\documentclass[epsfig, 11pt, onecolumn, oneside]{report}
\usepackage{epsfig}

\usepackage{color}
\usepackage{epsf,graphicx}

\usepackage{latexsym,amssymb,amsmath}
\usepackage{cite} 
\usepackage{setspace}
\usepackage{fancyhdr}
\usepackage{a4}

\newcommand{\newc}{\newcommand}
\newc{\gsim}{\lower.7ex\hbox{$\;\stackrel{\textstyle>}{\sim}\;$}}
\newc{\lsim}{\lower.7ex\hbox{$\;\stackrel{\textstyle<}{\sim}\;$}}
\newc{\gev}{\,{\rm GeV}}
\newc{\mev}{\,{\rm MeV}}
\newc{\ev}{\,{\rm eV}}
\newc{\kev}{\,{\rm keV}}
\newc{\tev}{\,{\rm TeV}}
\def\ln{\mathop{\rm ln}}

\newc{\mz}{M_Z}
\newc{\mpl}{M_*}
\newc{\mw}{m_{\rm weak}}
\newc{\nr}[1]{N^c_R{}_{#1}}

\def\be{\begin{equation}}
\def\ee{\end{equation}}
\def\ba{\begin{eqnarray}}
\def\ea{\end{eqnarray}}
\def\bitem{\begin{itemize}}
\def\eitem{\end{itemize}}
\newc{\ie}{{\it i.e.}}          \newc{\etal}{{\it et al.}}
\newc{\eg}{{\it e.g.}}          \newc{\etc}{{\it etc.}}
\newc{\cf}{{\it c.f.}}
\def\bar#1{\overline{#1}}

\def\inv{^{\raise.15ex\hbox{${\scriptscriptstyle -}$}\kern-.05em 1}}
\def\lbar{{\lower.35ex\hbox{$\mathchar'26$}\mkern-10mu\lambda}} 

\def\ee#1{\times 10^{#1} }

\def\to{\rightarrow}

\let\<=\langle
\let\>=\rangle

\let\+=\uparrow
\let\B=\bar

\let\be=\beta

\let\la=\lambda

\let\si=\sigma

\let\Up=\Upsilon

\let\nn=\nonumber
\let\ul=\underline
\let\ti=\tilde
\let\ve=\varepsilon
\let\ph=\phantom

\begin{document}

\baselineskip 0.6cm


\pagenumbering{roman}
\setcounter{page}{3}






\newpage

\thispagestyle{empty}



\vspace*{1cm}

\begin{center}

{\Large\bf M-Theory on Manifolds with $\boldsymbol{G_2}$ \bf Holonomy \\}
\vspace{2cm}

{\large
Adam Bruno Barrett\\
\vspace{0.5cm}
Balliol College\\
Oxford\\
\vspace{0.5cm}
\emph{and}\\
\vspace{0.5cm}
Rudolf Peierls Centre for Theoretical Physics\\
Department of Physics\\
University of Oxford}
\end{center}

\vspace{8cm}
\begin{center}
{\large
Thesis submitted for the Degree of Doctor of Philosophy
\\
 University of Oxford\\
\vspace{0.3cm}
$\cdot$ September 2006 $\cdot$}
\end{center}


\newpage


\begin{abstract}\noindent
We study M-theory on $G_2$ holonomy spaces that are constructed by
dividing a seven-torus by some discrete symmetry group. We classify possible
group elements that may be used in this construction and use them to
find a set of possible orbifold groups that lead to co-dimension four
singularities. We describe how to blow up such singularities,
and then derive the moduli K\"ahler potential for M-theory on the resulting
class of $G_2$ manifolds. To consider the singular limit it is necessary to
derive the supergravity action for M-theory on the orbifold
$\mathbb{C}^2/\mathbb{Z}_N$. We do this by coupling 11-dimensional
supergravity to a seven-dimensional Yang-Mills theory located on the
orbifold fixed plane. We show that the resulting action is
supersymmetric to leading non-trivial order in the 11-dimensional
Newton constant. Obtaining this action enables us to then reduce M-theory
on a toroidal $G_2$ orbifold with co-dimension four singularities, taking
explicitly into account the additional gauge fields at the singularities.
The four-dimensional effective theory has $\mathcal{N}=1$ supersymmetry with
non-Abelian $\mathcal{N}=4$ gauge theory sub-sectors. We present
explicit formulae for the K\"ahler potential, gauge-kinetic function
and superpotential. In the four-dimensional theory, blowing-up of the
orbifold is described by continuation along D-flat directions. Using
this interpretation, we demonstrate consistency of our results for singular
$G_2$ spaces with corresponding ones obtained for smooth $G_2$ spaces.
In addition, we consider the effects of switching on flux and Wilson lines
on singular loci of the $G_2$ space, and we discuss the relation to
$\mathcal{N}=4$ SYM theory.

\end{abstract}

\begin{center}
\phantom{ABC}
\vspace{8.0cm}
\begin{quote}
\it The eternal sand-glass of existence will ever be turned once
more, and thou with it, thou speck of dust!
\end{quote}
\end{center}
\hfill{\begin{flushright} \small From ``Joyful Wisdom'', Friedrich
Nietzsche\\ Translated by Thomas Common\\ (Frederick Ungar
Publishing Co, 1960)
\end{flushright}} \thispagestyle{empty}

\pagestyle{empty}
\pagenumbering{roman}
\setcounter{page}{3} \pagestyle{plain}

\tableofcontents

\chapter*{Acknowledgements}
\addcontentsline{toc}{chapter}
                 {\protect\numberline{Acknowledgements\hspace{-96pt}}}
I would like to begin by thanking my supervisor, Andr\'e Lukas, for
inputting limitless enthusiasm into our research and for always
being so generous with his time. At the beginning, I remember
wondering how I was ever going to understand anything of this
mystical field, let alone manage to make my own contribution to it.
I have been very lucky to have been shown a traversable path towards
these goals by someone who is undoubtedly an expert.

Next I would like to thank Lara Anderson, who has been my research
collaborator during the last year and a half. It has been a
rollercoaster ride, and I really appreciate Lara's unwaning optimism
during low periods, and also her putting
up patiently with all my tantrums!

I thank the Particle Physics and Astronomy Research Council for granting me 
a studentship, and the UK taxpayer for providing the money that came with it. 
This thesis could not have been written without this funding.

I have been really lucky to have been placed in such a cheerful office
environment, both during my first year at the University of Sussex,
and then subsequently at Oxford. I thank Thomas House, Eran Palti,
Babiker Hassanain, Tanya Elliott and Mario Serna for great
conversation, debate and jokes.

Finally, I thank all of my family for bringing me up in an
environment which has inspired me to embark upon this quest for
knowledge. I appreciate all the wisdom and philosophy, handed down
to me by my grandfather, Clement Krysler. The love, support and 
encouragement from my parents, Martyn and Annette, and my brother 
Alex has been invaluable throughout my years as a student.

\newpage

\doublespacing
\pagestyle{fancy}
\renewcommand{\sectionmark}[1]{\markright{\thesection\ #1}}

\pagenumbering{arabic}
\setcounter{page}{1}

\fancyhf{} \fancyhead[L]{\sl 1.~Introduction}
\fancyhead[R]{\rm\thepage}
\renewcommand{\headrulewidth}{0.3pt}
\renewcommand{\footrulewidth}{0pt}
\addtolength{\headheight}{3pt}
\fancyfoot[C]{}

\chapter{Introduction}
In comprising both general relativity and quantum field theory, it
is possible that M-theory will one day provide us with a unified
framework for describing all the forces of nature. Although M-theory
must be formulated in eleven spacetime dimensions, it has many
vacuum solutions with just four macroscopic dimensions, one of which
could correspond to the observed universe. Current thought favours
the idea that our world should be described by a quantum theory
that, when viewed from a four-dimensional perspective, is
$\mathcal{N}=1$ supersymmetric. This is because simple extensions of
the standard model that incorporate $\mathcal{N}=1$ supersymmetry
greatly increase its elegance. Firstly there is no hierarchy problem
in such models, and, furthermore, it is possible to achieve gauge coupling 
unification. Thus the main focus of research into solutions of
M-theory is on those with $\mathcal{N}=1$ unbroken supersymmetry.

M-theory is related to the various superstring theories by a network of
dualities. Superstring theories are formulated in ten dimensions, and much
work has been done towards obtaining realistic four-dimensional theories
using these as a starting point. Compactification of one of the heterotic
string theories on a six-dimensional Calabi-Yau space is one method for
obtaining a four-dimensional $\mathcal{N}=1$ supersymmetric theory. Another
route is to consider a solution of type IIA or IIB string theory whose
internal space is again a Calabi-Yau space, but that also involves D-branes
being present. The usual method for deriving properties of the
four-dimensional effective theory is to begin with the low-energy effective
supergravity action for the relevant superstring theory, and to work on
compactification of this. If one assumes that all the moduli of the internal
space are large compared to the Planck length then this method works well.

Although the fundamental degrees of freedom are currently
unknown, one can also study direct compactification of M-theory itself from
eleven to four dimensions by using the appropriate low-energy effective
supergravity. For a smooth spacetime this is simply 11-dimensional
 supergravity \cite{Witten:1995ex}. Compactifying this theory on a manifold of $G_2$ holonomy
leads to a four-dimensional theory with $\mathcal{N}=1$ supersymmetry.
Unfortunately this does not, however, provide a viable framework for particle
phenomenology, since the reduced theory contains only Abelian gauge
multiplets and massless uncharged chiral multiplets~\cite{Comp1},~\cite{Comp2}. To be
more specific, the four-dimensional spectrum consists of Abelian gauge
multiplets, which descend from the three-form of 11-dimensional supergravity,
and uncharged chiral multiplets that contain the metric moduli of the $G_2$
manifold, together with their associated axions. (There are no continuous
symmetries of a manifold of $G_2$ holonomy by which to obtain gauge fields.)

With the construction of M-theory on a manifold with a boundary,
$\mathcal{M}_{1,9}\times S^1/\mathbb{Z}_2$, Ho\v rava and
Witten~\cite{Hor-Wit} demonstrated for the first time that the
situation can be quite different for compactifications on singular
spaces.  In fact, they showed that new states in the form of two
10-dimensional $E_8$ super-Yang-Mills multiplets, located on the two
10-dimensional fixed planes of this orbifold, had to be added to the
theory for consistency, and they explicitly constructed the
corresponding supergravity theory by coupling 11-dimensional
supergravity in the bulk to these super-Yang-Mills theories. It soon
became clear that this theory allows for phenomenologically
interesting Calabi-Yau compactifications~\cite{Strong}--\cite{HetM}
and, as the strong-coupling limit of the heterotic string, should be
regarded as a promising avenue towards particle phenomenology from
M-theory.

More recently, it has been discovered that phenomenologically
interesting theories can also be obtained by M-theory
compactification on singular spaces with $G_2$
holonomy~\cite{Acharya}--\cite{He:2002fp}. In this context, certain
co-dimension four singularities within the $G_2$ space lead to
low-energy non-Abelian gauge fields~\cite{Acharya},~\cite{Gukov}. In
addition, chiral fermions, that are possibly charged under the gauge
multiplets, arise when the locus of such a singularity passes
through an isolated conical (co-dimension seven)
singularity~\cite{Gukov}--\cite{Cvetic}. Chapter~\ref{revchap} of
this thesis contains a review of M-theory on $G_2$ spaces, and also
presents some background material to this topic.
\\

There has been much work done on the subject of M-theory on $G_2$
spaces, in which calculations are carried out for some generic,
possibly compact manifold of $G_2$ holonomy or else for some
specific non-compact $G_2$ manifold. However, it is clear that
potentially realistic examples should rely on compact $G_2$ spaces
whose properties can be fairly well understood. An interesting set
of compact manifolds of $G_2$ holonomy, based on $G_2$ orbifolds,
has been constructed in~\cite{Joyce}. The method involves dividing a
seven-torus by some discrete symmetry group and then blowing up the
singularities of the resulting orbifold. It is an interesting task
to pursue this method to construct compact $G_2$ manifolds further
and, hence, to classify $G_2$ orbifolds and their associated
blown-up $G_2$ manifolds. In Chapter~\ref{Classification} of this
thesis we propose a method for such a classification, and present a
class of possible orbifold groups with co-dimension four fixed
points. We then construct an explicit class of manifolds, using
these orbifold groups. Many of the examples appear to be new.

Having obtained a class of manifolds of holonomy $G_2$ it would be
useful to be able to compare the four-dimensional effective theories
resulting from compactification on different members of the class.
An important ingredient for beginning such analysis is the
four-dimensional moduli K\"ahler potential. This has obvious
applications to various areas of study, for example, supersymmetry
breaking or the cosmological dynamics of moduli fields. For the
$G_2$ manifold based on the simplest $G_2$ orbifold~\cite{Joyce}
$\mathcal{T}^7/\mathbb{Z}_2^3$ the K\"ahler potential has been
calculated in Ref.~\cite{Lukas}. At the end of
Chapter~\ref{Classification} we generalise this result by deriving a
formula for the moduli K\"ahler potential valid for the manifolds of
our classification.
\\

Singular $G_2$ manifolds are readily obtained from our examples by
shrinking the blow-ups down to zero. Furthermore, by design, the
singularities that arise are co-dimension four $ADE$ singularities.
This means that using one of these singular manifolds as the
internal space for an M-theory universe results in non-Abelian gauge
fields being present in the low-energy physics. Chapters~\ref{117}
and~\ref{G2sing} of this thesis are concerned with obtaining the
explicit four-dimensional effective theory for such models.

It is well-known that the additional states which appear in M-theory
at $ADE$ spacetime singularities are seven-dimensional gauge
multiplets with the appropriate $ADE$ gauge group. In particular, an
$\mathrm{SU}(N)$ gauge multiplet is present for the case of a
singularity of type $A_{N-1}$, that is, when the structure of
spacetime around the singularity takes the form
$\mathcal{M}_{1,6}\times\mathbb{C}^2/\mathbb{Z}_N$, where
$\mathcal{M}_{1,6}$ is a smooth seven-dimensional space with
Minkowski signature. Chapter~\ref{117}, however, presents the first
explicit computation of the terms that couple these states to
11-dimensional supergravity. We construct 11-dimensional
supergravity on the general space $\mathcal{M}_{1,6}\times
\mathbb{C}^2/\mathbb{Z}_N$, coupled to seven-dimensional
$\mathrm{SU}(N)$ super-Yang-Mills theory located on the orbifold
fixed plane $\mathcal{M}_{1,6}\times \{ \boldsymbol{0} \}$. By
formulating the problem in a general context, our results are
applicable not just to compactifications of M-theory on $G_2$
spaces, but also to other problems such as M-theory on certain
singular limits of K3. This work is very much in the spirit of Ho\v
rava-Witten theory~\cite{Hor-Wit}, which couples 11-dimensional
supergravity on a manifold with a boundary, $\mathcal{M}_{1,9}\times
S^1/\mathbb{Z}_2$, to 10-dimensional super-Yang-Mills theory.

Having constructed the effective supergravity action for M-theory on
spaces of the form
$\mathcal{M}_{1,6}\times\mathbb{C}^2/\mathbb{Z}_N$,
Chapter~\ref{G2sing} of this thesis is concerned with
compactification of M-theory on a singular $G_2$ space. Central to
this is the reduction of the seven-dimensional $\mathrm{SU}(N)$
gauge-theories on the three-dimensional singular loci within the
$G_2$ space. In general, the shape of these loci can take various
forms. In this work, we focus on $G_2$ spaces based on the
construction adopted in Chapter~\ref{Classification}, for which
these loci are three-tori.

It is interesting to compare our results for M-theory on singular
$G_2$ spaces with those for compactification on the associated
smooth $G_2$ spaces obtained by blowing-up the singularities. It
turns out that blowing up of a singularity can in fact be described
by a Higgs effect induced by continuation along D-flat directions in
the four-dimensional effective theory. The gauge group is broken
from $\mathrm{SU}(N)$ to its maximal Abelian subgroup
$\mathrm{U}(1)^{N-1}$, and the fields that remain massless in this
scenario correspond exactly to the zero modes of M-theory on the
blown-up orbifold. Thus an explicit comparison of results can be
made. At the end of Chapter~\ref{G2sing}, we compare the K\"ahler
potentials that we have found for the singular and blown-up cases,
restricting the singular result to the Abelian matter fields. The
two results indeed turn out to be equivalent.
\\

In the final chapter we carry out some further analysis of M-theory
on the singular $G_2$ spaces. We examine the effects of introducing
backgrounds involving Wilson lines and flux. Wilson lines are
interesting for learning about possible patterns of gauge symmetry
breaking, whilst the inclusion of flux is useful since it can lead
to moduli stabilisation and low-energy supersymmetry breaking. We
compute the superpotential for Abelian flux of the $\mathrm{SU}(N)$
gauge fields on the internal singular locus $\mathcal{T}^3$. We show
that this can be obtained from a Gukov-type formula which involves
integration of the complexified Chern-Simons form of the gauge
theory over the internal three-torus. Our result provides a further
confirmation for the matching between the singular and smooth
theories, as the flux superpotential in both limits is of the same
form if the field identification suggested by the comparison of the
K\"ahler potentials is used.

We also consider one of the $\mathrm{SU}(N)$ gauge sectors of our
action with gravity switched off. Since the reduction of
seven-dimensional gauge fields on a three-torus does not in itself
break any supersymmetry, these subsectors have enhanced
$\mathcal{N}=4$ supersymmetry. This connection with $\mathcal{N}=4$
super-Yang-Mills is of particular interest because this theory has
special properties, and currently plays an important r\^{o}le in
various aspects of string theory, especially in the context of the
AdS/CFT conjecture~\cite{Maldacena}. One intriguing property is its
S-duality symmetry. We show that this translates into a T-duality on
the singular $\mathcal{T}^3$ locus, and speculate about a possible
extension of this S-duality to the full supergravity theory.
\\

The work presented in this thesis is drawn from three research
papers. Chapter~\ref{Classification} is based on
\begin{itemize}
  \item A.~B.~Barrett and A.~Lukas, ``Classification and Moduli
K\"{a}hler Potentials of $G_{2}$ Manifolds'', Phys.\ Rev.\ D \textbf{71}
(2005) 046004, [arXiv:hep-th/0411071],
\end{itemize}
Chapter~\ref{117} is based on
\begin{itemize}
  \item L.~B.~Anderson, A.~B.~Barrett and A.~Lukas, ``M-Theory
  on the Orbifold $\mathbb{C}^2/\mathbb{Z}_N$'', Phys.\ Rev.\ D
  \textbf{73}, (2006) 106011, [arXiv:hep-th/0602055],
\end{itemize}
and Chapters~\ref{G2sing} and \ref{Analysis} are based on
\begin{itemize}
  \item L.~B.~Anderson, A.~B.~Barrett, A.~Lukas and M.~Yamaguchi,
   ``Four-dimensional Effective M-theory on a Singular $G_2$ Manifold'', Phys.\ Rev.\ D \textbf{74} (2006) 086008, [arXiv:hep-th/0606285].
\end{itemize}
Before these however, comes Chapter~\ref{revchap}, which is a review
of the subject area of this thesis.

\chapter{Review of M-theory and Manifolds of $\boldsymbol{G_2}$ Holonomy} \label{revchap}
\fancyhf{} \fancyhead[L]{\sl \rightmark} \fancyhead[R]{\rm\thepage}
\renewcommand{\headrulewidth}{0.3pt}
\renewcommand{\footrulewidth}{0pt}
\addtolength{\headheight}{3pt} \fancyfoot[C]{} In this chapter we
review some of the background material to the work presented in this
thesis. The first section gives an introduction to $G_2$ holonomy
manifolds. This is followed in Section~\ref{Mtheoryrev} by a review
of M-theory on smooth spacetimes, which includes an outline of how
compactification on a general $G_2$ manifold works. Then in the
third section there is some discussion of M-theory on singular
spacetimes. There exist two lengthy review papers
\cite{Metzger},~\cite{Gukov} on the subject of M-theory on $G_2$
manifolds, and much of the material in this chapter has been drawn
from these. For the detailed mathematics of $G_2$ holonomy manifolds
we refer the reader to the book \cite{Joyce} on special holonomy
manifolds written by Joyce. Other material that has been useful for 
preparing this review includes a talk given by Duff
 \cite{Duff}, which contains an interesting discussion on the holonomy 
of internal spaces and its relation to supersymmetry, a paper by Lukas 
and Morris \cite{Lukas}, which contains a 
self-contained mini-review of the reduction of M-theory on a $G_2$ manifold, 
a paper by Witten \cite{Anomaly}, which discusses anomaly cancellation in 
M-theory on singular $G_2$ manifolds, and, for the review of Ho\v rava-Witten 
theory, the original paper, Ref.~\cite{Hor-Wit}.   

\section{Introduction to manifolds of $G_2$ holonomy}
\label{revgeometry}

We begin by introducing the exceptional Lie group $G_2$, and then
writing down the conditions for a seven-dimensional Riemannian
manifold to have holonomy group $G_2$. Let $\varphi_0$ be the
three-form on $\mathbb{R}^7$ given by
\begin{eqnarray}
\varphi_0 & = &
 \mathrm{d}x^1\wedge\mathrm{d}x^2\wedge\mathrm{d}x^3+
 \mathrm{d}x^1\wedge\mathrm{d}x^4\wedge\mathrm{d}x^5+
 \mathrm{d}x^1\wedge\mathrm{d}x^6\wedge\mathrm{d}x^7+
 \mathrm{d}x^2\wedge\mathrm{d}x^4\wedge\mathrm{d}x^6\nonumber \\& &
-\mathrm{d}x^2\wedge\mathrm{d}x^5\wedge\mathrm{d}x^7-
 \mathrm{d}x^3\wedge\mathrm{d}x^4\wedge\mathrm{d}x^7-
 \mathrm{d}x^3\wedge\mathrm{d}x^5\wedge\mathrm{d}x^6\, .
\label{structure}
\end{eqnarray}
Then $G_2$ is the subgroup of $\mathrm{SO}(7)$ that preserves this
form $\varphi_0$ under its usual action on three-forms. Thus, a
matrix $M\in \mathrm{SO}(7)$ is a member of the group $G_2$ if and
only if
\begin{equation}
(\varphi_0)_{ABC}={M^D}_A{M^E}_B{M^F}_C(\varphi_0)_{DEF}\,,
\end{equation}
where $A,B\ldots=1,\ldots, 7$. The group $G_2$ is spanned by 14 of
the 21 generators of $\mathrm{SO}(7)$. A seven-dimensional
Riemannian manifold has holonomy group $G_2$ if and only if it
admits a torsion-free $G_2$ structure and has finite first
fundamental group. A $G_2$ structure is a globally defined
three-form $\varphi$ which is everywhere locally isomorphic to
$\varphi_0$, and a torsion-free $G_2$ structure is one satisfying
$\mathrm{d}\varphi=\mathrm{d}\ast\varphi=0$. In this thesis, we will
often refer to seven-dimensional $G_2$ holonomy manifolds simply as
$G_2$ manifolds.

Next we list some properties of a general $G_2$ manifold
$\mathcal{X}$. The first property is that, associated with the
torsion-free $G_2$ structure $\varphi$, is a Ricci-flat metric $g$,
and this can be explicitly computed from $\varphi$ using the
equations
\begin{eqnarray}
 \label{met1}
g_{AB}&=&\mathrm{det}(h)^{-1/9}h_{AB}\,, \\
\label{met2}
h_{AB}&=&\frac{1}{144}\varphi_{ACD}\varphi_{BEF}\varphi_{GHI}
\hat{\epsilon}^{CDEFGHI}\,,
\end{eqnarray}
where $\hat{\epsilon}$ is the ``pure-number'' Levi-Civita
pseudo-tensor. These equations are derived by first noting that
$h_{AB}$ is, up to a rescaling, the unique rank two symmetric object
that can be constructed from $\varphi$, and then working out that
the rescaling \eqref{met1} is required in order to produce a tensor.
The factor of 144 is introduced so that the local metric associated
with the canonical local $G_2$ structure, $\varphi_0$ of
Eq.~\eqref{structure}, is the usual Kronecker delta. The second
property is that the manifold has two unfixed Betti numbers, namely
$b^2(\mathcal{X})$ and $b^3(\mathcal{X})$. If we assume the manifold
is connected then $b^0(\mathcal{X})=1$, and there are no one-cycles,
so $b^1(\mathcal{X})=0$. Thirdly, there exists precisely one
covariantly constant spinor $\psi$, and this is related to the $G_2$
structure by the bilinear relation
\begin{equation}
\varphi_{ABC}=i \bar{\psi} \gamma_{ABC} \psi\,,
\end{equation}
where $\gamma_{ABC}$ are anti-symmetrised products of gamma-matrices
on the manifold.

We now outline Joyce's method
\cite{joyce1},~\cite{joyce2},~\cite{Joyce} for constructing concrete
compact examples of $G_2$ manifolds. One takes, as the starting
point, some seven-torus $\mathcal{T}^7$. This space comes equipped
with a torsion-free $G_2$ structure that is simply given by
\begin{eqnarray} \label{revstructure}
\varphi & = &
R^1R^2R^3\mathrm{d}x^1\wedge\mathrm{d}x^2\wedge\mathrm{d}x^3
+R^1R^4R^5\mathrm{d}x^1\wedge\mathrm{d}x^4\wedge\mathrm{d}x^5+
R^1R^6R^7\mathrm{d}x^1\wedge\mathrm{d}x^6\wedge\mathrm{d}x^7 \nonumber \\
& &+R^2R^4R^6\mathrm{d}x^2\wedge\mathrm{d}x^4\wedge\mathrm{d}x^6
-R^2R^5R^7\mathrm{d}x^2\wedge\mathrm{d}x^5\wedge\mathrm{d}x^7-
R^3R^4R^7\mathrm{d}x^3\wedge\mathrm{d}x^4\wedge\mathrm{d}x^7 \nonumber \\
& & -R^3R^5R^6\mathrm{d}x^3\wedge\mathrm{d}x^5\wedge\mathrm{d}x^6\,,
\end{eqnarray}
where the $R^A$ are its seven radii. An orbifold with holonomy $G_2$
can be obtained from the seven-torus by dividing it by a discrete
symmetry group $\Gamma$, that is a subgroup of $G_2$, and renders
the first fundamental group finite. We shall refer to $\Gamma$ as
the orbifold group. To create a smooth manifold, one blows up the
singularities of the orbifold. Loosely speaking, this involves
removing a patch around the singularity and replacing it with a
smooth space of the same symmetry. Note that, following this
construction, the independent moduli will come from torus radii and
from the radii and orientation of cycles associated with the
blow-ups.

To illustrate this method of construction, it is useful to present
an example \cite{Joyce}. Let the starting point be the standard
seven-dimensional torus, $\mathcal{T}^7=\mathbb{R}^7/\mathbb{Z}^7$.
To create a $G_2$ holonomy orbifold we divide this space by the
group $\mathbb{Z}_2^3$, whose generators $\alpha$, $\beta$ and
$\gamma$ are taken to act on the torus coordinates according to
\begin{eqnarray}
\alpha : \left( x^1,\ldots,x^7 \right)& \mapsto &
\left(x^1,x^2,x^3,-x^4,-x^5,-x^6,-x^7\right)\,,\\
\beta : \left(x^1,\ldots,x^7 \right)&\mapsto
&\left(x^1,-x^2,-x^3,\frac{1}{2}-x^4,-x^5,x^6,x^7\right)\,,
\end{eqnarray}
\begin{eqnarray}
\gamma : \left(x^1,\ldots,x^7\right) &\mapsto &
\left(-x^1,-x^2,x^3,x^4,\frac{1}{2}-x^5,\frac{1}{2}-x^6,x^7\right)\,.
\end{eqnarray}
The orbifold has singularities corresponding to the fixed loci of
elements of the orbifold group. It can be shown that there are 12
singularities within the fundamental domain, and that these are all
of co-dimension four. An example of a singular locus is
$\left(x^1,x^2,x^3,0,0,0,0\right)$, which is fixed by the generator
$\alpha$. Near each singular point, the orbifold takes the
approximate form $\mathcal{T}^3\times\mathbb{C}^2/\mathbb{Z}_2$,
where $\mathcal{T}^3$ is a three-torus. Blowing up such
singularities involves the following. One first removes a
four-dimensional ball centred around each singularity times the
associated fixed three-torus. Then, secondly, one replaces the
resulting hole by $\mathcal{T}^3\times \mathcal{U}$, where
$\mathcal{U}$ is smoothed Eguchi-Hanson space, the blow-up of the
cone $\mathbb{C}^2/\mathbb{Z}_2$, as discussed in Chapter
\ref{Classification}. We shall not describe Eguchi-Hanson space in
detail in this section, but just mention that its homology consists
of a single two-cycle located at its centre, and that it approaches
the flat space $\mathbb{C}^2/\mathbb{Z}_2$ asymptotically at
infinity \cite{eguchi1}.

\section{Review of M-theory on smooth spacetimes} \label{Mtheoryrev}

The low-energy limit of M-theory on a smooth spacetime is
11-dimensional supergravity~\cite{Witten:1995ex}. In this review, we
describe the action and associated supersymmetry transformations of
11-dimensional supergravity, and present the field equations. We
then describe some simple solutions of M-theory in which
11-dimensional spacetime takes the form of a direct product of some
four-dimensional spacetime with a seven-dimensional compact
``internal'' manifold. We demonstrate how the amount of supersymmetry
preserved by a solution depends on the internal space. In
particular, we explain how, from a four-dimensional perspective, a
solution with a $G_2$ manifold as the internal space is
$\mathcal{N}=1$ supersymmetric. The section ends with an outline of
how the reduction to four-dimensions of M-theory on a $G_2$ manifold
works in general.

Our conventions for this section are as follows. We take spacetime
to have mostly positive signature, that is $(-+\ldots+)$, and use
indices $M,N,\ldots=0,1,\ldots,10$ to label 11-dimensional
coordinates $(x^M)$. When working with a product space of the form
$\mathcal{M}_{1,3}\times \mathcal{M}_7$, we use coordinates
$(x^\mu)$ on $\mathcal{M}_{1,3}$, where $\mu,\nu,\ldots=0,1,2,3$,
and $(x^A)$ on $\mathcal{M}_{7}$, where $A,B,\ldots=1,\ldots,7$.
Underlined versions of all index types denote tangent space indices.
\\

The field content of 11-dimensional supergravity consists of the
vielbein ${\hat{e}_M}^{\underline{M}}$ and associated metric
$\hat{g}_{MN}=\eta_{\underline{M}\underline{N}}
\hat{e}_M^{\phantom{M}\underline{M}}
\hat{e}_N^{\phantom{N}\underline{N}}$, the three-form field $C$,
with field strength $G=\mathrm{d}C$, and the gravitino $\Psi_M$, a
Majorana, spin-3/2 fermion.\footnote{Here we place a ``hat'' on the
metric and vielbein in order to reserve ``unhatted'' notation for
lower dimensional fields later on.} In this thesis we shall not
compute any four-fermi terms, or associated cubic fermion terms in
supersymmetry transformations, and so we shall neglect these
throughout. The action is given by \cite{Julia}
\begin{eqnarray} \label{rev11dsugra4}
\mathcal{S}_{11} & = &
\frac{1}{\kappa_{11}^2}\int_{\mathcal{M}_{1,10}}\mathrm{d}^{11}x\sqrt{-\hat{g}}
 \bigg( \frac{1}{2}\hat{R}-\frac{1}{2}\bar{\Psi}_M\Gamma^{MNP}\nabla_N\Psi_P
 -\frac{1}{96}G_{MNPQ}G^{MNPQ}
 \nn \\
 & & \hspace{3.5cm} -\frac{1}{192}\Big(\bar{\Psi}_M\Gamma^{MNPQRS}\Psi_S+12\bar{\Psi}^N
     \Gamma^{PQ}\Psi^R\Big)
 G_{NPQR}\bigg) \nn \\
& & -\frac{1}{12\kappa_{11}^2}\int_{\mathcal{M}_{1,10}}C\wedge
G\wedge G\,.
\end{eqnarray}
Here $\kappa_{11}$ is the 11-dimensional Newton constant,
$\Gamma^{M_1\ldots M_n}$ denote anti-symmetrised products of
11-dimensional gamma matrices, and $\nabla_M$ is the spinor
covariant derivative, defined in terms of the spin connection
$\omega$ by
\begin{equation}
\nabla_M=\partial_M+\frac{1}{4}\omega_M^{\phantom{M}\underline{M}\underline{N}}
\Gamma_{\underline{M}\underline{N}}\,.
\end{equation}
The transformation laws of local supersymmetry, parameterised by the
32 real component Majorana spinor $\eta$, read
\begin{eqnarray}
\delta \hat{e}_M^{\phantom{M}\underline{N}} & = & \bar{\eta}\Gamma^{\underline{N}}\Psi_{M} \,,\nn \\
\delta C_{MNP} & = & -3\bar{\eta}\Gamma_{[MN}\Psi_{P]}\,, \label{rev11dsusy1} \\
\delta \Psi_M & = & 2\nabla_M\eta+\frac{1}{144}\left(
\Gamma_M^{\phantom{M}NPQR}
 -8\delta_M^N\Gamma^{PQR}\right)\eta G_{NPQR} \,.  \nn
\end{eqnarray}

Vacuum solutions of M-theory are specified by a choice of spacetime,
$\mathcal{M}_{1,10}$, together with a set of field configurations
for the metric, three-form and gravitino that satisfy the equations
of motion of 11-dimensional supergravity. The equations of motion
for solutions with vanishing gravitino, $\Psi_M=0$, can be written
most succinctly as
\begin{eqnarray}
\mathrm{d}\ast G &=& -\frac{1}{2}G\wedge G\, ,  \label{reveom1}\\
R_{MN}&=&\frac{1}{12}\left(
G_{MPQR}{G_N}^{PQR}-\frac{1}{12}\hat{g}_{MN}G_{PQRS}G^{PQRS}
\right)\, . \label{reveom2}
\end{eqnarray}
In addition to these equations, the four-form field strength $G$
satisfies the Bianchi identity
\begin{equation}
\mathrm{d}G  = 0\,. \label{reveomn}
\end{equation}
In attempting to relate M-theory to observable physics, one often
studies vacua for which the 11-dimensional spacetime
$\mathcal{M}_{1,10}$ is a direct product of the form
\begin{equation}
\mathcal{M}_{1,10}=\mathcal{M}_{1,3}\times\mathcal{M}_{7}\,.
\end{equation}
Here $\mathcal{M}_{1,3}$ is a four-dimensional manifold to be
interpreted as the spacetime we perceive, whilst $\mathcal{M}_{7}$
is taken to be a seven-dimensional compact ``internal'' space, too
small to perceive, but with moduli large compared to the Planck
length so the supergravity approximation to M-theory is valid.

The simplest solutions with this geometry take the ``external'' space
to be four-dimensional Minkowksi space, thus
$\mathcal{M}_{1,3}=\mathbb{R}^{1,3}$, have vanishing gravitino and
three-form, $\Psi_M=0$, $C=0$, and a metric of the form
\begin{equation}
\mathrm{d}s^2=\eta_{\mu\nu}\mathrm{d}x^\mu\mathrm{d}x^\nu +
g_{AB}(x^C)\mathrm{d}x^A\mathrm{d}x^B.
\end{equation}
Here $\eta_{\mu\nu}=\mathrm{diag}(-1,1,1,1)$ is the Minkowski metric
on $\mathbb{R}^{1,3}$. By the Einstein equation \eqref{reveom2}, the
metric $g_{AB}(x^C)$ on the internal space $\mathcal{M}_7$ must be
Ricci-flat.

Another class of solutions, known as Freund-Rubin solutions
\cite{Freund}, take four-dimensional anti-de Sitter (AdS) space as
the external space. This space is curved but still maximally
symmetric. In such solutions, the presence of a non-vanishing Ricci
tensor in the Einstein equation \eqref{reveom2} is balanced by
switching on the external components $G_{\mu\nu\rho\sigma}$ of the
four-form field strength to create a gravitational source. The
internal space is also no longer Ricci-flat, but is still an
Einstein manifold, that is, a manifold with Ricci tensor
proportional to its metric.

Switching on internal flux, that is, purely internal components
$G_{ABCD}$ of the four-form field strength, modifies the geometry of
solutions to 11-dimensional supergravity. In general, it becomes
difficult to write down an explicit metric. However, if the flux is
sufficiently small, then the effect on the geometry can be ignored
for the purposes of deriving the four-dimensional effective field
theory. One can then study the effect of flux just on the
four-dimensional geometry.
\\

Given a solution of M-theory, it is interesting to ask whether it is
supersymmetric, and if so, how much supersymmetry it preserves. One
determines this by working out how many independent supersymmetry
transformations leave the solution invariant. Let us examine
supersymmetry of the Ricci-flat solutions. Since these solutions
have vanishing gravitino, all supersymmetry transformations
\eqref{rev11dsusy1} leave the metric and three-form invariant.
However, for a transformation to leave the gravitino invariant, the
spinor-parameter $\eta$ must satisfy the Killing spinor equation
\begin{equation}
\nabla_M \eta =0\,. \label{revkilling1}
\end{equation}
It is therefore the number of solutions there are to this equation
that determines the amount of preserved supersymmetry. On the direct
product spacetime
$\mathcal{M}_{1,10}=\mathbb{R}^{1,3}\times\mathcal{M}_7$, the
$\boldsymbol{32}$ spinor representation of $\mathrm{SO}(1,10)$
decomposes into the product representation
$\boldsymbol{4}\otimes\boldsymbol{8}$ of $\mathrm{SO}(1,3)\times
\mathrm{SO}(7)$. Thus the spinor-parameter $\eta$ can be written in
the form
\begin{equation}
\eta(x^{\mu},x^A)=\varepsilon(x^{\mu})\otimes\psi(x^A)\,,
\end{equation}
where $\varepsilon$ is a spinor on $\mathbb{R}^{1,3}$ and $\psi$ is
a spinor on $\mathcal{M}_7$. Under this decomposition, the Killing
spinor equation \eqref{revkilling1} reduces to
\begin{equation}
\nabla^{(7)}_A\psi=0\,,
\end{equation}
given that Minkowski space admits a basis of four constant spinors.
This implies that the number $\mathcal{N}$ of preserved
supersymmetries from a four-dimensional perspective is precisely the
number of covariantly constant spinors on $\mathcal{M}_7$.

In Section \ref{revgeometry} we mentioned the fact that $G_2$
manifolds admit precisely one covariantly constant spinor. Moreover,
such manifolds are Ricci-flat. Therefore, $G_2$ manifolds are viable
internal spaces for the class of solutions we have been considering,
and lead to $\mathcal{N}=1$ supersymmetry in four-dimensions.
Different choices of internal space result in other amounts of
supersymmetry being preserved. In general, the larger the holonomy
group of the space, the fewer covariantly constant spinors there
are. One computes the number of such spinors by decomposing the
$\boldsymbol{8}$ spinor representation of $\mathrm{SO}(7)$ into
irreducible representations of the holonomy group, and counting the
number of singlets obtained \cite{Freund:1983ir}. (For $G_2$ this
decomposition is given by
$\boldsymbol{8}\to\boldsymbol{7}\oplus\boldsymbol{1}$.) Taking an
internal space with trivial holonomy, such as a seven-torus, will
result in the maximal amount of supersymmetry, $\mathcal{N}=8$,
being preserved. An example of an intermediate case is the choice
$\mathcal{M}_7=S^1\times\mathcal{M}_{\mathrm{CY}_3}$, the product of
a circle with a Calabi-Yau three-fold. This space has holonomy
$\mathrm{SU}(3)$, and leads to $\mathcal{N}=2$ unbroken
supersymmetry, by virtue of the decomposition
$\boldsymbol{8}\to\boldsymbol{3}\oplus\bar{\boldsymbol{3}}
\oplus\boldsymbol{1}\oplus\boldsymbol{1}$.

Let us briefly discuss supersymmetry of Freund-Rubin type solutions.
For these solutions the analysis is similar to that above, except,
due to the presence of a non-vanishing four-form field strength, the
number $\mathcal{N}$ of preserved supersymmetries is determined by
the number of solutions to a Killing spinor equation of the form
\begin{equation}
\nabla^{(7)}\psi(x^A)=c\psi(x^A)\,,
\end{equation}
where $c$ is a constant. For $\mathcal{N}=1$ supersymmetry, one
chooses a so-called weak $G_2$ manifold
\cite{weak1}--\cite{thomas}. These manifolds are
similar to $G_2$ holonomy manifolds, except they have a $G_2$
structure which is not torsion-free.
\\

We now review the reduction to four-dimensions of M-theory on a
general $G_2$ manifold $\mathcal{X}$, following Ref.~\cite{Lukas}. 
From the field
equations~\eqref{reveom1}--\eqref{reveomn} one sees that the bosonic
zero modes are the Ricci-flat deformations of the metric and the
harmonic deformations of the three-form. Recall that the Ricci-flat
metric on a $G_2$ manifold is induced by the torsion-free $G_2$
structure via Eqs.~\eqref{met1} and \eqref{met2}. Thus Ricci-flat
deformations of the internal metric can be described by the
torsion-free deformations of the $G_2$ structure and, hence, by the
third cohomology $H^3(\mathcal{X},\mathbb{R})$. Consequently, the
number of independent metric moduli is given by the third Betti
number $b^3(\mathcal{X})$. To define these moduli explicitly, we
introduce to $\mathcal{X}$ an integral basis $\{C^I\}$ of
three-cycles, and a dual basis $\{\Phi_I\}$ of harmonic three forms
satisfying
\begin{equation} \label{dual}
\int_{C^I}\Phi_J=\delta^I_J\,,
\end{equation}
where $I,J,\ldots=1,\ldots,b^3(\mathcal{X})$.
 We can then expand the torsion-free $G_2$ structure $\varphi$ as
\begin{equation}
\varphi=\sum_Ia^I\Phi_I\,.
\end{equation}
Then, by equation \eqref{dual}, the $a^I$ can be computed in terms
of certain underlying geometrical parameters by performing the
period integrals
\begin{equation} \label{periodgen}
a^I=\int_{C^I}\varphi\,.
\end{equation}
Let us also introduce an integral basis $\{D^P\}$ of two-cycles,
where $P,Q,\ldots=1,\ldots b^2(\mathcal{X})$, and a dual basis
$\{\omega_P\}$ of two-forms satisfying
\begin{equation}
\int_{D^P}\omega_Q=\delta^P_Q\,.
\end{equation}
Then, the three-form field $C$ can be expanded in terms of the basis
$\{\Phi_I\}$ and $\{\omega_P\}$ as
\begin{equation} \label{Cexp}
C=\nu^I\Phi_I+A^P\wedge\omega_P\,.
\end{equation}
(Recall that there is never a one-form on a $G_2$ manifold, and note
that we are neglecting the purely external part of $C$, which is
non-dynamical.) The expansion coefficients $\nu^I$ represent
$b^3(\mathcal{X})$ axionic fields in the four-dimensional effective
theory, while the Abelian gauge fields $A^P$, with field strengths
$F^P$, are part of $b^2(\mathcal{X})$ Abelian vector multiplets. The
$\nu^I$ pair up with the metric moduli $a^I$ to form the bosonic
parts of $b^3(\mathcal{X})$ four-dimensional chiral superfields
\begin{equation} \label{sufielddefn}
T^I=a^I+i\nu^I\,.
\end{equation}

A key ingredient in the four-dimensional effective supergravity
theory is the K\"ahler metric $K_{I\bar{J}}$, which is precisely the
sigma-model metric for the chiral superfields \cite{Wess}. According
to Ref.~\cite{Beasley}, a general formula for this is
\begin{equation} \label{Kmetric}
K_{I\bar{J}}=\frac{1}{4\kappa_{4}^2V}\int_{\mathcal{X}} \Phi_I
\wedge \ast \Phi_J\,.
\end{equation}
Here the four-dimensional Newton constant $\kappa_4$ is related to
its 11-dimensional counterpart by
\begin{equation}
\kappa_{11}^2=\kappa_4^2v_7\,,
\end{equation}
where $v_7$ is a reference volume, and $V$ is the volume of the
$G_2$ space $\mathcal{X}$ as measured by the Ricci-flat internal
metric $g$,
\begin{equation}
v_7=\int_{\mathcal{X}} \mathrm{d}^7x\,, \hspace{0.5cm}
V=\int_{\mathcal{X}} \mathrm{d}^7x\sqrt{\mathrm{det} g}\,.
\end{equation}
Note that the Hodge star is also with respect to the
seven-dimensional metric $g$. The most straightforward way of
proving the formula~\eqref{Kmetric} for the K\"ahler metric is to
reduce the 11-dimensional three-form kinetic term by inserting the
expansion~\eqref{Cexp}. With a little more effort, it can also be
derived by reducing the 11-dimensional Einstein-Hilbert
term~\cite{Gutowski:2002dk}. Using general properties of $G_2$
manifolds, it was shown in Ref.~\cite{Beasley} that the K\"ahler
metric~\eqref{Kmetric} descends from the K\"ahler potential
\begin{equation} \label{Kformula1}
K=-\frac{3}{\kappa_4^2}\ln \left( \frac{V}{v_7} \right)\,.
\end{equation}
This formula tells us that the K\"ahler potential (and indeed the
K\"ahler metric) only depends on the metric moduli $a^I$, and not on
the axions $\nu^I$. In terms of superfields, this means that $K$ is
a function of the real parts $T^I+\bar{T}^I$ only.

It is interesting to note that, in string theory, there are classes
of internal manifolds for which a general, explicit formula for the
moduli K\"ahler potential exists in terms of certain topological
data. For example, Calabi-Yau three-folds, at large radius, have
their moduli K\"ahler potential determined by a cubic polynomial
with coefficients given by their triple intersection numbers
\cite{Candelas}. Key to this result is a quasi topological relation
between two-forms and their Hodge duals on the Calabi-Yau space
\cite{Strominger}. There appears to be no analogue for three-forms
on $G_2$ manifolds. The only known formulae for the moduli K\"ahler
metric and K\"ahler potential are those given in equations
\eqref{Kmetric} and \eqref{Kformula1}, and these cannot be evaluated
generically for all $G_2$ manifolds.

There is no superpotential for 11-dimensional supergravity on a
smooth $G_2$ manifold \cite{Comp1},~\cite{Beasley}, and so it is the
K\"ahler potential which determines most of the physics of the
reduced theory. Let us just conclude this discussion by writing down
a general form for the gauge-kinetic function. Reduction of the
Chern-Simons term of 11-dimensional
supergravity~\eqref{rev11dsugra4}, by inserting the gauge field part
of \eqref{Cexp}, leads to the four-dimensional term \cite{Beasley}
\begin{equation}
\int_{\mathbb{R}^{1,3}}c_{IPQ}\nu^IF^P\wedge F^Q\,,
\end{equation}
where the coefficients $c_{IPQ}$ are given by
\begin{equation}
c_{IPQ}\sim \int_{\mathcal{X}} \Phi_I \wedge \omega_P \wedge
\omega_Q\,.
\end{equation}
This implies that the gauge-kinetic function $f_{PQ}$, which couples
$F^P$ and $F^Q$, takes the form
\begin{equation} \label{gkfapp}
f_{PQ} \sim \sum_IT^Ic_{IPQ}\,.
\end{equation}

A quick look at the spectrum of the reduced theory tells us that it
is not going to be a viable model for realistic particle
phenomenology from M-theory. Firstly, there is no non-Abelian gauge
symmetry, and, secondly, the chiral multiplets are massless and
uncharged. This is in fact a general property of 11-dimensional
supergravity compactified on a seven-dimensional
manifold~\cite{Comp1},~\cite{Comp2}. If a realistic model of the
universe is to be built from M-theory, it is going to involve an
11-dimensional spacetime containing singularities. This brings us to
the next section.

\section{M-theory on singular spacetimes} \label{revMsing}
In general, low-energy M-theory on a singular spacetime involves
additional field content to that of 11-dimensional supergravity, and
this is localised at each singularity. In this section we take a
look at some examples of this effect. We begin with a review of Ho\v
rava and Witten's explicit construction of M-theory on a manifold
with a boundary, $\mathcal{M}_{1,9}\times S^1/\mathbb{Z}_2$. In this
case, the extra states fill out two 10-dimensional $E_8$
super-Yang-Mills (SYM) multiplets, one living on each of the two
10-dimensional orbifold fixed planes. This theory of Ho\v rava and
Witten provides the motivation and inspiration for the construction
of the M-theory action on a space with co-dimension four
singularities, which forms a key part of this thesis. Afterwards we
continue with an explanation of how non-Abelian gauge fields arise
on co-dimension four singular loci, and then we finish with a
discussion of chiral fermions from co-dimension seven singularities.
\\

The motivation for studying M-theory on a manifold of the form
$\mathcal{M}_{1,9}\times S^1/\mathbb{Z}_2$ came when Ho\v rava and
Witten realised that this is the strong coupling limit of
ten-dimensional $E_8\times E_8$ heterotic string
theory~\cite{Hor-Wit0}. Moreover, it was this identification that
established the precise field content at the singularities. The
action was derived in Ref.~\cite{Hor-Wit} by computing the unique
supersymmetric coupling of 10-dimensional $E_8$ vector multiplets
living on the boundary to an 11-dimensional supergravity multiplet
propagating in the bulk. An interesting feature of the action is
that it is not gauge-invariant at the classical level, but that
one-loop anomalies render the quantum theory gauge-invariant.
Furthermore, the cancellation mechanism for these two effects relies
crucially on the gauge group being $E_8$, and also the gauge
coupling taking a fixed value with respect to the gravitational
coupling.

The basic structure of the Ho\v rava-Witten action is
\begin{equation} \label{horwitaction}
\mathcal{S}_{11-10}=\mathcal{S}_{11}+\sum_{k=1}^{2}\mathcal{S}_{10}^{(k)}\,,
\end{equation}
where $\mathcal{S}_{11}$ is the action \eqref{rev11dsugra4} of
11-dimensional supergravity, and $\mathcal{S}_{10}^{(k)}$, $k=1,2$,
are two copies of the 10-dimensional SYM action, living on the two
respective orbifold-fixed planes $\mathcal{M}_{1,9}^{(k)}$. The
theory is constructed as an expansion in
$\zeta_{10}=\kappa_{11}/\la_{10}$, where $\lambda_{10}$ is the
Yang-Mills coupling. The anomaly analysis determines $\lambda_{10}$
in terms of the 11-dimensional Newton constant $\kappa_{11}$ as
\begin{equation}
\lambda_{10}^2=2\pi\left(4\pi\kappa_{11}^{2}\right)^{2/3}\,.
\end{equation}
The bulk action, 11-dimensional supergravity, appears at zeroth
order in this expansion, while the Yang-Mills theories arise at
higher order.

The fields of 11-dimensional supergravity have to be restricted in
accordance with the orbifold symmetry. Suppose we take the orbifold
$S^1/\mathbb{Z}_2$ to lie in the $x^{10}$ direction, use the range
$x^{10}\in[-1/2,1/2]$, with the endpoints identified, and take the
$\mathbb{Z}_2$ orbifold symmetry to act as $x^{10}\to -x^{10}$. Then
each bosonic field $\phi$ is either constrained to be even or odd
with respect to this coordinate, that is,
$\phi(x^{10})=\pm\phi(-x^{10})$. The constraint on spinors,
meanwhile, takes the form
$\Gamma_{10}\eta(x^{10})=\pm\eta(-x^{10})$.

With the conventions above, the orbifold planes
$\mathcal{M}_{1,9}^{(k)}$, $k=1,2$ are located respectively at
$x^{10}=0,1/2$. Let us discuss the 10-dimensional SYM theory located
at $\mathcal{M}_{1,9}^{(1)}$. The leading terms are order
$\zeta_{10}^2\sim\kappa_{11}^{2/3}$ relative to those of
11-dimensional supergravity, and are given by
\begin{eqnarray}
\mathcal{S}_{10}^{(1)}&=&\frac{1}{\la_{10}^2}\int_{\mathcal{M}_{1,9}^{(k)}}
\mathrm{d}^{10}x\sqrt{-\hat{g}}\left(
-\frac{1}{4}F_{\hat{M}\hat{N}}^aF_a^{\hat{M}\hat{N}}
-\frac{1}{2}\B{\la}^a\Gamma^{\hat{M}}\mathcal{D}_{\hat{M}}\la_a
-\frac{1}{8}\bar{\Psi}_{\hat{M}}\Gamma^{\hat{N}\hat{P}}
\Gamma^{\hat{M}}F^a_{\hat{N}\hat{P}}\la_a    \right. \nn \\
&&\hspace{3.6cm}\left.
+\frac{1}{48}\B{\la}^a\Gamma^{\hat{M}\hat{N}\hat{P}}\la_a
G_{\hat{M}\hat{N}\hat{P}10}\right)\,, \label{horwitbound}
\end{eqnarray}
where the indices $\hat{M},\hat{N},\ldots=0,1,\ldots,9$ label
coordinates $(x^{\hat{M}})$ on the orbifold plane. The localised
field content consists of the $E_8$ gauge vector
$F^a=\mathcal{D}A^a$ and the gaugino $\la_a$. The other fields are
projected from the bulk in the straightforward way. Supersymmetry of
the action \eqref{horwitaction} requires a local modification to the
Bianchi identity whereby
\begin{equation}
\mathrm{d}G_{10\hat{M}\hat{N}\hat{P}\hat{Q}}=-
\frac{3\kappa_{11}^2}{48\la_{10}^2}\delta(x^{10})F^a_{[\hat{M}\hat{N}}
F_{\hat{P}\hat{Q}]a}\,.
\end{equation}
To satisfy this, it is necessary for the three-form $C$ to transform
non-trivially under gauge transformations. From this, one can see
that the presence of the Chern-Simons $CGG$ term in the
11-dimensional supergravity action \eqref{rev11dsugra4} causes the
Ho\v rava-Witten action \eqref{horwitaction} to fail to be
gauge-invariant at the classical level. There is also a local
modification to the supersymmetry transformation law of the
three-form component $C_{10\hat{M}\hat{N}}$, and this is given by
\begin{equation}
\delta C_{10\hat{M}\hat{N}}=-\frac{\kappa_{11}^2}{2\la_{10}^2}
\delta(x^{10}) A_{[\hat{M}}^a\bar{\eta}\Gamma_{\hat{N}]}\la_a\,.
\label{revCmod}
\end{equation}
Finally, the supersymmetry transformation laws of the localised
fields are given by
\begin{eqnarray}
\delta A_{\hat{M}}^a&=&\bar{\eta}\Gamma_{\hat{M}}\la^a\,, \nn \\
\delta
\la^a&=&-\frac{1}{8}\Gamma^{\hat{M}\hat{N}}F^a_{\hat{M}\hat{N}}\eta\,.
\label{revbranesusy}
\end{eqnarray}

The Ho\v rava-Witten action
\eqref{horwitaction},~\eqref{rev11dsugra4},~\eqref{horwitbound} is
supersymmetric to order $\zeta_{10}^2\sim\kappa_{11}^{2/3}$ under
the transformations
\eqref{rev11dsusy1},~\eqref{revCmod},~\eqref{revbranesusy}. However,
at the next order, $\zeta_{10}^4\sim\kappa_{11}^{4/3}$, a
verification of supersymmetric invariance would involve the
cancellation of terms that are formally proportional to $\delta(0)$.
Indeed, one can see that such a term arises when one substitutes the
localised part of the three-form variation \eqref{revCmod} into the
localised part of the action \eqref{horwitbound}. Ho\v rava and
Witten interpret this occurrence as a problem with the classical
treatment of M-theory, and assume that in the quantum regime there
would be a built-in cut-off that would replace $\delta(0)$ by a
finite constant times $\kappa_{11}^{-2/9}$.  In this instance, the
gauge fields would propagate in a boundary layer, with a thickness
of order $\kappa_{11}^{2/9}$, and not precisely on the boundary of
spacetime.
\\

We now discuss, following Ref.~\cite{Gukov}, another type of
spacetime singularity that gives rise to non-Abelian gauge symmetry
in M-theory. This is the co-dimension four $ADE$ singularity. In the
neighbourhood of such a singularity, 11-dimensional spacetime takes
the form $\mathcal{M}_{1,6}\times\mathbb{C}^2/H$, where
$\mathcal{M}_{1,6}$ is a seven-dimensional smooth spacetime and $H$
is one of the finite subgroups of $\mathrm{SU}(2)$. The nomenclature
$ADE$ comes from the existence of a natural one-to-one correspondence
\cite{mckay},~\cite{mckay2}
between the set of possible groups $H$ and the Dynkin diagrams of the simply laced 
semi-simple Lie algebras,
$A_{N-1}=\mathfrak{su}(N)$, $D_M=\mathfrak{so}(2M)$, $E_6$, $E_7$
and $E_8$. (For each group $H$, a matrix can be constructed from the 
characters of its irreducible representations, and the set of such 
matrices exactly reproduces the set of Cartan matrices of the Dynkin diagrams.) If 
one chooses the standard action of $\mathrm{SU}(2)$ on
the coordinates of $\mathbb{C}^2$ then the generator of
$H_{A_{N-1}}$, which is isomorphic to $\mathbb{Z}_N$, is given by
\begin{equation}
\left( \begin{array}{cc} e^{2i\pi/N} & 0 \\
 0 & e^{-2i\pi/N}
 \end{array} \right)\,.
 \end{equation}
Generators for the $D$ and $E$ series are given, for example, in
Ref.~\cite{Gukov}. Non-Abelian gauge symmetry arises from $ADE$
singularities due to the presence of extra states on the orbifold
fixed singular locus $\mathcal{M}_{1,6}\times\{\boldsymbol{0}\}$.
The extra field content for orbifold group $H_G$ is a
seven-dimensional SYM gauge multiplet with gauge group $G$
\cite{Acharya}.

The emergence of non-Abelian gauge fields from $ADE$ singularities
was first discovered via the duality \cite{Witten:1995ex} between
M-theory on a K3 manifold and heterotic string theory on a
three-torus, $\mathcal{T}^3$. In the latter theory, non-Abelian
gauge symmetry arises at special points in moduli space, whilst
generic points lead to an Abelian unbroken gauge group,
$\mathrm{U}(1)^{16}$, the Cartan subgroup of $E_8\times E_8$ and
$\mathrm{SO}(32)$. This is seen by studying the possible Wilson
lines on the three-torus. Duality then tells us that there should be
points in the moduli space of M-theory on K3 that also exhibit
non-Abelian symmetry enhancement. It turns out that these points are
precisely the points in moduli space where the K3 develops orbifold
singularities. The manifold K3 is a compact four-manifold admitting
a metric of $\mathrm{SU}(2)$ holonomy. (In fact, up to
diffeomorphisms, it is the unique simply-connected manifold of this
type.) This special holonomy implies that, close to an orbifold
singularity, a K3 manifold must be described as $\mathbb{C}^2/H$,
where $H$ is a finite subgroup of $\mathrm{SU}(2)$. It is thus
inferred that non-Abelian gauge fields arise precisely at $ADE$
singularities. Since this is a local effect, the result can be
generalised from M-theory on a K3 manifold to M-theory on any space
containing $ADE$ singularities.

It can in fact also be argued directly from M-theory that
non-Abelian gauge fields appear at $ADE$ singularities. It is known
that, among the fundamental constituents of M-theory, there are
M2-branes, 2+1-dimensional objects. Since the M2-brane has tension,
its dynamics will push it to wrap the smallest volume two-cycle in
the space. This tension gives rise to massive states, and when, as
in Section~\ref{Mtheoryrev}, one assumes all the moduli of spacetime
are large compared to the Planck length, one is justified in
ignoring these states for the purposes of constructing low-energy
M-theory. However, singular spacetimes are formally limits of smooth
spacetimes in which some moduli are shrunk down to zero. If these
moduli describe a number of collapsing two-cycles then massless
membrane states arise in the singular limit and are therefore
present in the low-energy supergravity approximation to M-theory.
This is precisely the scenario for $ADE$ singularities, which are
blown up by asymptotically locally Euclidean manifolds containing
two-cycles.

As an example, let us illustrate this effect for the case of an
$A_1\sim\mathbb{Z}_2$ singularity. This is blown up by Eguchi-Hanson
space $\mathcal{U}$, which contains a single two-cycle $\gamma$ at
its centre. Consider M-theory on a smooth seven-space times
Eguchi-Hanson space, $\mathcal{M}_{1,6}\times\mathcal{U}$. We are
interested in the field content localised in the neighbourhood of
the two-cycle, as viewed from a seven-dimensional perspective. There
is precisely one $\mathrm{U}(1)$ gauge field $A$, and this descends
from the three-form zero mode $C_0=A\wedge\omega$, where $\omega$ is
the two-form dual to the two-cycle $\gamma$. This gauge field
combines with the three scalars parameterising the cycle $\gamma$ to
form the bosonic part of a $\mathrm{U}(1)$ vector multiplet in
seven-dimensions. These are the only massless localised states. In
addition, though, there are some massive states associated with the
M2-brane wrapping the two-cycle, and these appear as particles from
the seven-dimensional point of view. The world-volume $V$ of a
wrapped M2-brane takes the form $V=p\times(\pm\gamma)$, where $p$ is
a path in $\mathcal{M}_{1,6}$ and the sign $\pm$ indicates the
orientation of the membrane with respect to the cycle $\gamma$. The
membrane couples to the $C$-field via the following term in its
world-volume action,
\begin{equation}
\int_V C = \pm \int_pA\int_\gamma \omega = \pm \int_p A\,.
\end{equation}
Hence, the membrane states are charged under the $\mathrm{U}(1)$
gauge field $A$, and states corresponding to opposite orientations
of the membrane with respect to the cycle have opposite charges. In
the singular orbifold limit, when the two-cycle collapses, these
states give rise to two additional massless seven-dimensional vector
multiplets. Given their charges, these combine with the other vector
multiplet to form the field content of $A_1\sim\mathrm{SU}(2)$ SYM
theory.
\\

The final type of singularity we consider is the co-dimension seven
singularity. In general, these singularities lead to charged chiral
fermions \cite{Gukov}--\cite{Cvetic}. We take M-theory on a
spacetime of the form $\mathcal{M}_{1,3}\times\mathcal{M}_7$. Near a
co-dimension seven singularity, $\mathcal{M}_7$ looks like a cone,
its metric taking the structure
\begin{equation}
\mathrm{d}s^2=\mathrm{d}r^2 + r^2\mathrm{d}\Omega^2\,.
\end{equation}
Here $\mathrm{d}\Omega^2$ is a metric on some base six-manifold
$\mathcal{B}$, and the radial variable $r$ is non-negative, with the
singularity being located at $r=0$.

We shall argue for the presence of chiral fermions at such a
singularity by showing they are necessary for gauge anomaly
cancellation, and hence for consistency of M-theory. (One could also
use duality with heterotic string theory, and for this we refer the
reader to Ref.~\cite{B}.) Here, we follow Ref.~\cite{Anomaly}. We
consider the cubic anomaly that arises from the $\mathrm{U}(1)$
gauge fields $A^P$, $P=1,\ldots, b^2(\mathcal{M}_7)$, that appear in
the four-dimensional effective theory from the zero-mode expansion
\begin{equation}
C=\sum_P A^P \wedge\omega_P
\end{equation}
of the three-form field in terms of a basis $\{\omega_P\}$ of
two-forms on $\mathcal{M}_7$. The most general gauge transformation
takes $C\to C + \mathrm{d}\Lambda$, where $\Lambda$ is a two-form.
The anomaly arises from the variation, under this transformation, of
the Chern-Simons $CGG$ term of 11-dimensional supergravity
\eqref{rev11dsugra4}. It takes the form
\begin{equation} \label{anomaly1}
\Delta \sim \int_{\mathcal{M}_{1,3}\times\mathcal{M}_7}
\mathrm{d}\Lambda \wedge G \wedge G\,.
\end{equation}
To deal with the conical singularity, one formally regards
$\mathcal{M}_7$ as a manifold with a boundary, $\partial
\mathcal{M}_7 = \mathcal{B}$, and thus
\begin{equation} \label{anomaly2}
\Delta \sim  \left. \int_{\mathcal{M}_{1,3}\times\mathcal{B}}
\Lambda \wedge G \wedge G \right|_{r=0} \,.
\end{equation}
If we make the ansatz $\Lambda=\sum_P\Lambda^P\wedge\omega_P$ then
we find
\begin{equation}
\Delta \sim \sum_{P,Q,R} \int_{\mathcal{B}}
\omega_P\wedge\omega_Q\wedge\omega_R \int_{\mathcal{M}_{1,3}}
\Lambda^P F^Q \wedge F^R \,,
\end{equation}
where $F^P=\mathrm{d}A^P$ are the field strengths of the gauge
fields. The integrals over $\mathcal{B}$ are topological, and so
just numbers from a four-dimensional perspective. Hence, the
11-dimensional anomaly $\Delta$ can be cancelled by the usual chiral
anomaly of the form
\begin{equation}
\sum_{P,Q,R} q_Pq_Qq_R \int_{\mathcal{M}_{1,3}} \Lambda^P F^Q \wedge
F^R \,
\end{equation}
if there is an additional chiral superfield at the singularity, and
this carries the appropriate $\mathrm{U}(1)$ charges $\{q_P\}$.

If, on the internal space, the locus of a co-dimension four $ADE$
singularity passes through a conical singularity, one will in
general find localised chiral fermions that are charged under the
non-Abelian $ADE$ gauge group. In Ref.~\cite{Anomaly} this is shown
by demonstrating that certain anomalies must arise in the localised
seven-dimensional SYM theory.
\\

At the end of the previous section we saw that compactification of
M-theory on a smooth $G_2$ manifold did not lead to promising
particle phenomenology. However, with the results we have discussed
in this section, it is clear that the situation is potentially much
more interesting for M-theory compactified on a singular $G_2$
manifold. Specifically, if the $G_2$ space contained a co-dimension
four $ADE$ singularity, then there would be non-Abelian gauge
fields, and if it contained a conical (co-dimension seven)
singularity, then it would contain charged chiral fermions. If there
was an intersection of the loci of these two types of singularity,
then one could obtain chiral fermions that are charged under a
non-Abelian gauge group. This motivates the work in this thesis,
which focuses on orbifold-based $G_2$ spaces with co-dimension four
orbifold-fixed points.

\chapter{Classification and Moduli K\"ahler Potentials of
$\boldsymbol{G_2}$ Manifolds} \label{Classification}

\fancyhf{}
\fancyhead[L]{\sl 3.~Classification and Moduli K\"ahler
Potentials of $G_2$ Manifolds}
\fancyhead[R]{\rm\thepage}
\renewcommand{\headrulewidth}{0.3pt}
\renewcommand{\footrulewidth}{0pt}
\addtolength{\headheight}{3pt} \fancyfoot[C]{}

In this chapter we classify, and describe in detail, $G_2$ holonomy
manifolds of a certain type, constructed from the quotient of a
seven-torus, using the method that we outlined in
Chapter~\ref{revchap}. We also derive a formula that gives the
moduli K\"ahler potential for M-theory reduced on a $G_2$ manifold
of this type.

The opening two sections of this chapter are concerned with the
classification. This will be in terms of the orbifold group of the
manifold, and so, in Section~\ref{elements}, we draw up a list of
symmetries of seven-tori that may be used as generators of orbifold
groups. For a symmetry $\alpha$ to be suitable for the orbifolding
there must exist a $G_2$ structure $\varphi$ on the torus that is
preserved by $\alpha$. Then in Section~\ref{groups} we look at ways
of combining these symmetries to form orbifold groups that give
orbifolds of finite first fundamental group, and hence holonomy
group $G_2$. We find a straightforward way of checking for this
property. A summary of the results is as follows. There is only one
possible Abelian orbifold group, $\mathbb{Z}_2^3$, with three or
less generators. Further, we have been looking for viable examples
within the class of orbifold groups formed by three or less
generators with co-dimension four singularities subject to an
additional technical constraint on the allowed underlying lattices.
Within this class we have found all viable examples consisting of
ten distinct semi-direct product groups with three generators as
well as five exceptional cases built from three generators with a
more complicated algebra.

In Section~\ref{manifolds} we give a description of a general $G_2$
manifold with blown-up co-dimension four orbifold singularities of
$A$-type. We present a basis of its third homology, and write down
formulae for an ``almost Ricci-flat'' metric $g$ and its corresponding
$G_2$ structure $\varphi$ of small torsion. Recall that, following
Joyce's construction, the independent moduli will come from torus
radii and from the radii and orientation of cycles associated with
the blow-ups.

With all the machinery in place, in Section~\ref{ModKP} we compute
the moduli K\"ahler potential for our class of $G_2$ manifolds,
valid for sufficiently small blow-up moduli. Let us give an outline
of the calculation. Since the K\"ahler potential is given by a
simple formula \eqref{Kformula1} in terms of the volume $V$, the
first step in calculating it is to derive the volume in terms of
underlying geometrical parameters using the metric $g$. The second
step is to evaluate the period integrals \eqref{periodgen} of the
$G_2$ structure $\varphi$ to obtain the moduli $a^I$ in terms of the
same geometrical parameters. This enables us to re-write our
expression for the volume in terms of these moduli, and, hence, in
terms of the superfields $T^I$ for which the moduli are the real
part \eqref{sufielddefn}. We then substitute our final expression
for the volume into the formula~\eqref{Kformula1} for the K\"ahler
potential.

Note that ideally one would like to perform the calculation using a
torsion-free $G_2$ structure. However, such torsion-free structures
are not known explicitly on compact $G_2$ manifolds. Instead, as in
Ref.~\cite{Lukas}, we use our explicit $G_2$ structure with small
torsion and compute the K\"ahler potential in a controlled
approximation, knowing that there exists a ``nearby'' torsion-free
$G_2$ structure.

To keep the main text of this chapter more readable we have
collected some of the technical details in Appendices~\ref{a} and
\ref{c}. Appendix \ref{a} contains some results on $G_2$ structures
useful for the classification of Sections \ref{elements} and
\ref{groups}, whilst Appendix \ref{c} has some of the details of how
to blow up singularities, and describes some calculations on the
associated Gibbons-Hawking spaces. Also, we refer the reader to
Appendix~\ref{b} for a table listing the possible orbifold group
elements.

\section{Classification of orbifold group elements} \label{elements}
The most general group element $\alpha$ that we consider acts on
seven-dimensional vectors $\boldsymbol{x}$ by $\alpha :
\boldsymbol{x} \mapsto M_{(\alpha)}\boldsymbol{x} +
\boldsymbol{v}_{(\alpha)}$, where $M_{(\alpha)}$ is an orthogonal
matrix and $\boldsymbol{v}_{(\alpha)}$ is a shift vector. We shall
find all such $\alpha$ that give a consistent orbifolding of some
seven-torus, and that preserve some $G_{2}$ structure.
Mathematically, by a seven-torus, we mean the fundamental domain of
$\mathbb{R}^7/\Lambda$, where $\Lambda$ is some seven-dimensional
lattice. In the following we shall use bold Greek letters to denote
lattice vectors.

\fancyhf{} \fancyhead[L]{\sl \rightmark} \fancyhead[R]{\rm\thepage}
\renewcommand{\headrulewidth}{0.3pt}
\renewcommand{\footrulewidth}{0pt}
\addtolength{\headheight}{3pt} \fancyfoot[C]{}

Let us begin by considering the orbifolding. For consistency we require that
two points $\boldsymbol{x}$ and $\boldsymbol{y}$ in one unit cell are
equivalent under the orbifolding if and only if the corresponding two
points $\boldsymbol{x}+\boldsymbol{\lambda}$ and
$\boldsymbol{y}+\boldsymbol{\lambda}$ in another unit cell are
equivalent. Suppose
\begin{equation} \label{y1}
\boldsymbol{y}=  M_{(\alpha)}\boldsymbol{x} + \boldsymbol{v}_{(\alpha)}
                 + \boldsymbol{\mu}\,.
\end{equation}
Then there must exist a $ \boldsymbol{\nu}$ such that
\begin{equation}
\boldsymbol{y} + \boldsymbol{\lambda}=  M_{(\alpha)}(\boldsymbol{x}+
  \boldsymbol{\lambda}) + \boldsymbol{v}_{(\alpha)} + \boldsymbol{\nu}\,.
\end{equation}
Substituting for $\boldsymbol{y}$ using equation \eqref{y1}, we see
that this is the case if and only if
$M_{(\alpha)}\boldsymbol{\lambda}$ is a lattice vector. Since
$\boldsymbol{\lambda}$ is arbitrary, it follows that $\alpha$ gives
a consistent orbifolding precisely when $M_{(\alpha)}$ takes lattice
vectors to lattice vectors. Note that there is no constraint on $
\boldsymbol{v}_{(\alpha)}$.

We are interested in classifying orthogonal matrices that preserve a
seven-dimensional lattice. A useful starting point is to find out
the possible orders of such orthogonal group elements.  We first
quote a result from Ref.~\cite{Ono} in $n$-dimensions and then apply
it to $n=7$. Let $M\in \mathrm{O}(n)$ be of order $m$. Then all its
eigenvalues are $m^{\mathrm{th}}$ roots of unity, and so we can
choose an orthonormal basis for $\mathbb{R}^{n}$ with respect to
which $M$ takes the form
\begin{equation}
M= \left( \begin{array}{cccc}
M_{1} & \, & \, & \,  \\
\, & M_{2} & \, & \,  \\
\, & \, & \ddots & \, \\
\, & \, & \, & M_{k}  \\
\end{array} \right) \,,
\end{equation}
with $M_{j} \in \mathrm{O}(d_{j})$, the eigenvalues of $M_{j}$ being
primitive $m_{j}^{\mathrm{th}}$ roots of unity, $m_{j}\lvert m$, and
$m_{i} \neq m_{j}$ for $i \neq j$. The result is that there exists a
seven-dimensional lattice preserved by $M$ if and only if we can
write each $d_j$ in the form
\begin{equation} \label{euler}
d_{j}=n_{j} \Phi (m_{j})\,,
\end{equation}
where $n_{j} \in \mathbb{N}$ and $\Phi (m)$ is Euler's
function, the number of integers less than $m$ that are prime to
$m$. Furthermore, for an $M$ that satisfies \eqref{euler}, each primitive
$m_{j}^{\mathrm{th}}$ root of unity is
an eigenvalue of $M_{j}$ with geometric multiplicity precisely
$n_{j}$.

We now consider the values $m_{j}$ is allowed to take when
$n=7$. Since $d_{j}\le 7$, $M_{j}$ can only be a constituent block of
$M$ if $\Phi(m_{j}) \le 7$. There is a formula from number theory
\begin{equation} \label{5}
\Phi (a) = \prod_{i} (p_{i}-1)p_{i}^{r_{i}-1}\,,
\end{equation}
where now $a=\prod p_{i}^{r_{i}}, r_{i}\in \mathbb{N}$ is the prime
decomposition of $a$. It is then straightforward to show from
Eq.~\eqref{5} that the allowed values of $m_{j}$ are $m_{j}=1, \ldots,
10, 12, 14, 18$. It is also easy to see that this result is the same
as for $n=6$, a fact which we will use, since the $n=6$ case is the
relevant one for classifying orbifold-based Calabi-Yau spaces.

We now look for conditions on $M$ to belong in $G_{2}$. According to
\cite{Joyce} we must have $M\in \mathrm{SO}(7)$. It therefore has
eigenvalues 1 and complex conjugate pairs of modulus one. Hence we
can write $M$ in the canonical form
\begin{equation}
M= \left( \begin{array}{cccc}
1 & \, & \, & \,  \\
\, & S(\theta_{1}) & \, & \,  \\
\, & \, & S(\theta_{2}) & \, \\
\, & \, & \, & S(\theta_{3})  \\
\end{array} \right) \,,
\end{equation}
where
\begin{equation} \label{2theta}
 S(\theta_{i})= \left( \begin{array}{cc}
\cos\theta_{i} & -\sin\theta_{i} \\
\sin\theta_{i} & \cos\theta_{i} \\
\end{array} \right)\,.
\end{equation}
Accordingly, $M$ decomposes as
\begin{equation} \label{decomp}
M=(1) \oplus M^{\prime}\,,
\end{equation}
where $M^{\prime} \in \mathrm{SO}(6)$.

As an aside, let us just mention that naively one may have first
derived the decomposition \eqref{decomp} and then decided immediately
that this implies that the set of possible orders of $M$ is identical
to the set of possible orders of symmetries of six-dimensional
lattices. Although this does indeed turn out to be the case, it is not
obvious that the $M$ of \eqref{decomp} preserves a seven-dimensional
lattice if and only if the corresponding $M^{\prime}$ preserves a
six-dimensional lattice. Rather, this can be shown by
applying Eq.~\eqref{euler} which leads to the same
allowed values of $m_j$ for the cases $n=6$ and $n=7$.

Now $M\in G_{2}$ if and only if it leaves a $G_{2}$ structure
invariant. In other words $M$ must leave $\varphi_0$ defined by
equation \eqref{structure} invariant, or else there must exist an
$\mathrm{O}(7)$ transformation taking $\varphi_0$ to a three-form
$\tilde{\varphi}_0$ that \textit{is} left invariant by $M$. It is
convenient to recast $\varphi_0$ in complex form by taking
\begin{equation} \label{complex}
\begin{array}{cccc}
x_{0}=x_{1}\,,& z_{1}=\frac{1}{\sqrt{2}}(x_{2}+ix_{3})\,, &
z_{2}=\frac{1}{\sqrt{2}}(x_{4}+ix_{5})\,, &
z_{3}=\frac{1}{\sqrt{2}}(x_{6}+ix_{7})\,,
\end{array}
\end{equation}
to obtain
\begin{eqnarray} \label{form}
\varphi_0 & = & \mathrm{d}x_{0}\wedge
i(\mathrm{d}z_{1}\wedge\mathrm{d}\bar{z}_{1}+
\mathrm{d}z_{2}\wedge\mathrm{d}\bar{z}_{2}+\mathrm{d}z_{3}\wedge\mathrm{d}\bar{z}_{3})
+\sqrt{2}\mathrm{d}z_{1}\wedge\mathrm{d}z_{2}\wedge\mathrm{d}z_{3} \nonumber\\
& & +\sqrt{2}\mathrm{d}\bar{z}_{1}\wedge\mathrm{d}\bar{z}_{2}
    \wedge\mathrm{d}\bar{z}_{3}\,.
\end{eqnarray}
We can then see by inspection that $M$ preserves $\varphi_0$ if and
only if
\begin{equation}
\theta_{1}+\theta_{2}+\theta_{3}=0 \: \mathrm{mod} \: 2\pi \,.
\end{equation}
The following is also easily verified. Under a transformation only
containing reflections in coordinate axes, asking for the resulting
$\tilde{\varphi}_0$ to be left invariant by $M$ imposes one of the
following four conditions:
\begin{eqnarray}
 \theta_{1}+\theta_{2}+\theta_{3} & = & 0 \: \mathrm{mod} \: 2\pi\,, \label{one}\\
  -\theta_{1}+\theta_{2}+\theta_{3} & = & 0 \: \mathrm{mod} \: 2\pi\,, \label{two}\\
  \theta_{1}-\theta_{2}+\theta_{3} & = & 0 \: \mathrm{mod} \: 2\pi\,, \label{three}\\
  \theta_{1}+\theta_{2}-\theta_{3} & = & 0 \: \mathrm{mod} \: 2\pi\,. \label{four}
\end{eqnarray}
Hence, that $M$ satisfies one of \eqref{one}, \eqref{two},
\eqref{three} or \eqref{four} is sufficient for $M\in G_{2}$. It
turns out that this is also necessary. The proof of this is somewhat
technical, and an outline of it is given in Appendix \ref{a}. It is
useful to note that the condition we have on $M$ to belong in
$G_{2}$ is precisely the condition \cite{Bailin} on the $M^{\prime}$
of \eqref{decomp} to belong in $\mathrm{SU}(3)$ under some embedding
of $\mathrm{SU}(3)$ in $\mathrm{SO}(6)$.

By combining the above results we see that the classification is in
one-to-one correspondence with that of the possible orbifold group
elements of a Calabi-Yau space, which is given, for example, in
Refs.~\cite{Bailin} and \cite{Dixon}. The table in Appendix \ref{b}
gives this classification in terms of the rotation angles
$\theta_{i}$.

\section{Classification of orbifold groups} \label{groups}
Having obtained a class of possible generators, we now wish to find a
class of discrete symmetry groups from which compact manifolds of
$G_2$ holonomy may be constructed. Let us state the conditions for
$\Gamma$ to be a suitable orbifold group. There must exist both a
seven-dimensional lattice $\Lambda$ and a $G_2$ structure
$\varphi$ that are preserved by $\Gamma$, and the first fundamental
group $\pi_1$ of $(\mathbb{R}^7/\Lambda)/\Gamma$ must be finite.

It is useful to translate the condition on $\pi_1$ into an equivalent
condition that is more readily checked. An equivalent condition is
that there exist no non-zero vectors $\boldsymbol{n}$ with the
property that $M_{(\alpha)}\boldsymbol{n}=\boldsymbol{n}$ for each
$\alpha\in\Gamma$. That this condition is sufficient for $\pi_1$ to be
finite is shown in Ref. \cite{Joyce}. To show this is necessary, we demonstrate that if the condition is not satisfied then $\pi_1$ is infinite. 

Suppose then, that $\boldsymbol{n}$ is a non-zero vector which satisfies 
$M_{(\alpha)}\boldsymbol{n}=\boldsymbol{n}$ for each
$\alpha\in\Gamma$, and let us expand $\boldsymbol{n}$ as
\begin{equation} \label{nexp}
\boldsymbol{n}=\sum_{j}n_{j}\boldsymbol{\lambda}_{j}\,
\end{equation}
where $\{\boldsymbol{\lambda}_{j}\}$ is a basis of lattice vectors. Consider 
 the generators $\{\alpha_{1},\ldots,\alpha_{k}\}$ of
$\Gamma$. Applying $\alpha_{l}$ to both sides of \eqref{nexp},
we have
\begin{equation} \label{nexp2}
\boldsymbol{n}=\sum_{j,i}n_{j}a_{ji}^{(l)}\boldsymbol{\lambda}_{i}\,,
\end{equation}
where the $a_{ji}^{(l)}$ are matrices with integer coefficients. From
\eqref{nexp} and \eqref{nexp2} we obtain
\begin{equation} \label{system}
\sum_{j}n_{j}b_{ji}^{(l)}=0\,,
\end{equation}
for each $i$ and $l$, where $b_{ji}^{(l)}=a_{ji}^{(l)}-\delta_{ji}$.
Now since, by assumption, there exist non-zero solutions to
\eqref{system}, in constructing a particular solution we may choose
the value of at least one of the $n_i$s. Let us set the value of
this $n_i$ to unity. Then consider another $n_i$. If it is free then
let us set it also to unity. If it is constrained then it must be a
linear function of other $n_i$s with rational coefficients. We have
hence constructed a solution to \eqref{system} with each $n_i$
rational. Now, for our solution, write $n_i$ in the form
\begin{equation}
n_i=\frac{p_i}{q_i}\,,
\end{equation}
with $p_i$ and $q_i$ integers. Then
$\mathrm{lcm}\{q_i\}\boldsymbol{n}$ is a lattice vector (where lcm
stands for ``lowest common multiple'') and so the path that is a
straight line from the origin of the orbifold to
$w\,\mathrm{lcm}\{q_i\}\boldsymbol{n}$, with $w$ an integer, is a path
of winding number $w$. This establishes the result.  \\

Given the above result, it is clear from \eqref{decomp} that a group
$\Gamma$ must contain more than one generator if the resulting
orbifold $\mathcal{T}^{7}/\Gamma$ is to have finite first
fundamental group, and hence be suitable for constructing a manifold
of $G_2$ holonomy. We now attempt to construct Abelian groups of the
form $\mathbb{Z}_m\times\mathbb{Z}_n$ for which the corresponding
orbifold has finite first fundamental group. For now we take the
generator of the $\mathbb{Z}_n$ symmetry to be a straightforward
rotation
\begin{equation} \label{Rprime}
R= \left( \begin{array}{cccc}
1 & \, & \, & \,  \\
\, & S(\theta_{1}) & \, & \,  \\
\, & \, & S(\theta_{2}) & \, \\
\, & \, & \, & S(\theta_{3})  \\
\end{array} \right)\, ,
\end{equation}
with $S(\theta_i)$ as in \eqref{2theta} and
$(\theta_1,\theta_2,\theta_3)$ one of the triples of the table in
Appendix A.  We look for a second symmetry, also a pure rotation, with
corresponding matrix $P$ commuting with $R$ such that the group
generated by $P$ and $R$ is a suitable orbifold group. To find the
constraints on $P$ coming from commutativity we apply a generalisation
of Schur's Lemma, as follows.

Write the reducible representation $\rho$ of the group $G$ as
$\rho=n_1\rho_1\oplus\cdots\oplus n_r\rho_r$, where the $\rho_i$ are
irreducible
representations of $G$ of dimension $d_i$ and the integers $n_i$
indicate how often each $\rho_i$ appears in $\rho$. Then a matrix $P$ with
$[P,\rho(g)]=0$ for all $g\in G$ has the general form
\begin{equation} \label{Schur}
P=P_1\otimes \boldsymbol{1}_{d_1\times d_1}\oplus\cdots\oplus P_r\otimes
\boldsymbol{1}_{d_r\times d_r}\,,
\end{equation}
where the $P_i$ are $n_i\times n_i$ matrices.

The table in Appendix \ref{b} lists the $n_i$ and $d_i$ for the
representations defined by each $R$
we are considering. We can therefore simply go through this table and,
for each $r$, see if there are any $P$s that are suitable for our
construction. It turns out that there are in fact no suitable $P$s for
any of the $R$s. In fact every case fails because $\pi_1$ is not finite.
Below is one
example to provide an illustration.

There is the possibility that $R$ represents a $\mathbb{Z}_2$ symmetry of the lattice and is given by
\begin{equation}
R  = \mathrm{diag}(1,1,1,-1,-1,-1,-1)\,.
\end{equation}
Then we have $n_1=3$, $d_1=1$ and $n_2=4$, $d_2=1$. Applying the lemma \eqref{Schur},
$P$ must take the form
\begin{equation}
P= \left( \begin{array}{cc}
P_{1}^{3\times 3} & \,  \\
\, &  P_{2}^{4\times 4} \\
\end{array} \right)\,.
\end{equation}
Now (see Appendix \ref{a}), any $G_2$ structure $\varphi$ preserved
by $R$ has $\lvert\varphi_{123}\rvert=1$. Under $P$
$\varphi_{123}\mapsto\mathrm{det}(P_{1}^{3\times 3})\varphi_{123}$,
and so we must have $\mathrm{det}(P_{1}^{3\times 3})=1$. Hence
$P_{1}^{3\times 3}\in \mathrm{SO}(3)$, and leaves at least one
direction fixed. But $R$ fixes this direction too, since it fixes
all of the 1, 2 and 3 directions. We therefore rule out this case
since we can not
render $\pi_1$ finite by this construction.  \\

Let us now attempt to construct an orbifold group of the form
$\mathbb{Z}_m\times\mathbb{Z}_n\times\mathbb{Z}_p$. As above, we let
$R$ generate $\mathbb{Z}_p$, and use the same method as above to
find the possibilities for the other generators $P$ and $Q$. In
looking for $P$, most possibilities are still ruled out, but we can
now relax the condition that there are no non-zero fixed vectors of
the group generated by $R$ and $P$. There are then three cases we
need to consider. Firstly,
$\frac{1}{2\pi}(\theta_{1},\theta_{2},\theta_{3})=(\frac{1}{4},\frac{1}{4},
\frac{1}{2})$. In this case, we find that $P$ must take the form
\begin{equation}
P =  (-1)\oplus \left( \begin{array}{cc}
\cos\phi_{1} & \sin\phi_{1} \\
\sin\phi_{1} & -\cos\phi_{1} \\
\end{array} \right) \oplus
 \left( \begin{array}{cc}\cos\phi_{2} & \sin\phi_{2} \\
\sin\phi_{2} & -\cos\phi_{2} \\
\end{array} \right) \oplus
 \left( \begin{array}{cc} \cos\phi_{3} & \sin\phi_{3} \\
\sin\phi_{3} & -\cos\phi_{3} \\
\end{array} \right)\,.
\end{equation}
Here the $2\times 2$ blocks each represent the most general element
of $\mathrm{O}(2)-\mathrm{SO}(2)$. It is easily verified that such a
matrix can not commute with $R$, thus ruling out this case. (Note
that Eq.~\eqref{Schur} was a necessary but not sufficient condition
for commutativity.) The second case is when $n_1=3$, $d_1=1$ and
$n_2=2$, $d_2=2$, for which $p=3$, 4 or 6. On checking the
possibilities for this case we find that there is no way of forming
a group with all the correct properties. We are left with one
remaining case, the case with $p=2$, and this leads to the group
$\mathbb{Z}_2^{3}$. For this case, the matrices $P$, $Q$ and $R$ are
given essentially uniquely by
\begin{equation} \label{Z2}
R  = \mathrm{diag}(1,1,1,-1,-1,-1,-1)\,,
\end{equation}
\begin{equation} \label{Z3}
P  =  \mathrm{diag}(1,-1,-1,-1,-1,1,1)\,,
\end{equation}
\begin{equation} \label{Z4}
Q =  \mathrm{diag}(-1,-1,1,1,-1,-1,1)\,.
\end{equation}
\\

The next step is to consider not just pure rotations, but to now allow
the group elements to contain translations as well. Let us derive a
condition for commutativity. Let two generators of an Abelian orbifold
group be given by
\begin{equation}
\alpha_1 : \boldsymbol{x} \mapsto M_1\boldsymbol{x} + \boldsymbol{v}_1\,,
\end{equation}
\begin{equation}
\alpha_2 : \boldsymbol{x} \mapsto M_2\boldsymbol{x} + \boldsymbol{v}_2\,.
\end{equation}
Then commutativity requires us to be able to write
\begin{equation}
(\alpha_1\circ\alpha_2)\boldsymbol{x}=(\alpha_2\circ\alpha_1)\boldsymbol{x}
+\sum_jn_j\boldsymbol{\lambda}_j\,,
\end{equation}
where the $\boldsymbol{\lambda}_j$ form a
 basis of lattice vectors and the $n_j$ are
integers. This gives
\begin{equation}
[M_1,M_2]\boldsymbol{x}+M_1\boldsymbol{v}_2-M_2\boldsymbol{v}_1+
\boldsymbol{v}_1-\boldsymbol{v}_2=
\sum_jn_j\boldsymbol{\lambda}_j\,.
\end{equation}
Since $\boldsymbol{x}$ may vary continuously, we see
 that we must still have $[M_1,M_2]=0$,
and then we are left with
\begin{equation} \label{commute}
(M_1-I)\boldsymbol{v}_2-(M_2-I)\boldsymbol{v}_1=\sum_jn_j
\boldsymbol{\lambda}_j\,.
\end{equation}
We are now able to write down the most general Abelian orbifold
group, with at most three generators, from which a $G_2$ manifold
may be constructed. We apply the constraint \eqref{commute} to any
translations added to the matrix transformations of equations
\eqref{Z2}--\eqref{Z4}. The result is that the most general set of
generators act as follows on a vector $\boldsymbol{x}=(x_1,\ldots,
x_7)$ of the standard seven-torus:
\begin{equation}
\alpha : \boldsymbol{x}\mapsto
\bigg(x_1+\frac{m_1}{2},x_2+\frac{m_2}{2}, x_3+\frac{m_3}{2},
-x_4+a_4,-x_5+a_5,-x_6+a_6,-x_7+a_7\bigg)\,,
\end{equation}
\begin{equation}
\beta : \boldsymbol{x}\mapsto \bigg(x_1+\frac{n_1}{2},-x_2+b_2,
-x_3+b_3,
-x_4+b_4,-x_5+b_5,x_6+\frac{n_6}{2},x_7+\frac{n_7}{2}\bigg)\,,
\end{equation}
\begin{equation}
\gamma : \boldsymbol{x}\mapsto \bigg(-x_1+c_1,-x_2+c_2,
x_3+\frac{p_3}{2}, x_4+\frac{p_4}{2},
-x_5+c_5,-x_6+c_6,x_7+\frac{p_7}{2}\bigg)\,,
\end{equation}
where the $m_i$, $n_i$ and $p_i$ are integers and the $a_i$, $b_i$ and $c_i$ are unconstrained real numbers. The group generated is always $\mathbb{Z}_2^{3}$.
\\

Our objective was to find a class of orbifold groups and to achieve
this, it appears that commutativity is not the most suitable
constraint to impose, in spite of the systematic approach it gave
us. Since well-defined procedures to describe the metric on the
blow-ups are available for the cases with co-dimension four fixed
loci, we now focus on orbifold groups that only lead to these. We thus
restrict attention to generators that leave three directions of the
torus invariant, namely those whose rotation part is one of
$\mathbb{Z}_2$, $\mathbb{Z}_3$, $\mathbb{Z}_4$ or $\mathbb{Z}_6$ from
the table in Appendix \ref{b}. In fact, for this section we shall
consider pure rotations only. We show that a simple constraint on the
lattice itself enables us to come up with a substantial class of
possible orbifold groups. Let us insist that, for each generator
$\alpha$ of the orbifold group, there exists a partition of our basis
of lattice vectors into three sets, spanning the spaces $U$, $V$ and
$W$, of dimension 3, 2 and 2 respectively such that
\begin{equation} \label{latt1}
\Lambda =U\perp V\perp W\,,
\end{equation}
\begin{equation} \label{latt2}
\alpha U = U\,,\hspace{0.3cm} \alpha V = V\,, \hspace{0.3cm} \alpha
W = W\,,
\end{equation}
\begin{equation}\label{latt3}
\alpha \lvert_U = \mathrm{id}\,.
\end{equation}
This seems a sensible condition to impose, since it makes it easy to picture
how the orbifold group acts on the lattice. Basically, each generator rotates two
two-dimensional sub-lattices.

The classification of orbifold groups, subject to the above constraints and
containing three or fewer generators, now goes as follows. We take as the
first generator the
matrix $R$, in the canonical form of \eqref{Rprime} with $\theta_1=0$ and
$\theta_2=-\theta_3=2\pi / N_R$, where $N_R=2,$ 3, 4, or 6. We then use
coordinate freedom to choose a $G_2$ structure $\varphi$ that is the
standard one of Eq.~\eqref{structure} up to possible sign differences (see
Appendix \ref{a}).

We are then able to derive the other possible generators of the
orbifold group that are distinct up to redefinitions of the
coordinates. First we narrow the possibilities using the constraint of
$G_2$ structure preservation. In particular this imposes the condition
that the three fixed directions must correspond precisely to one of
the seven terms in the $G_2$ structure (see Appendix \ref{a}). Having
done this we look for a preserved lattice. It is straightforward to go
through all possibilities having imposed \eqref{latt1}, \eqref{latt2}
and \eqref{latt3}. We find the following distinct generators. Firstly
matrices in the canonical form of \eqref{Rprime} with
$\theta_1=-\theta_2=2\pi / N$ and $\theta_3=0$, with $N$ dividing $N_R$
or vice-versa (with the exception $N=2$, $N_R=3$). Secondly
$\mathbb{Z}_2$ symmetries not in canonical form, for example
\begin{equation} \label{Q}
Q_0 =  \mathrm{diag}(-1,1,-1,1,-1,1,-1)\,,
\end{equation}
and then, only for the cases $N_R=2$ or 4, $\mathbb{Z}_4$ symmetries not
in canonical form, for example
\begin{equation}
Q_1 : (x_1,x_2,x_3,x_4,x_5,x_6,x_7) \mapsto
(x_7,x_2,-x_5,x_4,x_3,x_6,-x_1)\,.
\end{equation}

We can now go about combining these generators together. It is easy to see that, as when we were considering Abelian groups, there are no orbifold groups built from just two generators that result in orbifolds with finite first fundamental group. Our class therefore consists solely of orbifold groups containing three generators. The main sub-class has generators $P$ and
$R$ in the canonical form of \eqref{Rprime}, with $P$ having
$\theta_1=-\theta_2=2\pi / N_P$ and $\theta_3=0$ and $R$ having
$\theta_1=0$ and $\theta_2=-\theta_3=2\pi / N_R$, and $Q_0$ as given in
\eqref{Q}. This class contains the following orbifold groups:
\begin{equation}
\begin{array}{l}
\mathbb{Z}_2 \times \mathbb{Z}_2 \times \mathbb{Z}_2\,, \\
\mathbb{Z}_2 \ltimes \left( \mathbb{Z}_2 \times \mathbb{Z}_3 \right)\,,
\end{array} \nn
\end{equation}
\begin{equation}
\begin{array}{l} 
\mathbb{Z}_2 \ltimes \left( \mathbb{Z}_2 \times \mathbb{Z}_4 \right)\, ,\\
\mathbb{Z}_2 \ltimes \left( \mathbb{Z}_2 \times \mathbb{Z}_6 \right) \,,\\
\mathbb{Z}_2 \ltimes \left( \mathbb{Z}_3 \times \mathbb{Z}_3 \right) \,,\\
\mathbb{Z}_2 \ltimes \left( \mathbb{Z}_3 \times \mathbb{Z}_6 \right) \,,\\
\mathbb{Z}_2 \ltimes \left( \mathbb{Z}_4 \times \mathbb{Z}_4 \right) \,,\\
\mathbb{Z}_2 \ltimes \left( \mathbb{Z}_6 \times \mathbb{Z}_6 \right) \,.\\
\end{array}
\end{equation}
The semi-direct product notation is defined by writing $G\ltimes H$ if
$G$ and $H$ are Abelian and $[g,h]\in H$ for any $g\in G$ and $h\in
H$. In fact, within our class of orbifold groups, the stronger condition
$[g,h]=h^2$ will be satisfied in each case of a semi-direct product. Note
that since we should really be thinking of orbifold group elements as
abstract group elements as opposed to matrices, the commutator is defined by
$[g,h]=g^{-1}h^{-1}gh$. Some
other semi-direct products are attained by using $R$ as above
and then taking $Q_0$ and $\mathrm{diag}(-1,-1,1,1,-1,-1,1)$ as the
other generators to obtain the following groups:
\begin{equation}
\mathbb{Z}_2^2 \ltimes \mathbb{Z}_N\,, \; \; N=3, \, 4 \,\,
\mathrm{or} \,\, 6\,.
\end{equation}

There are also some exceptional cases that contain a $\mathbb{Z}_4$
not in canonical form. These exceptional orbifold groups will contain
only $\mathbb{Z}_2$ and $\mathbb{Z}_4$ symmetries. We make the
observations that $\mathbb{Z}_2$ and $\mathbb{Z}_4$ symmetries either
commute or give semi-direct products and that two $\mathbb{Z}_4$
symmetries $\alpha$ and $\beta$ either commute or have the relation
$\alpha^2\beta\alpha^2=\beta^{-1}$. It is then straightforward to find
all possible
group algebras for our exceptional orbifold groups. Going through the
conditions for a $G_2$ orbifold, we find some of these algebras can be
realised and some can not. Those that can are constructed as
follows. Take $P$ and $R$ as before with $N_P=2$ and $N_R=4$ and use
either $Q_1$ from above or $Q_2$ given by
\begin{equation}
Q_2 : (x_1,x_2,x_3,x_4,x_5,x_6,x_7) \mapsto
(x_3,x_2,-x_1,x_4,-x_7,x_6,x_5)\,.
\end{equation}
These lead to the respective groups
\begin{equation}
\mathbb{E}_1 =: \langle P,\, Q_1,\, R\, \lvert\, P^2=1,\, Q_1^4=1,\,
R^4=1,\, [P,Q_1]=1,\, [P,R]=1,\, Q_1^2RQ_1^2=R^{-1}  \rangle\,,
\end{equation}
and
\begin{equation}
\mathbb{E}_2 =: \langle P,\, Q_2,\, R\, \lvert\, P^2=1,\, Q_2^4=1,\,
R^4=1,\, [P,Q_2]=Q_2^2,\, [P,R]=1,\, Q_2^2RQ_2^2=R^{-1}  \rangle\,.
\end{equation}
Then three more possibilities come about from using $R$ with $N=4$,
$Q_2$ and one of the $P_i$ given by
\begin{equation}
P_1 : (x_1,x_2,x_3,x_4,x_5,x_6,x_7) \mapsto
(-x_1,-x_2,x_3,-x_4,x_5,x_6,-x_7)\,,
\end{equation}
\begin{equation}
P_2 : (x_1,x_2,x_3,x_4,x_5,x_6,x_7) \mapsto
(x_1,-x_3,x_2,x_5,-x_4,x_6,x_7)\,,
\end{equation}
\begin{equation}
P_3 : (x_1,x_2,x_3,x_4,x_5,x_6,x_7) \mapsto
(x_1,-x_5,x_4,-x_3,x_2,x_6,x_7)\,.
\end{equation}
The groups we obtain can be described respectively by
\ba
\mathbb{E}_3 &=:& \langle P_1,\, Q_2,\, R\, \lvert\, P_1^2=1,\,
Q_2^4=1,\, R^4=1,\, [P_1,Q_2]=Q_2^2,\, [P_1,R]=R^2,\, \nn \\
&& \hspace{2.3cm} Q_2^2RQ_2^2=R^{-1}  \rangle\,, \\
\mathbb{E}_4 &=:& \langle P_2,\, Q_2,\, R\, \lvert\, P_2^4=1,\,
Q_2^4=1,\, R^4=1,\, P_2^2Q_2P_2^2=Q_2^{-1},\, [P_2,R]=1,\, \nn \\
&& \hspace{2.3cm} Q_2^2RQ_2^2=R^{-1}  \rangle\,, \\
\mathbb{E}_5 &=:& \langle P_3,\, Q_2,\, R\, \lvert\, P_3^4=1,\,
Q_2^4=1,\, R^4=1,\, P_3^2Q_2P_3^2=Q_2^{-1},\, P_3^2RP_3^2=R^{-1},\, \nn \\
&& \hspace{2.3cm} Q_2^2RQ_2^2=R^{-1}  \rangle\,.
\ea
\\

It is worth briefly summing up what we have found. We have found a
class of sixteen distinct groups, each composed of three pure
rotations, that may be used as orbifold groups to construct compact
manifolds of $G_2$ holonomy. If two simple constraints are imposed on
the manifolds, we have the complete class of such orbifold groups. The
first constraint states that the manifold is to have only co-dimension
four fixed points. The second constraint is on the lattice from which
the manifold is constructed. It states that the lattice must decompose
into an orthogonal sum of smaller lattices, with three constituents of
the sum being simple two-dimensional lattices, and the final
constituent being the trivial one-dimensional lattice. Furthermore the
action of each orbifold group generator must be to simply rotate two
two-dimensional sub-lattices.

There are two obvious ways of extending the classification we have
obtained. One is to remove the restriction on the lattice. Perhaps
this would give rise to some more complicated examples, in which
lattice vectors do not lie precisely in the planes of rotation of the
orbifold group generators. A second method would be to allow
co-dimension six \nopagebreak fixed points and thus allow any of the generators
listed in Appendix \ref{b}.


\section{Description of the manifolds and their $G_2$ structures} \label{manifolds}
In this section we discuss properties of a general smooth $G_2$
manifold $\mathcal{Y}^{\mathrm{S}}$, constructed from an orbifold
$\mathcal{Y}=\mathcal{T}^7/\Gamma$ with co-dimension four fixed
points. We assume that points on the torus that are fixed by one
generator of the orbifold group are not fixed by other generators.
Given an orbifold group, this can always be arranged by
incorporating appropriate translations into the generators, and thus
all of our previously found examples are relevant. Under this
assumption we have a well-defined blow-up procedure.

Let us introduce some notation. We use the index $\tau$ to label the
generators of the orbifold group $\Gamma$, and write $N_\tau$ for
the order of the generator $\alpha_\tau$. For $\Gamma$ to only have
co-dimension four fixed points the possible values of $N_\tau$ are
2, 3, 4 and 6 (see Appendix \ref{b}). Each generator will have a
certain number $M_\tau$ of fixed points associated with it. A
singular point on $\mathcal{Y}$ is therefore labelled by a pair
$(\tau,s)$, where $s=1,\ldots,M_\tau$.

Near a singular point $\mathcal{Y}$ takes the approximate form
$\mathcal{T}^3_{(\tau,s)}\times\mathbb{C}^2/\mathbb{Z}_{N_{\tau}}$,
where $\mathcal{T}^3_{(\tau,s)}$ is a three-torus. Note that this
means the singularities are $A$-type, according to the $ADE$
classification. Blowing up a singularity involves the following. One
firstly removes a four-dimensional ball centred around the
singularity times the associated fixed three-torus
$\mathcal{T}^3_{(\tau,s)}$. Secondly one replaces the resulting hole
by $\mathcal{T}^3_{(\tau,s)}\times \mathcal{U}_{(\tau,s)}$, where
$\mathcal{U}_{(\tau,s)}$ is the blow-up of
$\mathbb{C}^2/\mathbb{Z}_{N_{\tau}}$ as discussed in Appendix
\ref{c}.
\\

Before giving a more detailed description of what
$\mathcal{Y}^{\mathrm{S}}$ looks like, let us present a basis of
three-cycles. We will use this in the next section to compute the
moduli for M-theory on $\mathcal{Y}^{\mathrm{S}}$ in terms of
underlying geometrical parameters. (Recall that Ricci flat
deformations of the metric can be described by torsion-free
deformations of the $G_2$ structure, and, as we explained in
Chapter~\ref{revchap}, this means that the moduli can be computed by
performing period integrals of the $G_2$ structure over
three-cycles.) Localised on the blow-up labelled by $(\tau,s)$ there
are $3(N_\tau-1)$ three-cycles. These are formed by taking the
Cartesian product of one of the $(N_\tau-1)$ two-cycles on
$\mathcal{U}_{(\tau,s)}$ with one of the three one-cycles on
$\mathcal{T}^3_{(\tau,s)}$. Let us label these three-cycles by
$C(\tau,s,i,m)$, where $m=1,2,3$ labels the direction on
$\mathcal{T}^3_{(\tau,s)}$ and $i=1,\ldots,N_\tau-1$ labels the
two-cycles of $\mathcal{U}_{(\tau,s)}$. On the bulk, that is the
remaining parts of the torus, we can define three-cycles by setting
four of the coordinates $x^A$ to constants (chosen so there is no
intersection with any of the blow-ups). The number of these that
fall into distinct homology classes is then given by the number of
independent terms in the $G_2$ structure on the bulk. Let us explain
this statement. The bulk $G_2$ structure can always be chosen so as
to contain the seven terms of the standard $G_2$ structure
\eqref{structure}, with positive coefficients multiplying them. If
we write $r^A$ for the coefficient in front of the $A^{\mathrm{th}}$
term in Eq.~\eqref{structure}, then by the number of independent
terms we mean the number of $r^A$s that are not constrained by the
orbifolding. We then write $C^A$ for the cycle obtained by setting
the four coordinates on which the $A^{\mathrm{th}}$ term in
\eqref{structure} does not depend to constants, for example,
\begin{equation}
C^1=\{x^4,x^5,x^6,x^7=\mathrm{const}\}\,.
\end{equation}
A pair of $C^A$s for which the corresponding $r^A$s are
independent then belong to distinct homology classes. There is
therefore some subset $\mathcal{C}$ of $\{C^A\}$ such that the
collection $\{\mathcal{C},C(\tau,s,i,m)\}$ provides a basis for
$H_3(\mathcal{Y}^{\mathrm{S}},\mathbb{Z})$. We deduce the following
formula for the third Betti number of $\mathcal{Y}^{\mathrm{S}}$:
\begin{equation}
b^3(\mathcal{Y}^{\mathrm{S}})=b(\Gamma)+\sum_\tau M_\tau \cdot
3(N_\tau-1)\,,
\end{equation}
where $b(\Gamma)$ is the number of bulk three-cycles, a positive
integer less or equal than seven, and dependent on the orbifold
group $\Gamma$. For the class of orbifold groups obtained in the
previous section $b(\Gamma)$ takes values as given in
Table~\ref{tab:01}. A description of the derivation is given in the
discussion below on constructing the bulk $G_2$ structure.

As an aside, let us write down the formula for the other non-trivial
Betti number $b^2(\mathcal{Y}^{\mathrm{S}})$, valid for the set of
orbifold groups that do not permit any bulk two-cycles, which
includes all orbifold groups obtained in the previous section. This
is given by
\begin{equation}
b^2(\mathcal{Y}^{\mathrm{S}})=\sum_\tau M_\tau \cdot (N_\tau-1)\,,
\end{equation}
with each blow-up contributing the $(N_\tau-1)$ two-cycles on
$\mathcal{U}_{(\tau,s)}$.
\\

\begin{table}
\begin{center}
\begin{tabular}{c|c}
$\Gamma$ & $b(\Gamma)$ \\
\hline
$\mathbb{Z}_2\times\mathbb{Z}_2\times\mathbb{Z}_2$ & 7 \\
$\mathbb{Z}_2 \ltimes \left( \mathbb{Z}_2 \times \mathbb{Z}_3 \right)$ & 5 \\
$\mathbb{Z}_2 \ltimes \left( \mathbb{Z}_2 \times \mathbb{Z}_4 \right)$ & 5 \\
$\mathbb{Z}_2 \ltimes \left( \mathbb{Z}_2 \times \mathbb{Z}_6 \right)$ & 5\\
$\mathbb{Z}_2 \ltimes \left( \mathbb{Z}_3 \times \mathbb{Z}_3 \right)$ & 4\\
$\mathbb{Z}_2 \ltimes \left( \mathbb{Z}_3 \times \mathbb{Z}_6 \right)$ & 4\\
$\mathbb{Z}_2 \ltimes \left( \mathbb{Z}_4 \times \mathbb{Z}_4 \right)$ & 4\\
$\mathbb{Z}_2 \ltimes \left( \mathbb{Z}_6 \times \mathbb{Z}_6 \right)$ & 4\\
$\mathbb{Z}_2^2 \ltimes \mathbb{Z}_3$ & 5\\
$\mathbb{Z}_2^2 \ltimes \mathbb{Z}_4$ & 5\\
$\mathbb{Z}_2^2 \ltimes \mathbb{Z}_6$ & 5\\
$\mathbb{E}_1$ & 3\\
$\mathbb{E}_2$ & 3\\
$\mathbb{E}_3$ & 3\\
$\mathbb{E}_4$ & 2\\
$\mathbb{E}_5$ & 1\\
\end{tabular}
\end{center}
\caption{Bulk third Betti numbers of the orbifold groups}
\label{tab:01}
\end{table}

We now discuss the geometrical structure of
$\mathcal{Y}^{\mathrm{S}}$ in more detail, focusing in particular on
the blow-up regions. Metrics and $G_2$ structures will be presented
on each region of $\mathcal{Y}^{\mathrm{S}}$. Let us begin with the
bulk, which is the straightforward part. Assuming that, for a
constant metric, only the diagonal components survive the
orbifolding, which is certainly the case for the explicit examples
constructed in Section \ref{groups},
\begin{equation} \label{bulkmetric}
\mathrm{d}s^2=\sum_{A=1}^7(R^A\mathrm{d}x^A)^2\,.
\end{equation}
Here the $R^A$ are precisely the seven radii of the torus. The
corresponding $G_2$ structure $\varphi$ is obtained from the
standard flat $G_2$ structure $\varphi_0$ of Eq.~\eqref{structure}
by rescaling $x^A\to R^Ax^A$. This leads to
\begin{eqnarray} \label{structurex}
\varphi & = & R^1R^2R^3\mathrm{d}x^1\wedge\mathrm{d}x^2\wedge\mathrm{d}x^3
+R^1R^4R^5\mathrm{d}x^1\wedge\mathrm{d}x^4\wedge\mathrm{d}x^5+
R^1R^6R^7\mathrm{d}x^1\wedge\mathrm{d}x^6\wedge\mathrm{d}x^7 \nonumber \\
& &+R^2R^4R^6\mathrm{d}x^2\wedge\mathrm{d}x^4\wedge\mathrm{d}x^6
-R^2R^5R^7\mathrm{d}x^2\wedge\mathrm{d}x^5\wedge\mathrm{d}x^7-
R^3R^4R^7\mathrm{d}x^3\wedge\mathrm{d}x^4\wedge\mathrm{d}x^7 \nonumber \\
& & -R^3R^5R^6\mathrm{d}x^3\wedge\mathrm{d}x^5\wedge\mathrm{d}x^6\,.
\end{eqnarray}
Now, for the orbifolding to preserve the metric, some of the $R^A$
must be set equal to one another. It is straightforward to check
that if $\alpha_\tau$ involves a rotation in the $(A,B)$ plane by an
angle not equal to $\pi$, then we must set $R^A=R^B$. Following this
prescription it is easy to find the function $b(\Gamma)$ discussed
above.

On one of the blow-ups $\mathcal{T}^3\times \mathcal{U}$ (for convenience we
suppress $\tau$ and $s$ indices) we use coordinates $\xi^m$ on
$\mathcal{T}^3$ and four-dimensional coordinates $\zeta^{\hat{A}}$
on $\mathcal{U}$. We write $R^m$ to denote the three radii of $\mathcal{T}^3$,
which will be the three $R^A$ in the directions fixed by $\alpha$.
The $G_2$ structure can be written as
\begin{equation} \label{structure4}
\varphi=\sum_m\omega^m\left(w(\zeta),z(\zeta),\boldsymbol{b}_1,\ldots,
\boldsymbol{b}_{N_\tau}\right)\wedge R^m\mathrm{d}\xi^m-
R^1R^2R^3\mathrm{d}\xi^1\wedge\mathrm{d}\xi^2\wedge\mathrm{d}\xi^3\,.
\end{equation}
Here $w$ and $z$ are complex coordinates on $\mathcal{U}$ and the
$\boldsymbol{b}_i\equiv(\mathrm{Re}\:a_i,\mathrm{Im}\:a_i,b_i)$,
$i=1,\ldots,N$ are
a set of $N$ three-vectors, which parameterise the size of the blow-up
and its orientation with respect to the bulk. The $\omega^m$ are a
triplet of two-forms that constitute a ``nearly'' hyperk\"ahler
structure on $\mathcal{U}$, as discussed in Appendix \ref{c}. We will not need
to know explicitly the relation between the two sets of coordinates
$\zeta^{\hat{A}}$ and $w$ and $z$, although we keep in mind that
this relation will depend on the four radii $R^{\hat{A}}$ transverse
to the $R^m$. In terms of $w$ and $z$, we can write the $\omega^m$
as
\begin{equation} \label{om1}
\omega^1=\frac{i}{2}\partial\bar{\partial}\mathcal{K}\,,
\end{equation}
\begin{equation} \label{om2}
\omega^2=-\mathrm{Re}\left(\frac{\mathrm{d}w\wedge\mathrm{d}z}{w}\right)\,,
\: \: \:
\omega^3=-\mathrm{Im}\left(\frac{\mathrm{d}w\wedge\mathrm{d}z}{w}\right)\,,
\end{equation}
where $\mathcal{K}$ is the K\"ahler potential for $\mathcal{U}$, which
interpolates between that for Gibbons-Hawking space
\cite{GibbHawk},~\cite{Hitchin} in the central
region of the blow-up and that for flat space far away from the centre
of the blow-up. We clarify this statement in the discussion below, but
first we shall describe the central region of the blow-up, where $\mathcal{U}$
looks exactly like Gibbons-Hawking space. For a technical account of
this discussion, including how to write $\mathcal{K}$ explicitly, we
refer the reader to Appendix \ref{c}.

Gibbons-Hawking spaces
(or gravitational multi-instantons)~\cite{GibbHawk},~\cite{Hitchin} provide a
generalisation of the Eguchi-Hanson space~\cite{eguchi1},~\cite{eguchi2}
and their different
topological types are labelled by an integer $N$ (where the case $N=2$
corresponds to the Eguchi-Hanson case). While the Eguchi-Hanson space
contains a single two-cycle, the $N^{\rm th}$ Gibbons-Hawking space
contains a sequence $\gamma_1,\ldots,\gamma_{N-1}$ of such cycles at
the ``centre'' of the space.  Only neighbouring cycles $\gamma_i$ and
$\gamma_{i+1}$ intersect and in a single point and, hence, the
intersection matrix $\gamma_i\cdot\gamma_j$ equals the Cartan matrix of
$A_{N-1}$. Asymptotically, the $N^{\rm th}$ Gibbons-Hawking space has
the structure $\mathbb{C}^2/\mathbb{Z}_{N}$.  Accordingly, we take
$N=N_\tau$ when blowing up $\mathbb{C}^2/\mathbb{Z}_{N_{\tau}}$. The
metric on Gibbons-Hawking space can be written
\begin{equation}
\mathrm{d}s^2=\gamma\mathrm{d}z\mathrm{d}\bar{z} + \gamma^{-1}
\left( \frac{\mathrm{d}w}{w}+\bar{\delta}\mathrm{d}z \right) \left(
\frac{\mathrm{d}w}{w}+ \delta\mathrm{d}\bar{z} \right)\,,
\end{equation}
where
\begin{equation}
\gamma = \sum_i\frac{1}{r_i}\,,
\end{equation}
\begin{equation}
r_i=\sqrt{(x-b_i)^2+4\lvert z-a_i\rvert^2}\,,
\end{equation}
\begin{equation}
\delta =  \sum_i\frac{x-b_i-r_i}{2(\bar{z}-\bar{a}_i)r_i}\,,
\end{equation}
\begin{equation}
w\bar{w}=\prod_i(x-b_i+r_i)\,.
\end{equation}
Here $x$ is a real coordinate, given implicitly in terms of $w$ and
$z$ in the above equations. The sizes and orientations of the
two-cycles are determined by the $N_\tau$ points $\boldsymbol{b}_i$ in
the $(\mathrm{Re}\:z,\mathrm{Im}\:z,x)$ hyperplane. Concretely
$\gamma_i$ is parameterised by
\begin{equation}
z=a_i+\lambda (a_{i+1}-a_i)\,,
\end{equation}
\begin{equation}
w=e^{i\theta}h(\lambda)\,,
\end{equation}
for some function $h$, as $0\leq\lambda\leq 1$, $0\leq\theta\leq 2\pi$ \cite{Hitchin}.

We can add a periodic real coordinate $y$ to $x$, $z$ and $\bar{z}$ to form a well-defined coordinate system on the space. We can also define a radial coordinate $r$ given by
\begin{equation}
r= \sqrt{(x-\tilde{b})^2+4\lvert z-\tilde{a} \rvert^2}\,,
\end{equation}
where tildes denote mean values over the index $i$. It is precisely
this radial coordinate that the interpolating function $\epsilon$,
appearing in the K\"ahler potential $\mathcal{K}$, depends on. We
can now describe the interpolation more precisely. If we set all the
$a$ and $b$ parameters of Gibbons-Hawking space to zero we have flat
space. Therefore the method of constructing $\mathcal{K}$ is to
start with the K\"ahler potential for Gibbons-Hawking space, and to
then place a factor of $\epsilon$ next to every $a$ and $b$ that
appears. We can keep $\epsilon$ general, all we require are the
following properties:
\begin{equation}
\epsilon (r) = \left\{ \begin{array}{cc}
1 & \mathrm{if} \; r\leq r_0\,,  \\
0 & \mathrm{if} \; r\geq r_1\,, \\
\end{array} \right.
\end{equation}
where $r_0 $ and $r_1$ are two fixed radii satisfying $\lvert
a_i\rvert\ll r_0 < r_1$ and $\lvert b_i\rvert\ll r_0$ for each
$i$. Then, as already discussed, $\mathcal{U}$ is identical to Gibbons-Hawking
space for $r<r_0$. For $r>r_1$ $\mathcal{U}$ is identical to the flat space
$\mathbb{C}^2/\mathbb{Z}_N$, and we can match the $G_2$ structure
\eqref{structure4} to the bulk $G_2$ structure \eqref{structurex}.

Let us briefly discuss the torsion of the $G_2$ structure
\eqref{structure4}. By virtue of the blow-up $\mathcal{U}$ being hyperk\"ahler
in the regions $r<r_0$ and $r>r_1$, and the $\omega^m$ of equations
\eqref{om1} and \eqref{om2} forming the triplet of closed and
co-closed K\"ahler forms expected on such a space, the $G_2$
structure is torsion free in these regions. It departs from non-zero
torsion only in the ``collar'' region $r\in[r_0,r_1]$, where
$\omega^2$ and $\omega^3$ fail to be co-closed. However, for
sufficiently small blow-ups, $\lvert a_i\rvert\ll 1$, $\lvert
b_i\rvert\ll 1$, and a ``smooth'', slowly-varying interpolation
function $\epsilon$, the deviation from a torsion-free $G_2$
structure is small \cite{Lukas}. Consequently, we can use this $G_2$
structure to reliably perform M-theory calculations to leading
non-trivial order in the $a_i$s and $b_i$s.

Finally, we briefly discuss the metric on a blow-up. The metric can
be derived directly from the $G_2$ structure, using equations
\eqref{met1} and \eqref{met2}. Its structure is given by
\begin{equation} \label{met3}
\mathrm{d}s^2=\mathcal{G}_0\mathrm{d}\boldsymbol{\zeta}^2
+\sum_{m=1}^{3}\mathcal{G}_m(\mathrm{d}\xi^m)^2\,,
\end{equation}
where $\mathrm{d}\boldsymbol{\zeta}$ is the line element on the
appropriate smoothed Gibbons-Hawking space, which can be derived from
equations \eqref{Legendre} and \eqref{smoothF}, and the $\mathcal{G}$s
are conformal factors, that may depend on the blow-up moduli and the
interpolation function, but whose product must be equal to 1, since
they do not appear in the measure \eqref{measure}.

\section{Volume, periods and the K\"ahler potential} \label{ModKP}
In this section we derive the moduli K\"ahler potential for M-theory
on a $G_2$ manifold, $\mathcal{Y}^{\mathrm{S}}$, of the type we have
been describing in Section~\ref{manifolds}. Along the way we compute
the volume of the manifold and also the period integrals of the
$G_2$ structure over three-cycles.
\\

Our first task is to find the total volume of the manifold
$\mathcal{Y}^{\mathrm{S}}$ in terms of the geometrical parameters
from which it is constructed. These consist of the seven radii $R^A$
of the underlying seven-torus as well as $N_\tau$ triplets
$(\mathrm{Re}\:a_{(\tau,s,i)},\mathrm{Im}\:a_{(\tau,s,i)},b_{(\tau,s,i)})$
for each blow-up $\mathcal{T}^3_{(\tau,s)}\times
\mathcal{U}_{(\tau,s)}$. From the bulk metric \eqref{bulkmetric}, we
see that the bulk contribution to the volume is proportional to
$\prod_A R^A$. There will be a factor $c_\Gamma$ in front of this,
dependent on the orbifold group $\Gamma$. In certain simple cases
this is just the inverse of the order of $\Gamma$, and it is always
calculable by obtaining the fundamental domain of the orbifold. For
the purposes of most calculations, it is relatively unimportant,
since one will be able to absorb it into the normalisation of the
blow-up moduli fields. The volume of a blow-up $\mathcal{T}^3\times
\mathcal{U}$ has been computed in Appendix~\ref{c} and its
contribution to the overall volume of $\mathcal{Y}^{\mathrm{S}}$ can
be read off from the formula~\eqref{finalvol}. Putting everything
together we find, to lowest non-trivial order in the blow up
parameters,
\begin{equation} \label{volM}
V =   c_\Gamma\prod_AR^A  -   \frac{\pi^2}{6}\sum_{\tau,s}
N_\tau\big(  \mathrm{var}_i\{b_{(\tau,s,i)}\}+
2\,\mathrm{var}_i\{\mathrm{Re}\,a_{(\tau,s,i)}\} + 2\,
\mathrm{var}_i\{ \mathrm{Im}\,a_{(\tau,s,i)}\}\big)\prod_m
R^m_{(\tau)}\,.
\end{equation}
Here var refers to the variance, with the usual definition
\begin{equation}
\mathrm{var}_i\{X_i\}=\frac{1}{N}\sum_i\left(X_i-\frac{1}{N}\sum_j
X_j\right)^2\,.
\end{equation}
Note that this result is independent of the interpolation functions
$\epsilon_{(\tau,s)}$ of the blow-up.  \\

Having written down a $G_2$ structure of small
torsion~\eqref{structurex},~\eqref{structure4} on the manifold
$\mathcal{Y}^{\mathrm{S}}$, we can compute its period integrals over
the three-cycles we presented in Section~\ref{manifolds}. The bulk
periods
\begin{equation}
a^A=\int_{C^A}\varphi
\end{equation}
can be read off straight from~\eqref{structurex} and are given by
\begin{equation} \label{periodx}
\left. \begin{array}{cccc}
a^1=R^1R^2R^3\,, & a^2=R^1R^4R^5\,, & a^3=R^1R^6R^7\,, & a^4=R^2R^4R^6\,, \\
a^5=R^2R^5R^7\,, & a^6=R^3R^4R^7\,, & a^7=R^3R^5R^6\,. &  \, \\
\end{array} \right.
\end{equation}
To find the periods associated with the blow-ups, we firstly require
the period integrals
\begin{equation}
\int_{\gamma_i}\omega^m\,,
\end{equation}
$i=1,\ldots,(N-1)$, of the three hyperk\"ahler forms over the
two-cycles on a general Gibbons-Hawking space. Following
Ref.~\cite{Hitchin} we find
\begin{equation}
\int_{\gamma_i}\omega^1=\frac{\pi}{2}(b_i-b_{i+1})\,,
\end{equation}
\begin{equation}
\int_{\gamma_i}(\omega^2+i\omega^3)=\pi i(a_i-a_{i+1})\,.
\end{equation}
Having obtained these we can write down the periods
\begin{equation}
A(\tau,s,i,m)=\int_{C(\tau,s,i,m)}\varphi
\end{equation}
on each blow-up $\mathcal{T}^3_{(\tau,s)}\times
\mathcal{U}_{(\tau,s)}$. By inspection of the $G_2$
structure~\eqref{structure4} we find
\begin{eqnarray}\label{period1}
A(\tau,s,i,1) & = & \frac{\pi}{2} R^1_{(\tau)}\left( b_{(\tau,s,i)} - b_{(\tau,s,i+1)} \right)\,, \nonumber \\
A(\tau,s,i,2) & = & \frac{i\pi}{2}  R^2_{(\tau)}\left( a_{(\tau,s,i)}           - \bar{a}_{(\tau,s,i)}      - a_{(\tau,s,i+1)}  + \bar{a}_{(\tau,s,i+1)}  \right)\,, \\
A(\tau,s,i,3) & = & \frac{\pi}{2} R^3_{(\tau)}\left( a_{(\tau,s,i)}
+ \bar{a}_{(\tau,s,i)}      - a_{(\tau,s,i+1)}  -
\bar{a}_{(\tau,s,i+1)}  \right)\,. \nn
\end{eqnarray}
We remind the reader that $R^m_{(\tau)}$ denote the three radii of
$\mathcal{T}^3_{(\tau,s)}$, consistent with the notation of
equation~\eqref{structure4}.
\\

We are now ready to compute the K\"ahler potential. Using the
results \eqref{periodx} and \eqref{period1} for the periods, we can
rewrite the volume \eqref{volM} in terms of $a^A$ and
$A(\tau,s,i,m)$, which constitute the real, bosonic parts of
superfields. We denote these superfields by $T^A$ and
$U^{(\tau,s,i,m)}$ such that
\begin{equation}
\mathrm{Re}(T^A)=a^A\,, \; \; \;
\mathrm{Re}(U^{(\tau,s,i,m)})=A(\tau,s,i,m)\,.
\end{equation}
Note that in many cases some of the $T^A$s are identical to each
other and should be thought of as the same field. As discussed in
Section~\ref{manifolds}, this comes about from requiring some of the
radii $R^A$ to be equal for a consistent orbifolding of the base
torus. The number of distinct $T^A$ is $b(\Gamma)$, which for the
orbifold groups $\Gamma$ constructed in Section~\ref{groups} is
given in Table~\ref{tab:01}. To discover which $T^A$ are equal the
procedure is as follows. Let $x^A$ be coordinates in the bulk with
respect to which the $G_2$ structure is given by \eqref{structurex}.
Then a generator $\alpha_\tau$ of $\Gamma$ acts by simultaneous
rotations in two planes $(A,B)$ and $(C,D)$ say. If the order of
$\alpha_\tau$ is greater than two, then we identify $R^A$ with $R^B$
and $R^C$ with $R^D$.  We go through this process for all generators
of $\Gamma$, and then use \eqref{periodx} to determine which of the
$a^A$, and hence $T^A$, are equal.  From Eq.~\eqref{Kformula1} we
find for the K\"ahler potential
\begin{equation} \label{K}
K =  -\frac{1}{\kappa_4^2}\sum_{A=1}^{7}\ln
(T^A+\bar{T}^A)+\frac{2}{c_\Gamma\kappa_4^2}\sum_{\tau,s,m}\frac{1}{N_\tau}\frac{\sum_{i\leq
j}\left(\sum_{k=i}^{j}(U^{(\tau,s,k,m)}+\bar{U}^{(\tau,s,k,m)})\right)^2}{(T^{A(\tau,m)}+\bar{T}^{A(\tau,m)})(T^{B(\tau,m)}+\bar{T}^{B(\tau,m)})}
+ \frac{7}{\kappa_4^2}\ln 2\,.
\end{equation}
The index functions $A(\tau,m)$, $B(\tau,m)\in\{1,\ldots,7\}$
indicate by which two of the seven bulk moduli $T^A$ the blow-up
moduli $U^{(\tau,s,i,m)}$ are divided in the K\"ahler potential
\eqref{K}. Their values depend only on the generator index $\tau$
and the orientation index $m$. They may be calculated from the
formula
\begin{equation}
a^{A(\tau,m)}a^{B(\tau,m)}=\frac{\left(R^m_{(\tau)}\right)^2\prod_AR^A}{\prod_nR^n_{(\tau)}}\,.
\end{equation}
The $\tau$ dependence is only through the fixed directions of the
generator $\alpha_\tau$ and the possible values of the index
functions are given in Table~\ref{tab:two}.
\begin{table}
\begin{center}
\begin{tabular}{|c|c|c|c|}
\hline
$\mathrm{Fixed \, directions\, of\, } \alpha_\tau$ &$ m=1$ & $m=2$ &$ m=3$ \\
\hline
(1,2,3) & (2,3) & (4,5) & (6,7) \\
\hline
(1,4,5) & (1,3) & (4,6) & (5,7) \\
\hline
(1,6,7) & (1,2) & (4,7) & (5,6) \\
\hline
(2,4,6) & (1,5) & (2,6) & (3,7) \\
\hline
(2,5,7) & (1,4) & (2,7) & (3,6) \\
\hline
(3,4,7) & (1,7) & (2,4) & (3,5) \\
\hline
(3,5,6) & (1,6) & (2,5) & (3,4) \\
\hline
\end{tabular}
\caption{Values of the index functions $(A(\tau,m),B(\tau,m))$
specifying the bulk moduli $T^A$ by which the blow-up moduli
$U^{(\tau,s,i,m)}$ are divided in the K\"ahler potential.}
\label{tab:two}
\end{center}
\end{table}

We now state precisely and systematically the scenarios in which
\eqref{K} is valid. Firstly, all moduli must be larger than one (in
units where the Planck length is set to one) so that the
supergravity approximation to M-theory is valid. Secondly, all
blow-up moduli $U^{(\tau,s,i,m)}$ must be small compared to the bulk
moduli $T^A$ so that corrections of higher order in $U/T$ can be
neglected. The action of the generators of the orbifold group
$\Gamma$ on the base seven-torus $\mathcal{T}^7$ must lead to an
orbifold with co-dimension four singularities. Furthermore, no two
generators must fix the same point on the torus. This last
requirement is to ensure that the assumption of the structure
$\mathcal{T}^3\times\mathbb{C}^2/\mathbb{Z}_N$ around the
singularities is correct.
\\

Equation \eqref{K} gives the moduli K\"ahler potential for a large
class of compact manifolds of holonomy $G_2$. This class contains
manifolds constructed from a large number of different orbifolds,
based on at least sixteen distinct orbifold groups, namely those
constructed in Section~\ref{groups}. Moduli fields fall into two
categories. Firstly the fields $T^A$, which descend from the seven
radii of the manifold, and secondly the fields $U^{(\tau,s,i,m)}$
which descend from geometrical parameters describing the blow-up of
singularities of the manifold (namely the radii of two-cycles on the
blow-up and their orientation with respect to the bulk). Our formula
constitutes the first two terms in an expansion of the K\"ahler
potential in terms of the $U$s. The zeroth order term is simply a
consequence of the volume of the manifold being proportional to the
product of the seven radii. It is not surprising that the lowest
order correction terms arise at second order in the $U$s.
Heuristically, one can think of all $U$-dependent terms as being
associated with the volume subtracted from the manifold as a result
of the presence of two-cycles on the blow-ups. One expects these
terms to depend on the two-cycles through their area, and hence on
even powers of the $U$s. It had to be the case that second-order
terms in the $U$s are homogeneous of order minus two in the $T$s,
since the K\"ahler potential must be dimensionless, but it did not
have to necessarily turn out that the terms took on so simple a form
in the general case. This is an attractive feature of our result.

\chapter{M-theory on the Orbifold $\mathbb{C}^2/\mathbb{Z}_N$} \label{117}
\fancyhf{} \fancyhead[L]{\sl 4.~M-theory on the Orbifold
$\mathbb{C}^2/\mathbb{Z}_N$} \fancyhead[R]{\rm\thepage}
\renewcommand{\headrulewidth}{0.3pt}
\renewcommand{\footrulewidth}{0pt}
\addtolength{\headheight}{3pt} \fancyfoot[C]{} In the previous
chapter, we have discussed M-theory on a smooth manifold of $G_2$
holonomy. Our formula for the K\"ahler potential is useful for many
applications, for example in the context of moduli stabilisation. It
is however singular limits of the smooth $G_2$ manifolds which are
ultimately going to be interesting for phenomenology. That is, the
limit in which some or all of the blow-up moduli are shrunk down to
zero. With this in mind, the following two chapters of this thesis
are concerned with M-theory on spaces with co-dimension four $ADE$
singularities. Although much discussion has taken place on this
subject, prior to the work presented in this thesis, the explicit
supergravity action, valid for M-theory at low-energies, had not
been computed. The construction of this forms the basis of this
chapter.

The structure of a $G_2$ space close to a co-dimension four $ADE$
singularity is of the form $\mathcal{B}\times \mathbb{C}^2/H$, where
$H$ is one of the discrete $ADE$ subgroups of $\mathrm{SU}(2)$ and
$\mathcal{B}$ is a three-dimensional space. We will, for simplicity
focus on $A$-type singularities, that is $H=\mathbb{Z}_N$, which
lead to gauge fields with gauge group $\mathrm{SU}(N)$. This is in
keeping with the work in the previous chapter in which we considered
the blow-up of $A$-type singularities. Within the class of examples
of $G_2$ orbifolds constructed in Chapter~\ref{Classification}, the
possible values of $N$ are 2, 3, 4 and 6. However, in this chapter,
we keep $N$ general, given that there may be other constructions
which lead to more general values. In M-theory, the gauge fields are
located at the fixed point of $\mathbb{C}^2/\mathbb{Z}_N$ (the
origin of $\mathbb{C}^2$) times $\mathbb{R}^{1,3}\times\mathcal{B}$,
where $\mathbb{R}^{1,3}$ is the four-dimensional uncompactified
spacetime, and are, hence, seven-dimensional in nature. One would,
therefore, expect there to exist a supersymmetric theory which
couples 11-dimensional supergravity on the orbifold
$\mathbb{C}^2/\mathbb{Z}_N$ to seven-dimensional super-Yang-Mills
theory.

Although motivated by the application to $G_2$ compactifications, we
formulate this problem in a slightly more general context, seeking
to understand the general structure of low-energy M-theory on
orbifolds of $ADE$ type. Concretely, we will construct
11-dimensional supergravity on the general orbifold
$\mathcal{M}_{1,10}^N=\mathcal{M}_{1,6}\times
\mathbb{C}^2/\mathbb{Z}_N$ coupled to seven-dimensional $\rm{SU}(N)$
super-Yang-Mills theory located on the orbifold fixed plane
$\mathcal{M}_{1,6}\times \{\bf{0}\}$. For ease of terminology, we
will also refer to this orbifold plane, somewhat loosely as the
``brane''. This result can then be applied (see
Chapter~\ref{G2sing}) to compactifications of M-theory on $G_2$
spaces with $\mathbb{C}^2/\mathbb{Z}_N$ singularities, as well as to
other problems, (for example M-theory on certain singular limits of
K3). We stress that this construction is very much in the spirit of
Ho\v rava-Witten theory, which we reviewed in Chapter \ref{revchap}.

Let us briefly outline our method to construct this theory, which
relies on combining information from the known actions of
11-dimensional~\cite{Julia} and seven-dimensional
supergravity~\cite{Berghshoeff},~\cite{Park}. Firstly, we constrain
the field content of 11-dimensional supergravity (the ``bulk
fields'') to be compatible with the $\mathbb{Z}_N$ orbifolding
symmetry. We will call the Lagrangian for this constrained version
of 11-dimensional supergravity ${\cal L}_{11}$. As a second step
this action is truncated to seven dimensions, by requiring all
fields to be independent of the coordinates $y$ of the orbifold
$\mathbb{C}^2/\mathbb{Z}_N$ (or, equivalently, constraining it to
the orbifold plane at $y=0$). The resulting Lagrangian, which we
call ${\cal L}_{11}|_{y=0}$, is invariant under half of the original
32 supersymmetries and represents a seven-dimensional ${\cal N}=1$
supergravity theory which turns out to be coupled to a single
$\mathrm{U}(1)$ vector multiplet for $N>2$ or three $\mathrm{U}(1)$
vector multiplets for $N=2$. As a useful by-product, we obtain an
explicit identification of the (truncated) 11-dimensional bulk
fields with the standard fields of seven-dimensional
Einstein-Yang-Mills (EYM) supergravity. We know that the additional
states on the orbifold fixed plane should form a seven-dimensional
vector multiplet with gauge group ${\rm SU}(N)$.  In a third step,
we couple these additional states to the truncated seven-dimensional
bulk theory ${\cal L}_{11}|_{y=0}$ to obtain a seven-dimensional EYM
supergravity ${\cal L}_{\mathrm{SU}(N)}$ with gauge group
$\mathrm{U}(1)\times \mathrm{SU}(N)$ for $N>2$ or
$\mathrm{U}(1)^3\times \mathrm{SU}(N)$ for $N=2$. We note that,
given a fixed gauge group, the structure of ${\cal
L}_{\mathrm{SU}(N)}$ is essentially determined by seven-dimensional
supergravity. We further write this theory in a form which
explicitly separates the bulk degrees of freedom (which we have
identified with 11-dimensional fields) from the degrees of freedom
in the $\mathrm{SU}(N)$ vector multiplets. Given this preparation,
we prove in general that the action
\begin{equation}
\mathcal{S}_{11-7}=\int_{\mathcal{M}_{1,10}^N} \mathrm{d}^{11}x \left[ \mathcal{L}_{11} + \delta^{(4)}
(y^A)\left( \mathcal{L}_{\mathrm{\mathrm{SU}(N)}}-\kappa_{11}^{8/9}\mathcal{L}_{11}\right) \right]
\label{S117}
\end{equation}
is supersymmetric to leading non-trivial order in an expansion in
$\kappa_{11}$, the 11-dimensional Newton constant. Inserting the various
Lagrangians with the appropriate field identifications into this
expression then provides us with the final result.

The plan of this chapter is as follows. In Section~\ref{Bulk} we
discuss the constraints that arise on the fields of 11-dimensional
supergravity from putting this theory on the orbifold. We also lay
out our conventions for rewriting 11-dimensional fields according to
a seven plus four split of the coordinates. In
Section~\ref{reduction4} we examine our bulk Lagrangian constrained
to the orbifold plane and cast it in standard seven-dimensional
form. The proof that the action~\eqref{S117} is indeed
supersymmetric to leading non-trivial order is presented in
Section~\ref{construct}. Finally, in Section~\ref{results} we
present and discuss the explicit result for the coupled
11-/7-dimensional action and the associated supersymmetry
transformations. Some of the technical background material is placed
in the appendix. Appendix~\ref{spinors} covers our conventions for
spinors in eleven, seven and four dimensions, and describes how we
decompose 11-dimensional spinors. There are also some useful spinor
identities. In Appendix~\ref{Pauli} we have collected some useful
group-theoretical information related to the cosets
$SO(3,M)/SO(3)\times SO(M)$ of $d=7$ EYM supergravity which are used
in the reduction of the bulk theory to seven-dimensions. Finally,
Appendix~\ref{bigEYM} is a self-contained introduction to EYM
supergravity in seven dimensions.


\section{Eleven-dimensional supergravity on the orbifold} \label{Bulk}
\fancyhf{} \fancyhead[L]{\sl \rightmark} \fancyhead[R]{\rm\thepage}
\renewcommand{\headrulewidth}{0.3pt}
\renewcommand{\footrulewidth}{0pt}
\addtolength{\headheight}{3pt} \fancyfoot[C]{} In this section we
begin our discussion of M-theory on
$\mathcal{M}^N_{1,10}=\mathcal{M}_{1,6}\times\mathbb{C}^2/\mathbb{Z}_N$
by describing the bulk theory. We first lay out our conventions for
this chapter, and describe the decomposition of spinors in a four
plus seven split of the coordinates. Then we recall that fields
propagating on orbifolds are subject to certain constraints on their
configurations, and we list and explain the constraints that arise
in this instance.
\\

We take spacetime to have mostly positive signature, that is
$(-+\ldots+)$, and use indices $M,N,\ldots=0,1,\ldots,10$ to label
the 11-dimensional coordinates $(x^M)$. It is often convenient to
split these into four coordinates $y^{A}$, where
$A,B,\ldots=1,\ldots,4$,
in the directions of the orbifold $\mathbb{C}^2/\mathbb{Z}_N$ and seven
remaining coordinates $x^\mu$, where $\mu,\nu,\ldots=0,1,\ldots,6$,
on $\mathcal{M}_{1,6}$.  Frequently, we will also use complex
coordinates $(z^p,\bar{z}^{\bar{p}})$ on $\mathbb{C}^2/\mathbb{Z}_N$,
where $p,q,\ldots=1,2$, and $\bar{p},\bar{q},\ldots=\bar{1},\B{2}$
label holomorphic and anti-holomorphic coordinates, respectively.
The complex structure is chosen so that, explicitly,
\begin{equation}
z^1=\frac{1}{\sqrt{2}}(y^1+iy^2)\,,\hspace{0.3cm}
z^2=\frac{1}{\sqrt{2}}(y^3+iy^4)\,.
\end{equation}
Underlined versions of all the above index types denote the associated
tangent space indices.

All 11-dimensional spinors in this thesis are Majorana. Having
split coordinates into four- and seven-dimensional parts it is useful
to decompose 11-dimensional Majorana spinors accordingly as tensor
products of $\mathrm{SO}(1,6)$ and $\mathrm{SO}(4)$ spinors. To this end, we introduce a
basis of left-handed $\mathrm{SO}(4)$ spinors $\{\rho^i\}$ and their
right-handed counterparts $\{\rho^{\bar{\jmath}}\}$ with indices
$i,j,\ldots =1,2$ and $\bar{\imath},\bar{\jmath},\ldots
=\bar{1},\bar{2}$. Up to an overall rescaling, this basis can be
defined by the relations $\gamma^{\underline{A}}\rho^i =
\left(\gamma^{\underline{A}}\right)_{\B{\jmath}}^{\ph{j}i}\rho^{\B{\jmath}}$.
An 11-dimensional spinor $\psi$ can then be written as
\begin{equation}
\psi=\psi_i (x,y)\otimes\rho^i + \psi_{\bar{\jmath}}(x,y)\otimes\rho^{\bar{\jmath}}\,,
 \label{spinor7}
\end{equation}
where the 11-dimensional Majorana condition on $\psi$ forces
$\psi_i(x,y)$ and $\psi_{\bar{\jmath}}(x,y)$ to be
$\mathrm{SO}(1,6)$ symplectic Majorana spinors. In the following,
for any 11-dimensional Majorana spinor we will denote its associated
seven-dimensional symplectic Majorana spinors by the same symbol
with additional $i$ and $\bar{\imath}$ indices. A full account of
spinor conventions used in this chapter, together with a derivation
of the above decomposition
can be found in Appendix~\ref{spinors}.\\

With our 11-dimensional conventions the same as in Chapter
\ref{revchap}, we can refer the reader back there for the action
\eqref{rev11dsugra4} and supersymmetry transformations
\eqref{rev11dsusy1} of 11-dimensional supergravity. Here, let us
just present, once more, the field content. This consists of the
vielbein ${\hat{e}_M}^{\underline{M}}$ and associated metric
$\hat{g}_{MN}=\eta_{\underline{M}\underline{N}}
\hat{e}_M^{\phantom{M}\underline{M}}
\hat{e}_N^{\phantom{N}\underline{N}}$, the three-form field $C$,
with field strength $G=\mathrm{d}C$, and the gravitino $\Psi_M$. In
order for 11-dimensional supergravity to be consistent on the
orbifold $\mathcal{M}_{1,6}\times\mathbb{C}^2/\mathbb{Z}_N$ we need
to constrain these fields in accordance with the $\mathbb{Z}_N$
orbifold action. Let us discuss in detail how this works.

We denote by $R$ the $\mathrm{SO}(4)$ matrix of order $N$ that
generates the $\mathbb{Z}_N$ symmetry on our orbifold. This matrix
acts on the 11-dimensional coordinates as $(x,y)\mapsto (x,Ry)$,
which implies the existence of a seven-dimensional fixed plane
characterised by $\{ y=0\}$. For a field $X$ to be well-defined on
the orbifold it must satisfy
\begin{equation} \label{theta}
X(x,Ry)=\Theta(R)X(x,y)
\end{equation}
for some linear operator $\Theta(R)$ that represents the generator
of $\mathbb{Z}_N$. In constructing our theory we have to choose, for
each field, a representation $\Theta$ of $\mathbb{Z}_N$ for which we
wish to impose this constraint. For the theory to be well-defined,
these choices of representations must be such that the action
\eqref{rev11dsugra4} is invariant under the $\mathbb{Z}_N$ orbifold
symmetry. Concretely, what we do is choose how each index type
transforms under $\mathbb{Z}_N$. We take $R\equiv (R^{A}_{\ph{A}B})$
to be the transformation matrix acting on curved four-dimensional
indices $A,B,\ldots$ while the generator acting on tangent space
indices $\underline{A},\underline{B},\ldots$ is some other
$\mathrm{SO}(4)$ matrix
$T^{\underline{A}}_{\phantom{A}\underline{B}}$. It turns out that
this matrix must be of order $N$ for the four-dimensional components
of the vielbein to remain non-singular at the orbifold fixed plane.
Seven-dimensional indices $\mu ,\nu ,\ldots$ transform trivially.
Following the correspondence Eq.~\eqref{spinor7}, 11-dimensional
Majorana spinors $\psi$ are described by two pairs $\psi_i$ and
$\psi_{\bar{\imath}}$ of seven-dimensional symplectic Majorana
spinors.  We should, therefore, specify how the $\mathbb{Z}_N$
symmetry acts on indices of type $i$ and $\bar{\imath}$.
Supersymmetry requires that at least some spinorial degrees of
freedom survive at the orbifold fixed plane. For this to be the
case, one of these type of indices, $i$ say, must transform
trivially. Invariance of fermionic terms in the action
\eqref{rev11dsugra4} requires that the other indices, that is those
of type $\bar{\imath}$, be acted upon by a $\mathrm{U}(2)$ matrix
$S_{\bar{\imath}}^{\phantom{i}\bar{\jmath}}$ that satisfies the
equation
\begin{equation} \label{STconstraint}
{S_{\bar{\imath}}}^{\bar{k}}\left(\gamma^{\underline{A}}\right)_{\B{k}}^{\ph{k}j}=
{T^{\ul{A}}}_{\underline{B}}
{\left(\gamma^{\underline{B}}\right)_{\B{\imath}}}^{j}\,.
\end{equation}
Given this basic structure, the constraints satisfied by the fields are as follows
\begin{eqnarray}
\hat{e}_\mu^{\ph{\mu}\underline{\nu}}(x,Ry)&=&
\hat{e}_\mu^{\ph{\mu}\underline{\nu}}(x,y)\,, \label{cond1} \\
\hat{e}_{A}^{\ph{A}\underline{\nu}}(x,Ry)&=&
(R^{-1})_{A}^{\phantom{A}B}
\hat{e}_{B}^{\ph{B}\underline{\nu}}(x,y)\,,\\
\hat{e}_\mu^{\ph{\mu}\underline{A}}(x,Ry)&=&T^{\underline{A}}_{\phantom{A}\underline{B}}
\hat{e}_\mu^{\ph{\mu}\underline{B}}(x,y)\,,\\
\hat{e}_{A}^{\ph{A}\underline{B}}(x,Ry)&=&
(R^{-1})_{A}^{\phantom{A}C}
T^{\underline{B}}_{\phantom{B}\underline{D}}
\hat{e}_{C}^{\ph{C}\underline{D}}(x,y)\,, \label{cond4} \\
C_{\mu\nu\rho}(x,Ry) & = & C_{\mu\nu\rho}(x,y)\,, \\ \label{Ccond}
C_{\mu\nu A}(x,Ry) & = &
(R^{-1})_{A}^{\phantom{A}B}C_{\mu\nu B}(x,y)\,, \: \:\: \mathrm{etc.} \label{cond6}\\
\Psi_{\mu i}(x,Ry)&=&\Psi_{\mu i}(x,y)\,, \\ \Psi_{\mu
\bar{\imath}}(x,Ry)&=&S_{\bar{\imath}}^{\phantom{i}\bar{\jmath}}\Psi_{\mu
\bar{\jmath}}(x,y)\,,\\
\Psi_{Ai}(x,Ry)&=&(R^{-1})_{A}^{\phantom{A}B}\Psi_{B i}(x,y)\,,\\
\Psi_{A\bar{\imath}}(x,Ry)&=&(R^{-1})_{A}^{\phantom{A}B}
S_{\bar{\imath}}^{\phantom{i}\bar{\jmath}}\Psi_{B
\bar{\jmath}}(x,y)\,. \label{condn}
\end{eqnarray}
Furthermore, covariance of the supersymmetry transformation laws with
respect to $\mathbb{Z}_N$ requires
\begin{eqnarray}
 \eta_{i}(x,Ry)&=&\eta_{i}(x,y)\,, \\
\eta_{\bar{\imath}}(x,Ry)&=&S_{\bar{\imath}}^{\phantom{i}\bar{\jmath}}\eta_{\bar{\jmath}}(x,y)\,.
\end{eqnarray}
It is convenient to make the choice
\begin{equation}
({R^{A}}_{B})=\left( \begin{array}{cccc}
\cos(2\pi/N) & \sin(2\pi/N) & 0 & 0 \\
-\sin(2\pi/N) & \cos(2\pi/N) & 0 & 0 \\
0 & 0 & \cos(2\pi/N) & \sin(2\pi/N) \\
0 & 0 & -\sin(2\pi/N) & \cos(2\pi/N)
\end{array} \right)\, ,
\label{R}
\end{equation}
which, in complex coordinates $(z^p,\bar{z}^{\bar{p}})$, leads to
\begin{equation} \label{R4}
(R^p_{\ph{p}q})=e^{-2i\pi /N}\boldsymbol{1}_2\,, \hspace{0.3cm}
(R^{\B{p}}_{\ph{p}\B{q}})= e^{2i\pi /N}\boldsymbol{1}_2\,,
\hspace{0.3cm} (R^{\B{p}}_{\ph{p}q})=(R^p_{\ph{p}\B{q}})=0\, .
\end{equation}
Using this representation, the constraint~\eqref{cond4} implies
\begin{equation}
 {\hat{e}_p}^{\ul{A}}=e^{2i\pi /N}{T^{\ul{A}}}_{\ul{B}}\, {\hat{e}_p}^{\ul{B}}\, .
\end{equation}
Hence, for the vierbein ${\hat{e}_{\ul{A}}}^{\ul{B}}$ to be
non-singular $T$ must have two eigenvalues $e^{-2i\pi /N}$.
Similarly, the conjugate of the above equation shows that $T$ should
have two eigenvalues $e^{2i\pi /N}$. Therefore, in an appropriate
basis, we can use the representation
\begin{equation}
 ({T^{\ul{p}}}_{\ul{q}})=e^{-2i\pi /N}{\bf 1}_2\, ,\qquad
 ({T^{\ul{\bar p}}}_{\ul{\bar q}})=e^{2i\pi /N}{\bf 1}_2\, .\label{T}
\end{equation}
Given these representations for $R$ and $T$, the matrix $S$ is uniquely fixed
by Eq.~\eqref{STconstraint} to be
\begin{equation}
 ({S_{\B{\imath}}}^{\B{\jmath}})=e^{-2i\pi /N}{\bf 1}_2\, .\label{S}
\end{equation}
We will use the explicit form of $R$, $T$ and $S$ above to analyse the
degrees of freedom when we truncate fields to be $y$ independent.

When the 11-dimensional fields are taken to be independent of the
orbifold $y$ coordinates, the constraints~\eqref{cond1}--\eqref{condn}
turn into projection conditions, which force certain field components
to vanish. As we will see shortly, the surviving field components fit
into seven-dimensional ${\cal N}=1$ supermultiplets, a confirmation
that we have chosen the orbifold $\mathbb{Z}_N$ action on fields
compatible with supersymmetry. More precisely, for the case $N>2$, we
will find a seven-dimensional gravity multiplet and a single
$\mathrm{U}(1)$ vector multiplet. Hence, we expect the associated
seven-dimensional ${\cal N}=1$ Einstein-Yang-Mills (EYM) supergravity
to have gauge group $\mathrm{U}(1)$.  For $\mathbb{Z}_2$ the situation is
slightly more complicated, since, unlike for $N>2$, some of the field
components which transform bi-linearly under the generators are now
invariant.  This leads to two additional vector multiplets, so that
the associated theory is a seven-dimensional ${\cal N}=1$ EYM
supergravity with gauge group $\mathrm{U}(1)^3$.  In the following section, we
will write down this seven-dimensional theory, both for $N=2$ and
$N>2$, and find the explicit identifications of truncated
11-dimensional fields with standard seven-dimensional supergravity
fields.


\section{Truncating the bulk theory to seven dimensions} \label{reduction4}
In this section, we describe in detail how the bulk theory is
truncated to seven dimensions. We recall from the beginning of the
chapter that this constitutes one of the essential steps in the
construction of the theory. As a preparation, we explicitly write
down the components of the 11-dimensional fields that survive on the
orbifold plane and work out how these fit into seven-dimensional
super-multiplets. We then describe, for each orbifold, the
seven-dimensional EYM supergravity with the appropriate field
content. By an explicit reduction of the 11-dimensional theory and
comparison with this seven-dimensional theory, we find a list of
identifications between 11- and 7-dimensional fields which is
essential for our subsequent construction.\\

To discuss the truncated field content, we use the
representations~\eqref{R4}, \eqref{T}, \eqref{S} of $R$, $T$ and $S$
and the orbifold conditions \eqref{cond1}--\eqref{condn} for $y$
independent fields. For $N>2$ we find that the
surviving components are given by $\hat{g}_{\mu\nu}$,
$\hat{e}_p^{\ph{p}\underline{q}}$, $C_{\mu\nu\rho}$, $C_{\mu p\B{q}}$,
$\Psi_{\mu i}$, $(\Gamma^p\Psi_p)_i$ and
$(\Gamma^{\B{p}}\Psi_{\B{p}})_i$. Meanwhile, the spinor $\eta$ which
parameterises supersymmetry reduces to $\eta_i$, a single symplectic
Majorana spinor, which corresponds to seven-dimensional ${\cal N}=1$
supersymmetry. Comparing with the structure of seven-dimensional
multiplets (see Appendix~\ref{bigEYM} for a review of
seven-dimensional EYM supergravity), these field components fill out
the seven-dimensional supergravity multiplet and a single $\mathrm{U}(1)$
vector multiplet.  For the case of the $\mathbb{Z}_2$ orbifold, a
greater field content survives, corresponding in seven-dimensions to a
gravity multiplet plus three $\mathrm{U}(1)$ vector multiplets. The surviving
fields in this case are expressed most succinctly by $\hat{g}_{\mu\nu}$,
$\hat{e}_{A}^{\ph{A}\underline{B}}$, $C_{\mu\nu\rho}$, $C_{\mu AB}$,
$\Psi_{\mu i}$ and $\Psi_{A\B{\imath}}$. The spinor $\eta$ which
parameterises supersymmetry again reduces to $\eta_i$, a single
symplectic Majorana spinor.

These results imply that the truncated bulk theory is a
seven-dimensional $\mathcal{N}=1$ EYM supergravity with gauge group
$\mathrm{U}(1)^n$, where $n=1$ for $N>2$ and $n=3$ for $N=2$. In the
following, we discuss both cases and, wherever possible, use a
unified description in terms of $n$, which can be set to either $1$
or $3$, as appropriate.  The correspondence between 11-dimensional
truncated fields and seven-dimensional supermultiplets is as
follows. The gravity super-multiplet contains the purely
seven-dimensional parts of the 11-dimensional metric, gravitino and
three-form, that is, $\hat{g}_{\mu\nu}$, $\Psi_{\mu i}$ and
$C_{\mu\nu\rho}$, along with three vectors from $C_{\mu AB}$, a
spinor from $\Psi_{A\B{\imath}}$ and the scalar
$\mathrm{det}(\hat{e}_{A}^{\ph{A}\underline{B}})$. The remaining
degrees of freedom, that is, the remaining vector(s) from $C_{\mu
AB}$, the remaining spinor(s) from $\Psi_{A\B{\imath}}$ and the
scalars contained in $v_{A}^{\ph{A}\underline{B}}:=
\mathrm{det}(\hat{e}_{A}^{\ph{A}\underline{B}})^{-1/4}\hat{e}_{A}^{\ph{A}\underline{B}}$,
the unit-determinant part of $\hat{e}_{A}^{\ph{A}\underline{B}}$,
fill out $n$ seven-dimensional vector multiplets. The $n+3$ Abelian
gauge fields transform under the $\mathrm{SO}(3,n)$ global symmetry
of the $d=7$ EYM supergravity, while the vector multiplet scalars
parameterise the coset $\mathrm{SO}(3,n)/\mathrm{SO}(3)\times
\mathrm{SO}(n)$. Let us describe how such coset spaces are obtained
from the vierbein $v_{A}^{\ph{A}\underline{B}}$, starting with the
generic case $N>2$ with seven-dimensional gauge group
$\mathrm{U}(1)$, that is, $n=1$. In this case, the rescaled vierbein
$v_{A}^{\ph{A}\underline{B}}$ reduces to
$v_p^{\ph{p}\underline{q}}$, which represents a set of $2\times 2$
matrices with determinant one, identified by $\mathrm{SU}(2)$
transformations acting on the tangent space index. Hence, these
matrices form the coset $\mathrm{SL}(2,\mathbb{C})/\mathrm{SU}(2)$,
which is isomorphic to $\mathrm{SO}(3,1)/\mathrm{SO}(3)$, the
correct coset space for $n=1$. For the special $\mathbb{Z}_2$ case,
which implies $n=3$, the whole of $v_{A}^{\ph{A}\underline{B}}$ is
present and forms the coset space
$\mathrm{SL}(4,\mathbb{R})/\mathrm{SO}(4)$. This space is isomorphic
to $\mathrm{SO}(3,3)/\mathrm{SO}(3)^2$, which is indeed the correct
coset space for $n=3$.

We now briefly review seven-dimensional EYM supergravity with
gauge group $\mathrm{U}(1)^n$. A more general account of seven-dimensional
supergravity including non-Abelian gauge groups can be found in
Appendix \ref{bigEYM}. The seven-dimensional ${\cal N}=1$ supergravity multiplet
contains the vielbein $w_{\mu}^{\ph{\mu}\underline{\nu}}$ with associated
metric $h_{\mu\nu}$, the
gravitino $\psi_{\mu i}$, a triplet of vectors $A_{\mu i}^{\ph{\mu
i}j}$ with field strengths $F_i^{\ph{i}j}=\mathrm{d}A_i^{\ph{i}j}$, a
three-form $C$ with field strength $G=\mathrm{d}C$, a
spinor $\chi_i$, and a scalar $\si$. A seven-dimensional
vector multiplet contains a $\mathrm{U}(1)$ gauge field $A_\mu$ with field
strength $F=\mathrm{d}A$, a gaugino $\la_i$ and a triplet of scalars
$\phi_i^{\ph{i}j}$. Here, all spinors are symplectic Majorana spinors
and indices $i,j,\ldots=1,2$ transform under the $\mathrm{SU}(2)$
R-symmetry. For ease of notation, the three vector fields in the
supergravity multiplet and the $n$ additional Abelian gauge fields from
the vector multiplet are combined into a single $\mathrm{SO}(3,n)$ vector
$A_\mu^I$, where $I,J,\ldots =1,\ldots ,(n+3)$. The coset space
$\mathrm{SO}(3,n)/\mathrm{SO}(3)\times \mathrm{SO}(n)$ is described by a $(3+n)\times (3+n)$ matrix
$\ell_{I}^{\ph{I}\underline{J}}$, which depends on the $3n$ vector
multiplet scalars and satisfies the $\mathrm{SO}(3,n)$ orthogonality condition
\begin{equation}
\ell_{I}^{\ph{I}\underline{J}}\ell_{K}^{\ph{K}\underline{L}}\eta_{\underline{J}\underline{L}}=\eta_{IK}
\end{equation}
with
$(\eta_{IJ})=(\eta_{\underline{I}\underline{J}})=\rm{diag}(-1,-1,-1,+1,\ldots,+1)$. Here,
indices $I,J,\ldots=1,\ldots,(n+3)$ transform under $\mathrm{SO}(3,n)$. Their
flat counterparts $\underline{I},\underline{J}\ldots$ decompose into a
triplet of $\mathrm{SU}(2)$, corresponding to the gravitational directions and
$n$ remaining directions corresponding to the vector multiplets. Thus
we can write $\ell_{I}^{\ph{I}\underline{J}}\to
(\ell_{I}^{\ph{I}u},\ell_I^{\ph{I}\alpha})$, where $u=1,2,3$ and
$\alpha=1,\ldots,n$. The adjoint $\mathrm{SU}(2)$ index $u$ can be converted
into a pair of fundamental $\mathrm{SU}(2)$ indices by multiplication with the
Pauli matrices, that is,
\begin{equation}
\ell_{I\ph{i}j}^{\ph{I}i}=\frac{1}{\sqrt{2}}
\ell_I^{\ph{I}u}(\si_u)^i_{\ph{i}j}\,.
\end{equation}
The Maurer-Cartan forms $p$ and $q$ of the matrix $\ell$, defined by
\ba
p_{\mu\alpha \ph{i}j}^{\ph{\mu\alpha }i}&=&\ell^{I}_{\ph{I}\alpha}
\partial_\mu \ell_{I\ph{i}j}^{\ph{\mu}i}\,, \label{Maurer1}\\
q_{\mu \ph{i}j\ph{k}l}^{\ph{\mu}i\ph{j}k}&=&\ell^{Ii}_{\ph{Ii}j}
 \partial_\mu \ell_{I\ph{k}l}^{\ph{\mu}k}\,, \\
q_{\mu \phantom{i}j}^{\phantom{\mu}i}&=&\ell^{Ii}_{\ph{Ii}k}
\partial_\mu \ell_{I\ph{k}j}^{\ph{\mu}k}\,, \label{Maurern}
\ea
will be needed as well.

With everything in place, we can now write down our expression for $\mathcal{L}_7^{(n)}$,
the Lagrangian of seven-dimensional ${\cal N}=1$ EYM supergravity with gauge group
$\mathrm{U}(1)^n$ \cite{Park}. Neglecting four-fermi terms, it is given by
\begin{eqnarray} \label{7dsugra}
\mathcal{L}_7^{(n)}\!\!\! &=&\!\!\!\frac{1}{\kappa^2_7}\sqrt{-h}\left\{\frac{1}{2}
R_{(h)}-\frac{1}{2}\bar{\psi}_{\mu
}^{i}\Upsilon ^{\mu \nu \rho }\mathcal{D} _{\nu }\psi _{\rho i}-\frac{1}{4}e^{-2\si}
\left( \ell_{I\phantom{i}j}^{\phantom{I}i}\ell_{%
J\phantom{j}i}^{\phantom{J}j}+\ell_{I}^{\ph{I}\alpha}\ell_{J\alpha}\right)
F_{\mu \nu }^{I}F^{J\mu \nu } \right.  \notag \\
&&\hspace{0.45cm}-\frac{1}{96}e^{4\si}G_{\mu \nu \rho \sigma }G^{\mu \nu \rho \sigma }
-\frac{1}{2}\bar{\chi}^{i}\Upsilon ^{\mu }\mathcal{D} _{\mu }\chi _{i}-\frac{5}{%
2}\partial _{\mu }\si \partial ^{\mu }\si -\frac{1}{2}p_{\mu\alpha \phantom{i}j}^{\phantom{\mu\alpha}i}p_{%
\phantom{\mu\alpha j}i}^{\mu\alpha j}
-\frac{1}{2}\bar{\lambda}^{\alpha i}\Upsilon ^{\mu }\mathcal{D}
_{\mu }\lambda _{\alpha i}
\notag \\
&&\hspace{0.45cm}+\frac{\sqrt{5}}{2}\left( \bar{\chi}^{i}
\Upsilon ^{\mu \nu }\psi _{\mu i}+\bar{\chi}^{i}\psi _{i}^{\nu }\right)
\partial _{\nu }\si -\frac{1}{\sqrt{2}}\left( \bar{\lambda}^{\alpha i}
\Upsilon ^{\mu \nu }\psi _{\mu
j}+\bar{\lambda}^{\alpha i}\psi _{j}^{\nu }\right) p_{\nu\alpha \phantom{j}i}^{%
\phantom{\nu\alpha}j}  \notag \\
&&\hspace{0.45cm}+e^{2\si}G_{\mu \nu \rho \sigma }\left[
\frac{1}{192}\left( 12\bar{\psi}^{\mu i}\Upsilon ^{\nu \rho }\psi _{i}^{\si}+ \bar{\psi}_{\la}^{i}
\Upsilon ^{\lambda \mu \nu \rho \sigma \tau }\psi _{\tau
i}\right)+\frac{1}{48\sqrt{5}}\left( 4\bar{\chi}^{i}\Upsilon
^{\mu \nu \rho }\psi _{i}^{\si} \right.\right.\nn \\
&& \hspace{2.5cm} \left. \left. -\bar{\chi}^{i}\Upsilon ^{\mu \nu
\rho \sigma \tau }\psi _{\tau i}\right)
-\frac{1}{320}\bar{\chi}^{i}\Upsilon ^{\mu \nu \rho \sigma }\chi
_{i}+\frac{1}{192}\bar{\lambda}^{\alpha i}\Upsilon ^{\mu \nu
\rho\sigma }
\lambda _{\alpha i}\right] \nn \\
&&\hspace{0.45cm} -ie^{-\si}F_{\mu \nu }^{I}\ell_{I%
\phantom{j}i}^{\phantom{I}j}\left[ \frac{1}{4\sqrt{2}}\left( \bar{\psi}%
_{\rho }^{i}\Upsilon ^{\mu \nu \rho \sigma }\psi _{\sigma j}+2\bar{\psi}%
^{\mu i}\psi _{j}^{\nu }\right) +\frac{1}{2\sqrt{10}}\left( \bar{\chi}%
^{i}\Upsilon ^{\mu \nu \rho }\psi _{\rho j}-2\bar{\chi}^{i}\Upsilon ^{\mu
}\psi _{j}^{\nu }\right) \right.  \notag \\
&&\hspace{3.0cm}\left. +\frac{3}{20\sqrt{2}}\bar{\chi}^{i}\Upsilon
^{\mu \nu }\chi _{j}-\frac{1}{4\sqrt{2}}\bar{\lambda}^{\alpha
i}\Upsilon ^{\mu \nu }\lambda
_{\alpha j}\right]  \notag \\
&&\hspace{0.45cm}+e^{-\si }F_{\mu \nu }^{I}\ell_{I\alpha}\left[
\frac{1}{4}\left( 2\bar{\lambda}^{\alpha i}\Upsilon ^{\mu }\psi _{i}^{\nu }
-\bar{\lambda}^{\alpha i}\Upsilon ^{\mu \nu \rho }\psi _{\rho i}\right)
 +\frac{1}{2\sqrt{5}}\bar{\lambda}^{\alpha i}\Upsilon ^{\mu \nu }\chi _{i}\right]  \notag \\
&&\hspace{0.45cm}\left. -\frac{1}{96}\epsilon^{\mu\nu\rho\sigma\kappa\lambda\tau}C_{\mu\nu\rho
}F_{\sigma\kappa}^{\tilde{I}}F_{\tilde{I}}{}_{\lambda\tau} \right\}\, .
\end{eqnarray}
In this Lagrangian the covariant derivatives of symplectic Majorana spinors $\epsilon _{i}$
are defined by
\begin{equation}
\mathcal{D} _{\mu }\epsilon _{i}=\partial _{\mu }\epsilon _{i}+\frac{1}{2}q_{\mu
i}^{\phantom{\mu i}j}\epsilon _{j}+\frac{1}{4}\omega _{\mu }^{\phantom{\mu}%
\underline{\mu }\underline{\nu }}\Upsilon _{\underline{\mu }\underline{\nu }%
}\epsilon _{i}\,.
\end{equation}
The associated supersymmetry transformations, parameterised by the spinor $\varepsilon_i$,
are, up to cubic fermion terms, given by
\ba
\delta \sigma &=& \frac{1}{\sqrt{5}}\bar{\chi}^{i}\varepsilon _{i}\, ,  \nn\\
\delta w _{\mu }^{%
\underline{\nu}}&=&\B{\varepsilon }^{i}\Upsilon ^{\underline{\nu}}\psi
_{\mu i }\, , \nn\\
\delta \psi _{\mu i}&=&2\mathcal{D} _{\mu }\varepsilon _{i}-\frac{e^{2\sigma}}{80}
\left(\Upsilon _{\mu }^{\ph{\mu} \nu \rho \sigma \eta }-\frac{8}{3}\delta _{\mu }^{\nu
}\Upsilon ^{\rho \sigma \eta }\right) \varepsilon _{i}G_{\nu \rho \sigma \eta
}+\frac{ie^{-\sigma}}{5\sqrt{2}}\left(
\Upsilon _{\mu }^{\ph{\mu}\nu \rho }-8\delta _{\mu }^{\nu }\Upsilon ^{\rho
}\right) \varepsilon _{j}F_{\nu \rho }^{I}\ell_{I\phantom{i}i}^{\phantom{I}j}\, , \nn\\
\delta \chi _{i}&=& \sqrt{5}\Upsilon ^{\mu }\varepsilon _{i}\partial _{\mu }\si -
\frac{1}{24\sqrt{5}}\Upsilon ^{\mu \upsilon \rho \sigma }\varepsilon
_{i}G_{\mu \nu \rho \sigma }e^{2\si}\text{ }-\frac{i}{\sqrt{10}}\Upsilon ^{\mu\nu }
\varepsilon _{j}F_{\mu\nu }^{I}\ell_{I\phantom{i}i}^{\phantom{I}j}e^{-\si}\,, \nn\\
\delta C_{\mu \nu \rho }&=&\left( -3\B{\psi }^{i}_{\left[ \mu \right. }\Upsilon _{\left.
 \nu \rho \right] }\varepsilon _{i}-\frac{2}{\sqrt{5}}\B{\chi }^{i}\Upsilon _{\mu \nu \rho }
\varepsilon_{i}\right) e^{-2\si}\,,  \\
\ell_{I\phantom{i}j}^{\phantom{I}i}\delta A_{\mu }^{I}&=&\left[ i\sqrt{2}\left(
\B{\psi }_{\mu }^{i}\varepsilon _{j}-\frac{1}{2}%
\delta _{j}^{i}\B{\psi }_{\mu }^{k}\varepsilon _{k}\right) -\frac{2i}{%
\sqrt{10}}\left( \B{\chi }^{i}\Upsilon _{\mu }\varepsilon _{j}-\frac{1%
}{2}{}\delta _{j}^{i}\B{\chi }^{k}\Upsilon _{\mu }\varepsilon
_{k}\right) \right] e^{\si}\,,  \nn\\
\ell_{I}^{\ph{I}\alpha}\delta A_{\mu }^{I}&=&\B{\varepsilon }%
^{i}\Upsilon _{\mu }\lambda _{i}^\alpha e^{\si}\,,  \nn\\
\delta \ell_{I\phantom{i}j}^{\phantom{I}i}&=&-i\sqrt{2}\B{\varepsilon }^{i}
\lambda _{\alpha j}\ell_{I}^{\ph{I}\alpha}+\frac{i}{\sqrt{2}}\B{\varepsilon }^{k}
\lambda _{\alpha k}\ell_{I}^{\ph{I}\alpha}\delta _{j}^{i}\, ,  \nn\\
\delta \ell_{I}^{\ph{I}\alpha}&=&-i\sqrt{2}\B{\varepsilon }^{i}\lambda_{j}^\alpha
\ell_{I\phantom{j}i}^{\phantom{I}j}\, , \nn\\
\delta \lambda _{i}^\alpha &=&-\frac{1}{2}\Upsilon ^{\mu\nu
}\varepsilon _{i}F_{\mu\nu}^{I}\ell_{I}^{\ph{I}\alpha}e^{-\si}+\sqrt{2}i
\Upsilon ^{\mu }\varepsilon _{j}p_{\mu \phantom{\alpha i}j}^{\phantom{\mu}\alpha i}\, . \nn
\ea
\\
We now explain in detail how the truncated  bulk theory corresponds
to the above seven-dimensional EYM supergravity with gauge group $\mathrm{U}(1)^n$,
where $n=1$ for the $\mathbb{Z}_N$ orbifold with $N>2$ and $n=3$ for the
special $\mathbb{Z}_2$ case. It is convenient to choose the seven-dimensional
Newton constant $\kappa_7$ as $\kappa_7=\kappa_{11}^{5/9}$. The correspondence
between 11- and 7-dimensional Lagrangians can then be written as
\begin{equation} \label{equivalence}
\kappa_{11} ^{8/9}\mathcal{L}_{11}\lvert_{y=0}=\mathcal{L}_7^{(n)}\, .
\end{equation}
We have verified by explicit computation that this relation indeed
holds for appropriate identifications of the truncated 11-dimensional fields
with the standard seven-dimensional fields which appear in Eq.~\eqref{7dsugra}.
The relation is trivial for the three-form $C_{\mu\nu\rho}$. For the other
fields we have, for the generic $\mathbb{Z}_N$ orbifold
with $N>2$ and $n=1$
\begin{eqnarray}
\si &=& \frac{3}{20}\ln\det \hat{g}_{AB}\,,  \label{4id1}\\
h_{\mu \nu } &=&e^{\frac{4}{3}\si }\hat{g}_{\mu \nu }\,, \label{Weyl}\\
\psi_{\mu i} & = & \Psi_{\mu i}e^{\frac{1}{3}\si}-\frac{1}{5}\Up_\mu
 \left(\Gamma^{A}\Psi_{A}\right)_i e^{-\frac{1}{3}\si}\,, \\
\chi_i & = & \frac{3}{2\sqrt{5}}\left(\Gamma^{A}\Psi_{A}\right)_ie^{-\frac{1}{3}\si}\,, \label{4idn}\\
F^I_{\mu\nu}&=&\frac{i}{2}\mathrm{tr}\left( \si^IG_{\mu\nu}\right)\,, \\
\la_i & =&  -\frac{i}{2}\left( \Gamma^p\Psi_p - \Gamma^{\B{p}}\Psi_{\B{p}}\right)_ie^{-\frac{1}{3}\si}\,, \\
\ell_I^{\ph{I}\underline{J}}&=&\frac{1}{2}\mathrm{tr}\left( \B{\si}_I v \si^J v^{\dagger} \right)\,.
 \label{idn2}
\ea
Furthermore the seven-dimensional supersymmetry spinor $\ve_i$ is related to
its 11-dimensional counterpart $\eta$ by
\begin{equation} \label{spinorrel}
\varepsilon_i=e^{\frac{1}{3}\si}\eta_i\,.
\end{equation}
In these identities, we have defined the matrices $G_{\mu\nu}\equiv
(G_{\mu\nu p\B{q}})$, $v\equiv
(e^{5\si/6}\hat{e}^{p}_{\ph{p}\underline{q}})$ and made use of the
standard $\mathrm{SO}(3,1)$ Pauli matrices $\si^I$, defined in Appendix
\ref{Pauli}.
For the $\mathbb{Z}_2$ orbifold we have $n=3$ and, therefore, two
additional $\mathrm{U}(1)$ vector multiplets. Not surprisingly, field identification in
the gravity multiplet sector is unchanged from the generic case and still
given by Eqs.~\eqref{4id1}--\eqref{4idn}. It
is in the vector multiplet sector, where the additional states
appear, that we have to make a distinction. For the bosonic
vector multiplet fields we find
\ba
F^I_{\mu\nu}&=&-\frac{1}{4}\mathrm{tr}\left( t^IG_{\mu\nu}\right)\,,
\label{idn3} \\
\ell_I^{\ph{I}\underline{J}}&=&\frac{1}{4}\mathrm{tr}\left( \B{t}_I v
t^J v^{T} \right)\,, \label{idn4} \ea where now $G_{\mu\nu}\equiv
(G_{\mu\nu AB})$ and $v\equiv
(e^{5\si/6}\hat{e}^{A}_{\ph{A}\underline{B}})$. Here, $t^I$
are the six $\mathrm{SO}(4)$
generators, which are explicitly given in Appendix~\ref{Pauli}.


\section{General form of the supersymmetric bulk-brane action} \label{construct}
In this section, we present our general method of construction for the
full action, which combines 11-dimensional supergravity with the
seven-dimensional super-Yang-Mills theory on the orbifold plane in a
supersymmetric way. Main players in this construction will be the
constrained 11-dimensional bulk theory ${\cal L}_{11}$, as discussed
in Section~\ref{Bulk}, its truncation to seven dimensions, ${\cal
L}_7^{(n)}$, which has been discussed in the previous section and
corresponds to a $d=7$ EYM supergravity with gauge group $\mathrm{U}(1)^n$,
and
${\cal L}_{\mathrm{SU}(N)}$, a $d=7$ EYM supergravity with gauge group
$\mathrm{U}(1)^n\times \mathrm{SU}(N)$. The $\mathrm{SU}(N)$ gauge group in the latter theory
corresponds, of course, to the additional $\mathrm{SU}(N)$ gauge multiplet
which one expects to arise for M-theory on the orbifold
$\mathbb{C}^2/\mathbb{Z}_N$.

Let us briefly recall, from Chapter \ref{revchap}, the physical
origin of these $\mathrm{SU}(N)$ gauge fields on the orbifold fixed
plane. It is well-known~\cite{Comp1},~\cite{Lukas} that the $N-1$
Abelian $\mathrm{U}(1)$ gauge fields within $\mathrm{SU}(N)$ are
already massless for a smooth blow-up of the orbifold
$\mathbb{C}^2/\mathbb{Z}_N$ by an asymptotically locally Euclidean
(ALE) manifold. More precisely, they arise as zero modes of the
M-theory three-form on the blow-up ALE manifold. The remaining
vector fields arise from membranes wrapping the two-cycles of the
ALE space and become massless only in the singular orbifold limit,
when these two-cycles collapse. For our current purposes, the only
relevant fact is that all these $\mathrm{SU}(N)$ vector fields are
located on the orbifold fixed plane. This allows us to treat the
Abelian and non-Abelian parts of $\mathrm{SU}(N)$ on the same
footing,
despite their different physical origin.\\

We claim that the action for the bulk-brane system is given by
\begin{equation} \label{action1}
\mathcal{S}_{11-7}=\int_{\mathcal{M}^N_{1,10}} \mathrm{d}^{11}x \left[
 \mathcal{L}_{11} + \delta^{(4)}(y^{A})\mathcal{L}_{\mathrm{brane}} \right]\,,
\end{equation}
where
\begin{equation} \label{braneformula}
\mathcal{L}_{\mathrm{brane}}=\mathcal{L}_{\mathrm{SU}(N)}-\mathcal{L}_7^{(n)}\,.
\end{equation}
Here, as before, $\mathcal{L}_{11}$ is the Lagrangian for
11-dimensional supergravity~\eqref{rev11dsugra4} with fields
constrained in accordance with the orbifold $\mathbb{Z}_N$ symmetry,
as discussed in Section~\ref{Bulk}. The Lagrangian
$\mathcal{L}_7^{(n)}$ describes a seven-dimensional ${\cal N}=1$ EYM
theory with gauge group $\mathrm{U}(1)^n$. Choosing $n=1$ for
generic $\mathbb{Z}_N$ with $N>2$ and $n=3$ for $\mathbb{Z}_2$, this
Lagrangian corresponds to the truncation of the bulk Lagrangian
$\mathcal{L}_{11}$ to seven dimensions, as we have shown in the
previous section. This correspondence implies identifications
between truncated 11-dimensional bulk fields and the fields in
$\mathcal{L}_7^{(n)}$, which have also been worked out explicitly in
the previous section (see Eqs.~\eqref{4id1}--\eqref{idn2} for the
case $N>2$ and Eqs.~\eqref{4id1}--\eqref{4idn} and
\eqref{idn3}--\eqref{idn4} for $N=2$).  These identifications are
also considered part of the definition of the
Lagrangian~\eqref{action1}. The new Lagrangian
$\mathcal{L}_{\mathrm{SU}(N)}$ is that of seven-dimensional EYM
supergravity with gauge group $\mathrm{U}(1)^n\times
\mathrm{SU}(N)$, where, as usual, $n=1$ for generic $\mathbb{Z}_N$
with $N>2$ and $n=3$ for $\mathbb{Z}_2$. This Lagrangian contains
the ``old'' states in the gravity multiplet and the
$\mathrm{U}(1)^n$ gauge multiplet and the ``new'' states in the
$\mathrm{SU}(N)$ gauge multiplet. We will think of the former states
as being identified with the truncated 11-dimensional bulk states by
precisely the same relations we have used for $\mathcal{L}_7^{(n)}$.
The idea of this construction is, of course, that in
$\mathcal{L}_{\rm brane}$ the pure supergravity and
$\mathrm{U}(1)^n$ vector multiplet parts cancel between ${\mathcal
L}_{\mathrm{SU}(N)}$  and $\mathcal{L}_7^{(n)}$, so that we remain
with ``pure'' $\mathrm{SU}(N)$ theory on the orbifold plane. For
this to work out, we have to choose the seven-dimensional Newton
constant $\kappa_7$ within $\mathcal{L}_{\mathrm{SU}(N)}$ to be the
same as the one in $\mathcal{L}_7^{(n)}$, that is
\begin{equation}
\kappa_7=\kappa_{11}^{5/9}\, .
\end{equation}
The supersymmetry transformation laws for the action~\eqref{action1}
are schematically given by
\ba \label{susy1}
\delta_{11}&=&\delta_{11}^{\ph{11}11}+\kappa^{8/9}\delta^{(4)}(y^{A})\delta_{11}^{\ph{11}\mathrm{brane}}\,, \\
\delta_7&=&\delta_7^{\ph{7}\mathrm{SU}(N)}, \label{susy2}
\ea
where
\begin{equation} \label{susy3}
\delta_{11}^{\ph{11}\mathrm{brane}}=\delta_{11}^{\ph{11}\mathrm{SU}(N)}-\delta_{11}^{\ph{11}11}\,.
\end{equation}
Here $\delta_{11}$ and $\delta_7$ denote the supersymmetry
transformation of bulk fields and fields on the orbifold fixed
plane, respectively. A superscript 11 indicates a supersymmetry
transformation law of $\mathcal{L}_{11}$, as given in equations
\eqref{rev11dsusy1}, and a superscript $\mathrm{SU}(N)$ indicates a
supersymmetry transformation law of $\mathcal{L}_{\mathrm{SU}(N)}$,
as can be found by substituting the appropriate gauge group into
equations \eqref{7dsusy}. These transformation laws are
parameterised by a single 11-dimensional spinor, with the
seven-dimensional spinors in $\delta_{11}^{\ph{11}\mathrm{SU}(N)}$
and $\delta_{7}^{\ph{7}\mathrm{SU}(N)}$ being simply related to this
11-dimensional spinor by equation \eqref{spinorrel}. On varying
$\mathcal{S}_{11-7}$ with respect to these supersymmetry
transformations we find \ba \delta
\mathcal{S}_{11-7}&=&-\int_{\mathcal{M}_{1,10}^N} \mathrm{d}^{11}x
\delta^{(4)}(y^{A}) \left( 1-\kappa^{8/9}\delta^{(4)}(y^{A}) \right)
\delta_{11}^{\ph{11}\mathrm{brane}} \mathcal{L}_{\mathrm{brane}}\,.
\label{susycalc} \ea At first glance, the occurrence of
delta-function squared terms is concerning. However, as in Ho\v
rava-Witten theory, we can interpret the occurrence of these terms
as a symptom of attempting to treat in classical supergravity what
really should be treated in quantum M-theory \cite{Hor-Wit}. It is
presumed that in proper quantum M-theory, fields on the brane
penetrate a finite thickness into the bulk, and that there would be
some kind of built-in cut-off allowing us to replace
$\delta^{(4)}(0)$ by a finite constant times $\kappa_{11}^{-8/9}$.
If we could set this constant to one and formally substitute
\begin{equation}
\delta^{(4)}(0)=\kappa_{11}^{-8/9}
\end{equation}
then the above integral would vanish.

As in Ref.~\cite{Hor-Wit}, we can avoid such a regularisation if we
work only to lowest non-trivial order in $\kappa_{11}$, or, more
precisely to lowest non-trivial order in the parameter
$\zeta_7=\kappa_7/\la_7$, where $\la_7$ is the Yang-Mills gauge
coupling. Note that $\zeta_7$ has dimension of inverse energy. To
determine the order in $\zeta_7$ of various terms in the Lagrangian
we need to fix a convention for the energy dimensions of the fields.
We assign energy dimension 0 to bulk bosonic fields and energy
dimension 1/2 to bulk fermions. This is consistent with the way we
have written down 11-dimensional supergravity \eqref{rev11dsugra4}.
In terms of seven-dimensional supermultiplets this tells us to
assign energy dimension 0 to the gravity multiplet and the
$\mathrm{U}(1)$ vector multiplet bosons and energy dimension 1/2 to
the fermions in these multiplets. For the $\mathrm{SU}(N)$ vector
multiplet, that is for the brane fields, we assign energy dimension
1 to the bosons and 3/2 to the fermions. With these conventions we
can expand
\begin{equation} \label{expansion1}
\mathcal{L}_{\mathrm{SU}(N)}=\kappa_7^{-2}\left( \mathcal{L}_{(0)}+\zeta_7^2\mathcal{L}_{(2)}+
\zeta_7^4\mathcal{L}_{(4)}+\ldots \right)\,,
\end{equation}
where the $\mathcal{L}_{(m)}$, $m=0,2,4,\ldots$ are independent of $\zeta_7$.
The first term in this
series is equal to $\mathcal{L}_7^{(n)}$, and therefore the leading
order contribution to $\mathcal{L}_{\mathrm{brane}}$ is precisely the
second term of order $\zeta_7^2$ in the above series. It turns out that,
up to this order, the action $\mathcal{S}_{11-7}$ is supersymmetric
under \eqref{susy1} and \eqref{susy2}. To see this we also expand
the supersymmetry transformation in orders of $\zeta_7$, that is
\begin{equation}
\delta_{11}^{\ph{11}\mathrm{SU}(N)} = \delta_{11}^{(0)}+\zeta_7^2\delta_{11}^{(2)}+\zeta_7^4\delta_{11}^{(4)}+\ldots\, .
\end{equation}
Using this expansion and Eq.~\eqref{expansion1} one finds that the
uncanceled variation \eqref{susycalc} is, in fact, of order
$\zeta_7^4$. This means the action~\eqref{action1} is indeed supersymmetric
up to order $\zeta_7^2$ and can be used to write down a supersymmetric
theory to this order. This is exactly what we will do in the following
section. We have also checked explicitly that the terms of order $\zeta_7^4$
in Eq.~\eqref{susycalc} are non-vanishing, so that our method cannot
be extended straightforwardly to higher orders.

A final remark concerns the value of the Yang-Mills gauge coupling
$\lambda_7$.  The above construction does not fix the value of this
coupling, and our action is supersymmetric to order $\zeta_7^2$, as
well as gauge-invariant, for all values of $\lambda_7$.  This is in
contrast to Ho\v rava-Witten theory, for which the gauge coupling is
uniquely determined by demanding gauge invariance. (In this case,
there is no modification to the Bianchi identity to spoil manifest
gauge invariance.) One does, however, expect M-theory to somehow fix
$\lambda_7$ in terms of the 11-dimensional Newton constant
$\kappa_{11}$. Indeed, reducing M-theory on a circle to IIA string
theory, the orbifold seven-planes turn into D6 branes whose tension
is fixed in terms of the string tension~\cite{Polchinski}. By a
straightforward matching procedure this fixes the gauge coupling to
be~\cite{Friedmann}
\begin{equation} \label{couplerel}
\lambda_7^2=(4\pi)^{4/3}\kappa_{11}^{2/3}\, .
\end{equation}


\section{The explicit bulk/brane theory} \label{results}
In this section, we give a detailed description of M-theory on
$\mathcal{M}^N_{1,10}=\mathcal{M}_{1,6}\times\mathbb{C}^2/\mathbb{Z}_N$,
taking account of the additional states that appear on the brane. We begin
with a reminder of how the bulk fields, truncated to seven dimensions,
are identified with the fields that appear in the seven-dimensional
supergravity Lagrangians from which the theory is constructed.
Then we write down our full Lagrangian, and present the supersymmetry
transformation laws.\\

As discussed in the previous section, the full Lagrangian is
constructed from three parts: the Lagrangian of 11-dimensional
supergravity $\mathcal{L}_{11}$, with bulk fields constrained by the
orbifold action; $\mathcal{L}_7^{(n)}$, the Lagrangian of
seven-dimensional EYM supergravity with gauge group
$\mathrm{U}(1)^n$; and $\mathcal{L}_{\mathrm{SU}(N)}$, the
Lagrangian for seven-dimensional EYM supergravity with gauge group
$\mathrm{U}(1)^n\times \mathrm{SU}(N)$. The Lagrangian
$\mathcal{L}_{11}$ has been written down in
Eq.~\eqref{rev11dsugra4}, and the field constraints that go with it
have been discussed in Section \ref{Bulk}. We have dealt with
$\mathcal{L}_7^{(n)}$ in Section~\ref{reduction4}. The final piece,
$\mathcal{L}_{\mathrm{SU}(N)}$, is discussed in Appendix
\ref{bigEYM}, where we provide the reader with a general review of
seven-dimensional supergravity theories. Crucial to our construction
is the way in which we identify the fields in the supergravity and
$\mathrm{U}(1)^n$ gauge multiplets of the latter two Lagrangians
with the truncated bulk fields. Let us recall the structure of this
identification which has been worked out in
Section~\ref{reduction4}. The bulk fields truncated to seven
dimensions form a $d=7$ gravity multiplet and $n$ $\mathrm{U}(1)$
vector multiplets, where $n=1$ for the general $\mathbb{Z}_N$
orbifold with $N>2$ and $n=3$ for the $\mathbb{Z}_2$ orbifold. The
gravity multiplet contains the purely seven-dimensional parts of the
11-dimensional metric, gravitino and three-form, that is,
$\hat{g}_{\mu\nu}$, $\Psi_{\mu i}$ and $C_{\mu\nu\rho}$, along with
three vectors from $C_{\mu AB}$, a spinor from $\Psi_{A\B{\imath}}$
and the scalar $\mathrm{det}(\hat{e}_{A}^{\ph{A}\underline{B}})$.
Meanwhile, the vector multiplets contain the remaining vectors from
$C_{\mu AB}$, the remaining spinors from $\Psi_{A\B{\imath}}$ and
the scalars contained in $v_{A}^{\ph{A}\underline{B}}:=
\mathrm{det}(\hat{e}_{A}^{\ph{A}\underline{B}})^{-1/4}
\hat{e}_{A}^{\ph{A}\underline{B}}$, the unit-determinant part of
$\hat{e}_{A}^{\ph{A}\underline{B}}$. The gravity and $\mathrm{U}(1)$
vector fields naturally combine together into a single entity
$A_\mu^I$, $I=1,\ldots,(n+3)$, where the index $I$ transforms
tensorially under a global $\mathrm{SO}(3,n)$ symmetry. Meanwhile,
the vector multiplet scalars naturally combine into a single
$(3+n)\times (3+n)$ matrix $\ell$ which parameterises the coset
$\mathrm{SO}(3,n)/\mathrm{SO}(3)\times \mathrm{SO}(n)$. The precise
mathematical form of these identifications is given in equations
\eqref{4id1}--\eqref{idn2} for the general $\mathbb{Z}_N$ orbifold
with $N>2$, and equations \eqref{4id1}--\eqref{4idn} and
\eqref{idn3}--\eqref{idn4} for the $\mathbb{Z}_2$ orbifold.

In addition to those states which arise from projecting bulk states
to the orbifold fixed plane, the Lagrangian ${\cal
L}_{\mathrm{SU}(N)}$ also contains a seven-dimensional
$\mathrm{SU}(N)$ vector multiplet, which is genuinely located on the
orbifold plane. It consists of gauge fields $A_\mu^a$ with field
strengths $F^a=\mathcal{D}A^a$, gauginos $\la_i^a$, and
$\mathrm{SU}(2)$ triplets of scalars $\phi_{a\ph{i}j}^{\ph{a}i}$.
These fields are in the adjoint of $\mathrm{SU}(N)$ and we use
$a,b,\ldots=4,\ldots,(N^2+2)$ for $\mathfrak{su}(N)$ Lie algebra
indices. It is important to write ${\cal L}_{\mathrm{SU}(N)}$ in a
form where the $\mathrm{SU}(N)$ states and the
gravity/$\mathrm{U}(1)^n$ states are disentangled, since the latter
must be identified with truncated bulk states, as described above.
For most of the fields appearing in $\mathcal{L}_{\mathrm{SU}(N)}$,
this is just a trivial matter of using the appropriate notation. For
example, the vector fields in $\mathcal{L}_{\mathrm{SU}(N)}$ which
naturally combine into a single entity $A_\mu^{\ti{I}}$, where
$\ti{I}=1,\ldots,(3+n+N^2-1)$, and transforms as a vector under the
global $\mathrm{SO}(3,n+N^2-1)$ symmetry, can simply be decomposed
as $A_\mu^{\ti{I}}=(A_\mu^I,A_\mu^a)$, where $A_\mu^I$ refers to the
three vector fields in the gravity multiplet and the
$\mathrm{U}(1)^n$ vector fields and $A_\mu^a$ denotes the
$\mathrm{SU}(N)$ vector fields. For gauge group
$\mathrm{U}(1)^n\times \mathrm{SU}(N)$, the associated scalar fields
parameterise the coset $\mathrm{SO}(3,n+N^2-1)/\mathrm{SO}(3)\times
\mathrm{SO}(n+N^2-1)$. We obtain representatives $L$ for this coset
by expanding around the bulk scalar coset
$\mathrm{SO}(3,n)/\mathrm{SO}(3)\times \mathrm{SO}(n)$, represented
by matrices $\ell$, to second order in the $\mathrm{SU}(N)$ scalars
$\Phi\equiv (\phi_a^{\ph{a}u})$. For the details see
Appendix~\ref{bigEYM}. This leads to
\begin{equation}
L = \left( \begin{array}{ccc}
\ell+\frac{1}{2}\zeta_7^2\ell\Phi^T\Phi & m & \zeta_7\ell\Phi^T \\
\zeta_7\Phi & 0 & \boldsymbol{1}_{N^2-1}+\frac{1}{2}\zeta_7^2\Phi\Phi^T \\
\end{array} \right)\, .
\end{equation}
We note that the neglected $\Phi$ terms are of order $\zeta_7^3$ and higher
and, since we are aiming to construct the action only up to terms of order
$\zeta_7^2$, are, therefore, not relevant in the present context.

We are now ready to write down our final action.
As discussed in Section \ref{construct}, to order $\zeta_7^2\sim \lambda_7^{-2}$, it is given by
\begin{equation} \label{S11-7}
\mathcal{S}_{11-7}=\int_{\mathcal{M}^N_{1,10}} \mathrm{d}^{11}x \left[ \mathcal{L}_{11} +
         \delta^{(4)}(y^{A})\mathcal{L}_{\mathrm{brane}} \right]\,,
\end{equation}
where
\begin{equation}
\mathcal{L}_{\mathrm{brane}}=\mathcal{L}_{\mathrm{SU}(N)}-\mathcal{L}_7^{(n)}\,,
\end{equation}
and $n=3$ for the $\mathbb{Z}_2$ orbifold and $n=1$ for
$\mathbb{Z}_N$ with $N>2$. The bulk contribution,
$\mathcal{L}_{11}$, is given in equation \eqref{rev11dsugra4}, with
bulk fields subject to the orbifold
constraints~\eqref{cond1}--\eqref{condn}. On the orbifold fixed
plane, $\mathcal{L}_7^{(n)}$ acts to cancel all the terms in
$\mathcal{L}_{\mathrm{SU}(N)}$ that only depend on bulk fields
projected to the orbifold plane. Thus none of the bulk gravity terms
are replicated on the orbifold plane. To find
$\mathcal{L}_{\mathrm{brane}}$ explicitly we need to expand ${\cal
L}_{\mathrm{SU}(N)}$ in powers of $\zeta_7$, using, in particular,
the above expressions for the gauge fields $A_\mu^{\tilde{I}}$ and
the coset matrices $L$, and extract the terms of order $\zeta_7^2$.
The further details of this calculation are provided in Appendix
\ref{bigEYM}. The result is \ba \label{braneaction}
\mathcal{L}_{\mathrm{brane}}\!\!\! &=&\!\!\! \frac{1}{\lambda_7^2}
 \sqrt{-h} \left\{ -\frac{1}{4}e^{-2\si}F^a_{\mu\nu}F_a^{\mu\nu}
-\frac{1}{2}\mathcal{D}_\mu\phi_{a\ph{i}j}^{\ph{a}i}
\mathcal{D}^\mu\phi_{\ph{aj}i}^{aj}
 -e^{-2\si}\ell_{I\ph{i}j}^{\ph{I}i}\phi_{a\ph{j}i}^{\ph{a}j}F^I_{\mu\nu}F^{a\mu\nu}\right.  \nn \\
&&\hspace{0.45cm} -\frac{1}{2}\B{\la}^{ai}\Up^\mu\mathcal{D}_\mu\la_{ai}-\frac{1}{2}e^{-2\si}\ell_{I\ph{i}j}^{\ph{I}i}
{\phi^{aj}}_i\ell_{J\ph{k}l}^{\ph{J}k}\phi_{a\ph{l}k}^{\ph{a}l}
F^I_{\mu\nu}F^{J\mu\nu}-\frac{1}{2}p_{\mu\alpha\ph{i}j}^{\ph{\mu\alpha}i}
\phi_{a\ph{j}i}^{\ph{a}j}p^{\mu\alpha k}_{\ph{\mu\alpha k}l}\phi^{al}_{\ph{al}k} \nn \\
&&\hspace{0.45cm}+\frac{1}{4}\phi_{a\ph{i}k}^{\ph{a}i}
\mathcal{D}_\mu\phi_{\ph{ak}j}^{ak}\left( \B{\psi}_\nu^j
\Up^{\nu\mu\rho}\psi_{\rho i} + \B{\chi}^j\Up^\mu\chi_i+ \B{\la}^{\alpha j}
\Up^\mu\la_{\alpha i} \right) \nn \\
&&\hspace{0.45cm}+\frac{1}{192}e^{2\si }G_{\mu \nu \rho
\sigma }\bar{\lambda}^{ai}\Upsilon ^{\mu
\nu \rho \sigma }\lambda_{ai}-\frac{1}{2\sqrt{2}}\left( \bar{\lambda}^{\alpha i}
\Upsilon ^{\mu \nu }\psi _{\mu j}+\bar{\lambda}^{\alpha i}\psi _{j}^{\nu }\right)
\phi_{a\ph{j}i}^{\ph{a}j}\phi^{ak}_{\ph{ak}l} p_{\nu\alpha \phantom{l}k}^{%
\phantom{\nu\alpha}l} \nn \\
&& \hspace{0.45cm}-\frac{1}{\sqrt{2}}\left( \bar{\lambda}^{ai}
\Upsilon ^{\mu \nu }\psi_{\mu j}  +\bar{\lambda}^{ai}\psi_{j}^{\nu }\right)
\mathcal{D}_\nu \phi_{ a\phantom{j}i}^{\phantom{ a}j}+\frac{i}{4\sqrt{2}}e^{-\si}F^I_{\mu\nu}
\ell_{I\ph{j}i}^{\ph{I}j}\B{\la}^{ai}\Up^{\mu\nu}\la_{aj}  \notag \\
&&\hspace{0.45cm}-\frac{i}{2}e^{-\si}\left(F_{\mu\nu}^I
\ell_{I\ph{k}l}^{\ph{I}k}\phi^{al}_{\ph{al}k}\phi^{\ph{a}j}_{a\ph{j}i}+
2F^a_{\mu\nu}\phi_{a\ph{j}i}^{\ph{a}j}\right)\left[ \frac{1}{4\sqrt{2}}\left( \bar{\psi}_{\rho
}^{i}\Upsilon ^{\mu \nu \rho \sigma }\psi _{\sigma j}+2\bar{\psi}^{\mu
i}\psi _{j}^{\nu }\right) \right.  \notag \\
&&\hspace{1.7cm}\left. +\frac{3}{20\sqrt{2}}\bar{\chi}^{i}\Upsilon ^{\mu \nu
}\chi _{j} -\frac{1}{4\sqrt{2}}\bar{\lambda}^{\alpha i}\Upsilon ^{\mu \nu
}\lambda _{\alpha j} +\frac{1}{2\sqrt{10}}\left( \bar{\chi}^{i}\Upsilon
^{\mu \nu \rho }\psi _{\rho j}-2\bar{\chi}^{i}\Upsilon ^{\mu }\psi _{j}^{\nu
}\right) \right]  \nn \\
&&\hspace{0.45cm} +e^{-\si}F_{a\mu\nu}\left[ \frac{1}{4}\left( 2\bar{\lambda}^{ai}
\Upsilon ^{\mu }\psi _{i}^{\nu }-\bar{%
\lambda}^{ai}\Upsilon ^{\mu \nu \rho }\psi _{\rho i}\right) +\frac{1}{2%
\sqrt{5}}\bar{\lambda}^{ai}\Upsilon ^{\mu \nu }\chi _{i}\right] \nn \\
&&\hspace{0.45cm}+\frac{1}{4}e^{2\si}f_{bc}^{\ph{bc}a}f_{dea}
\phi^{bi}_{\ph{bi}k}\phi^{ck}_{\ph{ck}j}\phi^{dj}_{\ph{dj}l}
\phi^{el}_{\ph{el}i} -\frac{1}{2}e^{\si}f_{abc}\phi^{bi}_{\ph{bi}k}\phi^{ck}_{\ph{ck}j}
\left(\bar{\psi}_{\mu}^{j}\Upsilon ^{\mu }\lambda_{i}^a +\frac{2}{\sqrt{5}}\bar{\chi}^{j}\lambda _{i}^{%
\phantom{i}a}\right) \notag \\
&&  \hspace{0.45cm}-\frac{i}{\sqrt{2}}e^{\si }f_{ab}^{\ph{ab}c}\phi_{c\ph{i}j}^{\ph{c}i}
\bar{\lambda}^{aj}\lambda _{i}^b+\frac{i}{60\sqrt{2}}e^{\si }f_{ab}^{\ph{ab}c}
\phi^{al}_{\ph{al}k}\phi^{bj}_{\ph{bj}l}\phi_{c\ph{k}j}^{\ph{c}k}\left(
5\bar{\psi}_{\mu }^{i}\Upsilon ^{\mu \nu }\psi _{\nu i}+2\sqrt{5}\bar{\psi}%
_{\mu }^{i}\Upsilon ^{\mu }\chi _{i}\right.  \notag \\
&& \hspace{4.2cm}\left.\left. +3\bar{\chi}^{i}\chi _{i}-5\B{\la}^{\alpha i}
\la_{\alpha i}\right)-\frac{1}{96}\epsilon ^{\mu \nu \rho \sigma \kappa \lambda
 \tau }C_{\mu\nu \rho} F_{\si \kappa}^aF_{a\la \tau }\right\}.
\ea
Here $f_{ab}^{\ph{ab}c}$ are the structure constants of $\mathrm{SU}(N)$.
The covariant derivatives that appear are given by
\ba
\mathcal{D}_\mu A_{\nu a}&=&\partial_\mu A_{\nu a} -\Gamma^\rho_{\mu\nu}A_{\rho a}+
f_{ab}^{\phantom{ab}c}A_\mu^bA_{c\nu}\,, \\
\mathcal{D} _{\mu }\la_{ai}&=&\partial _{\mu }\la_{ai}+\frac{1}{2}
q_{\mu i}^{\phantom{\mu i}j}\la_{aj}+\frac{1}{4}\omega _{\mu }^{\phantom{\mu}%
\underline{\mu }\underline{\nu }}\Upsilon _{\underline{\mu }\underline{\nu }%
}\la_{ai}+ f_{ab}^{\phantom{ab}c}A_\mu^b\la_{ci}\,, \\
\mathcal{D}_\mu\phi_{a\ph{i}j}^{\ph{a}i}&=&\partial_\mu \phi_{a\ph{i}j}^{\ph{a}i}
 -q_{\mu \ph{i}j\ph{k}l}^{\ph{\mu}i\ph{j}k}\phi_{a\ph{l}k}^{\ph{a}l}+
f_{ab}^{\phantom{ab}c}A_\mu^b\phi_{c\ph{i}j}^{\ph{c}i}\,, \ea with
the Christoffel and spin connections $\Gamma$ and $\omega$ taken in
the seven-dimensional Einstein frame, (with respect to the metric
$h$, as given by \eqref{Weyl}). Finally, the quantities $p$ and $q$
are the Maurer-Cartan forms of the bulk scalar coset matrix
$\ell_I^{\ph{I}\underline{J}}$ as given by equations
\eqref{Maurer1}--\eqref{Maurern}. Once again, the identities for
relating the seven-dimensional gravity and $\mathrm{U}(1)$ vector
multiplet fields to 11-dimensional bulk fields are given in
equations \eqref{4id1}--\eqref{idn2} for the generic $\mathbb{Z}_N$
orbifold with $N>2$, and equations \eqref{4id1}--\eqref{4idn} and
\eqref{idn3}--\eqref{idn4} for the $\mathbb{Z}_2$ orbifold. We
stress that these identifications are part of the definition of the
theory.

The leading order brane corrections to the supersymmetry
transformation laws \eqref{rev11dsusy1} of the bulk fields are
computed using equations \eqref{susy1} and \eqref{susy2}. They are
given by \ba \delta^{\mathrm{brane}}\psi_{\mu i}&=&
\frac{\kappa_7^2}{\lambda_7^2}
                                       \left\{ \frac{1}{2}\left( \phi_{ak}^{\ph{ak}j}
\mathcal{D}_\mu\phi^{a\ph{i}k}_{\ph{a}i}-\phi^{a\ph{i}k}_{\ph{a}i}
\mathcal{D}_\mu\phi_{ak}^{\ph{ak}j}\right)\ve_j -\frac{i}{15\sqrt{2}}
\Upsilon _{\mu }\varepsilon _{i}f_{ab}^{\ph{ab}c}\phi^{al}_{\ph{al}k}\phi^{bj}_{\ph{bj}l}
\phi_{c\ph{k}j}^{\ph{c}k}e^{\si} \right. \nn \\
&& \left. \hspace{0.9cm} +\frac{i}{10\sqrt{2}}\left(
\Upsilon _{\mu }^{\ph{\mu}\nu\rho }-8\delta _{\mu }^{\nu }\Upsilon ^{\rho
}\right) \varepsilon _{j}\left( F_{\nu\rho }^{I}\ell_{I\phantom{k}l}^{%
\phantom{I}k}\phi^{al}_{\ph{al}k}\phi^{\ph{a}j}_{a\ph{j}i}+2F^a_{\nu\rho}
\phi^{\ph{a}j}_{a\ph{j}i}\right)e^{-\si} \right\}\,, \nn \\
\delta^{\mathrm{brane}}\chi_i&=&\frac{\kappa_7^2}{\lambda_7^2}
\left\{ -\frac{i}{2\sqrt{10}}\Up^{\mu\nu}\ve_j\left( F_{\mu\nu}^I
\ell_{I\ph{k}l}^{\ph{I}k}\phi^{al}_{\ph{al}k}\phi^{\ph{a}j}_{a\ph{j}i}+
2F^a_{\mu\nu}\phi^{\ph{a}j}_{a\ph{j}i}\right)e^{-\si} \right. \nn \\
&&\hspace{0.9cm}\left.+\frac{i}{3\sqrt{10}}\varepsilon _{i}f_{ab}^{\ph{ab}c}
\phi^{al}_{\ph{al}k}\phi^{bj}_{\ph{bj}l}\phi_{c\ph{k}j}^{\ph{c}k}e^{\si} \right\}\,, \nn \\
\ell_{I\ph{i}j}^{\ph{I}i}\delta^{\mathrm{brane}}A^I_\mu&=&
\frac{\kappa_7^2}{\lambda_7^2}\left\{ \left( \frac{i}{\sqrt{2}}
 \B{\psi }_{\mu }^{k}\varepsilon _{l} -\frac{i}{%
\sqrt{10}}\B{\chi }^{k}\Upsilon _{\mu }\varepsilon _{l} \right)
\phi^{al}_{\ph{al}k}\phi^{\ph{a}i}_{a\ph{i}j} e^{\si}-\B{\ve}^k
\Up_\mu\la^a_k\phi_{a\ph{i}j}^{\ph{a}i}e^\si \right\}, \label{bulksusycorr}\\
\ell_{I}^{\ph{I}\alpha}\delta^{\mathrm{brane}}A^I_\mu&=&0\,, \nn \\
\delta^{\mathrm{brane}}\ell_{I\ph{i}j}^{\ph{I}i}&=&
\frac{\kappa_7^2}{\lambda_7^2}\left\{ \frac{i}{\sqrt{2}}
\bigg[ \B{\ve}^k\la_{\alpha l}\phi^{al}_{\ph{al}k}
\phi^{\ph{a}i}_{a\ph{i}j}\ell_I^{\ph{I}\alpha}+\B{\ve}^l
\la_{ak}\phi^{ai}_{\ph{ai}j}\ell_{I\ph{k}l}^{\ph{I}k} \right. \nn \\
&&\left. \hspace{0.9cm}  -\left( \B{\ve}^i\la_{aj}-\frac{1}{2}\delta^i_j\B{\ve}^m\la_{am}\right)
 \phi^{al}_{\ph{al}k}\ell_{I\ph{k}l}^{\ph{I}k} \bigg] \right\}\,,\nn \\
\delta^{\mathrm{brane}}\ell_{I}^{\ph{I}\alpha}&=&
\frac{\kappa_7^2}{\lambda_7^2}\left\{ -\frac{i}{\sqrt{2}}\B{\ve}^i\la^\alpha_j
\phi^{aj}_{\ph{aj}i}\phi^{\ph{a}l}_{a\ph{l}k}\ell_{I\ph{k}l}^{\ph{I}k}\right\}\,, \nn \\
\delta^{\mathrm{brane}}\la_i^\alpha &=&\frac{\kappa_7^2}{\lambda_7^2}
\left\{ \frac{i}{\sqrt{2}}\Up^\mu\ve_j\phi_{ai}^{\ph{ai}j}
p_{\mu \ph{\alpha k}l}^{\ph{\mu}\alpha k}\phi^{al}_{\ph{al}k}\right\}\,, \nn
\ea
where $\ve_i$ is the 11-dimensional supersymmetry spinor $\eta$
projected onto the orbifold plane, as given in \eqref{spinorrel}. We note that
not all of the bulk fields receive corrections to their
supersymmetry transformation laws. The leading order supersymmetry
transformation laws of the $\mathrm{SU}(N)$ multiplet fields are found using equation
\eqref{susy2} and take the form
\ba
\delta A_\mu^a&=& \B{\ve}^i\Up_\mu\la_i^ae^\si-\left( i\sqrt{2}\psi_\mu^i\ve_j
 -\frac{2i}{\sqrt{10}}\B{\chi}^i\Up_\mu\ve_j \right)\phi^{aj}_{\ph{aj}i}e^\si\,,\nn \\
\delta \phi_{a\ph{i}j}^{\ph{a}i}&=&-i\sqrt{2}\left( \B{\ve}^i\la_{aj}-
\frac{1}{2}\delta^i_j\B{\ve}^k\la_{ak}\right)\,, \label{susybranexmfn} \\
\delta\la^a_i&=&-\frac{1}{2}\Up^{\mu\nu}\ve_i\left( F^I_{\mu\nu}
\ell^{\ph{I}j}_{I\ph{j}k}\phi^{ak}_{\ph{ak}j}+F^a_{\mu\nu}\right)e^{-\si}-i\sqrt{2}
\Up^\mu\ve_j\mathcal{D}_\mu\phi^{a\ph{i}j}_{\ph{a}i}-i\ve_j
f^a_{\ph{a}bc}\phi^{bj}_{\ph{bj}k}\phi^{ck}_{\ph{ck}i}\,. \nn
\ea
\\

To make some of the properties of our result more transparent, it is
helpful to extract the bosonic part of the action. This bosonic part
will also be sufficient for many practical applications. We recall
that the full Lagrangian \eqref{braneaction} is written in the
seven-dimensional Einstein frame to avoid the appearance of
$\sigma$--dependent pre-factors in many terms. The bosonic part,
however, can be conveniently formulated in terms of $\hat{g}_{\mu\nu}$, the
seven-dimensional part of the 11-dimensional bulk metric
$\hat{g}_{MN}$. This requires performing the Weyl-rescaling~\eqref{Weyl}.
It also simplifies the notation if we rescale the scalar $\si$ as
$\tau=10\si/3$.
Let us now write down the purely bosonic part of our action, subject
to these small modifications. We find
\begin{equation}
\mathcal{S}_{11-7,{\rm bos}}=\mathcal{S}_{11,{\rm bos}} + \mathcal{S}_{7,{\rm bos}}\, ,
\end{equation}
where $\mathcal{S}_{11,{\rm bos}}$ is the bosonic part of
11-dimensional supergravity~\eqref{rev11dsugra4}, with fields
subject to the orbifold constraints \eqref{cond1}--\eqref{cond6}.
Further, $\mathcal{S}_{7,{\rm bos}}$ is the bosonic part of
Eq.~\eqref{braneaction}, subject to the above modifications, for
which we obtain \ba \label{4boseaction} \mathcal{S}_{7,{\rm bos}}&=&
\frac{1}{\lambda_7^2} \int_{y=0} \mathrm{d}^7x \sqrt{-\hat{g}}
\left( -\frac{1}{4}H_{ab}F^a_{\mu\nu}F^{b\mu\nu}
-\frac{1}{2}H_{aI}F^a_{\mu\nu}
F^{I\mu\nu} -\frac{1}{4}(\delta H)_{IJ}F^I_{\mu\nu}F^{J\mu\nu}  \right. \nn \\
&&\hspace{3.2cm} \left. -\frac{1}{2}e^\tau\mathcal{D}_\mu
\phi_{a\ph{i}j}^{\ph{a}i}\mathcal{D}^\mu\phi_{\ph{aj}i}^{aj}
-\frac{1}{2}\left( \delta K
\right)^{\ph{I}j\ph{iJ}l}_{I\ph{j}iJ\ph{l}k}
\partial_\mu{\ell^{Ii}}_j\partial^\mu{\ell^{Jk}}_l+\frac{1}{4}D^{ai}_{\ph{ai}j}D^{\ph{a}j}_{a\ph{j}i}\right) \nn \\
&&-\frac{1}{4\lambda_7^2}\int_{y=0} C\wedge F^a \wedge F_a\,,
\ea
where
\ba
H_{ab}&=&\delta_{ab}\,, \label{4H1} \\
H_{aI}&=&2\ell_{I\ph{i}j}^{\ph{I}i}\phi_{a\ph{j}i}^{\ph{a}j}\,, \label{H2} \\
(\delta H)_{IJ}&=&2\ell_{I\ph{i}j}^{\ph{I}i}\phi_{a\ph{j}i}^{\ph{a}j}
\ell_{J\ph{k}l}^{\ph{J}k}\phi_{a\ph{l}k}^{\ph{a}l}\,,  \label{4H3} \\
\left( \delta K \right)^{\ph{I}j\ph{iJ}l}_{I\ph{j}iJ\ph{l}k}&=&
 e^\tau \phi_{a\ph{j}i}^{\ph{a}j}\phi^{al}_{\ph{al}k}\delta_{\alpha\beta}
{\ell_I}^\alpha {\ell_J}^\beta\,, \\
D^{ai}_{\ph{ai}j}&=&e^\tau f^a_{\ph{a}bc}\phi^{bi}_{\ph{bi}k}\phi^{ck}_{\ph{ck}j}\,,
\ea
and the Maurer-Cartan forms $p$ and $q$ of the matrix of scalars $\ell$ are defined by
\ba
p_{\mu\alpha \ph{i}j}^{\ph{\mu\alpha }i}&=&\ell^{I}_{\ph{I}\alpha}\partial_\mu
 \ell_{I\ph{i}j}^{\ph{\mu}i}\,, \\
q_{\mu \ph{i}j\ph{k}l}^{\ph{\mu}i\ph{j}k}&=&\ell^{Ii}_{\ph{Ii}j}\partial_\mu
 \ell_{I\ph{k}l}^{\ph{\mu}k}\,.
\ea
The bosonic fields localised on the orbifold plane are the $\mathrm{SU}(N)$
gauge vectors $F^a=\mathcal{D}A^a$ and the $\mathrm{SU}(2)$ triplets of scalars
$\phi_{a\ph{i}j}^{\ph{a}i}$. All other fields are projected from the
bulk onto the orbifold plane, and there are algebraic equations relating
them to the 11-dimensional fields in $\mathcal{S}_{11}$.
As discussed above, these relations are trivial for the metric
$\hat{g}_{\mu\nu}$
and the three-form $C_{\mu\nu\rho}$, whilst the scalar $\tau$ is given by
\begin{equation} \label{5id1}
\tau = \frac{1}{2}\ln\det \hat{g}_{AB}\, ,
\end{equation}
and can be interpreted as an overall scale factor of the orbifold
$\mathbb{C}^2/\mathbb{Z}_N$. For the remaining fields, the ``gravi-photons''
$F^I_{\mu\nu}$ and the ``orbifold moduli'' $\ell_I^{\ph{I}\underline{J}}$,
we have to distinguish between the generic $\mathbb{Z}_N$ orbifold with
$N>2$ and the $\mathbb{Z}_2$
orbifold. For $\mathbb{Z}_N$ with $N>2$ we have four $\mathrm{U}(1)$ gauge fields,
so that $I=1,\ldots ,4$, and $\ell_I^{\ph{I}\underline{J}}$ parameterises
the coset $\mathrm{SO}(3,1)/\mathrm{SO}(3)$. They are identified with 11-dimensional fields through
\ba
F^I_{\mu\nu}&=&\frac{i}{2}\mathrm{tr}\left( \si^IG_{\mu\nu}\right)\,, \\
\ell_I^{\ph{I}\underline{J}}&=&\frac{1}{2}\mathrm{tr}\left( \B{\si}_I v \si^J v^{\dagger} \right)\,,\label{5idZN}
\ea
where $G_{\mu\nu}\equiv (G_{\mu\nu p\B{q}})$,
$v\equiv (e^{\tau/4}\hat{e}^{p}_{\ph{p}\underline{q}})$ and $\si^I$
are the $\mathrm{SO}(3,1)$ Pauli matrices as given in Appendix \ref{Pauli}.
For the $\mathbb{Z}_2$ case, we have six $\mathrm{U}(1)$ vector fields, so that
$I=1,\ldots ,6$, and $\ell_I^{\ph{I}\underline{J}}$ parameterises the
coset $\mathrm{SO}(3,3)/\mathrm{SO}(3)^2$. The field identifications now read
\ba
F^I_{\mu\nu}&=&-\frac{1}{4}\mathrm{tr}\left( t^IG_{\mu\nu}\right)\,, \label{5idZ2}\\
\ell_I^{\ph{I}\underline{J}}&=&\frac{1}{4}\mathrm{tr}\left( \B{t}_I v t^J v^{T} \right)\,, \label{5idn}
\ea
where this time $G_{\mu\nu}\equiv (G_{\mu\nu AB})$,
$v\equiv (e^{\tau/4}\hat{e}^{A}_{\ph{A}\underline{B}})$, and $t^I$
are the generators of $\mathrm{SO}(4)$, as given in Appendix \ref{Pauli}.\\

Let us discuss a few elementary properties of the bosonic
action~\eqref{4boseaction} on the orbifold plane, starting with the
gauge-kinetic functions \eqref{4H1}--\eqref{4H3}. The first observation
is, that the gauge-kinetic function for the $\mathrm{SU}(N)$ vector fields is
trivial (to the order we have calculated), which confirms the result
of Ref.~\cite{Friedmann}. On the other hand, we find non-trivial gauge
kinetic terms between the $\mathrm{SU}(N)$ vectors and the gravi-photons, as
well as between the gravi-photons. We also note the appearance of the
Chern-Simons term $C\wedge F^a\wedge F_a$, which has been
predicted~\cite{Anomaly} from anomaly cancellation in configurations
which involve additional matter fields on conical singularities, but,
in our case, simply follows from the structure of seven-dimensional
supergravity without any further assumption. We note that, while there
is no seven-dimensional scalar field term which depends only on
orbifold moduli, the scalar field kinetic terms in~\eqref{4boseaction}
constitute a complicated sigma model which mixes the orbifold moduli
and the scalars in the $\mathrm{SU}(N)$ vector multiplets. A further interesting
feature is the presence of the seven-dimensional D-term potential in
Eq.~\eqref{4boseaction}. Introducing the matrices $\phi_a\equiv
(\phi^{\ph{a}i}_{a\ph{i}j})$ and $D^a\equiv (D^{ai}_{\ph{ai}j})$
this potential can be written as
\begin{equation}
 V=\frac{1}{4\lambda_7^2}{\rm tr}\left( D^aD_a\right)\, ,
\end{equation}
where
\begin{equation}
D^a=\frac{1}{2}e^\tau f^a_{\ph{a}bc}[\phi^b,\phi^c]\, .
\end{equation}
The flat directions, $D^a=0$, of this potential, which correspond to
unbroken supersymmetry, as can be seen from Eq.~\eqref{susybranexmfn},
can be written as
\begin{equation}
\phi^a=\upsilon^a\sigma^3
\end{equation}
with vacuum expectation values $\upsilon^a$. The $\upsilon^a$
correspond to elements in the Lie algebra of $\mathrm{SU}(N)$ which
can be diagonalised into the Cartan sub-algebra. Generic such
diagonal matrices break $\mathrm{SU}(N)$ to $\mathrm{U}(1)^{N-1}$,
while larger unbroken groups are possible for non-generic choices.
Looking at the scalar field masses induced from the D-term in such a
generic situation, we have one massless scalar for each of the
non-Abelian gauge fields, which is absorbed as their longitudinal
degree of freedom. For each of the $N-1$ unbroken Abelian gauge
fields, we have all three associated scalars massless, as must be
the case from supersymmetry. This situation corresponds exactly to
what happens when the orbifold singularity is blown up. We can,
therefore, see that within our supergravity construction blowing-up
is encoded by the D-term. Moreover, the Abelian gauge fields in
$\mathrm{SU}(N)$ correspond to (a truncated version of) the massless
vector fields which arise from zero modes of the M-theory three-form
on a blown-up orbifold, while the $3(N-1)$ scalars in the Abelian
vector fields correspond to the blow-up moduli. In the next chapter
we shall explore this further, using this interpretation of the
Higgs effect to continue the K\"ahler potential for M-theory on a
singular $G_2$ manifold to that on a smooth $G_2$ manifold with
small blow-ups.
\begin{equation}
\ast \nn
\end{equation}

In this chapter, we have constructed the effective supergravity action for
M-theory on the orbifold $\mathcal{M}_{1,6}\times\mathbb{C}^2/\mathbb{Z}_N$,
by coupling 11-dimensional supergravity, constrained in accordance with
the orbifolding, to $\mathrm{SU}(N)$ super-Yang-Mills theory located on the
seven-dimensional fixed plane of the orbifold. We have found that the
orbifold-constrained fields of 11-dimensional supergravity, when restricted
to the orbifold plane, fill out a seven-dimensional supergravity multiplet
plus a single $\mathrm{U}(1)$ vector multiplet for $N>2$ and three $\mathrm{U}(1)$ vector
multiplets for $N=2$. The seven-dimensional action on the orbifold plane,
which has to be added to 11-dimensional supergravity, couples these bulk
degrees of freedom to genuine seven-dimensional states in the $\mathrm{SU}(N)$
multiplet. We have obtained this action on the orbifold plane by ``up-lifting''
information from the known action of ${\cal N}=1$ Einstein-Yang-Mills
supergravity and identifying 11- and 7-dimensional degrees of freedom appropriately.
The resulting 11-/7-dimensional theory is given as an expansion in the
parameter $\zeta_7=\kappa_{11}^{5/9}/\lambda_7$, where $\kappa_{11}$ is the 11-dimensional
Newton constant and $\lambda_7$ is the seven-dimensional $\mathrm{SU}(N)$ coupling.
The bulk theory appears at zeroth order in $\zeta_7$, and we have determined the
complete set of leading terms on the orbifold plane which
are of order $\zeta_7^2$. At order $\zeta_7^4$ we encounter a singularity due to
a delta function square, similar to what happens in Ho\v{r}ava-Witten
theory~\cite{Hor-Wit}. As in Ref.~\cite{Hor-Wit}, we assume that this singularity
will be resolved in full M-theory, when the finite thickness of the
orbifold plane is taken into account, and that it does not invalidate the
results at order $\zeta_7^2$.

While we have focused on the $A$-type orbifolds
$\mathbb{C}^2/\mathbb{Z}_N$, we expect our construction to work
analogously for the other four-dimensional orbifolds of $ADE$ type. Our
result represents the proper starting point for compactifications of
M-theory on $G_2$ spaces with singularities of the type
$\mathcal{B}\times\mathbb{C}^2/\mathbb{Z}_N$, where $\mathcal{B}$ is a
three-dimensional
manifold. In particular, our result enables us to proceed with
compactification on a toroidal $G_2$ orbifold of the type constructed
in Chapter~\ref{Classification}, for which $\mathcal{B}=\mathcal{T}^3$,
a three-torus,
at each singularity. This will form the basis of the next chapter.

\chapter{Four-dimensional Effective M-theory on a
Singular $\boldsymbol{G_2}$ Manifold} \label{G2sing} 
\fancyhf{}
\fancyhead[L]{\sl 5.~Four-dimensional Effective M-theory on a Singular $G_2$ Manifold} 
\fancyhead[R]{\rm\thepage}
\renewcommand{\headrulewidth}{0.3pt}
\renewcommand{\footrulewidth}{0pt}
\addtolength{\headheight}{3pt} 
\fancyfoot[C]{} 
Compactification on
$G_2$ spaces with $A$-type singularities, which leads to
four-dimensional $\mathrm{SU}(N)$ gauge multiplets, can be seen as a
first step towards successful particle phenomenology from M-theory
on $G_2$ spaces. In this chapter we carry out such a
compactification, using the results of the previous chapter to
explicitly describe the non-Abelian gauge fields which arise at the
singularities. More specifically we calculate the four-dimensional
effective theory for M-theory on a $G_2$ orbifold $\mathcal{Y}$,
using the
action~\eqref{S11-7},~\eqref{rev11dsugra4},~\eqref{braneaction} for
11-dimensional supergravity coupled to seven-dimensional
super-Yang-Mills theory. The orbifold $\mathcal{Y}$ is of the kind
constructed in Chapter~\ref{Classification} by dividing a
seven-torus $\mathcal{T}^7$ by a discrete symmetry group $\Gamma$,
with co-dimension four fixed points. We assume that points on the
torus that are fixed by one generator of $\Gamma$ are not fixed by
other generators. This ensures the approximate form of $\mathcal{Y}$
near an orbifold fixed locus is
$\mathcal{T}^3\times\mathbb{C}^2/\mathbb{Z}_{N}$, and hence that,
according to the $ADE$ classification, the singularities of
$\mathcal{Y}$ are all of type $A_{N-1}\sim\mathbb{Z}_N$, for some $N$.
For concrete examples we refer the reader back to
Chapter~\ref{Classification}, where a list of sixteen possible
orbifold groups was drawn up. Note that the class of orbifolds we
consider corresponds precisely to the class of smooth $G_2$
manifolds for which we computed the moduli K\"ahler potential in
Section~\ref{ModKP}; we are dealing now with the limit in which all
blow-up moduli are taken to zero. This enables a direct comparison
of results for the singular and smooth cases, and this is carried
out at the end of the chapter.

Performing the compactification involves the reduction of the
seven-dimensional $\mathrm{SU}(N)$ gauge
theories on the three-dimensional singular loci within the $G_2$
space. For our $G_2$ orbifold $\mathcal{Y}$
these will always be three-tori, $\mathcal{T}^3$. Hence, while the full
four-dimensional theory is $\mathcal{N}=1$ supersymmetric, the gauge
sub-sectors associated with each singularity have enhanced
$\mathcal{N}=4$ supersymmetry. We explore the consequences of this
in Chapter~\ref{Analysis}.

This chapter is divided into four sections. In the first section
we lay out our conventions and then
describe the background and zero modes for M-theory on $\mathcal{M}_{1,10}
=\mathbb{R}^{1,3}\times\mathcal{Y}$. Zero modes can be split into
``bulk fields'' which
descend from 11-dimensional supergravity and ``matter fields''
which descend from the seven-dimensional super-Yang-Mills theories at the
singularities. Each are discussed in turn. In Section~\ref{red5} we
calculate, by reduction, the bosonic terms of the four-dimensional
effective theory. Since this is an $\mathcal{N}=1$ supergravity theory,
it is instructive to find the associated K\"ahler potential and
superpotential. This is carried out in Section~\ref{n=1}. Finally, in
Section~\ref{compare} we compare our results with those for smooth $G_2$
spaces. For most of the calculation we will focus on one singularity
for simplicity, and only introduce a sum over all singularities into
the final results in Section~\ref{n=1}.

Let us summarise the conventions for this chapter, and warn the reader that
notation for index types differs slightly
from that of Chapter~\ref{117}. As usual, we take 11-dimensional spacetime
to have mostly positive signature, that is $(-+\ldots+)$, and use indices
$M,N,\ldots=0,1,\ldots,10$ to label 11-dimensional coordinates $(x^M)$. We
now use $\mu,\nu,\ldots=0,1,2,3$ to label four-dimensional coordinates
on $\mathbb{R}^{1,3}$, while points on the internal $G_2$ space $\mathcal{Y}$
are labelled by coordinates $(x^A)$, where $A,B,\ldots=1,\ldots,7$.
Without loss of generality, the singularity of $\mathcal{Y}$ that we
consider explicitly is located
at $x^4=x^5=x^6=x^7=0$. Near this singularity, we split coordinates
according to
\begin{equation} \label{7coordid}
(x^A)\in \mathcal{Y} \to (x^m,y^{\hat{A}})\in \mathcal{T}^3
\times\mathbb{C}^2/\mathbb{Z}_N\, ,
\end{equation}
where $m,n,\ldots=1,2,3$ and $\hat{A},\hat{B},\ldots=1,\ldots,4$.
The action of the generator of the $\mathbb{Z}_N$ symmetry is
$y^{\hat{A}} \to {R^{\hat{A}}}_{\hat{B}} y^{\hat{B}}$, where
${R^{\hat{A}}}_{\hat{B}}$ is the same matrix as that given by
Eq.~\eqref{R}. Finally, to describe the seven-dimensional gauge
theories living on the full spacetime singular locus
$\mathbb{R}^{1,3}\times \mathcal{T}^3$, it is useful to introduce
coordinates $(x^{\hat{\mu}})$, where
$\hat{\mu},\hat{\nu},\ldots=0,1,\ldots,6$.

\section{Background solution and zero modes}
 We begin by
discussing the M-theory background on
$\mathcal{M}_{1,10}=\mathbb{R}^{1,3}\times \mathcal{Y}$. Throughout
our calculations we take the expectation values of fermions to
vanish, and this means that we only need concern ourselves with the
bosonic equations of motion. For the bulk 11-dimensional
supergravity~\eqref{rev11dsugra4} they are given by
Eqs.~\eqref{reveom1}--\eqref{reveomn}, and we choose the standard
background solution discussed in Chapter \ref{revchap}. Thus we take
the metric $\langle\hat{g}\rangle$ to be Ricci flat, whilst for the
three-form, we set $\langle C \rangle=0$. For $\mathcal{Y}$ being a
toroidal $G_2$ orbifold, the Ricci flatness condition implies
$\langle\hat{g}\rangle$ has constant components. In addition, these
components should be constrained in accordance with the orbifold
symmetry. Truncating these 11-dimensional fields to our particular
$\mathbb{Z}_N$ singularity, this background leads to constant
seven-dimensional fields $\langle \tau \rangle$ and $\langle
{\ell_I}^{\ul{J}}\rangle$, and vanishing $\langle A_{\hat{\mu}}^I
\rangle$, according to the identifications
\eqref{5id1}--\eqref{5idZN} for $N>2$ and
\eqref{5id1},~\eqref{5idZ2},~\eqref{5idn} for $N=2$. Substituting
this background into the field equations for the localised fields
gives \ba
\mathcal{D}F&=&0\, , \\
\mathcal{D}_{\hat{\mu}}F^{a\hat{\mu}\hat{\nu}}&=&e^{\langle\tau\rangle}
     {f^a}_{bc}{\phi^{bi}}_j\mathcal{D}^{\hat{\nu}}{\phi^{cj}}_i\, ,\\
\mathcal{D}_{\hat{\mu}}\mathcal{D}^{\hat{\mu}}{{\phi_{a}}^i}_j&=&
     -f_{abc}{\phi^{bi}}_k{D^{ck}}_j\, .
\ea
A valid background is thus obtained by setting the genuine
seven-dimensional fields to zero, that is, $\langle A_{\hat{\mu}}^a\rangle=0$,
and $\langle {{\phi_{a}}^i}_j \rangle =0$. With
these fields switched off, the singularity
causes no modification to the background for the bulk fields.

\fancyhf{} \fancyhead[L]{\sl \rightmark} \fancyhead[R]{\rm\thepage}
\renewcommand{\headrulewidth}{0.3pt}
\renewcommand{\footrulewidth}{0pt}
\addtolength{\headheight}{3pt} \fancyfoot[C]{}
We now discuss supersymmetry of the background. Substitution of our
background into the supersymmetry transformation
laws~\eqref{rev11dsusy1},~\eqref{bulksusycorr},~\eqref{susybranexmfn}
makes every term vanish except for the $\nabla_M \eta$ term in the
variation of the gravitino. Hence, the existence of precisely one
Killing spinor on a $G_2$ space guarantees that our background is
supersymmetric, with $\mathcal{N}=1$ supersymmetry from a
four-dimensional point of view. \\

Let us now discuss the zero modes of these background solutions,
both for the bulk and the localised fields. We begin with the bulk
zero modes. All the orbifold examples discussed in
Chapter~\ref{Classification} restrict the internal metric to be
diagonal (and do not allow any invariant two-forms). If we focus on
examples of this type, then the bulk zero modes are the same as for
the blown-up case studied in Section~\ref{ModKP}. Doing this, the
11-dimensional metric can be written as
\begin{equation} \label{metric1}
\mathrm{d}s^2=\left( \prod_{A=1}^7 R^A \right)^{-1} g_{\mu\nu}dx^\mu dx^\nu +
              \sum_{A=1}^{7}\left(R^A\mathrm{d}x^A\right)^2\, ,
\end{equation}
where the $R^A$ are precisely the seven radii of the underlying
seven-torus. The factor in front of the first part has been
chosen so that $g_{\mu\nu}$ is the four-dimensional metric in the
Einstein frame. There exists a $G_2$ structure $\varphi$ associated with each
internal Ricci-flat metric. For the seven-dimensional
part of the above metric and an appropriate choice of
coordinates it is given by\footnote{There are some sign differences between
corresponding terms in this $G_2$ structure and the one given
by Eq.~\eqref{structurex}. These come about from taking $x^7\to -x^7$
in~\eqref{structurex}, and this change is necessary to satisfy invariance
under our current choice of action~\eqref{R} of the
$\mathbb{Z}_N$ symmetry. In the notation of
Chapter~\ref{Classification},
the rotation angles are $(\theta_1,\theta_2,\theta_3)=(0,-2\pi/N,-2\pi/N)$,
and the sum of these is not in general 0 mod $2\pi$.
The reason for making this non-standard choice of action in
Chapter~\ref{117} was to obtain simple
expressions for the bulk fields constrained to the orbifold plane.}
\begin{eqnarray} \label{structure3}
\varphi & = & R^1R^2R^3\mathrm{d}x^1\wedge\mathrm{d}x^2\wedge\mathrm{d}x^3+
R^1R^4R^5\mathrm{d}x^1\wedge\mathrm{d}x^4\wedge\mathrm{d}x^5-
R^1R^6R^7\mathrm{d}x^1\wedge\mathrm{d}x^6\wedge\mathrm{d}x^7 \nonumber \\
 & &+R^2R^4R^6\mathrm{d}x^2\wedge\mathrm{d}x^4\wedge\mathrm{d}x^6 +
R^2R^5R^7\mathrm{d}x^2\wedge\mathrm{d}x^5\wedge\mathrm{d}x^7+
R^3R^4R^7\mathrm{d}x^3\wedge\mathrm{d}x^4\wedge\mathrm{d}x^7 \nonumber \\
 & & -R^3R^5R^6\mathrm{d}x^3\wedge\mathrm{d}x^5\wedge\mathrm{d}x^6\,.
\end{eqnarray}
It is the coefficients of $\varphi$ that define the metric moduli $a^A$ in the reduced theory, and these become the real bosonic parts of chiral
superfields. In this instance we use the index range $A=0,1,\ldots,6$ and set
\begin{equation} \label{period2}
\left. \begin{array}{cccc}
a^0=R^1R^2R^3\,, & a^{1}=R^1R^4R^5\,, & a^{2}=R^1R^6R^7\,, & a^{3}=R^2R^4R^6\,, \\
a^{4}=R^2R^5R^7\,, & a^{5}=R^3R^4R^7\,, & a^{6}=R^3R^5R^6\,. &  \, \\
\end{array} \right.
\end{equation}
Since there are no one-forms on a $G_2$ space, and our assumption
above states that there are no two-forms on $\mathcal{Y}$, the three-form field $C$ expands
purely in terms of three-forms, and takes the same form as $\varphi$, thus
\ba \label{Cexp1}
C&=& \nu^0\mathrm{d}x^1\wedge\mathrm{d}x^2\wedge\mathrm{d}x^3+
\nu^{1}\mathrm{d}x^1\wedge\mathrm{d}x^4\wedge\mathrm{d}x^5-
\nu^{2}\mathrm{d}x^1\wedge\mathrm{d}x^6\wedge\mathrm{d}x^7 +
 \nu^{3}\mathrm{d}x^2\wedge\mathrm{d}x^4\wedge\mathrm{d}x^6 \nn \\
 && + \nu^{4}\mathrm{d}x^2\wedge\mathrm{d}x^5\wedge\mathrm{d}x^7+
\nu^{5}\mathrm{d}x^3\wedge\mathrm{d}x^4\wedge\mathrm{d}x^7-
\nu^{6}\mathrm{d}x^3\wedge\mathrm{d}x^5\wedge\mathrm{d}x^6\,.
\ea
The $\nu^A$ become axions in the reduced theory, and pair up with the metric moduli
to form the superfields
\begin{equation}
\label{TA}
T^A=a^A+i\nu^A\,.
\end{equation}
We recall from Chapter~\ref{Classification} that not all of the
$T^A$ are independent, and we refer the reader back to Section
\ref{ModKP} for a description of the procedure for determining which
of the $T^A$ are constrained to be equal.

We now discuss a convenient re-labelling of the metric moduli, adapted to the
structure of the singularity. Under the identification of coordinates \eqref{7coordid}, the
metric modulus $a^0$ can be viewed as the volume modulus of the three-torus locus
$\mathcal{T}^3$ of the singularity. The other moduli, meanwhile, are each a product
of one radius of the torus $\mathcal{T}^3$ with two radii of $\mathcal{Y}$ transverse to
the singularity. It is sometimes useful to change the notation for
these moduli to the form $a^{mi}$ where $m=1,2,3$ labels the radius on
$\mathcal{T}^3$ that $a^{mi}$ depends on, and $i=1,2$. Thus
\begin{equation}
\begin{array}{cccccc}
a^{11}=a^1\,, & a^{12}=a^2\,, & a^{21}=a^3\,, & a^{22}=a^4\,, &
a^{31}=a^5\,, & a^{32}=a^6\,.
\end{array}
\end{equation}
We will also sometimes make the analogous change of notation for $\nu^A$
and $T^A$.\\

Having listed the bulk moduli, we now turn to the zero modes
associated with the singularity. The decomposition of
seven-dimensional fields works as follows. We take the straightforward
basis $(\mathrm{d}x^m)$ of harmonic one-forms on the three-torus, so
$A^a_{\hat{\mu}}$ simply decomposes into a four-dimensional vector $A^a_{\mu}$ plus the
three scalar fields $A^a_m$ under the reduction. The seven-dimensional
scalars $\phi_{au}$ simply become four-dimensional scalars. Setting
\ba
{b_a}^m&=&-A_{ma}\,, \\
{\rho_a}^1&=&\sqrt{a^{11}a^{12}}{\phi_a}^3\,, \\
{\rho_a}^2&=&-\sqrt{a^{21}a^{22}}{\phi_a}^2\,, \\
{\rho_a}^3&=&\sqrt{a^{31}a^{32}}{\phi_a}^1\,,
\ea
we can define the complex fields
\begin{equation} \label{cdefn}
{\mathcal{C}_a}^m={\rho_a}^m + i {b_a}^m\, .
\end{equation}
As we will see, the fields ${\mathcal{C}_a}^m$ are indeed the correct
four-dimensional chiral matter superfields. \\

The moduli in the above background solutions are promoted to four-dimensional fields,
as usual, and we will call the corresponding bulk fields $\hat{g}^{(0)}$ and $C^{(0)}$,
in the following. In a pure bulk theory, this would be a standard procedure and
the reduction to four dimension would proceed without further complication.
However, in the presence of localised fields there is a subtlety which we
will now discuss. Allowing the moduli to fluctuate introduces localised
stress energy on the seven-dimensional orbifold plane and this excites the
heavy modes of the theory which we would like to truncate in the reduction.
This phenomenon is well-known from Ho\v rava-Witten theory and can be dealt
with by explicitly integrating out the heavy modes, thereby generating
higher-order corrections to the effective theory~\cite{Lukas:1998ew}.
As we will now argue, in our case these corrections are always of higher
order. More precisely, we will compute the four-dimensional effective theory
up to second order in derivatives and up to order $\kappa_{11}^{4/3}\sim
\kappa_{11}^2/\lambda_7^2$, relative
to the leading gravitational terms. Let  $\hat{g}^{(1)}$ and $C^{(1)}$
be the first order corrections to the metric and the three-form which originate from
integrating out the localised stress energy on the orbifold plane, so that
we can write for the corrected fields
\ba
\hat{g}^{(\mathrm{B})}&=&\hat{g}^{(0)}+ \kappa_{11}^{4/3} \hat{g}^{(1)}\, ,
\label{backrfields1}\\
C^{(\mathrm{B})}&=&C^{(0)}+\kappa_{11}^{4/3} C^{(1)}\, .\label{backrfields2}
\ea
We note that these corrections are already suppressed by $\kappa_{11}^{4/3}$ relative
to the pure background fields. Therefore, when inserted into the orbifold
action~\eqref{braneaction}, the resulting corrections are of order $\kappa_{11}^{8/3}$
or higher and will, hence, be neglected. Inserted into the bulk action, the
fields~\eqref{backrfields1} and \eqref{backrfields2} lead to order
$\kappa_{11}^{4/3}$ corrections which can be written as
\begin{equation}
\delta\mathcal{S}_7=\kappa_{11}^{4/3}\hat{g}_{MN}^{(1)}\left. \frac{\delta \mathcal{S}_{11}}
{\delta \hat{g}_{MN}}\right|_{\hat{g}=\hat{g}^{(0)},\, C=C^{(0)}} +
\kappa_{11}^{4/3}C_{MNP}^{(1)}\left. \frac{\delta \mathcal{S}_{11}}{\delta C_{MNP}}
\right|_{\hat{g}=\hat{g}^{(0)},\, C=C^{(0)}}\, .
\label{dS}
\end{equation}
Let us analyse the properties of the terms contained in this
expression. The functional derivatives in the above expression
vanish for constant moduli fields since the background
configurations $\hat{g}^{(0)}$ and $\hat{C}^{(0)}$ are exact
solutions of the 11-dimensional bulk equations in this case. Hence,
allowing the moduli fields to be functions of the external
coordinates, the functional derivatives must contain at least two
four-dimensional derivatives. All terms in $\hat{g}^{(1)}$ and
$\hat{C}^{(1)}$ with four-dimensional derivatives will, therefore,
generate higher-dimensional derivative terms in four dimensions and
can be neglected. The only terms which are not of this type arise
from the D-term potential and covariant derivatives on the orbifold
plane and they appear within $\hat{g}^{(1)}$. These terms are of
order $\kappa_{11}^{4/3}$, and should in principle be kept. However,
they are of fourth order in the matter fields $\mathcal{C}^{am}$,
and contain two four-dimensional derivatives acting on bulk moduli.
They can, therefore, be thought of as corrections to the moduli
kinetic terms. As we will see, the K\"ahler potential of the
four-dimensional theory can be uniquely fixed without knowing these
correction terms explicitly.

\section{Calculation of the four-dimensional effective theory}~\label{red5}
We will now reduce our theory to four dimensions, starting with the
lowest order in the $\kappa$-expansion, that is, with the bulk
theory. The reduction of the bulk theory leads to a well-defined
four-dimensional supergravity theory in its own right. Recall that
the superpotential and D-term vanish when one reduces 11-dimensional
supergravity on a $G_2$ space. Also, we have no gauge fields to
consider since our $G_2$ orbifolds do not admit two-forms. Thus we
need only specify the K\"ahler potential $K_0$ to determine the
four-dimensional effective theory. Some of the technicalities of
computing this for a general $G_2$ space were laid out in
Chapter~\ref{revchap}. The K\"ahler potential is proportional to the
logarithm of the volume of the $G_2$ space, which for the case of
the $G_2$ orbifold $\mathcal{Y}=\mathcal{T}^7/\Gamma$ is
proportional to the product of the seven radii $R^A$ in
Eq.~\eqref{metric1}. To obtain the precise formula we use
equation~\eqref{Kformula1} and find
\begin{equation} \label{kahler1}
K_0=-\frac{1}{\kappa_4^2}\sum_{A=0}^6 \ln \left( T^A + \bar{T}^A
\right)+\frac{7}{\kappa_4^2}\ln 2\, .
\end{equation}
Here the four-dimensional Newton constant $\kappa_4$ is related to its
11-dimensional counterpart by
\begin{equation} \label{kappa114}
\kappa_{11}^2=\kappa_4^2v_7\, , \qquad
v_7=\int_{\mathcal{Y}}\mathrm{d}^7x \, .
\end{equation}
As one should expect, this K\"ahler potential $K_0$ coincides exactly with
the bulk contribution for the case of the
blown-up orbifold (see Eq.~\eqref{K}).
\\

Next we perform the reduction
of the seven-dimensional Yang-Mills
theory on the singular locus $\mathcal{T}^3$.
It is useful in this instance
to consider together both the case of the generic $\mathbb{Z}_N$,
$N>2$, singularity, and $\mathbb{Z}_2$.
The difference between the two cases was that for the $\mathbb{Z}_2$
case, some extra bulk field components survived the orbifolding.
Therefore, we can apply the seven-dimensional theory valid for the
$\mathbb{Z}_2$ case to the completely general case, keeping in mind that
for $N>2$ the $\mathbb{Z}_N$ orbifold symmetry
imposes some constraints on fields that are projected from the bulk.
(In any case, for many orbifold groups, there will be more constraints
arising from the other generators.)
Ultimately all constraints will manifest themselves as the usual ones on the
moduli superfields $T^A$, whereby some of them must be set equal to one
another, following the prescription in the discussion on bulk zero modes
in the previous section.

We need to express the truncated bulk fields $F^I_{\hat{\mu}\hat{\nu}}$,
${\ell_I}^{\ul{J}}$ and $\tau$ in terms of the bulk metric moduli
$a^A$ and the bulk axions $\nu^A$. This is done by using the
formulae~\eqref{metric1},~\eqref{period2},~\eqref{Cexp1}
for the 11-dimensional fields in terms of $a^A$ and $\nu^A$, together
with the field identifications~\eqref{5id1},~\eqref{5idZ2},~\eqref{5idn}
between 11-dimensional and seven-dimensional fields.
We find that the only non-vanishing
components of $F_{\hat{\mu}\hat{\nu}}^I$ are some of the mixed
components $F^I_{\mu m}$, and these are given by
\ba
F^1_{\mu 3} = -\frac{1}{2}\left( \partial_\mu \nu^{31}
+ \partial_\mu \nu^{32} \right)\,, &&
F^2_{\mu 2} = \frac{1}{2}\left( \partial_\mu \nu^{21}
+ \partial_\mu \nu^{22} \right)\,,\nn \\
F^3_{\mu 1} = -\frac{1}{2}\left( \partial_\mu \nu^{11}
+ \partial_\mu \nu^{12} \right)\,,
 && F^4_{\mu 1} = \frac{1}{2}\left( -\partial_\mu \nu^{11}
+\partial_\mu \nu^{12} \right)\,,\\
F^5_{\mu 2} = \frac{1}{2}\left( \partial_\mu \nu^{21}
- \partial_\mu \nu^{22} \right)\,, &&
F^6_{\mu 3} = \frac{1}{2}\left( \partial_\mu \nu^{31} - \partial_\mu \nu^{32} \right)\, .\nn
\ea
For the coset matrix $\ell$, which is symmetric, we find the non-zero components
\ba
{\ell_1}^1={\ell_6}^6=\frac{a^{31}+a^{32}}{2\sqrt{a^{31}{a^{32}}}}, && {\ell_1}^6={\ell_6}^1=\frac{a^{31}-a^{32}}{2\sqrt{a^{31}{a^{32}}}}\,,\nn \\
{\ell_2}^2={\ell_5}^5=\frac{a^{21}+a^{22}}{2\sqrt{a^{21}{a^{22}}}}, && {\ell_2}^5={\ell_5}^2=\frac{-a^{21}+a^{22}}{2\sqrt{a^{21}{a^{22}}}}\,, \\
{\ell_3}^3={\ell_4}^4=\frac{a^{11}+a^{12}}{2\sqrt{a^{11}{a^{12}}}}, && {\ell_3}^4={\ell_4}^3=\frac{-a^{11}+a^{12}}{2\sqrt{a^{11}{a^{12}}}}\,. \nn
\ea
Finally, we have the following relation for the orbifold scale factor $\tau$:
\begin{equation}
e^{\tau} = (a^0)^{-2/3}\prod_{m=1}^3\left(a^{m1}a^{m2}\right)^{1/3}.
\end{equation}
We now present the results of our reduction of bosonic terms at the
singularity. We neglect terms of the form $\mathcal{C}^n(\partial
T)^2$, where $n\geq 2$, and thus neglect the back-reaction term
$\delta\mathcal{S}_7$ in Eq.~\eqref{dS} completely. From the bosonic
part of the seven-dimensional action~\eqref{4boseaction} we get the following
terms, divided up into
scalar kinetic terms, gauge-kinetic terms and scalar potential:
\ba
\mathcal{L}_{4,\mathrm{kin}} &=&
 -\frac{1}{2{\lambda_4}^2}\sqrt{-g}\sum_{m=1}^3\left\{ \frac{1}{a^{m1}a^{m2}}
 \left(\mathcal{D}_\mu\rho_a^m\mathcal{D}^\mu\rho^{am}
 +\mathcal{D}_\mu b_a^m\mathcal{D}^\mu b^{am}\right)\right. \nn \\
 && \hspace{3.1cm}- \frac{1}{3} \sum_{A=0}^6 \frac{1}{a^{m1}a^{m2}a^A}
\partial_\mu a^A\left( \rho_a^m\mathcal{D}^\mu\rho^{am}+b_a^m\mathcal{D}^\mu b^{am}
 \right) \nn \\
&&  \hspace{3.1cm}- \frac{1}{(a^{m1})^2a^{m2}}\rho_{a}^m\left( \partial_\mu
    \nu^{m1}\mathcal{D}^\mu b^{am} + \partial_\mu a^{m1} \mathcal{D}^\mu \rho^{am} \right)
    \nn \\
&& \hspace{3.1cm}\left. -\frac{1}{a^{m1}(a^{m2})^2}\rho_a^m\left(
    \partial_\mu \nu^{m2}\mathcal{D}^\mu b^{am} + \partial_\mu a^{m2}
    \mathcal{D}^\mu \rho^{am} \right) \right\},\label{kinactual} \\
\mathcal{L}_{4,\mathrm{gauge}}&=&-\frac{1}{4\lambda_4^2}\sqrt{-g}\left(a^0
    F_{\mu\nu}^aF^{\mu\nu}_a-\frac{1}{2}\nu^0\epsilon^{\mu\nu\rho\si}
    F^a_{\mu\nu}F_{a\rho\si}\right), \label{gauge}\\
\mathcal{V}&=&\frac{1}{4\lambda_4^2a^0}\sqrt{-g}{f^a}_{bc}f_{ade}
    \sum_{m,n,p=1}^3 \epsilon_{mnp}\frac{1}{a^{n1}a^{n2}a^{p1}a^{p2}}
    \left( \rho^{bn}\rho^{dn}\rho^{cp}\rho^{ep}+\rho^{bn}\rho^{dn}b^{cp}b^{ep}\right. \nn \\
    &&\hspace{6.5cm}\left. +b^{bn}b^{dn}\rho^{cp}\rho^{ep}+b^{bn}b^{dn}b^{cp}b^{ep}\right).
    \label{scalarpot}
\ea
The four-dimensional gauge coupling $\lambda_4$ is related to the seven-dimensional analogue by
\begin{equation} \label{la74}
\lambda_4^{-2}=v_3\lambda_7^{-2}\, ,\qquad v_3=\int_{\mathcal{T}^3}\mathrm{d}^3x\, ,
\end{equation}
where $v_3$ is the reference volume for the three-torus.
Note that the above matter field action is suppressed relative to the
gravitational
action by a factor $\zeta_4^2=\kappa_4^2/\lambda_4^2\sim \kappa_{11}^{4/3}$,
as mentioned earlier.
\\

\section{Finding the superpotential and K\"ahler potential}\label{n=1}
The above reduced action must be the bosonic part of a
four-dimensional $\mathcal{N}=1$ supergravity and we would now like
to determine the associated K\"ahler potential and superpotential.
We start by combining the information from the expression
\eqref{kahler1} for the bulk K\"ahler potential $K_0$ descending
from 11-dimensional supergravity, with the matter field terms
\eqref{kinactual}--\eqref{scalarpot} descending from the
singularity, to obtain the full K\"ahler potential. In general, one
cannot expect the definition~\eqref{TA} of the moduli in terms of
the underlying geometrical fields to remain unchanged in the
presence of additional matter fields. We, therefore, begin by
writing the most general form for the correct superfield
$\tilde{T}^A$ in the presence of matter fields as
\begin{equation}
\ti{T}^A=T^A + F^A\left( T^B, \bar{T}^B, \mathcal{C}_m^a, \bar{\mathcal{C}}_m^a \right)
 \,.\label{Tcorr}
\end{equation}
Analogously, the most general form of the K\"ahler potential in the presence
of matter can be written as
\begin{equation}
K=K_0+K_1\left( T^A, \bar{T}^A, \mathcal{C}_m^a, \bar{\mathcal{C}}_m^a \right)\,. \label{Kcorr}
\end{equation}
Given this general form for the superfields and the K\"ahler potential, we can work
out the resulting matter field kinetic terms by taking second derivatives of $K$ with
respect to $\ti{T}^A$ and $\mathcal{C}_a^m$. Neglecting terms of order $C^n(\partial T)^2$,
as we have done in the reduction to four dimensions, we find
\ba \label{kingeneric}
\mathcal{L}_{4,\mathrm{kin}}&=&-\sqrt{-g}\left\{ \sum_{m,n=1}^3
\frac{\partial^2K_1}{\partial \mathcal{C}_a^m \partial \bar{\mathcal{C}}_b^n}
\mathcal{D}_\mu\mathcal{C}_a^m\mathcal{D}^\mu\bar{\mathcal{C}}_b^n+
\left( 2\sum_{m=1}^3\sum_{A=0}^6\frac{\partial^2K_1}{\partial \mathcal{C}_a^m
 \partial \bar{T}^{A}}\mathcal{D}_\mu\mathcal{C}_a^m\partial^\mu \bar{T}^A+
 \mathrm{c.c.} \right)\right. \nn \\
&& \hspace{1.6cm} + \left. \left( \sum_{A,B=0}^6 \frac{\partial^2K_0}{\partial T^A
   \partial \bar{T}^B}\partial_\mu T^A \partial^\mu \bar{F}^B + \mathrm{c.c.} \right)\right\}\, .
\ea
By matching kinetic terms \eqref{kinactual} from the reduction with
the kinetic terms in the above equation \eqref{kingeneric} we can
uniquely determine the expressions for the superfields $\ti{T}^A$
and the K\"ahler potential. They are given respectively by
\begin{equation}
\ti{T}^A = T^A -\frac{1}{24\lambda_4^2}\left( T^A+\bar{T}^A \right) \sum_{m=1}^3
 \frac{\mathcal{C}_a^m\bar{\mathcal{C}}^{am}}{(T^{m1}+\bar{T}^{m1})(T^{m2}+\bar{T}^{m2})}\,,
\end{equation}
\begin{equation}
K= \frac{7}{\kappa_4^2}\ln 2 - \frac{1}{\kappa_4^2}\sum_{A=0}^6
\ln (\ti{T}^A + \bar{\ti{T}}^A ) + \frac{1}{4\lambda_4^2}
\sum_{m=1}^3\frac{(\mathcal{C}_a^m+\bar{\mathcal{C}}_a^m)(\mathcal{C}^{am}+
\bar{\mathcal{C}}^{am})}{(\ti{T}^{m1}+\bar{\ti{T}}^{m1})(\ti{T}^{m2}+\bar{\ti{T}}^{m2})}\,.
\label{1singkahler}
\end{equation}
\\

We now come to the computation of the gauge-kinetic function $f_{ab}$
and the superpotential $W$. The former is straightforward to read off
from the gauge-kinetic part \eqref{gauge} of the reduced action and is given by
\begin{equation}
f_{ab}=\frac{1}{\lambda_4^2}\ti{T}^0\delta_{ab}.
\end{equation}
To find the superpotential, we can compare the scalar potential \eqref{scalarpot} of
the reduced theory to the standard supergravity formula \cite{Wess} for the scalar potential
\begin{equation}
\mathcal{V}=\frac{1}{\kappa_4^4}\sqrt{-g}e^{\kappa_4^2K}\left( K^{X\bar{Y}}\mathcal{D}_X W
\mathcal{D}_{\bar{Y}} \bar{W} - 3 \kappa_4^2 \lvert W \rvert^2\right) +
\sqrt{-g}\frac{1}{2\kappa_4^4}(\mathrm{Re}f)^{-1ab}D_a D_b\, ,
\end{equation}
taking into account the above results for the K\"ahler potential and the gauge
kinetic function. This leads to the superpotential and D-terms
\ba
W &=& \frac{\kappa_4^2}{24\lambda_4^2}f_{abc}\sum_{m,n,p=1}^3
\epsilon_{mnp}\mathcal{C}^{am}\mathcal{C}^{bn}\mathcal{C}^{cp}\,,
\label{1singsuperpot} \\
D_a &=& \frac{2i\kappa_4^2}{\lambda_4^2}f_{abc}\sum_{m=1}^{3} \frac{\mathcal{C}^{bm}\bar{\mathcal{C}}^{cm}}{(\ti{T}^{m1}+\bar{\ti{T}}^{m1}) (\ti{T}^{m2}+\bar{\ti{T}}^{m2})}\,.
\ea
It can be checked that these D-terms are consistent with the gauging of
the $\mathrm{SU}(N)$ K\"ahler potential isometries, as they should be.
\\

We are now ready to write down our formulae for the quantities that
specify the four-dimensional effective supergravity for M-theory on
$\mathcal{Y}=\mathcal{T}^7/\Gamma$, including the contribution from all singularities.
To do this we simply introduce a sum over the singularities.

Let us present the notation we need to write down these results. As
in Chapter \ref{Classification}, we introduce a label $(\tau,s)$ for
each singularity, the index $\tau$ labelling the generators of the
orbifold group, and $s$ labelling the $M_\tau$ fixed points
associated with the generator $\alpha_\tau$. We write $N_\tau$ for
the order of the generator $\alpha_\tau$. For the sixteen example
orbifold groups, these integer numbers can be computed from
information provided in Chapter~\ref{Classification}. Thus, near a
singular point, $\mathcal{Y}$ takes the approximate form
$\mathcal{T}^3_{(\tau,s)}\times \mathbb{C}^2/\mathbb{Z}_{N_\tau}$,
where $\mathcal{T}^3_{(\tau,s)}$ is a three-torus. The matter fields
at the singularities we denote by $(\mathcal{C}^{(\tau,s)})_a^m$ and
it is understood that the index $a$ transforms in the adjoint of
$\mathrm{SU}(N_\tau)$. The gauge couplings depend only on the type
of singularity, that is on the index $\tau$, and are denoted by
$\la_{(\tau)}$. M-theory determines the values of these gauge
couplings, and they can be derived using equations
\eqref{couplerel}, \eqref{kappa114} and \eqref{la74}. We find
\begin{equation}
\la_{(\tau)}^2=\left( 4\pi \right)^{4/3}
\frac{v_7^{1/3}}{v_3^{(\tau)}}\kappa_4^{2/3}\,
\end{equation}
in terms of the reference volumes $v_7$ for $\mathcal{Y}$ and
$v_3^{(\tau)}$ for $\mathcal{T}^3_{(\tau,s)}$.

The respective formulae for the moduli superfields, K\"ahler potential, superpotential and D-term potential are
\begin{eqnarray}
\ti{T}^A \!&\!=\!&\! T^A -\left( T^A+\bar{T}^A \right) \sum_{\tau,s,m}
\frac{1}{24\la_{(\tau)}^2}\frac{(\mathcal{C}^{(\tau,s)})_a^m(\bar{\mathcal{C}}^{(\tau,s)})^{am}}{(T^{B(\tau,m)}+\bar{T}^{B(\tau,m)})(T^{C(\tau,m)}+\bar{T}^{C(\tau,m)})}\,,\label{superfield}\\
K\!&\!=\!&\! - \frac{1}{\kappa_4^2}\sum_{A=0}^6 \ln (\ti{T}^A + \bar{\ti{T}}^A )
+\frac{7}{\kappa_4^2}\ln 2 \nn \\
&& +
\sum_{\tau,s,m}\frac{1}{4\lambda_{(\tau)}^2}\frac{\left[(\mathcal{C}^{(\tau,s)})_a^m+(\bar{\mathcal{C}}^{(\tau,s)})_a^m\right]\left[(\mathcal{C}^{(\tau,s)})^{am}+(\bar{\mathcal{C}}^{(\tau,s)})^{am}\right]}{(\ti{T}^{B(\tau,m)}+\bar{\ti{T}}^{B(\tau,m)})(\ti{T}^{C(\tau,m)}+\bar{\ti{T}}^{C(\tau,m)})}\,,
\label{fullkahler} \\
 \label{fullsuperpot}
W \!&\!=\!&\! \frac{1}{24}\sum_{\tau,s,m,n,p}\frac{\kappa_4^2}{\lambda_{(\tau)}^2}f_{abc} \epsilon_{mnp}(\mathcal{C}^{(\tau,s)})^{am}(\mathcal{C}^{(\tau,s)})^{bn}(\mathcal{C}^{(\tau,s)})^{cp}\,,\\
 \label{fullD}
D_a \!&\!=\!&\! 2i\sum_{\tau,s,m}\frac{\kappa_4^2}{\lambda_{(\tau)}^2}f_{abc} \frac{(\mathcal{C}^{(\tau,s)})^{bm}(\bar{\mathcal{C}}^{(\tau,s)})^{cm}}{(\ti{T}^{B(\tau,m)}+\bar{\ti{T}}^{B(\tau,m)})(\ti{T}^{C(\tau,m)}+\bar{\ti{T}}^{C(\tau,m)})}\,.
\end{eqnarray}
The index functions $B(\tau,m)$, $C(\tau,m)\in \{0,1,\ldots,6\}$ indicate
by which two of the seven moduli the matter fields are divided by in
equations \eqref{superfield}, \eqref{fullsuperpot} and
\eqref{fullD}. Their values depend only on the generator index $\tau$
and the R-symmetry index $m$. They may be calculated from the formula
\begin{equation}
a^{B(\tau,m)}a^{C(\tau,m)}=\frac{\left( R^m_{(\tau)} \right) ^2
\prod_A R^A}{\prod_n R^n_{(\tau)}}\, ,
\end{equation}
where $R^m_{(\tau)}$ denote the radii of the three-torus
$\mathcal{T}^3_{(\tau,s)}$. The possible
values of the index functions are given in Table~\ref{tableG2sing}.
\begin{table}[t]
\begin{center}
\begin{tabular}{|c|c|c|c|c|}
\hline
$\mathrm{Fixed \, directions\, of\, } \alpha_\tau$ &
$(B(\tau,1),C(\tau,1))$ & $(B(\tau,2),C(\tau,2))$ & $(B(\tau,3),C(\tau,3))$ & $h(\tau)$ \\
\hline
(1,2,3) & (1,2) & (3,4) & (5,6) & 0\\
\hline
(1,4,5) & (0,2) & (3,5) & (4,6) & 1\\
\hline
(1,6,7) & (0,1) & (3,6) & (4,5) & 2\\
\hline
(2,4,6) & (0,4) & (1,5) & (2,6) & 3\\
\hline
(2,5,7) & (0,3) & (1,6) & (2,5) & 4\\
\hline
(3,4,7) & (0,6) & (1,3) & (2,4) & 5\\
\hline
(3,5,6) & (0,5) & (1,4) & (2,3) & 6\\
\hline
\end{tabular}
\caption{Values of the index functions $(B(\tau,m),C(\tau,m))$ and
$h(\tau)$ that appear in the superfield definitions,
K\"ahler potential, D-term potential and gauge-kinetic functions.}
\label{tableG2sing}
\end{center}
\end{table}

There is a universal gauge-kinetic function for each $\mathrm{SU}(N_\tau )$
gauge theory given by
\begin{equation}
f_{(\tau)}=\frac{1}{\la_{(\tau)}^2}\ti{T}^{h(\tau)}\,,
\end{equation}
where $\ti{T}^{h(\tau)}$ is the modulus that corresponds to the volume
of the fixed three-torus $\mathcal{T}^3_{(\tau,s)}$ of the symmetry
$\alpha_\tau$. The value of $h(\tau)$ in terms of the fixed directions
of $\alpha_\tau$ is given in Table~\ref{tableG2sing}.
\\

\section{Comparison with results for smooth $G_2$ spaces}\label{compare}
In Chapter~\ref{Classification} we described how to construct a
smooth $G_2$ manifold $\mathcal{Y}^{\mathrm{S}}$ by blowing up the
singularities of the $G_2$ orbifold $\mathcal{Y}$. The moduli
K\"ahler potential for M-theory on this space was computed in
Section~\ref{ModKP}. Let us present the formula for this again,
focusing on the contribution from a single singularity:
\begin{equation} \label{Ksing}
K =  -\frac{1}{\kappa_4^2}\sum_{A=0}^{6}\ln (T^A+\bar{T}^A) +
     \frac{2}{N c_\Gamma\kappa_4^2}\sum_m\frac{\sum_{i\leq j}
     \left(\sum_{k=i}^{j}(U^{km}+\bar{U}^{km})\right)^2}{(T^{m1}+
     \bar{T}^{m1})(T^{m2}+\bar{T}^{m2})} + \frac{7}{\kappa_4^2}\ln2\, .
\end{equation}
Here, as usual, $T^A$ are the bulk moduli and $U^{im}$ are the blow-up
moduli. As for the formula for the singular manifold, the index $m$
can be thought of as labelling the directions on the three-torus
transverse to the particular blow-up. Each blow-up modulus is
associated with a two-cycle within the blow-up of a given singularity,
and the index $i,j,\ldots =1,\ldots, (N-1)$ labels these. Finally,
$c_\Gamma$ is a constant, dependent on the orbifold group.

In computing the K\"ahler potential \eqref{Ksing}, the M-theory action
was taken to be 11-dimensional supergravity, and so the result is
valid when all the moduli, including blow-up moduli, are large
compared to the Planck length. Therefore, the above result for the
K\"ahler potential cannot be applied to the orbifold limit, where
$\mathrm{Re}(U^{im})\rightarrow 0$. However, the corresponding singular result
\eqref{1singkahler} can be used to consider the case of
small blow-up moduli. To do this
explicitly, it is useful to recall from Chapter~\ref{117}
that the geometrical procedure
of blowing up a singularity can be described within the $\mathrm{SU}(N)$
gauge theory as a Higgs effect, whereby VEVs are assigned to the
(real parts of the) Abelian matter fields $\mathcal{C}^{im}$ along
the D-flat directions. This generically breaks $\mathrm{SU}(N)$ to
its maximal Abelian subgroup $\mathrm{U}(1)^{N-1}$ and leaves only the
$3(N-1)$ chiral multiplets $\mathcal{C}^{im}$ massless.
This field content corresponds precisely to the zero modes of M-theory
on the blown-up geometry, with the Abelian gauge fields arising as zero
modes of the M-theory three-form on the $N-1$ two-spheres of the blow-up
and the chiral multiplets corresponding to its moduli. Note that the
non-Abelian components of the matter fields $\mathcal{C}^{am}$
correspond to membrane states that are massless only in the singular
limit.

With the above interpretation of the blowing up procedure, by
switching off the non-Abelian components of $\mathcal{C}^{am}$ in
equation~\eqref{1singkahler}, one obtains a formula for the moduli
K\"ahler potential for M-theory on $\mathcal{Y}^{\mathrm{S}}$ with
small blow-up
moduli. At first glance this is slightly different from the
smooth result~\eqref{Ksing} which contains a double-sum over the Abelian
gauge directions. However, we can show that they are actually equivalent.
First, we identify the bulk moduli $T^A$ in~\eqref{Ksing} with $\ti{T}^A$ in
\eqref{1singkahler}.
One obvious way of introducing a double-sum into the singular result
\eqref{1singkahler} is to introduce a non-standard basis $X_i$ for the Cartan
sub-algebra of $\mathrm{SU}(N)$, which introduces a metric
\begin{equation}
 \label{kij}
 \kappa_{ij}=\mathrm{tr}(X_iX_j)\, .
\end{equation}
Neglecting an overall rescaling of the fields, identification of the smooth
and singular results for $K$ then requires the identity
\begin{equation} \label{identity}
\sum_{i,j}\kappa_{ij}(\mathcal{C}^{im}+\bar{\mathcal{C}}^{im})(\mathcal{C}^{jm}+
\bar{\mathcal{C}}^{jm})=\sum_{i\leq j}\left(\sum_{k=i}^{j}(U^{(k,m)}+\bar{U}^{(k,m)})\right)^2
\end{equation}
to hold. So far we have been assuming the canonical choice $\kappa_{ij}=\delta_{ij}$,
which is realised by the standard generators
\begin{equation}
X_1= \frac{1}{\sqrt{2}}\mathrm{diag}(1,-1,0,\ldots,0)\,, \hspace{0.3cm} X_2= \frac{1}{\sqrt{6}}\mathrm{diag}(1,1,-2,0,\ldots,0)\, , \hspace{0.3cm}  \cdots \, , \nn
\end{equation}
\begin{equation}
X_{N-1}= \frac{1}{\sqrt{N(N-1)}}\mathrm{diag}(1,\ldots,1,-(N-1))\,.
\end{equation}
Clearly, the relation~\eqref{identity} cannot be satisfied with a holomorphic
relation between fields for this choice of generators. Instead, from the
RHS of Eq.~\eqref{identity} we need the metric $\kappa_{ij}$ to be
\begin{equation} \label{killingmetric}
\kappa_{ij}= \left\{ \begin{array}{c}
 (N-j)i\, , \hspace{0.3cm} i\leq j\, ,\\
 (N-i)j\, , \hspace{0.3cm} i>j\, .
\end{array} \right.
\end{equation}
From Eq.~\eqref{identity} this particular metric $\kappa_{ij}$ is
positive definite and, hence, there is always a choice of generators
$X_i$ which reproduces this metric via Eq.~\eqref{kij}. For the simplest case $N=2$,
there is only one generator $X_1$ and the above statement becomes trivial.
For the $N=3$ case, a possible choice for the two generators $X_1$ and $X_2$ is
\begin{equation}
X_1=\mathrm{diag}(0,-1,1)\,, \hspace{0.3cm} X_2=\mathrm{diag}(1,0,-1)\,.
\end{equation}
Physically, these specific choices of generators tell us how the Abelian
group $\mathrm{U}(1)^{N-1}$ which appears in the smooth case is embedded into the
$\mathrm{SU}(N)$ group which is present in the singular limit.

In addition to the agreement of K\"ahler potentials for small and large
blow-up moduli, there is also consistency in the vanishing superpotential
for both cases. To see this for small blow-up moduli,
one observes that the right-hand side
of~\eqref{1singsuperpot} vanishes when restricted to the Abelian fields
$\mathcal{C}^{im}$.

The precise matching of the K\"ahler potentials in the
small and large blow-up limits is perhaps somewhat
surprising given that there does not
seem to be a general reason why the results for
large blow-up moduli should not
receive corrections when the supergravity approximation breaks down as one
moves towards the singular limit. At any rate, our result allows us to deal
with M-theory compactifications close to and at the singular limit of
co-dimension four $A$-type singularities. This opens up a whole range of
applications, for example, in the context of wrapped branes and their
associated low-energy physics.


\chapter{Further Analysis of M-theory on a $\boldsymbol{G_2}$ Orbifold}
\label{Analysis}

In this final chapter, we study some further properties of M-theory
on the $G_2$ orbifold $\mathcal{Y}=\mathcal{T}^7/\Gamma$. The first two
sections are concerned with
the effects of more complicated backgrounds. In Section~\ref{wilson} we
explore explicit patterns for the breaking
of gauge symmetry via Wilson lines. Then in Section~\ref{FLUX}
we consider flux. We show
explicitly that the superpotential for Abelian flux of the
$\mathrm{SU}(N)$ gauge fields on $\mathcal{T}^3$ is
obtained from a Gukov-type
formula consisting of the integral of the complexified Chern-Simons
form. In addition, we show that
the superpotential~\eqref{1singsuperpot} for the fluctuating parts of the
gauge fields descends from the same Gukov formula,
an observation first made in Ref.~\cite{AcharyaMod}.
The flux superpotential we find provides a further confirmation for the
matching between M-theory on singular and smooth $G_2$ manifolds;
it takes the same form in both limits if the field identification suggested
by the comparison of K\"ahler potentials is used.

In Section~\ref{N=4}, we consider one of the $\mathcal{N}=4$
super-Yang-Mills sectors of the four-dimensional
effective action with gravity switched off. We show how S-duality, in this
context, translates into a T-duality on the
singular $\mathcal{T}^3$ locus.
Furthermore, we speculate about a possible extension of this S-duality
to the full supergravity theory. Finally, we discuss how singular
and blown-up geometries correspond respectively
to the conformal and Coulomb phases of the  $\mathcal{N}=4$ theory.
\\

\section{Wilson lines}\label{wilson}
In this section we discuss breaking of the $\mathrm{SU}(N)$ gauge symmetry
through inclusion of Wilson lines in the internal three-torus $\mathcal{T}^3$.
As preparation, let us briefly recall the main features of Wilson-line
breaking~\cite{GSW},~\cite{Witten:1985xc},~\cite{Breit},~\cite{Cvetic}. A Wilson line
is a configuration
of the (internal) gauge field  $A^{a}$ with vanishing associated field
strength. For a non-trivial Wilson-line to be possible, the first fundamental group, $\pi_1$,
of the internal space needs to be non-trivial, a condition satisfied in our case,
as $\pi_1(\mathcal{T}^3)=\mathbb{Z}^3$. Practically, a Wilson line around a non-contractible loop $\gamma$ can be
described by
\begin{equation}
U_{\gamma}=P\exp\left(  -i\oint_{\gamma}X_{a}A^{a}{}_{m}dx^{m}\right)
\end{equation}
where $X_{a}$ are the generators of the Lie algebra of the gauge
group, $G$. This expression induces a group homomorphism,
$\gamma\mapsto U_{\gamma}$, between the fundamental group and the
gauge group of our theory.

We can explicitly determine the possible symmetry breaking patterns by
examining particular embeddings (that is, choices of representation) of
the fundamental group into the gauge group. For convenience, we will
focus on gauge groups $\mathrm{SU}(N)$, where $N=2,3,4,6$, since these are
the gauge groups that arise from the explicit constructions of $G_2$
orbifolds in Chapter~\ref{Classification}. We write a generic group element
of $\pi _{1}(\mathcal{T}^3)=\mathbb{Z}^{3}$
as a triple of integers $(n,m,p)$, and take addition as the group
multiplication.
Then a possible embedding of $\mathbb{Z}^3$ in, for example, $\mathrm{SU}(4)$ can be defined by
\begin{equation}
(n,m,p)\mapsto\left(
\begin{array}
[c]{ccc}%
e^{in}{\bf 1}_{2\times2} &  & \\
& 1 & \\
&  & e^{-2 in}%
\end{array}
\right)\,,
\end{equation}
which will clearly break the symmetry to $\mathrm{SU}(2)\times \mathrm{U}(1)\times \mathrm{U}(1)$. There
is, however, a great deal of redundancy in these choices of embedding and the
homomorphisms we define are clearly not unique. For example, we could have
instead chosen the map so as to take $(n,m,p)$ to an element in the
subgroup
$\mathrm{SU}(2)\times \mathrm{SU}(2)\times \mathrm{U}(1)$, say
\begin{equation}
(n,m,p)\mapsto \left(
\begin{array}
[c]{ccc}%
(-1)^{n}{\bf 1}_{2\times2} &  & \\
& e^{ im} & \\
&  & e^{- im}%
\end{array}
\right)\, ,
\end{equation}
which would also break $\mathrm{SU}(4)$ to $\mathrm{SU}(2)\times \mathrm{U}(1)\times \mathrm{U}(1)$. It is interesting to note the types of reduced
symmetry possible with Wilson lines. For example, the embedding
\begin{equation}
(n,m,p)\mapsto\left(
\begin{array}
[c]{ccc}%
e^{in}{\bf 1}_{2\times2} &  & \\
& e^{\frac{-2in}{3}}{\bf 1}_{3\times3} & \\
&  & 1
\end{array}
\right)\,
\end{equation}
of $\mathbb{Z}^{3}$ into $\mathrm{SU}(6)$ breaks this gauge group to
$\mathrm{SU}(3)\times \mathrm{SU}(2)\times \mathrm{U}(1)\times
\mathrm{U}(1)$, which contains the symmetry group of the Standard
Model. (One should note, however, that even in this case, our theory
does not contain the particle content of the Standard Model.)

Having given a number of examples, we now classify in general, which unbroken
subgroups of $\mathrm{SU}(N)$ are possible (using the group-theoretical tools
provided in Ref.~\cite{Groups}). It is clear that the generic unbroken
subgroup is the Cartan sub-algebra
$\mathrm{U}(1)^{N-1}$. However, as we have already demonstrated in the above
examples, certain choices of embedding leave a larger symmetry group intact.
These special choices are interesting, although there always exist small deformations
of such choices that lead to embeddings that break to the generic
$\mathrm{U}(1)^{N-1}$ unbroken subgroup. To see an example of this, consider
the mapping of $\mathbb{Z}^{3}$ into $\mathrm{SU}(3)$ defined by
\begin{equation}
(n,m,p)\mapsto\left(
\begin{array}
[c]{ccc}%
e^{i\alpha m+ip} &  & \\
& e^{i\alpha n+ip} & \\
&  & e^{-i\alpha(n+m)-2ip}%
\end{array}
\right)\,,
\end{equation}
where $\alpha$ is a parameter that may be varied. For general values of
$\alpha$ this embedding breaks $\mathrm{SU}(3)$ to $\mathrm{U}(1)^{2}$,
but for the special case of
$\alpha=0$ there is a larger unbroken gauge group, namely
$\mathrm{SU}(2)\times \mathrm{U}(1)$.

We find that Wilson lines can
break the $\mathrm{SU}(N)$ symmetry group to any subgroup of rank $N-1$, and
the results for all possible unbroken gauge groups are summarised in
Table~\ref{tab:chap6}.
\begin{table}[ptb]
\begin{center}%
\begin{tabular}
[c]{|c|c|}\hline
Gauge Group & Residual Gauge Groups from Wilson lines\\\hline
$\mathrm{SU}_{2}$ & $\mathrm{U}_1$\\\hline
$\mathrm{SU}_{3}$ & $\mathrm{SU}_{2}\times \mathrm{U}_1$, $\mathrm{U}_1^{2}$\\\hline
$\mathrm{SU}_{4}$ & $\mathrm{SU}_{3}\times \mathrm{U}_1$, $\mathrm{SU}_{2}\times \mathrm{U}_1^{2}$, $\mathrm{SU}_{2}^{2}\times
\mathrm{U}_1$, $\mathrm{U}_1^{3}$\\\hline
$\mathrm{SU}_{6}$ & $\mathrm{SU}_{5}\times \mathrm{U}_1$, $\mathrm{SU}_{4}\times \mathrm{U}_1^{2}$, $\mathrm{SU}_{2}\times
\mathrm{SU}_{3}\times \mathrm{U}_1^{2}$, $\mathrm{SU}_{2}^{2}\times \mathrm{U}_1^{3}$, $\mathrm{SU}_{2}\times \mathrm{U}_1%
^{4}$,\\
& $\mathrm{SU}_{3}\times \mathrm{U}_1^{3}$, $\mathrm{SU}_{2}\times \mathrm{SU}_{4}\times \mathrm{U}_1$, $\mathrm{SU}_{2}%
^{3}\times \mathrm{U}_1^{2}$, $\mathrm{SU}_{3}^{2}\times \mathrm{U}_1$, $\mathrm{U}_1^{5}$\\\hline
\end{tabular}
\end{center}
\caption{The symmetry group reductions in the presence of Wilson lines}
\label{tab:chap6}
\end{table}
Note that the Cartan subgroups are included as the last entries for
each of the gauge groups.

It is worth noting briefly that we can view this symmetry breaking by
Wilson lines in an alternate light in four dimensions. Rather than
consider a seven-dimensional compactification and Wilson lines, we
could obtain the same results by turning on VEVs for certain
directions of the scalar fields in our four-dimensional theory.\footnote{In fact,
the scalars which directly correspond to Wilson lines in seven dimensions
are the axionic, Abelian parts of the fields ${\cal C}^{am}$.} For
example, if we give generic VEVs to all the Abelian directions of the
scalar fields in Eq.~\eqref{scalarpot} we can break the symmetry to a
purely Abelian gauge group. This corresponds to a generic embedding in
the Wilson line picture. Likewise, we can obtain the larger symmetry
groups listed in Table~\ref{tab:chap6} by giving non-generic VEVs to the
scalar fields.

\section{$G$- and $F$-flux}\label{FLUX}
The discussion of Wilson lines can be thought of as describing
non-trivial background configurations for which we still maintain the
condition $F=0$ on the field strength. However, to gain a better understanding
of the possible vacua and their effects, we need to consider the
contributions of flux both from bulk and seven-dimensional field strengths.
Let us start with a bulk flux $G_{\mathcal{Y}}$ for the internal
part\footnote{We do not discuss flux in the external part of $G$.} of the M-theory
four-form field strength $G$. For M-theory compactifications on smooth $G_2$
spaces this was discussed in Ref.~\cite{Beasley}. In our case, all we have
to do is modify this discussion to include possible effects of the
singularities and their associated seven-dimensional gauge theories.
However, inspection of (the bosonic part of) the seven-dimensional gauge field
action~\eqref{4boseaction} shows that a non-vanishing internal $G_{\mathcal{Y}}$ will not
generate any additional contributions to the four-dimensional scalar potential,
apart from the ones descending from the bulk. Hence, we can use the
standard formula~\cite{Beasley}
\begin{equation}
W=\frac{1}{4}\int_{\mathcal{Y}}\left(  \frac{1}{2}C+i\varphi\right)  \wedge G_{\mathcal{Y}}\, ,
\label{Wfluxgen}
\end{equation}
where $C$
is the three-form of 11-dimensional supergravity and $\varphi$ is the $G_{2}%
$-structure of $\mathcal{Y}$. For a completely singular $G_2$ space, where the torus moduli
$T^A$ are the only bulk moduli, this formula leads to a flux superpotential
\begin{equation}
 W\sim n_AT^A\, ,
\end{equation}
with flux parameters $n_A$, which has to be added to the ``matter field''
superpotential~\eqref{1singsuperpot}. If some of the singularities are
blown up we also have blow-up moduli $U^{im}$ and the flux superpotential contains
additional terms, thus
\begin{equation}
 W\sim n_AT^A + n_{im}U^{im} \, . \label{Wflux}
\end{equation}

We now turn to a discussion of the seven-dimensional
super-Yang-Mills theory at the singularity. First, it is natural to
ask whether the matter field superpotential~\eqref{1singsuperpot}
can also be obtained from a Gukov-type formula, analogous to
Eq.~\eqref{Wfluxgen}, but with an integration over the
three-dimensional internal space on which the gauge theory is
compactified. To this end, we begin by defining the complexified
internal gauge field
\begin{equation}
\mathcal{C}_{a}=\rho_{am}\mathrm{d}x^{m}+ib_{am}\mathrm{d}x^{m}\, .
\end{equation}
It is worthwhile to note at this stage, that writing the real parts of these fields
(which are scalar fields in the original seven-dimensional theory) as forms is, in fact,
an example of the procedure referred to as `twisting'~\cite{Bershadsky:1995sp},~\cite{Acharya}.
In this particular case, the twisting amounts to identifying the
$R$-symmetry index ($m=1,2,3$) of our original seven-dimensional supergravity
with the tangent space indices of the three-dimensional compact space,
$\mathcal{T}^{3}$.
A plausible guess for the Gukov-formula for the seven-dimensional gauge theory is
an expression proportional to the integral of the complexified Chern-Simons form
\begin{equation}
\omega_{CS}=\left(  \mathcal{F}^{a}\wedge\mathcal{C}_{a}-\frac{1}{3}%
f_{abc}\mathcal{C}^{a}\wedge\mathcal{C}^{b}\wedge\mathcal{C}^{c}\right)
\end{equation}
over the three-dimensional internal space \cite{AcharyaMod}. Here, $\mathcal{F}$ is the
complexified field strength
\begin{equation}
\mathcal{F}^{a}=\mathrm{d}\mathcal{C}^{a}+f^{a}{}_{bc}\mathcal{C}^{b}\wedge
\mathcal{C}^{c}\,.
\end{equation}
Indeed, if we specialise to the case of vanishing flux, that is
$\mathrm{d}\mathcal{C}^a=0$, our matter field
superpotential~\eqref{1singsuperpot} is exactly reproduced by the
formula
\begin{equation}
W=\frac{\kappa_{4}^{2}}{16\lambda_{4}^{2}}\frac{1}{v_{3}}\int_{\mathcal{T}^{3}}%
\omega_{CS}\, . \label{NewGukov}%
\end{equation}

To see that Eq.~\eqref{NewGukov} also correctly incorporates the contributions of
$F$-flux, we can look at the following simple example of an Abelian $F$-flux.
Let the Abelian parts of the gauge field strength, $F^i$, be expanded in a basis of the harmonic
two-forms, $\omega_{m}=\frac{1}{2}\epsilon_{mnp}\mathrm{d}x^n\wedge \mathrm{d}x^p$, on the internal
three-torus $\mathcal{T}^3$, as
\begin{equation}
F^i=f^{im}\omega_m\, ,\label{fluxAnsatz}
\end{equation}
where $f^{im}$ are flux parameters.  Substituting this expression into the seven-dimensional
bosonic action~\eqref{4boseaction} and performing a compactification on $\mathcal{T}^3$ we find
a scalar potential which, taking into account the K\"ahler potential~\eqref{1singkahler},
can be reproduced from the superpotential
\begin{equation}
W=\frac{\kappa_{4}^{2}}{8\lambda_{4}^{2}}f_{im}\mathcal{C}^{im}\, .
\label{Wfluxabelian}
\end{equation}
This superpotential is exactly reproduced by the Gukov-type
formula~\eqref{NewGukov} which, after substituting the flux
ansatz~\eqref{fluxAnsatz}, specialises to its Abelian part. Hence,
the formula~\eqref{NewGukov} correctly reproduces the matter field
superpotential as well as the superpotential for Abelian $F$-flux.
The explicit Gukov formula for multiple singularities is analogous
to Eq.~\eqref{NewGukov}, with an additional sum to run over all
singularities as in Eq.~\eqref{fullsuperpot}. We also note that the
$F$-flux superpotential~\eqref{Wfluxabelian} is consistent with the
blow-up part of the $G$-flux superpotential~\eqref{Wflux} when the
identification of the Abelian scalar fields $\mathcal{C}^{im}$ with
the blow-up moduli $U^{im}$ is taken into account.

\section{Relation to $\mathcal{N}$=4 supersymmetric Yang-Mills theory}
\label{N=4}
In the previous two sections we have explored aspects and modifications of
the four-dimensional effective theory of M-theory
on the $G_2$ orbifold $\mathcal{Y}=\mathcal{T}^7/\Gamma$. We shall now
take a step back and look at the theory without flux, rephrasing it in order to
provide us with several new insights. The full theory is
clearly $\mathcal{N}=1$ supersymmetric, by
virtue of our choice to compactify on a $G_{2}$ holonomy
space. However, if we neglect the gravity sector (that is, in
particular hold constant the moduli $T^A$), the remaining theory is an
$\mathcal{N}=4$ super-Yang-Mills (SYM) theory. This is to be expected since we
are compactifying the seven-dimensional SYM theory on a three-torus.
In this section we make this connection more explicit by matching the
Yang-Mills part of our four-dimensional effective theory with
$\mathcal{N}=4$ SYM theory in its standard form. This connection is of
particular interest since $\mathcal{N}=4$ SYM theory is of central
importance in many current aspects of string theory, particularly in
the context of the AdS/CFT conjecture \cite{Maldacena}.

In $\mathcal{N}=1$ language, the field content of $\mathcal{N}=4$
SYM theory consists of
Yang-Mills multiplets  $(A_{\mu}^{a},\lambda^{a})$, where $a$ is a gauge index,
and a triplet of chiral multiplets $(A^a_m+iB^a_m,\chi^a_m)$ per gauge multiplet,
where $A^a_m$ and $B^a_m$ are real scalars, $\chi^a_m$ are Weyl fermions and
$m,n,\ldots = 1,2,3$. The field strength of the gauge field $A_\mu^a$
is given by $G=\mathrm{d}A$. Using this notation, the bosonic part of
the Lagrangian for $\mathcal{N}=4$ SYM theory is given
by~\cite{Gliozzi}--\cite{AdS}
\begin{eqnarray}
\mathcal{L}_{\mathcal{N}=4} \!&\!  =\! &\! -\frac{1}{4g^{2}}G_{\mu\nu}^{a}G_{a}^{\mu\nu
}+\frac{\theta}{64\pi^{2}}\epsilon^{\mu\nu\rho\sigma}G_{\mu\nu}^{a}%
G_{a\rho\sigma}-\frac{1}{2}\left(  \mathcal{D}_{\mu}A_{m}^{a}\mathcal{D}^{\mu}A_{a}^{m}%
-\frac{1}{2}\mathcal{D}_{\mu}B_{m}^{a}\mathcal{D}^{\mu}B_{a}^{m}\right)  \nonumber\\
&& +\frac{g^{2}}{4}\mathrm{tr}\left( [A_{m},A_{n}][A^{m}%
,A^{n}]+[B_{m},B_{n}][B^{m},B^{n}]+2[A_{m},B_{n}][A^{m},B^{n}]\right)\, .  \label{N4YM}%
\end{eqnarray}

Let us consider the contribution~\eqref{kinactual}--\eqref{scalarpot}
of one singularity, and hence one Yang-Mills sector, to M-theory reduced on
the $G_2$ orbifold $\mathcal{Y}$. If we switch off bulk fields,
that is, set the four-dimensional metric $g$ and the geometrical moduli $T^A$
to constants, then subject to an appropriate identification of the remaining
fields, these terms precisely reproduce $\mathcal{N}=4$
SYM theory~\eqref{N4YM}. By inspection these field identifications are given
by
\begin{align}
A_{a}^{m} &  =\frac{1}{\lambda_{4}\sqrt{a^{m1}a^{m2}}}\rho_{a}^{m}\, ,\label{Aa}\\
B_{a}^{m} &  =\frac{1}{\lambda_{4}\sqrt{a^{m1}a^{m2}}}b_{a}^{m}\, ,\\
G_{\mu\nu}^{a} &  =F_{\mu\nu}^{a}\, .
\end{align}
The $\mathcal{N}=4$ coupling constants are related to the $\mathcal{N}=1$
constants by%
\begin{equation}
g^{2} =\frac{\lambda_{4}^{2}}{a^{0}}\, ,\qquad
\theta =\frac{8\pi^{2}\nu^{0}}{\lambda_{4}^{2}}\,.\label{4couplings}%
\end{equation}

With this identification in place, it is interesting to consider the
Montonen-Olive and S-duality conjecture~\cite{MontonenOlive} in the
context of our theory. This duality acts on the complex coupling
\begin{equation}
\tau\equiv\frac{\theta}{2\pi}-\frac{4\pi i}{g^{2}}%
\end{equation}
by the standard $\mathrm{SL}(2,\mathbb{Z})$ transformation
\begin{equation} \label{mod}
\tau\rightarrow\frac{a\tau+b}{c\tau+d}%
\end{equation}
with $ad-bc=1$ and $a,b,c,d\in$ $\mathbb{Z}$. Note that these
transformations contain in particular $\tau\rightarrow
-\frac{1}{\tau}$, an interchange of strong and weak coupling.
Specifically, the S-duality conjecture is the statement that an
$\mathcal{N}=4$ Yang-Mills theory with parameter $\tau$ as defined
above and gauge group $G$, is identical to the theory with coupling
parameter transformed as in \eqref{mod} and the dual gauge group,
$\widehat{G}$. Note that here ``dual group'' refers to the Langlands
dual group, (which for $G=\mathrm{SU}(N)$ is given by $\widehat
{G}=\mathrm{SU}(N)/\mathbb{Z}_{N}$)~\cite{Dual}.

When we consider the above transformations within the context of our theory,
several interesting features emerge immediately. With the field and coupling
identifications (\ref{Aa})--(\ref{4couplings}) we have
\begin{equation}
\tau=-\frac{4\pi i\ti{T}^{0}}{\lambda_{4}^{2}}\,.
\end{equation}
Therefore, the shift symmetry $\tau\rightarrow\tau+b$ is equivalent to
an axionic shift of $\ti{T}^0$, and $\tau\rightarrow-\frac{1}{\tau}$ is given by
$\ti{T}^0\rightarrow\frac{1}{\ti{T}^0}$. Since $\mathrm{Re}(T^0)=a^{0}$ describes the
volume of the torus, $\mathcal{T}^{3}$, S-duality in the present context
is really a form of T-duality.

Bearing in mind this behavior in the Yang-Mills sector, we turn now to the
gravity sector. In a toroidal compactification of M-theory, the T-duality
transformation of $\ti{T}^0$ would be part of the U-duality group \cite{U-duality}
and would, therefore, be an exact symmetry. One may speculate that this
is still the case for our compactification on a $G_2$ orbifold and we proceed
to analyse the implications of such an assumption. Examining the
structure of our four-dimensional effective theory~\eqref{kinactual}--\eqref{scalarpot},
we see that the expressions for $K$, $W$ and
$D$ are indeed invariant under axionic shifts of $\ti{T}^0$. However, it is not
so clear what happens for $\ti{T}^0\rightarrow\frac{1}{\ti{T}^0}$.
An initial inspection of Eqs.~\eqref{kinactual}--\eqref{scalarpot}
shows that while the K\"ahler potential changes by%
\begin{equation}
\delta K\sim\ln\left(  \ti{T}^0\bar{\ti{T}}^{0}\right)\, ,
\end{equation}
the kinetic terms and superpotential will remain unchanged.
In order for the whole supergravity theory to be invariant
we need the supergravity function $\mathcal{G}=K+\ln |W|^2$
to be invariant. However, as stands, with the $\ti{T}^0$ independent
superpotential~\eqref{1singsuperpot} this is clearly not the case.
One should, however, keep in mind that this superpotential is valid only
in the large radius limit and can, therefore, in principle be subject to
modifications for small $\mathrm{Re}(T^0)$. Such a possible modification which
would make the supergravity function $\mathcal{G}$ invariant and reproduce the
large-radius result~\eqref{1singsuperpot} for large $\mathrm{Re}(T^0)$ is given by
\begin{equation}
W\rightarrow h(\ti{T}^0)W\,,
\end{equation}
where
\begin{equation}
h(  \ti{T}^0)  = \frac{1}{\eta^{2}(i\ti{T}^0)\left(
j(i\ti{T}^0)-744\right)  ^{1/12}}\,
\end{equation}
and $\eta$ and $j$ are the usual Dedekind $\eta$-function and Jacobi $j$-function.
For large $\mathrm{Re}(T^0)$ the function $h$ can be expanded as
\begin{equation}
 h(\ti{T}^0)=1+2e^{-2\pi\ti{T}^0}+\dots\, .
\end{equation}
Recalling that $\mathrm{Re}(T^0)$ measures the volume of the singular
locus $\mathcal{T}^3$, the above expansion suggests that the function $h$
may arise from membrane instantons wrapping this three-torus.
It would be interesting to verify this by an explicit membrane instanton
calculation along the lines of Ref.~\cite{Harvey:1999as}.

It is well known that there are two dynamical phases in $\mathcal{N}=4$
Yang-Mills theory in four-dimensions \cite{AdS}. A supersymmetric ground
state of the $\mathcal{N}=4$ theory is attained when the full scalar
potential in Eq.~\eqref{N4YM} vanishes. This is equivalent to the condition
\begin{equation}
\left[  Z^{am},Z^{bn}\right]  =0 \label{comm}
\end{equation}
with $Z^{am}=A^{am}+iB^{am}$. There are two classes of solutions to
this equation.  The first, the ``superconformal phase'', corresponds
to the case where $\left\langle Z^{am}\right\rangle =0$ for all
$a,m$. The gauge symmetry is unbroken for this regime, as is the
superconformal symmetry. In the present context, this phase
corresponds to the neighbourhood of a $\mathbb{C}^{2}/\mathbb{Z}_{N}$
singularity in which the full $\mathrm{SU}(N)$ symmetry is present.
As a result of the $\mathcal{N}=4$ supersymmetry in the gauge sector,
this phase will not be destabilised by low-energy gauge dynamics and,
hence, the theory will not be driven away from the orbifold point by
such effects. However, one can also expect a non-perturbative moduli
superpotential from membrane instantons~\cite{Harvey:1999as} whose
precise form for small blow-up cycles is unknown. It would be
interesting to investigate whether such membrane instanton corrections
can stabilise the system at the orbifold point or whether they drive
it away towards the smooth limit.

The second phase, called the ``Coulomb phase'' (or spontaneously
broken phase) corresponds to the flat directions of the potential
where Eq.~\eqref{comm} is satisfied for $\left\langle
  Z^{am}\right\rangle \neq0$.  The dynamics depend upon the amount of
unbroken symmetry. For generic breaking, $\mathrm{SU}(N)$ is reduced
to $\mathrm{U}(1)^{N-1}$. If this breaking is achieved through
non-trivial VEVs in the $A^{am}$ directions it corresponds,
geometrically, to blowing up the singularity in the internal $G_2$
space.


\chapter*{Final Remarks}
\addcontentsline{toc}{chapter}
                 {\protect\numberline{Final Remarks\hspace{-96pt}}}
\fancyhf{} \fancyhead[L]{\sl Final Remarks}
\fancyhead[R]{\rm\thepage}
\renewcommand{\headrulewidth}{0.3pt}
\renewcommand{\footrulewidth}{0pt}
\addtolength{\headheight}{3pt} \fancyfoot[C]{} In this thesis we
have been following a programme aimed at developing interesting
phenomenology from M-theory, in its explicit supergravity
approximation, compactified on spaces of $G_2$ holonomy. We have
focused on $G_2$ spaces constructed from orbifolds with co-dimension
four fixed points, studying singular limits as well as cases in
which singularities are blown-up. This has been motivated by the
fact that M-theory on spaces with co-dimension four $ADE$
singularities produces non-Abelian gauge
symmetry~\cite{Acharya},~\cite{Gukov}.

In Chapter~\ref{Classification} we found a class of sixteen distinct
orbifold groups, many of which are new examples, that may be used in
Joyce's construction of $G_2$ spaces. We then went on to construct,
for members of the class, $G_2$ structures with small torsion, and
their associated ``almost Ricci-flat'' metrics. This enabled us to
obtain a formula for the moduli K\"ahler potential for M-theory on
the general smooth manifold. In order to consider the singular
limit, it was necessary to construct, for the first time, the
explicit coupling of 11-dimensional supergravity to
seven-dimensional super-Yang-Mills theory that describes M-theory at
a co-dimension four $ADE$ singularity. This formed the basis of
Chapter~\ref{117}. In Chapter~\ref{G2sing} we used our result to
construct the explicit four-dimensional effective supergravity
action for M-theory on the general $G_2$ space in its singular
limit. Finally, in Chapter~\ref{Analysis} we explored some further
properties of the compactification of the seven-dimensional
super-Yang-Mills sector on a three-dimensional singular locus within
the $G_2$ space.

Whilst we have focused on a specific class of $G_2$ manifolds, the
methods we have developed in this thesis will work more generally,
and could be applied to other examples. In this way, many more
four-dimensional theories could be obtained and compared with each
other for their relative phenomenological merits.

In Chapter~\ref{Classification} we found a large number of orbifold
group elements that lead to co-dimension six singularities. One
could attempt to generalise the blow-up procedure and moduli
K\"ahler potential calculation to the case of manifolds with
orbifold groups containing some of these elements. There is also the
possibility of a more complicated orbifold fixed point structure.
For example, if there exist points on the torus that are fixed by
more than one generator of the orbifold group, there can be several
topologically distinct ways of blowing up the associated
singularity~\cite{Joyce}.

The construction of an explicit action for M-theory in the
neighbourhood of a co-dimension four $ADE$ singularity is one of the
main achievements of this thesis. It provides a basic starting point
for compactification of M-theory on many kinds of internal space
besides those we have considered. For example, it would be used for
any internal space that has a singular K3 manifold as a factor. It
would be particularly interesting to perform a compactification on a
$G_2$ manifold whose singular loci are different from
$\mathcal{T}^3$. This would allow a reduction of the $\mathcal{N}=4$
supersymmetry in the gauge theory sub-sector to $\mathcal{N}=1$,
giving rise to richer infrared gauge dynamics.

One remarkable feature of our results is the matching up of the
K\"ahler potentials in the limits of small and large blow-up moduli.
Moduli fields move along geodesics in moduli space, and the single
formula could be used to study evolution of moduli fields through
diverse regions of moduli space. Moduli stabilisation has been
achieved for M-theory on our class of $G_2$ manifolds, in the large
blow-up regime \cite{decarlos}. This involved switching on flux and
computing the effect of instantons arising from M5 branes wrapping
three-cycles on the $G_2$ manifold. Vacua were found with either
vanishing or negative cosmological constant. Using the matching of
flux superpotentials for small and large blow-up moduli, it would be
interesting to find out if there are vacua for which some of the
blow-up moduli are fixed at small values, and if there are, whether
a positive cosmological constant can be obtained.

Ultimately, the most fascinating extension of the work in this
thesis would be to attempt to include conical, or co-dimension
seven, singularities onto the manifolds, thus supporting charged
chiral matter. To consider such models, one would use experience
gained from the work in this
thesis to attempt to construct a supersymmetric coupling of four-dimensional
chiral multiplets to 11-dimensional supergravity. In this case, the
occurrence of anomaly in-flow from the bulk would give rise to an additional
problem. Namely, the theory would have to be built in such a way that chiral
anomalies at the singularity cancel this anomaly in-flow.

\appendix

\chapter{Some Results about $\boldsymbol{G_2}$ Structures} \label{a}
\fancyhf{} \fancyhead[L]{\sl A.~Some Results about $G_2$ Structures}
\fancyhead[R]{\rm\thepage}
\renewcommand{\headrulewidth}{0.3pt}
\renewcommand{\footrulewidth}{0pt}
\addtolength{\headheight}{3pt} \fancyfoot[C]{}
In this appendix we
derive some results about $G_2$ structures that are used in
Chapter~\ref{Classification} to obtain our classification
of possible orbifold groups of $G_2$ manifolds.  \\

Let $M$ be an element of $\mathrm{SO}(7)$ of the form
\begin{equation} \label{a1}
M=\left( \begin{array}{cccc}
1 & \, & \, & \,  \\
\, & S(\theta_{1}) & \, & \,  \\
\, & \, & S(\theta_{2}) & \, \\
\, & \, & \, & S(\theta_{3})  \\
\end{array} \right) ,
\end{equation}
where
\begin{equation}
 S(\theta_{i})= \left( \begin{array}{cc}
\cos\theta_{i} & -\sin\theta_{i} \\
\sin\theta_{i} & \cos\theta_{i} \\
\end{array} \right).
\end{equation}
Then $M$ is in $G_2$ (for some embedding of $G_2$ into
$\mathrm{SO}(7)$) if and only if one of the following conditions
hold on the $\theta_i$:
\begin{eqnarray}
 \theta_{1}+\theta_{2}+\theta_{3} & = & 0 \: \mathrm{mod} \: 2\pi, \label{1}\\
  -\theta_{1}+\theta_{2}+\theta_{3} & = & 0 \: \mathrm{mod} \: 2\pi, \label{2}\\
  \theta_{1}-\theta_{2}+\theta_{3} & = & 0 \: \mathrm{mod} \: 2\pi, \label{3}
\end{eqnarray}
\ba
  \theta_{1}+\theta_{2}-\theta_{3} & = & 0 \: \mathrm{mod} \: 2\pi. \label{4}
\end{eqnarray}

\textit{Proof:} That this is sufficient has already been demonstrated
in Chapter~\ref{Classification}. Now let us assume that this is not
necessary and we will find a contradiction. We shall attempt to
construct a three-form $\varphi$ that defines a $G_2$ structure and
that is left invariant by some $M$ not
satisfying any of Eqs.~\eqref{1}--\eqref{4}.

When expressed in terms of the coordinates $x_{0},$ $z_{1},$ $z_{2}$
and $z_{3}$, as in \eqref{complex}, each non-vanishing component of a
three-form imposes a definite constraint on the angles $\theta_{i}$ if
it is to be left invariant by $M$. For example a three-form with a
non-vanishing coefficient of
$\mathrm{d}x_{0}\wedge\mathrm{d}z_{1}\wedge\mathrm{d}z_{2}$ imposes
the constraint $\theta_1+\theta_2=0$. We shall use this property
whilst attempting to construct our $G_{2}$ structure.

The $G_{2}$ structure $\varphi_0$ given in \eqref{structure}
satisfies the tensorial identity
\begin{equation}
(\varphi_0)_{ABE}(\varphi_0)_{CD}^{\phantom{CD}E}=(\phi_0)_{ABCD}
+\delta_{AC}\delta_{BD}-\delta_{AD}\delta_{BC},
\end{equation}
where $\phi_0$ is the four-form dual to $\varphi_0$, and
$A,B,\ldots=1,\ldots,7$ label coordinates $(x^A)$ on $\mathbb{R}^7$.
Therefore, so must the $G_{2}$ structure $\varphi$ that we are
attempting to construct. From this equation we obtain
\begin{equation} \label{constraint1}
\sum_{A=1}^{7}\varphi_{ABC}^{2}=1,
\end{equation}
for each $B,C\in 1,\ldots,7$. Now in order for invariance of $\varphi$ to
not impose any of the constraints \eqref{1}--\eqref{4} on $M$,
when expressed in the coordinates of
\eqref{complex} it must not contain any of the following terms:
\begin{equation}
\mathrm{d}z_{1}\wedge\mathrm{d}z_{2}\wedge\mathrm{d}z_{3}, \: \: \mathrm{d}\bar{z}_{1}\wedge\mathrm{d}z_{2}\wedge\mathrm{d}z_{3}, \: \: \mathrm{d}z_{1}\wedge\mathrm{d}\bar{z}_{2}\wedge\mathrm{d}z_{3}, \: \: \mathrm{d}z_{1}\wedge\mathrm{d}z_{2}\wedge\mathrm{d}\bar{z}_{3}.
\end{equation}
For this to be the case, all of the following must be zero:
\begin{equation}
\varphi_{246}, \: \varphi_{247}, \: \varphi_{256},
\: \varphi_{257}, \: \varphi_{346}, \: \varphi_{347}, \: \varphi_{356}, \: \varphi_{357}.
\end{equation}
Bearing this in mind, and taking $(B,C)=(4,6)$ in equation
\eqref{constraint1} we observe that at least one of the following must
be non-zero:
\begin{equation}
\varphi_{146}, \: \varphi_{546}, \: \varphi_{746}. \:
\end{equation}
By changing coordinates to those of \eqref{complex}, we can spot the
constraints this imposes on the $\theta_{i}$. We can repeat this
process for $(B,C)=(2,4)$ and $(B,C)=(2,7)$, and put all the
constraints together to obtain the result that $M$ only leaves
$\varphi$ invariant if at least two of the $\theta_{i}$ are zero. By
assumption $\varphi$ is left invariant by some $M$ not satisfying any
of the constraints~\eqref{1}--\eqref{4}, and so
assume, by the above, and without loss of generality, that this $M$
has $\theta_{1}=\theta_{2}=0,$ and $\theta_{3}\neq0$ mod $2\pi$. Then
it is easily verified that only the following components (and
components with permuted indices of those below) of $\varphi$ may be
non-zero:
\begin{equation}
\begin{array}{l}
\varphi_{123}, \: \varphi_{145}, \: \varphi_{167}, \: \varphi_{124}, \: \varphi_{125}, \:\varphi_{134}, \: \varphi_{135}, \: \varphi_{234}, \\  \varphi_{235}, \: \varphi_{245},  \: \varphi_{345}, \: \varphi_{267}, \: \varphi_{367}, \: \varphi_{467}, \: \varphi_{567}.
\end{array}
\end{equation}
Now, using \eqref{constraint1} we find that
\begin{equation} \label{constraint2}
\lvert\varphi_{467}\rvert=\lvert\varphi_{567}\rvert=\lvert\varphi_{267}\rvert=\lvert\varphi_{367}\rvert=1,
\end{equation}
by using for example $(B,C)=(4,6)$. We then invoke the identity
\begin{equation}
\varphi_{ACD}\varphi^{BCD}=6\delta_{A}^{B},
\end{equation}
which implies
\begin{equation}
\varphi_{7CD}\varphi^{7CD}=6.
\end{equation}
However \eqref{constraint2} gives
\begin{eqnarray}
\varphi_{7CD}\varphi^{7CD} & \geq & 2(\varphi_{746}\varphi^{746}+\varphi_{756}\varphi^{756}+\varphi_{726}\varphi^{726}+\varphi_{736}\varphi^{736}) \nonumber\\
&= & 8.
\end{eqnarray}
We therefore have a contradiction, and hence our result.
\\

Now let $M$ be as in \eqref{a1}, and let it satisfy one of the conditions \eqref{1}--\eqref{4} so that it preserves some $G_2$ structure, but now let us assume that $\theta_1=0$ for simplicity. Then any $G_2$ structure preserved by $M$ may be brought to the standard form of equation \eqref{structure}, up to possible sign differences, by some redefinition of coordinates that preserves the structure of $M$ up to some redefinition of $\theta_2$ and $\theta_3$.

\textit{Proof:} Following the method used to prove the previous result, we can draw up a list of the components of $\varphi$ that may be non-zero if it is to be preserved by $M$. These are
\begin{equation}
\begin{array}{l}
\varphi_{123}, \: \varphi_{145}, \: \varphi_{167}, \: \varphi_{146}, \: \varphi_{147}, \:\varphi_{156}, \: \varphi_{157}, \: \varphi_{245}, \: \varphi_{345}, \: \varphi_{267}, \\  \varphi_{367}, \: \varphi_{246}, \: \varphi_{247}, \: \varphi_{256}, \: \varphi_{257}, \: \varphi_{346}, \: \varphi_{347},\: \varphi_{356}, \: \varphi_{357}.
\end{array}
\end{equation}
Now using \eqref{constraint1} and our freedom to choose the orientation of the 1, 2 and 3 directions we see that we have $\varphi_{123}=1$. We can then show similarly that, without loss, $\varphi_{145}=1$. Then $\varphi_{167}=\pm 1$ by consistency of the following identity, with $(A, B, D, E, F)=(2,6,2,3,6)$:
\begin{equation}
\varphi_{AB}^{\phantom{AB}C}\phi_{CDEF}=\delta_{AD}\varphi_{BEF}+\delta_{AE}\varphi_{BFD}+\delta_{AF}\varphi_{BDE}-\delta_{BD}\varphi_{AEF}-\delta_{BE}\varphi_{AFD}-\delta_{BF}\varphi_{ADE}.
\end{equation}
Finally, repeated use of \eqref{constraint1} and remaining coordinate freedom enables us to establish the result.
\\

There is a useful corollary of the above result. Let $M$ be a
rotation matrix with three independent preserved directions $A$, $B$
and $C$. Then $M$ preserves a given $G_2$ structure $\varphi$ only
if $\lvert\varphi_{ABC}\rvert=1$.

\chapter{Table of Possible Orbifold Group Elements
of $\boldsymbol{G_{2}}$ Manifolds} \label{b}

\fancyhf{} \fancyhead[L]{\sl B.~Table of Possible Orbifold Group
Elements of $G_{2}$ Manifolds} \fancyhead[R]{\rm\thepage}
\renewcommand{\headrulewidth}{0.3pt}
\renewcommand{\footrulewidth}{0pt}
\addtolength{\headheight}{3pt} \fancyfoot[C]{}

\begin{table}[h]
\begin{center}
\begin{tabular}{|c|c|cc|cc|cc|cc|}
\hline
$\mathrm{Symmetry}$ & $\frac{1}{2\pi}(\theta_{1},\theta_{2},\theta_{3})$  & $n_{1}$ & $d_{1}$ & $n_{2}$ & $d_{2}$ & $n_{3}$ & $d_{3}$ & $n_{4}$ & $d_{4}$  \\ \hline
$\mathbb{Z}_2$& $(0,\frac{1}{2},\frac{1}{2})$  & 3 & 1 & 4 &1&-&-&-&- \\
$\mathbb{Z}_3$ &$ (0,\frac{1}{3},\frac{1}{3})$  & 3 & 1 & 2 &2&-&-&-&- \\
$\mathbb{Z}_3^\ast$ & $(\frac{1}{3},\frac{1}{3},\frac{1}{3})$ & 1 & 1 & 3 &2&-&-&-&- \\
$\mathbb{Z}_4$ & $(0,\frac{1}{4},\frac{1}{4})$  & 3 & 1 & 2 &2&-&-&-&- \\
$\mathbb{Z}_4^\ast$ &$ (\frac{1}{4},\frac{1}{4},\frac{1}{2})$  & 1 & 1 & 2 &2&2&1&-&- \\
$\mathbb{Z}_6$ &$ (0,\frac{1}{6},\frac{1}{6})$  & 3 & 1 & 2 &2&-&-&-&- \\
$\mathbb{Z}_6^\ast$ &$ (\frac{1}{6},\frac{1}{6},\frac{1}{3}) $ & 1 & 1 & 2 &2&1&2&-&- \\
$\mathbb{Z}_6^\dagger$ &$ (\frac{1}{6},\frac{1}{3},\frac{1}{2}) $ & 1 & 1 & 1 & 2 & 1& 2 & 2 & 1 \\
$\mathbb{Z}_7$ &$ (\frac{1}{7},\frac{2}{7},\frac{3}{7})$  & 1 & 1 & 1 & 2 & 1& 2 & 1 & 2 \\
$\mathbb{Z}_8$ &$ (\frac{1}{8},\frac{1}{4},\frac{3}{8}) $ & 1 & 1 & 1 & 2 & 1& 2 & 1 & 2 \\
$\mathbb{Z}_8^\ast$ &$ (\frac{1}{8},\frac{3}{8},\frac{1}{2})$  & 1 & 1 & 1 & 2 & 1& 2 & 2 & 1 \\
$\mathbb{Z}_{12}$ & $(\frac{1}{12},\frac{1}{3},\frac{5}{12}) $ & 1 & 1 & 1 & 2 & 1& 2 & 1 & 2 \\
$\mathbb{Z}_{12}^\ast$ & $(\frac{1}{12},\frac{5}{12},\frac{1}{2}) $ & 1 & 1 & 1 & 2 & 1& 2 & 2 & 1 \\
\hline
\end{tabular}
\caption{List of orbifold group elements}\label{tab:appb}
\end{center}
\end{table}
\noindent The possible generators $\alpha$ of orbifold groups take
the form $\alpha : \boldsymbol{x} \mapsto M_{(\alpha)}\boldsymbol{x}
+ \boldsymbol{v}_{(\alpha)}$, where $M_{(\alpha)}$ is an orthogonal
matrix, which can be put into block diagonal form
\begin{equation}
\left( \begin{array}{cccc}
1 & \, & \, & \,  \\
\, & S(\theta_{1}) & \, & \,  \\
\, & \, & S(\theta_{2}) & \, \\
\, & \, & \, & S(\theta_{3})  \\
\end{array} \right)\, ,
\end{equation}
where
\begin{equation}
 S(\theta_{i})= \left( \begin{array}{cc}
\cos\theta_{i} & -\sin\theta_{i} \\
\sin\theta_{i} & \cos\theta_{i} \\
\end{array} \right)\,.
\end{equation}
Table~\ref{tab:appb} lists the possibilities
for the $\theta_i$ (up to signs). The symmetries are labelled as
$\mathbb{Z}_N$, where $N$ is the order. The matrix $M_{(\alpha)}$
corresponding to each symmetry defines a
representation $\rho$ of $\mathbb{Z}_N$. In the table,
the $n_i$ and $d_i$ label how the representation $\rho$ decomposes into
irreducibles according to $\rho=n_1\rho_1\oplus\cdots\oplus n_r\rho_r$,
as described in Chapter~\ref{Classification}.


\chapter{Blow-up and some Calculations on Gibbons-Hawking Space} \label{c}
\fancyhf{} \fancyhead[L]{\sl C.~Blow-up and some Calculations on
Gibbons-Hawking Space} \fancyhead[R]{\rm\thepage}
\renewcommand{\headrulewidth}{0.3pt}
\renewcommand{\footrulewidth}{0pt}
\addtolength{\headheight}{3pt} \fancyfoot[C]{}
In this appendix we
begin by deriving a general volume formula valid on regions of $G_2$
manifolds that take the form $\mathcal{T}^3\times \mathcal{U}$,
where $\mathcal{T}^3$ is some three-torus and $\mathcal{U}$ is a
four-dimensional hyperk\"ahler space. We then describe, for
arbitrary $N$, how to blow up
$\mathcal{T}^3\times\mathbb{C}^2/\mathbb{Z}_N$, and use our formula
to compute volumes on blow-ups, as induced by a given $G_2$
structure. We consider in turn cases in which $\mathcal{U}$
approaches flat space asymptotically and in which $\mathcal{U}$
becomes exactly flat for sufficiently large radius. We follow and
generalise results from Ref.~\cite{Lukas}.
\\

Let us briefly recall the definition of a hyperk\"ahler space. A hyperk\"ahler space is a $4n$-dimensional Riemannian manifold admitting a triplet $J^m$ of covariantly constant complex structures satisfying the algebra
\begin{equation}
J^mJ^n=-\boldsymbol{1}\delta^{mn}+\epsilon^{mn}_{\phantom{mn}p}J^p\,.
\end{equation}
Associated with the $J^m$ via \suppressfloats[t]
\begin{equation}
\omega^m_{\phantom{m}\hat{A}\hat{B}}=(J^m)_{\hat{A}}^{\phantom{A}\hat{C}}g_{\hat{C}\hat{B}}
\end{equation}
we have a triplet $\omega^m$ of covariantly constant so-called K\"ahler forms.

If we let $\mathcal{U}$ be a hyperk\"ahler space, then we can write down the following $G_2$ structure on $\mathcal{T}^3\times \mathcal{U}$:
\begin{equation} \label{struc2}
\varphi=\sum_m\omega^m\wedge\mathrm{d}\xi^m-\mathrm{d}\xi^1\wedge\mathrm{d}\xi^2\wedge\mathrm{d}\xi^3\,,
\end{equation}
where $\xi^m$ are coordinates on the torus $\mathcal{T}^3$. This is torsion free by virtue of $\mathrm{d}\omega^m=\mathrm{d}\ast\omega^m=0$.

The volume element $\sqrt{\mathrm{det}(g)}$ on $\mathcal{T}^3\times
\mathcal{U}$ may be found from the $G_2$ structure via the equations
\begin{equation} \label{appmet1}
g_{AB}=\mathrm{det}(h)^{-1/9}h_{AB}\,, \: \: \:
\sqrt{\mathrm{det}(g)}=\mathrm{det}(h)^{1/9}\,,
\end{equation}
where
\begin{equation} \label{appmet2}
h_{AB}=\frac{1}{144}\varphi_{ACD}\varphi_{BEF}\varphi_{GHI}\hat{\epsilon}^{CDEFGHI}\,
\end{equation}
and $\hat{\epsilon}$ is the ``pure-number'' Levi-Civita pseudo-tensor.

We now follow a general method of construction of hyperk\"ahler spaces \cite{Hitch1} to derive a formula for the measure. The triplet of K\"ahler forms is given by
\begin{equation}
\omega^1=\frac{i}{2}\partial\bar{\partial}\mathcal{K}\,,
\end{equation}
\begin{equation}
\omega^2=\mathrm{Re}(\mathrm{d}u\wedge\mathrm{d}z)\,, \: \: \:
\omega^3=\mathrm{Im}(\mathrm{d}u\wedge\mathrm{d}z)\,,
\end{equation}
where $\mathcal{K}$ is the K\"ahler potential for $\mathcal{U}$. These give
\begin{equation} \label{det}
\mathrm{det}(h)=\frac{1}{4^9}(\mathcal{K}_{,u\bar{z}}\mathcal{K}_{,z\bar{u}}-\mathcal{K}_{,u\bar{u}}\mathcal{K}_{,z\bar{z}})^3\,.
\end{equation}
Here $u$ and $z$ are complex coordinates on $\mathcal{U}$ and $\mathcal{K}_{,u\bar{z}}\equiv\partial^2\mathcal{K}/\partial u\partial\bar{z}$ etc. We can reduce this to a simpler expression if we write $\mathcal{K}$ as the Legendre transform of a real function $\mathcal{F}(x,z,\bar{z})$ with respect to the real coordinate $x$:
\begin{equation} \label{Legendre}
\mathcal{K}(u,\bar{u},z,\bar{z})=\mathcal{F}(x,z,\bar{z})-(u+\bar{u})x\,.
\end{equation}
Here $x$ is a function of $z$, $\bar{z}$, $u$ and $\bar{u}$ determined by
\begin{equation} \label{u}
\frac{\partial \mathcal{F}}{\partial x}=u+\bar{u}\,.
\end{equation}
We can then re-express \eqref{det} in terms of partial derivatives of $\mathcal{F}$ and obtain the rather neat result that
\begin{equation} \label{measure}
\sqrt{\mathrm{det}(g)}\equiv \frac{1}{4}\,.
\end{equation}
Note that this is entirely general and valid on any region of a $G_2$
manifold on which the $G_2$ structure can be written as in
Eq.~\eqref{struc2}, and on which the K\"ahler potential can be
expressed as in \eqref{Legendre}. Eq.~\eqref{measure} gives the measure
for integrating over the coordinates $u$, $z$, $\bar{u}$, $\bar{z}$
and $\xi^m$. However it will be more convenient in what follows for us
to substitute $u$ and $\bar{u}$ for the real coordinates $x$ and $y$,
with $x$ as in \eqref{u} and $y$ given by $y=i(\bar{u}-u)$. Then,
assuming for convenience unit volume for $\mathcal{T}^3$, the volume of
$\mathcal{T}^3\times \mathcal{U}$ over some compact subspace $\mathcal{U}_0\subset \mathcal{U}$ is given by
\begin{equation} \label{integral}
\mathrm{vol}(\mathcal{U}_0)=\frac{1}{8}\int_{\mathcal{U}_0}\lvert\mathcal{F}_{,xx}\mathrm{d}z\mathrm{d}\bar{z}\mathrm{d}x\mathrm{d}y\rvert\,.
\end{equation}
\\

As suggested by the above, the blow-up of
$\mathcal{T}^3\times\mathbb{C}^2/\mathbb{Z}_N$ will take the form $\mathcal{T}^3\times \mathcal{U}$,
where $\mathcal{U}$ is an appropriate hyperk\"ahler space. More specifically,
$\mathcal{U}$ will belong to the family of spaces referred to as Gibbons-Hawking
spaces or ``gravitational multi-instantons''~\cite{GibbHawk},~\cite{Hitchin}. Note that
these are generalised versions of Eguchi-Hanson space~\cite{eguchi1},~\cite{eguchi2}, which
corresponds to the case of $N=2$. For each $N$, we may take $\mathcal{U}$ to be
the $N$-centred Gibbons-Hawking space, for which the function
$\mathcal{F}$ is given by
\begin{equation} \label{potential}
\mathcal{F}=\sum_{i=1}^{N}\bigg(r_i-x_i\mathrm{ln}(x_i+r_i)+\frac{x_i}{2}\ln(4z_i\bar{z}_i)\bigg)\,,
\end{equation}
where
\begin{equation}
x_i=x-b_i\,, \: \: z_i=z-a_i\,,
\end{equation}
\begin{equation}
r_i=\sqrt{x_i^2+4\lvert z_i\rvert^2}\,.
\end{equation}
We can derive the metric from $\mathcal{F}$ by using the expression \eqref{Legendre} for the K\"ahler potential. In addition the change of coordinate
\begin{equation}
u=-\ln w + \sum_i\frac{1}{2}\ln\big(2(z-a_i)\big)\,,
\end{equation}
brings the metric into a familiar form \cite{Hitchin}
\begin{equation}
\mathrm{d}s^2=\gamma\mathrm{d}z\mathrm{d}\bar{z} + \gamma^{-1} \left( \frac{\mathrm{d}w}{w}+\bar{\delta}\mathrm{d}z \right) \left( \frac{\mathrm{d}w}{w}+ \delta\mathrm{d}\bar{z} \right)\,,
\end{equation}
where
\begin{equation}
\gamma = \sum_i\frac{1}{r_i}\,,
\end{equation}
\begin{equation}
\delta =  \sum_i\frac{x-b_i-r_i}{2(\bar{z}-\bar{a}_i)r_i}\,,
\end{equation}
\begin{equation}
w\bar{w}=\prod_i(x-b_i+r_i)\,.
\end{equation}

Since we would like to calculate the effect of blow-up on the volume of a ball around the origin of $\mathbb{C}^2/\mathbb{Z}_N$, we wish to relate the coordinates $\{ z,\bar{z}, x, y\}$ to the ordinary Cartesian coordinates for flat space.  Let us do this from first principles. Consider the ``blown-down'' version of $\mathcal{U}$, which is actually flat space. This is constructed from
\begin{equation}
\mathcal{F}=N\left( r-x\ln (x+r)+\frac{1}{2}x\ln (4z\bar{z}) \right)\,,
\end{equation}
where
\begin{equation} \label{r}
r=\sqrt{x^2+4\lvert z\rvert^2}\,.
\end{equation}
Using \eqref{Legendre} and \eqref{u} we find the K\"ahler potential is simply
\begin{equation} \label{coord2}
\mathcal{K}=Nr\,.
\end{equation}
Hence, $Nr$ corresponds to the square of the usual radius in flat space. We can also derive the relation
\begin{equation}
u+\bar{u}=\frac{N}{2}\ln \left( \frac{r-x}{r+x} \right)\,,
\end{equation}
which leads to
\begin{equation}
r=2\lvert z \rvert \cosh \left( \frac{u+\bar{u}}{N} \right)\,,
\end{equation}
\begin{equation}
x=-2\lvert z \rvert \sinh \left( \frac{u+\bar{u}}{N} \right)\,.
\end{equation}
Now flat space Cartesian coordinates $z_1$ and $z_2$ satisfy
\begin{equation} \label{flat}
\mathcal{K}=\lvert z_1 \rvert^2 + \lvert z_2 \rvert^2
\end{equation}
and so identifying \eqref{coord2} and \eqref{flat} we can come up with the following holomorphic transformation relating the two sets of coordinates.
\begin{equation} \label{trans}
z_1=\sqrt{Nz}e^{\frac{u}{N}}\,, \: \: z_2=\sqrt{Nz}e^{-\frac{u}{N}}\,.
\end{equation}
The coordinates $z_1$ and $z_2$ are unrestricted, and from this fact and \eqref{trans} we can infer the ranges of the coordinates $\{ x, y, z, \bar{z} \}$. We find that $x$ and $z$ are unrestricted, whilst $y$ is periodic with period $4\pi$.

We would like to compute the volume of a ball around the origin of $\mathcal{U}$. First however, we need to define a radial coordinate analogous to $r$ in equation \eqref{r}. The most sensible choice is to define
\begin{equation} \label{r1}
r\equiv \sqrt{(x-\tilde{b})^2+4\lvert z-\tilde{a} \rvert^2}\,,
\end{equation}
where tildes denote mean values over the index $i$. Having done this, we can perform the integration \eqref{integral} over the region $0\leq r\leq R$. Since we will ultimately be interested in the small blow-up limit, let us assume that $R$ is much larger than the $\lvert a_i \rvert$ and $\lvert b_i \rvert$, and derive an answer that is correct to lowest non-trivial order in these blow-up moduli. From \eqref{integral} and \eqref{potential}, the contribution from one term in the sum is given by
\begin{equation}
V=\frac{1}{8}\int\frac{1}{\sqrt{(x-b)^2+4(z-a)(\bar{z}-\bar{a})}}\mathrm{d}x\mathrm{d}y\lvert \mathrm{d}z\mathrm{d}\bar{z}\rvert\,,
\end{equation}
dropping the subscript $i$ for convenience. We make the change of variables $z=u+iv$, $a=u_0+iv_0$ and then $x^\prime=x$, $u^\prime=2u$, $v^\prime=2v$, $c=2u_0$, $d=2v_0$ and carry out the $y$ integration to obtain
\begin{equation}
V=\frac{\pi}{4}\int\frac{1}{\sqrt{(x^\prime-b)^2+(u^\prime-c)^2+(v^\prime-d)^2}}\mathrm{d}u^\prime\mathrm{d}v^\prime\mathrm{d}x^\prime\,,
\end{equation}
with the range $(x^\prime-\tilde{b})^2+(u^\prime-\tilde{c})^2+(v^\prime-\tilde{d})^2\leq R^2$. It is now straightforward to obtain
\begin{equation}
V=\frac{\pi}{8}\int\left( \sqrt{R^2+f(\theta , \phi )^2 - \boldsymbol{b}^2}+ f(\theta , \phi ) \right)^2\sin\theta\mathrm{d}\theta\mathrm{d}\phi\,,
\end{equation}
where $\boldsymbol{b}=(b-\tilde{b},\;c-\tilde{c},\;d-\tilde{d})$ and
\begin{equation}
f(\theta , \phi )=(b-\tilde{b})\sin\theta\cos\phi+(c-\tilde{c})\sin\theta\sin\phi+(d-\tilde{d})\cos\theta\,.
\end{equation}
Finally, we can do this to lowest non-trivial order in $\boldsymbol{b}$ and substitute back for $a$ to find
\begin{equation} \label{vol1}
V=\frac{\pi^2}{2}\left( R^2 - \frac{1}{3}\left( (b-\tilde{b})^2+4\lvert a-\tilde{a} \rvert^2\right)\right)+\mathcal{O}(\lvert\boldsymbol{b}\rvert^3)\,.
\end{equation}
When we sum \eqref{vol1} over all moduli we obtain the result that, correct to second order in the $a_i$ and $b_i$,
\begin{equation} \label{blowvol}
\mathrm{vol}_U(r=0,R)=\frac{\pi^2}{2}\bigg(NR^2-\frac{N}{3}(\mathrm{var}_i\{b_i\} + 2\,\mathrm{var}_i\{\mathrm{Re}\, a_i\} + 2\,\mathrm{var}_i\{\mathrm{Im}\,a_i\}) \bigg)\,.
\end{equation}
Here var refers to the variance, with the usual definition
\begin{equation}
\mathrm{var}_i\{X_i\}=\frac{1}{N}\sum_i(X_i-\tilde{X})^2\,.
\end{equation}
\\

The Gibbons-Hawking space that we have been discussing approaches flat space asymptotically. However, what we really need for our construction of $G_2$ manifolds are smoothed versions of this space which become exactly flat for sufficiently large radius. We now describe how our previous results generalise to a space $\mathcal{U}$ which interpolates between Gibbons-Hawking space at small radius and flat space at large radius.

The smoothed version of $\mathcal{F}$ is given by
\begin{equation} \label{smoothF}
\mathcal{F}=\sum_{i=1}^{N}\bigg(r_i-x_i\mathrm{ln}(x_i+r_i)+\frac{x_i}{2}\ln(4z_i\bar{z}_i)\bigg)\,,
\end{equation}
where now
\begin{equation}
x_i=x-\epsilon b_i\,, \: \: z_i=z-\epsilon a_i\,,
\end{equation}
\begin{equation}
r_i=\sqrt{x_i^2+4\lvert z_i\rvert^2}\,.
\end{equation}
Here $\epsilon$ is the smoothing function, dependent on the radius
\begin{equation}
r\equiv \sqrt{(x-\tilde{b})^2+4\lvert z-\tilde{a} \rvert^2}\,,
\end{equation}
and satisfying
\begin{equation}
\epsilon (r) = \left\{ \begin{array}{cc}
1 & \mathrm{if} \; r\leq r_0\,,  \\
0 & \mathrm{if} \; r\geq r_1\,. \\
\end{array} \right.
\end{equation}
Further, $r_0 $ and $r_1$ are two characteristic radii satisfying
$\lvert a_i\rvert\ll r_0 < r_1$ and $\lvert b_i\rvert\ll r_0$ for each
$i$ while $\mathcal{U}$ describes Gibbons-Hawking space for $r<r_0$ and the flat
space $\mathbb{C}^2/\mathbb{Z}_N$ for $r>r_1$.

Although this space interpolates between two hyperk\"ahler spaces it
is not a hyperk\"ahler space by itself. Accordingly the forms
$\omega^2$ and $\omega^3$ are no longer co-closed in the ``collar''
region $r\in[r_0,r_1]$. However, this space can be thought of as being
close to hyperk\"ahler as long as the blow-up moduli are sufficiently
small compared to one and the function $\epsilon$ is slowly varying
\cite{Lukas}. Analogously, the $G_2$ structure on $\mathcal{T}^3\times \mathcal{U}$ is not
actually torsion free, but has small torsion under the same
assumptions.

Let us now work out the volume of the region $r\leq\sigma$, for
$\sigma > r_1$. Considering first the contribution up to some radius
$R$, much larger than the $\lvert a_i \rvert$ and $\lvert b_i \rvert$
but smaller than $r_0$ so that $\epsilon$ is identically 1 on the
region of integration, we have exactly the same result as before
\eqref{blowvol}.  The corresponding contribution coming from a shell
$\rho_1<r<\rho_2$ in which $\epsilon=0$ is
\begin{equation}
V=\frac{N\pi^2}{2}\left( \rho_2^2-\rho_1^2 \right)\,.
\end{equation}
Finally we discuss what happens in the ``collar'' region
$r_0<r<r_1$. Here $\mathcal{U}$ looks locally like Gibbons-Hawking space, except
that as one moves outward, away from the origin, the modulus
$\boldsymbol{b}$ is decreasing. Therefore local contributions to the
volume from $\boldsymbol{b}$ become smaller as one moves away from the
origin. We make the observation that the contribution to \eqref{vol1}
from $\boldsymbol{b}$ is independent of $R$. Hence, the volume of a
shell with radii much larger than $\lvert \boldsymbol{b} \rvert$ but
smaller than $r_0$ is independent of $\boldsymbol{b}$ to second order.
We deduce that at second order there can be no contribution from
$\boldsymbol{b}$ to the volume of the collar region. Hence
\eqref{vol1} also holds for $R>r_0$. Hence, the result is identical to
the unsmoothed case, namely that
\begin{equation} \label{finalvol}
\mathrm{vol}_{U}(r=0,\sigma) = \frac{\pi^2}{2}\left(N\sigma^2-\frac{N}{3}\left(\mathrm{var}_i\{b_i\} + 2\mathrm{var}_i\{\mathrm{Re}\: a_i\} + 2\mathrm{var}_i\{\mathrm{Im}\:a_i\}\right) \right) + \mathcal{O}(\lvert\boldsymbol{b}_i\rvert^3)\,.
\end{equation}
Note that this expression is independent of the precise form of the smoothing function $\epsilon$.


\chapter{Spinor Conventions} \label{spinors}
\fancyhf{} \fancyhead[L]{\sl D.~Spinor Conventions}
\fancyhead[R]{\rm\thepage}
\renewcommand{\headrulewidth}{0.3pt}
\renewcommand{\footrulewidth}{0pt}
\addtolength{\headheight}{3pt} \fancyfoot[C]{}
In this appendix, we
provide the conventions used in Chapter~\ref{117}
 for gamma matrices and
spinors in eleven, seven and four dimensions and the relations between
them. This split of eleven dimensions into seven plus four arises
naturally from the orbifolds
$\mathcal{M}_{1,6}\times\mathbb{C}^{2}/\mathbb{Z}_{N}$ which we
consider in Chapter~\ref{117}. We need to work out the appropriate spinor
decomposition for this product space and, in particular, write
11-dimensional Majorana spinors as a product of seven-dimensional
symplectic Majorana spinors with an appropriate basis of
four-dimensional spinors. We denote 11-dimensional coordinates by
$(x^M)$, with indices $M,N,\ldots = 0,1,\ldots , 10$. They are split up
as $x^{M}=(x^{\mu},y^{A})$ with seven-dimensional coordinates $x^\mu$,
where $\mu ,\nu ,\ldots = 0,1,\ldots ,6$, on $\mathcal{M}_{1,6}$ and
four-dimensional coordinates $y^A$, where $A,B,\ldots = 1,\ldots,4$, on
$\mathbb{C}^2/\mathbb{Z}_N$.\\

We begin with gamma matrices and spinors in 11-dimensions. The gamma-matrices,
$\Gamma^{M}$, satisfy the standard Clifford algebra
\begin{equation}
\{\Gamma^{M},\Gamma^{N}\}=2\hat{g}^{MN}\,,\label{clifford}%
\end{equation}
where $\hat{g}_{MN}$ is the metric on the full space $\mathcal{M}_{1,6}%
\times\mathbb{C}^{2}/\mathbb{Z}_{N}$. We define the Dirac conjugate of an
11-dimensional spinor $\Psi$ to be
\begin{equation}
\bar{\Psi}=i\Psi^{\dagger}\Gamma^{0}\,.
\end{equation}
The 11-dimensional charge conjugate is given by
\begin{equation}
\Psi^{C}=B^{-1}\Psi^{\ast}\,,
\end{equation}
where the charge conjugation matrix $B$ satisfies~\cite{Tanii}
\begin{equation}
B\Gamma^{M}B^{-1}=\Gamma^{M\ast}\,,\hspace{0.5cm}B^{\ast}B=\boldsymbol{1} _{32}\,.
\end{equation}
In this work, all spinor fields in 11-dimensions are taken to satisfy the
Majorana condition, $\Psi^{C}=\Psi$, thereby reducing $\Psi$ from 32 complex
to 32 real degrees of freedom.\\

Next, we define the necessary conventions for $\mathrm{SO}(1,6)$ gamma matrices and
spinors in seven dimensions. The gamma matrices, denoted by $\Upsilon^\mu$,
satisfy the algebra
\begin{equation}
 \{\Upsilon^\mu ,\Upsilon^\nu\}=2\hat{g}^{\mu\nu}\, ,
\end{equation}
where $\hat{g}_{\mu\nu}$ is the metric on $\mathcal{M}_{1,6}$. The Dirac conjugate
of a general eight complex component spinor $\psi$ is defined by
\begin{equation}
\bar{\psi}=i\psi^{\dagger}\Upsilon^{0}\, .
\end{equation}
In seven dimensions, the charge conjugation matrix $B_8$ has the
following properties \cite{Tanii}
\begin{equation}
B_{8}\Upsilon^{\mu}B_{8}^{-1}=\Upsilon^{\mu\ast},\hspace{0.5cm}B_{8}^{\ast
}B_{8}=-\boldsymbol{1} _{8}\, .
\end{equation}
The second of these relations implies that charge conjugation, defined by
\begin{equation}
 \psi^c = B_8^{-1}\psi^\ast
\end{equation}
squares to minus one. Hence, one cannot define seven-dimensional
$\mathrm{SO}(1,6)$ Majorana spinors.  However, the supersymmetry algebra in seven
dimensions contains an $\mathrm{SU}(2)$ R-symmetry and spinors can be naturally
assembled into $\mathrm{SU}(2)$ doublets $\psi^i$, where $i,j,\ldots =
1,2$. Indices $i,j,\ldots$ can be lowered and raised with the
two-dimensional Levi-Civita tensor $\epsilon_{ij}$ and $\epsilon^{ij}$,
normalised so that $\epsilon^{12}=\epsilon_{21}=1$. With these conventions
a symplectic Majorana condition
\begin{equation}
\psi_{i}=\epsilon_{ij}B_{8}^{-1}\psi^{\ast j}\,,\label{symplectic majorana}%
\end{equation}
can be imposed on an $\mathrm{SU}(2)$ doublet $\psi^i$ of spinors, where we have
defined $\psi^{\ast i}\equiv(\psi_{i})^{\ast}$. All seven-dimensional
spinors in this paper are taken to be such symplectic Majorana spinors.
Further, in computations with seven-dimensional spinors, the following
identities are frequently useful,
\begin{align}
\bar{\chi}^{i}\Upsilon^{\mu_{1}\ldots\mu_{n}}\psi^{j} &  =(-1)^{n+1}\bar{\psi
}^{j}\Upsilon^{\mu_{n}\ldots\mu_{1}}\chi^{i}\,,\\
\bar{\chi}^{i}\Upsilon^{\mu_{1}\ldots\mu_{n}}\psi_{i} &  =(-1)^{n}\bar{\psi
}^{i}\Upsilon^{\mu_{n}\ldots\mu_{1}}\chi_{i}\,.
\end{align}
\\

Finally, we need to fix conventions for four-dimensional Euclidean gamma
matrices and spinors. Four-dimensional gamma matrices, denoted by
$\gamma^A$, satisfy
\begin{equation}
 \{\gamma^A,\gamma^B\}=2\hat{g}^{AB}\, ,
\end{equation}
with the metric $\hat{g}_{AB}$ on $\mathbb{C}^2/\mathbb{Z}_N$. The chirality
operator, defined by
\begin{equation}
\gamma=\gamma^{\underline{7}}\gamma^{\underline{8}}\gamma^{\underline{9}}
       \gamma^{\underline{10}}\, ,
\end{equation}
satisfies $\gamma^2 = {\bf 1}_4$. The four-dimensional charge conjugation
matrix $B_4$ satisfies the properties
\begin{equation}
B_4\gamma^AB_4^{-1}=\gamma^{A\ast}\, ,\qquad B_4^\ast B_4=-{\bf 1}_4\, .\label{B4}
\end{equation}
It will often be more convenient to work with complex coordinates
$(z^p,\bar{z}^{\bar{p}})$ on $\mathbb{C}^2/\mathbb{Z}_N$, where
$p,q,\ldots = 1,2$ and $\bar{p},\bar{q},\ldots = \bar{1},\bar{2}$.
In these coordinates, the Clifford algebra takes the well-known
``harmonic oscillator'' form
\begin{equation}
\left\{  \gamma^{p},\gamma^{q}\right\}  =0\,, \qquad
\left\{  \gamma^{\bar{p}},\gamma^{\bar{q}}\right\}  =0\, ,\qquad
\left\{  \gamma^{p},\gamma^{\bar{q}}\right\}     =2\hat{g}^{p\bar{q}}\, ,
\end{equation}
with creation and annihilation ``operators'' $\gamma^p$ and $\gamma^{\bar{p}}$,
respectively. In this new basis, complex conjugation of gamma matrices~\eqref{B4}
is described by
\begin{equation}
B_{4}\gamma^{\bar{p}}B_{4}^{-1}=\gamma^{p\ast}\; , \qquad B_{4}\gamma
^{p}B_{4}^{-1}=\gamma^{\bar{p}\ast}\, . \label{B4c}
\end{equation}
 A basis of spinors can be obtained by starting with the
``vacuum state'' $\Omega$, which is annihilated by $\gamma^{\bar{p}}$, that
is $\gamma^{\bar{p}}\Omega =0$, and applying creation operators to it.
This leads to the three further states
\begin{equation}
\rho^{\underline{p}}=\frac{1}{\sqrt{2}}\gamma^{\underline{p}}\Omega\, ,\qquad
\bar{\Omega}=\frac{1}{2}\gamma^{\ul{1}}\gamma^{\ul{2}}\Omega\, .
\end{equation}
In terms of the gamma matrices in complex coordinates, the chirality operator
$\gamma$ can be expressed as
\begin{equation}
\gamma=-1+\gamma^{\bar{\ul 1}}\gamma^{\ul 1}+\gamma^{\bar{\ul 2}}\gamma^{\ul 2}-\gamma
^{\bar{\ul 1}}\gamma^{\ul 1}\gamma^{\bar{\ul 2}}\gamma^{\ul 2}\, .
\end{equation}
Hence, the basis $(\Omega ,\rho^{\ul p},\bar{\Omega})$ consists of chirality
eigenstates satisfying
\begin{equation}
\gamma\Omega=-\Omega\, , \qquad \gamma\bar{\Omega}=-\bar{\Omega}\, , \qquad
\gamma\rho^{\underline{p}} = \rho^{\underline{p}}\, .
\end{equation}
For ease of notation, we will write the left-handed states as
$(\rho^i )=(\rho^{\ul{1}},\rho^{\ul{2}})$, where $i ,j ,\ldots =1,2$
and the right-handed states as $(\rho^{\bar{\imath}})=(\Omega ,\bar{\Omega})$
where $\bar{\imath},\bar{\jmath},\ldots =\bar{1},\bar{2}$. Note, it follows
from Eq.~\eqref{B4c} that
\begin{equation}
 B_4^{-1}\Omega^\ast = \bar{\Omega}\, ,\qquad
 B_4^{-1}\rho^{\ul{1}\ast} = \rho^{\ul{2}}\, .
\end{equation}
Hence $\rho^i$ and $\rho^{\bar{\imath}}$ each form a Majorana
conjugate pair of spinors with definite chirality.\\

We should now discuss the four plus seven split of 11-dimensional gamma
matrices and spinors. It is easily verified that the matrices
\begin{equation}
\Gamma^{\mu}    =\Upsilon^{\mu}\otimes\gamma\, ,\qquad
\Gamma^{A}    =\boldsymbol{1}  _{8}\otimes\gamma^{A}\,, \label{gammas}
\end{equation}
satisfy the Clifford algebra~\eqref{clifford} and, hence, constitute a
valid set of 11-dimensional gamma-matrices. Further, it is clear that
an 11-dimensional charge conjugation matrix $B$ can be obtained from
its seven- and four-dimensional counterparts $B_8$ and $B_4$ by
\begin{equation}
 B=B_8\otimes B_4\, .
\end{equation}
A general 11-dimensional Dirac spinor $\Psi$ can now be expanded in terms
of the basis $(\rho^i,\rho^{\bar{\imath}})$ of four-dimensional spinors
as
\begin{equation}
\Psi =\psi_{i}(x,y)\otimes\rho^{i}+\psi_{\bar{\jmath}}(x,y)\otimes\rho
^{\bar{\jmath}}\,,
\end{equation}
where $\psi_i$ and $\psi_{\bar{\jmath}}$ are four independent seven-dimensional
Dirac spinors. Given the properties of the four-dimensional spinor basis under
charge conjugation, a Majorana condition on the 11-dimensional spinor $\Psi$
simply translates into $\psi_i$ and $\psi_{\bar{\jmath}}$ each being
symplectic $\mathrm{SO}(1,6)$ Majorana spinors.


\chapter{Some group-theoretical Properties}
\label{Pauli} \fancyhf{} \fancyhead[L]{\sl E.~Some group-theoretical
Properties} \fancyhead[R]{\rm\thepage}
\renewcommand{\headrulewidth}{0.3pt}
\renewcommand{\footrulewidth}{0pt}
\addtolength{\headheight}{3pt} \fancyfoot[C]{}
In this appendix we
summarise some group-theoretical properties related to the coset
spaces $\mathrm{SO}(3,n)/$ $\mathrm{SO}(3)\times \mathrm{SO}(n)$ of
seven-dimensional EYM supergravity, which we encounter in
Chapter~\ref{117}. We focus on the parameterisation of these coset
spaces in terms of 11-dimensional metric components, which is an
essential ingredient in re-writing 11-dimensional supergravity,
truncated on the orbifold $\mathbb{C}^2/\mathbb{Z}_N$, into standard
seven-dimensional EYM supergravity language.

We begin with the generic $\mathbb{C}^2/\mathbb{Z}_N$ orbifold, where
$N>2$ and $n=1$, so the relevant coset space is $\mathrm{SO}(3,1)/\mathrm{SO}(3)$. In
this case, it is convenient to use complex coordinates
$(z^p,\bar{z}^{\bar{p}})$, where $p,q,\ldots = 1,2$ and
$\bar{p},\bar{q},\ldots = \bar{1},\bar{2}$, on the orbifold.  After
truncating the 11-dimensional metric to be independent of the orbifold
coordinates, the surviving degrees of freedom of the orbifold part of
the metric can be described by the components ${\hat{e}_p}^{\ul{p}}$ of the
vierbein, see Eqs.~\eqref{cond1}--\eqref{condn}. Extracting the
overall scale factor from this, we have a determinant one object
${v_p}^{\ul{p}}$, together with identifications by $\mathrm{SU}(2)$ gauge
transformations acting on the tangent space index. Hence, ${v_p}^{\ul{p}}$ should
be thought of as parameterising the coset $\mathrm{SL}(2,\mathbb{C})/\mathrm{SU}(2)$. This
space is indeed isomorphic to $\mathrm{SO}(3,1)/\mathrm{SO}(3)$. To work this out explicitly,
it is useful to introduce the map $f$ defined by
\begin{equation}
 f(u) = u_I\sigma^I
\end{equation}
which maps four-vectors $u_I$, where $I,J,\ldots =1,\ldots ,4$,
into hermitian matrices $f(u)$. Here the matrices $\sigma^I$ and their
conjugates $\bar{\sigma}^I$ are given by
\begin{equation}
(\sigma^{I})=(\sigma^{u},\boldsymbol{1}  _{2})\, ,\qquad
(\bar{\sigma}^{I})=(-\sigma^{u},\boldsymbol{1}  _{2})\, ,
\end{equation}
where the $\sigma^{u}$, $u=1,2,3$, are the standard Pauli matrices.
They satisfy the following useful identities
\begin{align}
\mathrm{tr}\left(  \sigma^{I}\bar{\sigma}^{J}\right)   &  =2\eta^{IJ}\,,\\
\mathrm{tr}\left(  \bar{\sigma}^{I}\sigma^{(J}\bar{\sigma}^{\lvert K\rvert
}\sigma^{L)}\right)   &  =2\left(  \eta^{IJ}\eta^{KL}+\eta^{IL}\eta^{JK}%
-\eta^{IK}\eta^{JL}\right)\,  ,
\end{align}
where $I,J,\ldots$ indices are raised and lowered with the Minkowski metric
$(\eta_{IJ})=\rm{diag}(-1,-1,-1,+1)$. A key property of the map $f$ is that
\begin{equation}
 u_Iu^I={\rm det}(f(u))
\end{equation}
for four-vectors $u_I$. This property is crucial in demonstrating that
the map $F$ defined by
\begin{equation}
 F(v)u = f^{-1}\left( vf(u)v^\dagger\right)
\end{equation}
is a group homeomorphism $F:\mathrm{SL}(2,\mathbb{C})\rightarrow \mathrm{SO}(3,1)$.
Solving explicitly for the $\mathrm{SO}(3,1)$ images ${\ell_I}^{\ul{J}}={(F(v))_I}^{\ul{J}}$
one finds
\begin{equation}
 {\ell_I}^{\ul{J}}=\frac{1}{2}{\rm tr}\left(\bar{\sigma}_Iv\sigma^Jv^\dagger\right)\,.
\end{equation}
This map induces the desired map $\mathrm{SL}(2,\mathbb{C})/\mathrm{SU}(2)\rightarrow \mathrm{SO}(3,1)/\mathrm{SO}(3)$
between the cosets.\\

The structure is analogous, although slightly more involved, for the orbifold
$\mathbb{C}^2/\mathbb{Z}_2$, where $n=3$ and the relevant coset space is
$\mathrm{SO}(3,3)/\mathrm{SO}(3)^2$. In this case, it is more appropriate to work with real
coordinates $y^A$ on the orbifold, where $A,B,\ldots =1,\ldots,4$. The
orbifold part of the truncated 11-dimensional metric, rescaled to
determinant one, is then described by the vierbein ${v_A}^{\ul{A}}$ in real
coordinates, which parameterises the coset $\mathrm{SL}(4,\mathbb{R})/\mathrm{SO}(4)$.
The map $f$ now identifies $\mathrm{SO}(3,3)$ vectors $u$ with elements of the $\mathrm{SO}(4)$ Lie algebra
according to
\begin{equation}
 f(u)=u_IT^I\, ,
\end{equation}
where $T^I$, with $I,J,\ldots = 1,\ldots ,6$ is a basis of
anti-symmetric $4\times 4$ matrices. We would like to choose these
matrices so that the first four, $T^1,\ldots ,T^4$ correspond to the
Pauli matrices $\sigma^1,\ldots , \sigma^4$ of the previous $N>2$
case, when written in real coordinates.  This ensures that our
result for $N=2$ indeed exactly reduces to the one for $N>2$ when
the additional degrees of freedom are ``switched off'' and,
hence, the action for both cases can be written in a uniform language.
It turns out that such a choice of matrices is given by
\begin{align}
T^{1}  &  =\left(
\begin{array}
[c]{cccc}%
0 & 0 & 0 & -1\\
0 & 0 & 1 & 0\\
0 & -1 & 0 & 0\\
1 & 0 & 0 & 0
\end{array}
\right) \, ,\hspace{0.3cm}T^{2}=\left(
\begin{array}
[c]{cccc}%
0 & 0 & 1 & 0\\
0 & 0 & 0 & 1\\
-1 & 0 & 0 & 0\\
0 & -1 & 0 & 0
\end{array}
\right) \, ,\hspace{0.3cm}\\
T^{3}  &  =\left(
\begin{array}
[c]{cccc}%
0 & -1 & 0 & 0\\
1 & 0 & 0 & 0\\
0 & 0 & 0 & 1\\
0 & 0 & -1 & 0
\end{array}
\right) \, ,\hspace{0.3cm} T^{4}=\left(
\begin{array}
[c]{cccc}%
0 & -1 & 0 & 0\\
1 & 0 & 0 & 0\\
0 & 0 & 0 & -1\\
0 & 0 & 1 & 0
\end{array}
\right)\, .\hspace{0.3cm}%
\end{align}
The two remaining matrices can be taken as
\begin{equation}
T^{5}=\left(
\begin{array}
[c]{cccc}%
0 & 0 & 1 & 0\\
0 & 0 & 0 & -1\\
-1 & 0 & 0 & 0\\
0 & 1 & 0 & 0
\end{array}
\right) \, ,\hspace{0.3cm}T^{6}=\left(
\begin{array}
[c]{cccc}%
0 & 0 & 0 & 1\\
0 & 0 & 1 & 0\\
0 & -1 & 0 & 0\\
-1 & 0 & 0 & 0
\end{array}
\right)  \,.
\end{equation}
Note that $T^{1,2,3}$ and $T^{4,5,6}$ form the two sets of $\mathrm{SU}(2)$ generators
within the $\mathrm{SO}(4)$ Lie algebra. We may introduce a ``dual'' to the six $T^{I}$ matrices,
analogous to the definition of the $\bar{\sigma}^{I}$ matrices of the $N>2$ case, which will
prove useful in many calculations. We define
\begin{equation}
\left(  \overline{T}^{I}\right)  ^{AB}=-\frac{1}{2}\left(  T^{I}\right)
_{CD}\epsilon^{ABCD}%
\end{equation}
which has the simple form
\begin{equation}
(\overline{T}^{I})=(T^{u},-T^{\alpha})\,,
\end{equation}
where $u,v,\ldots =1,2,3$ and $\alpha ,\beta ,\ldots =4,5,6$. Indices $I,J,\ldots$ are raised
and lowered with the metric $(\eta_{IJ})={\rm diag}(-1,-1,-1,+1,+1,+1)$.
The matrices $T^{I}$ satisfy the following useful identities
\begin{align}
\mathrm{tr}\left(  T^{I~}\overline{T}^{J}\right)   &  =4\eta^{IJ}\,,\\
\left(  T^{I}\right)  _{AB}\left(  T^{J}\right)  _{CD}\eta_{IJ}  &
=2\epsilon_{ABCD}\,,\\
\left(  T^{u}\right)  _{AB}\left(  T^{v}\right)  _{CD}\delta_{uv}  &
=\delta_{AC}\delta_{BD}-\delta_{AD}\delta_{BC}-\epsilon_{ABCD}\,,\\
(T^{\alpha})_{AB}(T^{\beta})_{CD}\delta_{\alpha\beta}  &  =\delta_{AC}%
\delta_{BD}-\delta_{AD}\delta_{BC}+\epsilon_{ABCD}\,.
\end{align}
Key property of the map $f$ is
\begin{equation}
 (u_Iu^I)^2={\rm det}(f(u))
\end{equation}
for any $\mathrm{SO}(3,3)$ vector $u_I$. This property can be used to show
that the map $F$ defined by
\begin{equation}
 F(v)u = f^{-1}\left( vf(u)v^T\right)
\end{equation}
is a group homomorphism $F:\mathrm{SL}(4,\mathbb{R})\rightarrow \mathrm{SO}(3,3)$.
Solving for the $\mathrm{SO}(3,3)$ images ${\ell_I}^{\ul{J}}={(F(v))_I}^{\ul{J}}$
one finds
\begin{equation}
 {\ell_I}^{\ul{J}}=\frac{1}{4}{\rm tr}\left( \bar{T}_IvT^Jv^T\right)\, .
\end{equation}
This induces the desired map between the cosets $\mathrm{SL}(4,\mathbb{R})/\mathrm{SO}(4)$
and $\mathrm{SO}(3,3)/\mathrm{SO}(3)^2$.


\chapter{Einstein-Yang-Mills Supergravity in seven Dimensions} \label{bigEYM}
\fancyhf{} \fancyhead[L]{\sl \rightmark} \fancyhead[R]{\rm\thepage}
\renewcommand{\headrulewidth}{0.3pt}
\renewcommand{\footrulewidth}{0pt}
\addtolength{\headheight}{3pt} \fancyfoot[C]{}
In this final
appendix we give a self-contained summary of minimal, ${\cal N}=1$
Einstein-Yang-Mills (EYM) supergravity in seven dimensions. This
theory may be formulated in two equivalent ways. In one
formulation~\cite{Berghshoeff},~\cite{Han} the gravity multiplet
contains a two-form field, whilst in the dual formulation this is
replaced by a three-form field~\cite{Park}. We shall consider the
latter version in this appendix, in light of the presence of a
three-form field in 11-dimensional supergravity. In
Chapter~\ref{117}, when we constrain the fields of this theory to a
seven-dimensional orbifold plane, the degrees of freedom fill out a
seven-dimensional gravity supermultiplet (in addition to a number of
vector multiplets dependent on the orbifold symmetry). Thus the
version with the three-form is the one that is best suited for our
application to M-theory on singular spaces.

There exists an $\mathrm{SU}(2)$ rigid R-symmetry in
seven-dimensional supergravity, and this may be gauged. The massive
theories obtained in this way were first constructed in
Refs.~\cite{Townsend}--\cite{Giani}. The seven-dimensional
supergravities we obtain by truncating M-theory are not massive and,
for this reason, we will not consider such theories with gauged
R-symmetry. The seven-dimensional pure supergravity theory can also
be coupled to $M$ vector
multiplets~\cite{Berghshoeff},~\cite{Park},~\cite{Avramis}--\cite{Nicolai},
transforming under a Lie group $G=\mathrm{U}(1)^n\times H$, where
$H$ is semi-simple, in which case the vector multiplet scalars
parameterise the coset space $\mathrm{SO}(3,M)/\mathrm{SO}(3)\times
\mathrm{SO}(M)$. In this appendix, we will first review
seven-dimensional ${\cal N}=1$ EYM supergravity with such a gauge
group $G$. This theory is used in Chapter~\ref{117} to construct the
complete action for low-energy M-theory on an orbifold of the form
$\mathcal{M}_{1,6}\times\mathbb{C}^{2}/\mathbb{Z}_{N}$. The
truncation to seven-dimensions of M-theory on such an orbifold leads
to a $d=7$ EYM supergravity with gauge group $\mathrm{U}(1)^n\times
\mathrm{SU}(N)$, where $n=1$ for $N>2$ and $n=3$ for $N=2$. Here,
the $\mathrm{U}(1)^n$ part of the gauge group originates from
truncated bulk states, while the $\mathrm{SU}(N)$ non-Abelian part
corresponds to the additional states which arise on the orbifold
fixed plane.  Since we are constructing the coupled
11-/7-dimensional theory as an expansion in $\mathrm{SU}(N)$ fields,
the crucial building block is a version of $d=7$ EYM supergravity
with gauge group $\mathrm{U}(1)^n\times \mathrm{SU}(N)$, expanded
around the supergravity and $\mathrm{U}(1)^n$ part. This expanded
version of the theory is presented in the second part of this
appendix.

\section{General action and supersymmetry transformations} \label{full}
The field content of gauged $d=7$, ${\cal N}=1$ EYM supergravity
consists of two types of multiplets. The first, the gravitational
multiplet, contains a graviton $g_{\mu\nu}$ with associated vielbein ${e_\mu}^{\ul{\nu}}$, a gravitino $\psi_\mu^i$,
a symplectic Majorana spinor $\chi^i$, an $\mathrm{SU}(2)$ triplet of Abelian
vector fields ${{A_\mu}^i}_j$ with field strengths ${F^i}_j=\mathrm{d}{A^i}_j$, a three form field $C_{\mu\nu\rho}$ with field strength $G=\mathrm{d}C$,
and a real scalar $\sigma$. So, in summary we have
\begin{equation} \label{gravmult}
 \left( g_{\mu\nu}\,,\;C_{\mu\nu\rho}\,,\;{{A_\mu}^i}_j\,,\;\sigma\, ,\;\psi^i_\mu \,,\;\chi^i\right).
\end{equation}
Here, $i,j,\ldots = 1,2$ are $\mathrm{SU}(2)$ R-symmetry indices. The second type is the vector multiplet, which contains
gauge vectors $A_\mu^a$ with field strengths $F^a={\cal D}A^a$, gauginos $\lambda^{ai}$ and $\mathrm{SU}(2)$ triplets of real scalars ${\phi^{ai}}_j$. In summary, we have
\begin{equation} \label{vecmult}
 \left( A_\mu^a\,, \;{\phi^{ai}}_j\,, \;\lambda^{ai}\right),
\end{equation}
where $a,b,\ldots = 4,\ldots ,(M+3)$ are Lie algebra indices of the gauge group $G$.

It is sometimes useful to combine all vector fields, the three Abelian ones in the gravity
multiplet as well as the ones in the vector multiplets, into a single $\mathrm{SO}(3,M)$
vector
\begin{equation}
 (A_\mu^{\tilde{I}})=\left({{A_\mu}^i}_j\,,\; A_\mu^a \right),
\end{equation}
where $\tilde{I},\tilde{J},\ldots = 1,\ldots ,(M+3)$. Under this combination, the corresponding field strengths are given by
\begin{equation}
F_{\mu\nu}^{\tilde{I}}=2\partial_{\lbrack\mu}A_{\nu]}^{\tilde{I}}+{f_{\tilde
{J}\tilde{K}}}^{\ti{I}}A_{\mu}^{\tilde{J}}A_{\nu}^{\tilde{K}}\,,
\end{equation}
where ${f_{bc}}^a$ are the structure constants for $G$ and all other components of ${f_{\tilde{J}\tilde{K}}}^{\ti{I}}$ vanish.

The coset space $\mathrm{SO}(3,M)/\mathrm{SO}(3)\times \mathrm{SO}(M)$ is described by a
$(3+M)\times (3+M)$ matrix
$L_{\tilde{I}}^{\phantom{I}\underline{\tilde{J}}}$, which depends on
the $3M$ vector multiplet scalars and satisfies the $\mathrm{SO}(3,M)$
orthogonality condition
\begin{equation}
L_{\tilde{I}}^{\phantom{I}  \underline{\tilde{J}}}L_{\tilde{K}}^{\phantom{K}
\underline{\tilde{L}}}\eta_{\underline{\tilde{J}}\underline{\tilde{L}}}%
=\eta_{\tilde{I}\tilde{K}}
\end{equation}
with
$(\eta_{\tilde{I}\tilde{J}})=(\eta_{\underline{\tilde{I}}\underline
{\tilde{J}}})=\mathrm{diag}(-1,-1,-1,+1,\ldots,+1)$. Here, indices
$\ti{I},\ti{J},\ldots = 1,\ldots,(M+3)$ transform under
$\mathrm{SO}(3,M)$. Their flat counterparts $\ti{\ul{I}},\ti{\ul{J}},\ldots$
decompose into a triplet of $\mathrm{SU}(2)$, corresponding to the
gravitational directions and $M$ remaining directions corresponding to
the vector multiplets. Thus we can write $L_{\tilde{I}}^{\phantom{I}
\underline{\tilde{J}}}\to\left(
{L_{\tilde{I}}}^u,L_{\tilde{I}}{}^{a}\right)$, where $u=1,2,3$. The
adjoint $\mathrm{SU}(2)$ index $u$ can be converted into a pair of fundamental
$\mathrm{SU}(2)$ indices by multiplication with the Pauli matrices, that is,
\begin{equation}
L_{\tilde{I}}{}^{i}{}_{j}=\frac{1}{\sqrt{2}}L_{\tilde{I}}{}^{u}\left(
\sigma_{u}\right) ^{i}{}_{j}\,.
\end{equation}
There are obviously many ways in
which one can parameterise the coset space $\mathrm{SO}(3,M)/\mathrm{SO}(3)\times \mathrm{SO}(M)$
in terms of the physical vector multiplet scalar degrees of freedom
${{\phi_a}^i}_j$. A simple parameterisation of this coset in terms
of $\Phi\equiv({\phi_{a}}^u)$ is given by
\begin{equation}
L_{\tilde{I}}^{\phantom{I}  \underline{\tilde{J}}}=\left(  \exp\left[
\begin{array}[c]{cc}%
0 & \Phi^T\\
\Phi & 0
\end{array}
\right]  \right)  _{\tilde{I}}^{\phantom{I}  \underline{\tilde{J}}}\,.%
\end{equation}
In the final paragraph of this appendix, when we expand seven-dimensional
supergravity, we will use a different parameterisation, which is better
adapted to this task. The Maurer-Cartan form of the matrix $L$, defined by
$L^{-1}\mathcal{D}L$, is needed to write down the theory. The
components $P$ and $Q$ are given explicitly by
\begin{align}
P_{\mu a\phantom{i} j}^{\phantom{\mu a} i} &  =
L_{\phantom{I} a}^{\tilde{I}}\left( \delta_{\tilde{I}}^{\tilde{K}}
 \partial_{\mu}+f_{\tilde{I}\tilde{J}}{}^{\tilde{K}}A_{\mu}^{\ti{J}}\right)
 L_{\tilde{K}\phantom{i} j}^{\phantom{K} i}\,,\\
Q_{\mu\phantom{i} j}^{\phantom{\mu} i} &  =
L_{\phantom{Ii} k}^{\tilde{I}i}\left( \delta_{\tilde{I}}^{\tilde{K}}
 \partial_{\mu}+f_{\tilde{I}\tilde{J}}{}^{\tilde{K}}A_{\mu}^{\ti{J}}\right)
 L_{\tilde{K}\phantom{k} j}^{\phantom{K} k}\,.
\end{align}
The final ingredients needed are the following projections of the structure constants
\begin{align}
D  &  =if_{ab}^{\phantom{ab}  c}L_{\phantom{ai}  k}^{ai}L_{\phantom{bj}
i}^{bj}L_{c\phantom{k}  j}^{\phantom{c}  k}\,,\nonumber\\
D_{\phantom{ai}  j}^{ai}  &  =if_{bc}^{\phantom{bc}  d}L_{\phantom{bi}
k}^{bi}L_{\phantom{ck}  j}^{ck}L_{d}^{\phantom{d}  a}\,,\nonumber\\
D_{ab\phantom{i}  j}^{\phantom{ab}  i}  &  =f_{cd}^{\phantom{cd}  e}
L_{a}^{\phantom{a}  c}L_{b}^{\phantom{b}  d}L_{e\phantom{i}  j}^{\phantom{e}
i}\,.
\end{align}

It is worth mentioning that invariance of the theory under the gauge
group $G$ and the R-symmetry group $\mathrm{SU}(2)$ requires that the
Maurer-Cartan forms $P$ and $Q$ transform covariantly. It can be shown
that this is the case, if and only if the ``extended'' set of
structure constants $f_{\tilde{I}\tilde{J}}{}^{\tilde{K}}$ satisfy the
condition
\begin{equation}
f_{\tilde{I}\tilde{J}}{}^{\tilde{L}}\eta_{\tilde{L}}{}_{\tilde{K}
}=f_{[\tilde{I}\tilde{J}}{}^{\tilde{L}}\eta{}_{\tilde{K}]}{}_{\tilde{L}
}\,.\label{gauge condition}
\end{equation}
For any direct product factor of the total gauge group, this condition
can be satisfied in two ways. Either, the structure constants are
trivial, or the metric $\eta_{\tilde{I}\tilde{J}}$ is the Cartan-Killing
metric of this factor. In our particular case, the condition~\eqref{gauge condition}
is satisfied by making use of both these possibilities.
The structure constants vanish for the ``gravitational'' part of the
gauge group and the $\mathrm{U}(1)^n$ part within $G$. For the semi-simple part $H$
of $G$, one can always choose a basis, so its Cartan-Killing metric is
simply the Kronecker delta.

With everything in place, we now write down the Lagrangian for the
theory. Setting coupling constants to one, and neglecting four-fermi
terms, it is given by \cite{Park}
\begin{align}
e^{-1}\mathcal{L}_{\mathrm{YM}}  &  =\frac{1}{2}R-\frac{1}{2}\bar{\psi}_{\mu
}^{i}\Upsilon^{\mu\nu\rho}\hat{\mathcal{D}}_{\nu}\psi_{\rho i}-\frac{1}%
{96}e^{4\sigma}G_{\mu\nu\rho\sigma}G^{\mu\nu\rho\sigma}-\frac{1}{2}\bar{\chi
}^{i}\Upsilon^{\mu}\hat{\mathcal{D}}_{\mu}\chi_{i}-\frac{5}{2}\partial_{\mu
}\sigma\partial^{\mu}\sigma\nonumber\label{EYM}\\
&\hspace{0.4cm}  +\frac{\sqrt{5}}{2}\left(  \bar{\chi}^{i}\Upsilon^{\mu\nu}\psi_{\mu i}%
+\bar{\chi}^{i}\psi_{i}^{\nu}\right)  \partial_{\nu}\sigma+e^{2\sigma}%
G_{\mu\nu\rho\sigma}\left[  \frac{1}{192}\left(  \bar{\psi}_{\lambda}%
^{i}\Upsilon^{\lambda\mu\nu\rho\sigma\tau}\psi_{\tau i}+12\bar{\psi}^{\mu
i}\Upsilon^{\nu\rho}\psi_{i}^{\sigma}\right)  \right. \nonumber\\
&  \hspace{4.6cm}\left.  +\frac{1}{48\sqrt{5}}\left(  4\bar{\chi}^{i}%
\Upsilon^{\mu\nu\rho}\psi_{i}^{\sigma}-\bar{\chi}^{i}\Upsilon^{\mu\nu
\rho\sigma\tau}\psi_{\tau i}\right)  -\frac{1}{320}\bar{\chi}^{i}\Upsilon
^{\mu\nu\rho\sigma}\chi_{i}\right] \nonumber
\end{align}
\begin{align}
&\hspace{0.4cm}  -\frac{1}{4}e^{-2\sigma}\left(  L_{\tilde{I}\phantom{i}  j}^{\phantom{I}
i}L_{\tilde{J}\phantom{j}  i}^{\phantom{J}  j}+L_{\tilde{I}}^{a}L_{\tilde{J}%
a}\right)  F_{\mu\nu}^{\tilde{I}}F^{\tilde{J}\mu\nu}-\frac{1}{2}\bar{\lambda
}^{ai}\Upsilon^{\mu}\hat{\mathcal{D}}_{\mu}\lambda_{ai}-\frac{1}{2}%
P_{\mu\phantom{ia}  j}^{\phantom{\mu}  ai}P_{\phantom{\mu}  a\phantom{j}
i}^{\mu\phantom{a}  j}\nonumber\\
&\hspace{0.4cm}  -\frac{1}{\sqrt{2}}\left(  \bar{\lambda}^{ai}\Upsilon^{\mu\nu}\psi_{\mu
j}+\bar{\lambda}^{ai}\psi_{j}^{\nu}\right)  P_{\nu a\phantom{j}
i}^{\phantom{\nu a}  j}+\frac{1}{192}e^{2\sigma}G_{\mu\nu\rho\sigma}%
\bar{\lambda}^{ai}\Upsilon^{\mu\nu\rho\sigma}\lambda_{ai}\nn\\
&\hspace{0.4cm}  -ie^{-\sigma}F_{\mu\nu}^{\tilde{I}}L_{\tilde{I}\phantom{j}  i}%
^{\phantom{I}  j}\left[  \frac{1}{4\sqrt{2}}\left(  \bar{\psi}_{\rho}%
^{i}\Upsilon^{\mu\nu\rho\sigma}\psi_{\sigma j}+2\bar{\psi}^{\mu i}\psi
_{j}^{\nu}\right)  +\frac{3}{20\sqrt{2}}\bar{\chi}^{i}\Upsilon^{\mu\nu}%
\chi_{j}-\frac{1}{4\sqrt{2}}\bar{\lambda}^{ai}\Upsilon
^{\mu\nu}\lambda_{aj}\right. \nonumber\\
&  \hspace{3.1cm}\left.  +\frac{1}{2\sqrt{10}}\left(  \bar{\chi}^{i}\Upsilon
^{\mu\nu\rho}\psi_{\rho j}-2\bar{\chi}^{i}\Upsilon^{\mu}\psi_{j}^{\nu}\right)
\right] \nonumber\\
&\hspace{0.4cm}  +e^{-\sigma}F_{\mu\nu}^{\tilde{I}}L_{\tilde{I}a}\left[  \frac{1}{4}\left(
2\bar{\lambda}^{ai}\Upsilon^{\mu}\psi_{i}^{\nu}-\bar{\lambda}^{ai}%
\Upsilon^{\mu\nu\rho}\psi_{\rho i}\right)  +\frac{1}{2\sqrt{5}}\bar{\lambda
}^{ai}\Upsilon^{\mu\nu}\chi_{i}\right] \nonumber\\
&\hspace{0.4cm}  +\frac{5}{180}e^{2\sigma}\left(  D^{2}-9D_{\phantom{ai}  j}^{ai}%
D_{a\phantom{j}  i}^{\phantom{a}  j}\right)  -\frac{i}{\sqrt{2}}e^{\sigma
}D_{ab\phantom{i}  j}^{\phantom{ab}  i}\bar{\lambda}^{aj}\lambda_{i}%
^{b}+\frac{i}{2}e^{\sigma}D_{a\phantom{i}  j}^{\phantom{a}  i}\left(
\bar{\psi}_{\mu}^{j}\Upsilon^{\mu}\lambda_{i}^{a}+\frac{2}{\sqrt{5}}\bar{\chi
}^{j}\lambda_{i}^{\phantom{i}  a}\right) \nonumber\\
&\hspace{0.4cm}  +\frac{1}{60\sqrt{2}}e^{\sigma}D\left(  5\bar{\psi}_{\mu}^{i}\Upsilon
^{\mu\nu}\psi_{\nu i}+2\sqrt{5}\bar{\psi}_{\mu}^{i}\Upsilon^{\mu}\chi
_{i}+3\bar{\chi}^{i}\chi_{i}-5\bar{\lambda}^{ai}\lambda_{ai}\right)
\nonumber\\
&\hspace{0.4cm}  -\frac{1}{96}\epsilon^{\mu\nu\rho\sigma\kappa\lambda\tau}C_{\mu\nu\rho
}F_{\sigma\kappa}^{\tilde{I}}F_{\tilde{I}}{}_{\lambda\tau}.
\end{align}
The covariant derivatives that appear here are given by
\ba
\mathcal{D}_\mu\psi_{\nu i}&=&\partial_{\mu}\psi_{\nu i}+\frac{1}{2}Q_{\mu
i}{}^{j}\psi_{\nu j}-\Gamma^\rho_{\mu\nu}\psi_{\rho i} +\frac{1}{4}\omega_{\mu}^{\phantom{\mu}  \underline
{\mu}\underline{\nu}}
\Upsilon_{\underline{\mu}\underline{\nu}}\psi_{\nu i}\,, \\
\mathcal{D}_{\mu}\chi_{i}&=&\partial_{\mu}\chi_{i}+\frac{1}{2}Q_{\mu
i}{}^{j}\chi_{j}+\frac{1}{4}\omega_{\mu}^{\phantom{\mu}  \underline
{\mu}\underline{\nu}}\Upsilon_{\underline{\mu}\underline{\nu}}\chi_{i}\,,\\
\mathcal{D}_{\mu}\la_{ai}&=&\partial_{\mu}\la_{ai}+\frac{1}{2}Q_{\mu
i}{}^{j}\la_{aj}+\frac{1}{4}\omega_{\mu}^{\phantom{\mu}  \underline
{\mu}\underline{\nu}}\Upsilon_{\underline{\mu}\underline{\nu}}\la_{ai}+{f_{ab}}^cA_\mu^b\la_{ci}\,.
\ea
The associated supersymmetry transformations, parameterised by the spinor $\varepsilon_i$, are, up to cubic fermion terms, given by
\begin{align}
\delta\sigma &  =\frac{1}{\sqrt{5}}\bar{\chi}^{i}\varepsilon_{i}\, ,
\nonumber\\
{\delta e_{\mu}}^{\underline{\nu}}  &  =\bar{\varepsilon}^{i}\Upsilon
^{\underline{\nu}}\psi_{\mu i}\,,\nonumber \\
\delta\psi_{\mu i}  &  =2\mathcal{D}_{\mu}\varepsilon_{i}-\frac{1}%
{80}\left(  \Upsilon_{\mu}^{\phantom{\mu}  \nu\rho\sigma\eta}-\frac{8}%
{3}\delta_{\mu}^{\nu}\Upsilon^{\rho\sigma\eta}\right)  \varepsilon_{i}%
G_{\nu\rho\sigma\eta}e^{2\sigma}\nonumber\\
&\hspace{0.4cm}  +\frac{i}{5\sqrt{2}}\left(  \Upsilon_{\mu}^{\phantom{\mu}  \nu\rho}%
-8\delta_{\mu}^{\nu}\Upsilon^{\rho}\right)  \varepsilon_{j}F_{\nu\rho
}^{\tilde{I}}L_{\tilde{I}\phantom{i}  i}^{\phantom{I}  j}e^{-\sigma}%
-\frac{1}{15\sqrt{2}}e^{\sigma}\Upsilon_{\mu}\varepsilon_{i}D\,,\nonumber\\
\delta\chi_{i}  &  =\sqrt{5}\Upsilon^{\mu}\varepsilon_{i}\partial_{\mu}%
\sigma-\frac{1}{24\sqrt{5}}\Upsilon^{\mu\upsilon\rho\sigma}\varepsilon
_{i}G_{\mu\nu\rho\sigma}e^{2\sigma}\text{ }-\frac{i}{\sqrt{10}}\Upsilon
^{\mu\nu}\varepsilon_{j}F_{\mu\nu}^{\tilde{I}}L_{\tilde{I}\phantom{i}
i}^{\phantom{I}  j}e^{-\sigma}+\frac{1}{3\sqrt{10}}e^{\sigma}\varepsilon
_{i}D\,,\nonumber\\
\delta C_{\mu\nu\rho}  &  =\left(  -3\bar{\psi}_{\left[  \mu\right.  }%
^{i}\Upsilon_{\left.  \nu\rho\right]  }\varepsilon_{i}-\frac{2}{\sqrt{5}}%
\bar{\chi}^{i}\Upsilon_{\mu\nu\rho}\varepsilon_{i}\right)
e^{-2\sigma}\,, \label{7dsusy} \\
L_{\tilde{I}\phantom{i}  j}^{\phantom{I}  i}\delta A_{\mu}^{\tilde{I}}  &
=\left[  i\sqrt{2}\left(  \bar{\psi}_{\mu}^{i}\varepsilon_{j}-\frac{1}%
{2}\delta_{j}^{i}\bar{\psi}_{\mu}^{k}\varepsilon_{k}\right)  -\frac{2i}%
{\sqrt{10}}\left(  \bar{\chi}^{i}\Upsilon_{\mu}\varepsilon_{j}-\frac{1}{2}%
{}\delta_{j}^{i}\bar{\chi}^{k}\Upsilon_{\mu}\varepsilon_{k}\right)  \right]
e^{\sigma}\,,\nonumber
\end{align}
\begin{align}
L_{\tilde{I}}^{\phantom{I}  a}\delta A_{\mu}^{\tilde{I}}  &  =\bar
{\varepsilon}^{i}\Upsilon_{\mu}\lambda_{i}^{a}e^{\sigma}\, ,\nonumber\\
\delta L_{\tilde{I}\phantom{i}  j}^{\phantom{I}  i}  &  =-i\sqrt{2}%
\bar{\varepsilon}^{i}\lambda_{aj}L_{\tilde{I}}^{\phantom{I}  a}+\frac{i}
{\sqrt{2}}\bar{\varepsilon}^{k}\lambda_{ak}L_{\tilde{I}}^{a}\delta_{j}
^{i}\,,\nonumber\\
\delta L_{\tilde{I}}^{\phantom{I}  a}  &  =-i\sqrt{2}\bar{\varepsilon}
^{i}\lambda_{j}^{a}L_{\tilde{I}\phantom{j}  i}^{\phantom{I}  j}\,,\nonumber\\
\delta\lambda_{i}^{a}  &  =-\frac{1}{2}\Upsilon^{\mu\nu}\varepsilon_{i}
F_{\mu\nu}^{\tilde{I}}L_{\tilde{I}}^{a}e^{-\sigma}+\sqrt{2}i\Upsilon^{\mu
}\varepsilon_{j}P_{\mu\phantom{aj}  i}^{\phantom{\mu }  aj}\text{ }-e^{\sigma
}\varepsilon_{j}D^{aj}{}_{i}\,. \nn
\end{align}

\section{A perturbative expansion}\label{bigEYM2}
In this final section we expand the EYM supergravity of Section
\ref{full} around its supergravity and $\mathrm{U}(1)^n$ part. The parameter
for the expansion is $\zeta_7:=\kappa_7/\la_7$, where $\kappa_7$ is
the coupling for gravity and $\mathrm{U}(1)^n$ and $\la_7$ is the
coupling for $H$, the non-Abelian part of the gauge group. To
determine the order in $\zeta$ of each term in the Lagrangian, we need to
fix a convention for the energy dimensions of the fields. Within the
gravity and $\mathrm{U}(1)$ vector multiplets, we assign energy dimension 0 to
bosonic fields and energy dimension 1/2 to fermionic fields.  For the
$H$ vector multiplet, we assign energy dimension 1 to the bosons and
3/2 to the fermions. With these conventions we can write
\begin{equation}
\mathcal{L}_{\mathrm{YM}}=\kappa_7^{-2}\left( \mathcal{L}_{(0)}+\zeta_7^2\mathcal{L}_{(2)}+\zeta_7^4\mathcal{L}_{(4)}+\ldots \right)\,,
\end{equation}
where the $\mathcal{L}_{(m)}$, $m=0,2,4,\ldots$ are independent of
$\zeta_7$. The first term in this series is the Lagrangian for EYM
supergravity with gauge group $\mathrm{U}(1)^n$, whilst the second term
contains the leading order non-Abelian gauge multiplet terms. We will
write down these first two terms and provide truncated supersymmetry
transformation laws suitable for the theory at this order.

In order to carry out the expansion, it is necessary to cast the field
content in a form where the $H$ vector multiplet fields and the
gravity/$\mathrm{U}(1)^n$ vector multiplet fields are disentangled. To this
end, we decompose the single Lie algebra indices
$a,b,\ldots=4,\ldots,(M+3)$ used in Section \ref{full} into indices
$\alpha,\beta,\ldots=4,\ldots,(3+n)$ that label the $\mathrm{U}(1)$ directions
and redefined indices $a,b,\ldots=(n+4),\ldots,(M+3)$ that are Lie algebra
indices of $H$. This makes the disentanglement straightforward for
most of the fields. For example, vector fields, which naturally
combine into the single entity $A_\mu^{\ti{I}}$, can simply be
decomposed as $A_\mu^{\ti{I}}=(A_\mu^I,A_\mu^a)$, where $A_\mu^I$,
$I=1,\ldots,(n+3)$, refers to the three vector fields in the gravity
multiplet and the $\mathrm{U}(1)^n$ vector fields, and $A_\mu^a$ denotes the
$H$ vector fields. Similarly, the $\mathrm{U}(1)$ gauginos are denoted by
$\la_{\alpha i}$, whilst the $H$ gauginos are denoted by
$\la_{ai}$. The situation is somewhat more complicated for the vector
multiplet scalar fields, which, as discussed, all together combine
into the single coset $\mathrm{SO}(3,M)/\mathrm{SO}(3)\times \mathrm{SO}(M)$, parameterised by
the $\mathrm{SO}(3,M)$ matrix $L$. It is necessary to find an explicit form for
$L$, which separates the $3n$ scalars in the $\mathrm{U}(1)^n$ vector
multiplets from the $3(M-n)$ scalars in the $H$ vector multiplet. To
this end, we note that, in the absence of the $H$ states, the $\mathrm{U}(1)^n$
states parameterise a $\mathrm{SO}(3,n)/\mathrm{SO}(3)\times \mathrm{SO}(n)$ coset, described by
$(3+n)\times (3+n)$ matrices
${\ell_I}^{\underline{I}}=({\ell_I}^u,{\ell_I}^\alpha )$. Here,
$\ell\equiv (\ell_I^{\ph{I}u})$ are $(3+n)\times 3$ matrices where the
index $u=1,2,3$ corresponds to the three ``gravity'' directions and
$m\equiv ( \ell_I^{\ph{I}\alpha})$ are $(3+n)\times n$ matrices with
$\alpha =4,\ldots ,(n+3)$ labeling the $\mathrm{U}(1)^n$ directions.  Let us
further denote the $\mathrm{SU}(N)$ scalars by $\Phi\equiv
(\phi_a^{\ph{a}u})$. Then we can construct approximate representatives
$L$ of the large coset $\mathrm{SO}(3,M)/\mathrm{SO}(3)\times \mathrm{SO}(M)$ by expanding, to
the appropriate order in $\Phi$, around the small coset
$\mathrm{SO}(3,n)/\mathrm{SO}(3)\times \mathrm{SO}(n)$ represented by $\ell$ and $m$. Neglecting
terms of cubic and higher order in $\Phi$, this leads to
\begin{equation}\label{L}
L = \left( \begin{array}{ccc}
\ell+\frac{1}{2}\zeta_7^2\ell\Phi^T\Phi & m & \zeta_7\ell\Phi^T \\
\zeta_7\Phi & 0 & \boldsymbol{1}_{M-n}+\frac{1}{2}\zeta_7^2\Phi\Phi^T \\
\end{array} \right)\, .
\end{equation}
We note that the neglected $\Phi$ terms are of order $\zeta_7^3$ and higher
and, since we are aiming to construct the action only up to terms of order
$\zeta^2$, are, therefore, not relevant in the present context.

For the expansion of the action it is useful to re-write the coset parameterisation~\eqref{L}
and the associated Maurer-Cartan forms $P$ and $Q$ in component form. We find
\begin{align}
{L_{I}{}^{i}}_{j}  &  ={\ell_{I}{}^{i}}_{j}+\frac{1}{2}\zeta_7^2\ell_{I}{}^{k}{}_{l}\phi_{\phantom{bi}  k}^{al}\phi_{a\phantom{i}j}^{\ph{a}i}\, ,\\
{L_I}^\alpha &=\zeta_7 {\ell_I}^\alpha\,,  \\
L_{I}{}^{a}  &  =\zeta_7\ell_{I}{}^{i}{}_{j}\phi_{\phantom{aj}i}^{aj}\,,\\
L_{a}{}^{i}{}_{j}  &  =\zeta_7\phi_{a}{}_{\phantom{b}j}^{i}\, ,\\
{L_a}^\alpha & = 0\, ,
\end{align}
\begin{align}
L_{a}{}^{b}  &  ={\delta_{a}}^{b}+\frac{1}{2}\zeta_7^2\phi_{a}{}_{\phantom{i}j}^{i}
\phi^{bj}_{\phantom{bj}i}\, , \\
{P_{\mu\alpha}{}^{i}}_{j}  &  = {p_{\mu\alpha}{}^{i}}_{j}+\frac{1}{2}\zeta_7^2{p_{\mu\alpha}}^{k}{}_{l}\phi^{a}{}^{l}{}_{k}{\phi_{a}{}^{i}}_{j}\, ,\\
{{P_{\mu a}}^i}_j  & = -\zeta_7\mathcal{D}_{\mu}{\phi_{a}{}^{i}}_{j}\, ,\\
Q_{\mu}{}^{i}{}_{j}  &  =q_{\mu}{}^{i}{}_{j}+\frac{1}{2}\zeta_7^2\left( \phi^{a}{}^{i}{}%
_{k}\mathcal{D}_{\mu}\phi_{a}{}^{k}{}_{j}-\phi_{a}{}^{k}{}_{j}\mathcal{D}_{\mu}\phi^{a}{}^{i}{}_{k}\right)\, ,
\end{align}
where $p$ and $q$ are the Maurer-Cartan forms associated with the small coset matrix $\ell$. Thus
\begin{align}
p_{\mu\alpha\phantom{i}  j}^{\phantom{\mu\alpha }  i}  &  =\ell_{\phantom{I}
\alpha}^{I}\partial_{\mu}\ell_{I\phantom{i}  j}^{\phantom{\mu}  i}\,
, \\
q_{\mu\phantom{i}  j\phantom{k}  l}^{\phantom{\mu}  i\phantom{j}  k}  &
=\ell_{\phantom{Ii}  j}^{Ii}\partial_{\mu}\ell_{I\phantom{k}  l}%
^{\phantom{\mu}  k}\, ,\\
q_{\mu\phantom{i}  j}^{\phantom{\mu}  i}  &  =\ell_{\phantom{Ii}  k}%
^{Ii}\partial_{\mu}\ell_{I\phantom{k}  j}^{\phantom{\mu}  k}\, . %
\end{align}
Furthermore, $q$ is now taken to be the $\mathrm{SU}(2)_{\mathrm{R}}$
connection, and the covariant derivatives that appear in the
expanded theory are given by
\ba
\mathcal{D}_{\mu}\phi_{a\phantom{i}j}^{\phantom{a}i}&=&\partial_{\mu
}\phi_{a\phantom{i}  j}^{\phantom{a}i}-q_{\mu\phantom{i}j\phantom{k}
l}^{\phantom{\mu}i\phantom{j}k}\phi_{a\phantom{l}  k}^{\phantom{a}
l}+f_{ab}^{\phantom{ab}c}A_{\mu}^{b}
\phi_{c\phantom{i}j}^{\phantom{c}i}\,, \\
\mathcal{D} _{\mu }\la_{ai}&=&\partial _{\mu }\la_{ai}+\frac{1}{2}
q_{\mu i}^{\phantom{\mu i}j}\la_{aj}+\frac{1}{4}\omega_{\mu }^{\phantom{\mu}%
\underline{\mu }\underline{\nu }}\Upsilon _{\underline{\mu }\underline{\nu }%
}\la_{ai}+ f_{ab}^{\phantom{ab}c}A_\mu^b\la_{ci}\,, \\
\mathcal{D}_\mu\psi_{\nu i}&=&\partial_{\mu}\psi_{\nu i}+\frac{1}{2}q_{\mu
i}{}^{j}\psi_{\nu j}-\Gamma^\rho_{\mu\nu}\psi_{\rho i} +\frac{1}{4}\omega_{\mu}^{\phantom{\mu}  \underline
{\mu}\underline{\nu}}
\Upsilon_{\underline{\mu}\underline{\nu}}\psi_{\nu i}\,, \\
\mathcal{D}_{\mu}\chi_{i}&=&\partial_{\mu}\chi_{i}+\frac{1}{2}q_{\mu
i}{}^{j}\chi_{j}+\frac{1}{4}\omega_{\mu}^{\phantom{\mu}  \underline
{\mu}\underline{\nu}}\Upsilon_{\underline{\mu}\underline{\nu}}\chi_{i}\,,\\
\mathcal{D}_{\mu}\la_{\alpha i}&=&\partial_{\mu}\la_{\alpha i}
+\frac{1}{2}q_{\mu i}{}^{j}\la_{\alpha j}
+\frac{1}{4}\omega_{\mu}^{\phantom{\mu}  \underline{\mu}\underline{\nu}}
\Upsilon_{\underline{\mu}\underline{\nu}}\la_{ \alpha i}\,.
\ea

Using the expressions above, it is straightforward to perform the expansion of $\mathcal{L}_{\mathrm{YM}}$ up to order $\zeta_7^2\sim \la_7^{-2}$. It is given by
\begin{align} \label{truncatedL}
\mathcal{L}_{\mathrm{YM}}\!  &  =\!\frac{1}{\kappa_{7}^{2}}\sqrt{-g}\left\{  \frac{1}{2}R-\frac{1}{2}\bar{\psi}_{\mu}^{i}\Upsilon^{\mu\nu\rho
}\mathcal{D}_{\nu}\psi_{\rho i}-\frac{1}{4}e^{-2\sigma}\left(
\ell_{I\phantom{i}  j}^{\phantom{I}  i}\ell_{J\phantom{j}  i}^{\phantom{J}
j}+\ell_{I}^{\phantom{I}  \alpha}\ell_{J\alpha}\right)  F_{\mu\nu}^{I}%
F^{J\mu\nu}\right. \nonumber\\
&  \hspace{1.0cm}-\frac{1}{96}e^{4\sigma}G_{\mu\nu\rho\sigma}G^{\mu\nu\rho\sigma}-\frac{1}{2}\bar{\chi}^{i}\Upsilon^{\mu}\hat
{\mathcal{D}}_{\mu}\chi_{i}-\frac{5}{2}\partial_{\mu}\sigma\partial^{\mu
}\sigma+\frac{\sqrt{5}}{2}\left(  \bar{\chi}^{i}\Upsilon^{\mu\nu}\psi_{\mu
i}+\bar{\chi}^{i}\psi_{i}^{\nu}\right)  \partial_{\nu}\sigma\nonumber\\
&  \hspace{1.0cm}-\frac{1}{2}\bar{\lambda}^{\alpha i}\Upsilon^{\mu}%
\mathcal{D}_{\mu}\lambda_{\alpha i}-\frac{1}{2}p_{\mu\alpha\phantom{i}
j}^{\phantom{\mu\alpha}  i}p_{\phantom{\mu\alpha j}  i}^{\mu\alpha j}%
-\frac{1}{\sqrt{2}}\left(  \bar{\lambda}^{\alpha i}\Upsilon^{\mu\nu}\psi_{\mu
j}+\bar{\lambda}^{\alpha i}\psi_{j}^{\nu}\right)  p_{\nu\alpha\phantom{j}
i}^{\phantom{\nu\alpha}  j}\nonumber
\end{align}
\begin{align}
&  \hspace{1.0cm}+e^{2\sigma}G_{\mu\nu\rho\sigma}\left[  \frac{1}%
{192}\left(  12\bar{\psi}^{\mu i}\Upsilon^{\nu\rho}\psi_{i}^{\sigma}+\bar
{\psi}_{\lambda}^{i}\Upsilon^{\lambda\mu\nu\rho\sigma\tau}\psi_{\tau
i}\right)  +\frac{1}{48\sqrt{5}}\left(  4\bar{\chi}^{i}\Upsilon^{\mu\nu\rho
}\psi_{i}^{\sigma}\right.  \right. \nonumber\\
&  \hspace{3.2cm}\left.  \left.  -\bar{\chi}^{i}\Upsilon^{\mu\nu\rho\sigma
\tau}\psi_{\tau i}\right)  -\frac{1}{320}\bar{\chi}^{i}\Upsilon^{\mu\nu
\rho\sigma}\chi_{i}+\frac{1}{192}\bar{\lambda}^{\alpha i}\Upsilon^{\mu\nu
\rho\sigma}\lambda_{\alpha i}\right] \nonumber\\
&  \hspace{1.0cm}-ie^{-\sigma}F_{\mu\nu}^{I}\ell_{I\phantom{j}  i}%
^{\phantom{I}  j}\left[  \frac{1}{4\sqrt{2}}\left(  \bar{\psi}_{\rho}%
^{i}\Upsilon^{\mu\nu\rho\sigma}\psi_{\sigma j}+2\bar{\psi}^{\mu i}\psi
_{j}^{\nu}\right)  +\frac{1}{2\sqrt{10}}\left(  \bar{\chi}^{i}\Upsilon^{\mu
\nu\rho}\psi_{\rho j}-2\bar{\chi}^{i}\Upsilon^{\mu}\psi_{j}^{\nu}\right)
\right. \nonumber\\
&  \hspace{3.6cm}\left.  +\frac{3}{20\sqrt{2}}\bar{\chi}^{i}\Upsilon^{\mu\nu
}\chi_{j}-\frac{1}{4\sqrt{2}}\bar{\lambda}^{\alpha i}\Upsilon^{\mu\nu}%
\lambda_{\alpha j}\right] \nonumber\\
&  \hspace{1.0cm}+e^{-\sigma}F_{\mu\nu}^{I}\ell_{I\alpha}\left[  \frac{1}%
{4}\left(  2\bar{\lambda}^{\alpha i}\Upsilon^{\mu}\psi_{i}^{\nu}-\bar{\lambda
}^{\alpha i}\Upsilon^{\mu\nu\rho}\psi_{\rho i}\right)  +\frac{1}{2\sqrt{5}%
}\bar{\lambda}^{\alpha i}\Upsilon^{\mu\nu}\chi_{i}\right] \nonumber\\
&  \hspace{1.0cm}\left.  -\frac{1}{96}\epsilon^{\mu\nu\rho\sigma\kappa
\lambda\tau}C_{\mu\nu\rho}F_{\sigma\kappa}^{I}F_{I\lambda\tau}\right\} \nn \\
& +\frac{1}{\la_7^{2}}\sqrt{-g}\left\{
-\frac{1}{4}e^{-2\sigma}F_{\mu\nu}^{a}F_{a}^{\mu\nu}-\frac{1}{2}%
\mathcal{D}_{\mu}\phi_{a\phantom{i}  j}^{\phantom{a}  i}\hat
{\mathcal{D}}^{\mu}\phi_{\phantom{aj}  i}^{aj}-\frac{1}{2}\bar{\lambda}^{ai}\Upsilon^{\mu}\mathcal{D}_{\mu}\lambda_{ai}\right. \nonumber\\
& \hspace{1.0cm} \left.  -e^{-2\sigma}\ell_{I\phantom{i}  j}^{\phantom{I}  i}%
\phi_{a\phantom{j}  i}^{\phantom{a}  j}F_{\mu\nu}^{I}F^{a\mu\nu}-\frac{1}%
{2}e^{-2\sigma}\ell_{I\phantom{i}  j}^{\phantom{I}  i}\phi_{a\phantom{j}
i}^{\phantom{a}  j}\ell_{J\phantom{k}  l}^{\phantom{J}  k}{\phi^{al}}_k
F_{\mu\nu}^{I}F^{J\mu\nu}\right. \nonumber\\
& \hspace{1.0cm} -\frac{1}{2}p_{\mu\alpha\phantom{i}  j}^{\phantom{\mu\alpha}  i}%
\phi_{a\phantom{j}  i}^{\phantom{a}  j}p_{\phantom{\mu\alpha k}  l}^{\mu\alpha
k}\phi_{\phantom{al}  k}^{al}+\frac{1}{4}\phi_{a\phantom{i}  k}^{\phantom{a}
i}\mathcal{D}_{\mu}{\phi^{ak}}_j\bar{\lambda
}^{\alpha j}\Upsilon^{\mu}\lambda_{\alpha i}\nonumber\\
& \hspace{1.0cm} -\frac{1}{\sqrt{2}}\left(  \bar{\lambda}^{\alpha i}\Upsilon^{\mu\nu}%
\psi_{\mu j}+\bar{\lambda}^{\alpha i}\psi_{j}^{\nu}\right)  \phi
_{a\phantom{j}  i}^{\phantom{a}  j}\phi_{\phantom{ak}  l}^{ak}p_{\nu
\alpha\phantom{l}  k}^{\phantom{\nu\alpha}  l}-\frac{1}{\sqrt{2}}\left(
\bar{\lambda}^{ai}\Upsilon^{\mu\nu}\psi_{\mu j}+\bar{\lambda}^{ai}\psi
_{j}^{\nu}\right)  \mathcal{D}_{\nu}\phi_{a\phantom{j}  i}%
^{\phantom{ a}  j}\nonumber\\
& \hspace{1.0cm} +\frac{1}{192}e^{2\sigma}G_{\mu\nu\rho\sigma}\bar{\lambda}%
^{ai}\Upsilon^{\mu\nu\rho\sigma}\lambda_{ai}+\frac{i}{4\sqrt{2}}e^{-\sigma
}F_{\mu\nu}^{I}\ell_{I\phantom{j}  i}^{\phantom{I}  j}\bar{\lambda}%
^{ai}\Upsilon^{\mu\nu}\lambda_{aj} \nn \\
& \hspace{1.0cm} -\frac{i}{2}e^{-\sigma}\left(  F_{\mu\nu}^{I}\ell_{I\phantom{k}
l}^{\phantom{I}  k}\phi_{\phantom{al}  k}^{al}\phi_{a\phantom{i}
j}^{\phantom{a}  i}+2F_{\mu\nu}^{a}\phi_{a\phantom{j}  i}^{\phantom{a}
j}\right)  \left[  \frac{1}{4\sqrt{2}}\left(  \bar{\psi}_{\rho}^{i}%
\Upsilon^{\mu\nu\rho\sigma}\psi_{\sigma j}+2\bar{\psi}^{\mu i}\psi_{j}^{\nu
}\right)  \right. \nonumber \\
& \hspace{1.0cm} \left.  +\frac{3}{20\sqrt{2}}\bar{\chi}^{i}\Upsilon^{\mu\nu}\chi
_{j}-\frac{1}{4\sqrt{2}}\bar{\lambda}^{\alpha i}\Upsilon^{\mu\nu}%
\lambda_{\alpha j}+\frac{1}{2\sqrt{10}}\left(  \bar{\chi}^{i}\Upsilon^{\mu
\nu\rho}\psi_{\rho j}-2\bar{\chi}^{i}\Upsilon^{\mu}\psi_{j}^{\nu}\right)
\right] \nonumber\\
& \hspace{1.0cm} +e^{-\sigma}F_{\mu\nu}^{a}\left[  \frac{1}{4}\left(  2\bar{\lambda}%
^{ai}\Upsilon^{\mu}\psi_{i}^{\nu}-\bar{\lambda}^{ai}\Upsilon^{\mu\nu\rho}%
\psi_{\rho i}\right)  +\frac{1}{2\sqrt{5}}\bar{\lambda}^{ai}\Upsilon^{\mu\nu
}\chi_{i}\right] \nonumber\\
& \hspace{1.0cm} +\frac{1}{4}e^{2\sigma}f_{bc}^{\phantom{bc}  a}f_{dea}\phi_{\phantom{bi}
k}^{bi}\phi_{\phantom{ck}  j}^{ck}\phi_{\phantom{dj}  l}^{dj}\phi
_{\phantom{el}  i}^{el}-\frac{1}{2}e^{\sigma}f_{abc}\phi_{\phantom{bi}
k}^{bi}\phi_{\phantom{ck}  j}^{ck}\left(  \bar{\psi}_{\mu}^{j}\Upsilon^{\mu
}\lambda_{i}^{a}+\frac{2}{\sqrt{5}}\bar{\chi}^{j}\lambda_{i}^{\phantom{i}
a}\right) \nonumber\\
& \hspace{1.0cm} -\frac{i}{\sqrt{2}}e^{\sigma}f_{ab}^{\phantom{ab}  c}\phi_{c\phantom{i}
j}^{\phantom{c}  i}\bar{\lambda}^{aj}\lambda_{i}^{b}+\frac{i}{60\sqrt{2}%
}e^{\sigma}f_{ab}^{\phantom{ab}  c}\phi_{\phantom{al}  k}^{al}\phi
_{\phantom{bj}  l}^{bj}\phi_{c\phantom{k}  j}^{\phantom{c}  k}\left(
5\bar{\psi}_{\mu}^{i}\Upsilon^{\mu\nu}\psi_{\nu i}+2\sqrt{5}\bar{\psi}_{\mu
}^{i}\Upsilon^{\mu}\chi_{i}\right. \nonumber\\
& \hspace{1.0cm} \left.  \left.  +3\bar{\chi}^{i}\chi_{i}-5\bar{\lambda}^{\alpha i}%
\lambda_{\alpha i}\right)  -\frac{1}{96}\epsilon^{\mu\nu\rho\sigma
\kappa\lambda\tau}C_{\mu\nu\rho}F_{\sigma\kappa}^{a}F_{a\lambda\tau}\right\}\,.
\end{align}
The associated supersymmetry transformations have an expansion similar
to that of the Lagrangian. Thus, the supersymmetry transformation of a
field $X$ takes the form
\begin{equation}
\delta X = \delta^{(0)}X+\zeta_7^2\delta^{(2)}X+\zeta_7^4\delta^{(4)}X+\ldots\,.
\end{equation}
We give the first two terms of this series for the gravity and $\mathrm{U}(1)$
vector multiplet fields, and just the first term for the $H$ vector
multiplet fields. These terms are precisely those required to prove
that the Lagrangian given in Eq.~\eqref{truncatedL} is supersymmetric
to order $\zeta_7^2\sim \la_7^{-2}$. They are
\begin{align}
\delta\sigma &  =\frac{1}{\sqrt{5}}\bar{\chi}^{i}\varepsilon_{i}\,,\nonumber\\
\delta {e_{\mu}}^{\underline{\nu}}  &  =\bar{\varepsilon}^{i}%
\Upsilon^{\underline{\nu}}\psi_{\mu i}\,,\nonumber\\
\delta\psi_{\mu i}  &  =2\mathcal{D}_{\mu}\varepsilon_{i}-\frac{1}%
{80}\left(  \Upsilon_{\mu}^{\phantom{\mu}  \nu\rho\sigma\eta}-\frac{8}%
{3}\delta_{\mu}^{\nu}\Upsilon^{\rho\sigma\eta}\right)  \varepsilon_{i}%
G_{\nu\rho\sigma\eta}e^{2\sigma}+\frac{i}{5\sqrt{2}}\left(
\Upsilon_{\mu}^{\phantom{\mu}  \nu\rho}-8\delta_{\mu}^{\nu}\Upsilon^{\rho
}\right)  \varepsilon_{j}F_{\nu\rho}^{I}\ell_{I\phantom{i}  i}^{\phantom{I}
j}e^{-\sigma}\nonumber\\
&\hspace{0.4cm}  +\frac{\kappa_{7}^{2}}{\la_7^{2}}\left\{  \frac{1}{2}\left(
\phi_{ak}^{\phantom{ak}  j}\mathcal{D}_{\mu}\phi_{\phantom{a}
i}^{a\phantom{i}  k}-\phi_{\phantom{a}  i}^{a\phantom{i}  k}\hat{\mathcal{D}%
}_{\mu}\phi_{ak}^{\phantom{ak}  j}\right)  \varepsilon_{j}-\frac{i}{15\sqrt
{2}}\Upsilon_{\mu}\varepsilon_{i}f_{ab}^{\phantom{ab}  c}\phi_{\phantom{al}
k}^{al}\phi_{\phantom{bj}  l}^{bj}\phi_{c\phantom{k}  j}^{\phantom{c}
k}e^{\sigma}\right. \nonumber\\
&  \left.  \hspace{1.6cm}+\frac{i}{10\sqrt{2}}\left(  \Upsilon_{\mu
}^{\phantom{\mu}  \nu\rho}-8\delta_{\mu}^{\nu}\Upsilon^{\rho}\right)
\varepsilon_{j}\left(  F_{\nu\rho}^{I}\ell_{I\phantom{k}  l}^{\phantom{I}
k}\phi_{\phantom{al}  k}^{al}\phi_{a\phantom{j}  i}^{\phantom{a}  j}%
+2F_{\nu\rho}^{a}\phi_{a\phantom{j}  i}^{\phantom{a}  j}\right)  e^{-\sigma
}\right\}  \,,\nonumber\\
\delta\chi_{i}  &  =\sqrt{5}\Upsilon^{\mu}\varepsilon_{i}\partial_{\mu}%
\sigma-\frac{1}{24\sqrt{5}}\Upsilon^{\mu\upsilon\rho\sigma}\varepsilon
_{i}G_{\mu\nu\rho\sigma}e^{2\sigma}\text{ }-\frac{i}{\sqrt{10}%
}\Upsilon^{\mu\nu}\varepsilon_{j}F_{\mu\nu}^{I}\ell_{I\phantom{i}
i}^{\phantom{I}  j}e^{-\sigma}\nonumber\\
&\hspace{0.4cm}  +\frac{\kappa_{7}^{2}}{\la_7^{2}}\left\{  -\frac{i}{2\sqrt{10}%
}\Upsilon^{\mu\nu}\varepsilon_{j}\left(  F_{\mu\nu}^{I}\ell_{I\phantom{k}
l}^{\phantom{I}  k}\phi_{\phantom{al}  k}^{al}\phi_{a\phantom{j}
i}^{\phantom{a}  j}+2F_{\mu\nu}^{a}\phi_{a\phantom{j}  i}^{\phantom{a}
j}\right)  e^{-\sigma} \right. \nn \\
&\hspace{1.6cm} \left. +\frac{i}{3\sqrt{10}}\varepsilon_{i}f_{ab}^{\phantom{ab}  c}%
\phi_{\phantom{al}  k}^{al}\phi_{\phantom{bj}
l}^{bj}\phi_{c\phantom{k}
j}^{\phantom{c}  k}e^{\sigma}\right\}  \,,\nonumber\\
\delta C_{\mu\nu\rho}  &  =\left( -3\bar{\psi}_{\left[ \mu\right.
}^{i}\Upsilon_{\left.  \nu\rho\right]  }\varepsilon_{i}-\frac{2}{\sqrt{5}}%
\bar{\chi}^{i}\Upsilon_{\mu\nu\rho}\varepsilon_{i}\right)  e^{-2\sigma
}\,,\nonumber\\
\ell_{I\phantom{i}  j}^{\phantom{I}  i}\delta A_{\mu}^{I}  &  =\left[
i\sqrt{2}\left(  \bar{\psi}_{\mu}^{i}\varepsilon_{j}-\frac{1}{2}\delta_{j}%
^{i}\bar{\psi}_{\mu}^{k}\varepsilon_{k}\right)  -\frac{2i}{\sqrt{10}}\left(
\bar{\chi}^{i}\Upsilon_{\mu}\varepsilon_{j}-\frac{1}{2}{}\delta_{j}^{i}%
\bar{\chi}^{k}\Upsilon_{\mu}\varepsilon_{k}\right)  \right]  e^{\sigma
}\nonumber\\
&\hspace{0.4cm}  +\frac{\kappa_{7}^{2}}{\la_7^{2}}\left\{  \left(  \frac{i}%
{\sqrt{2}}\bar{\psi}_{\mu}^{k}\varepsilon_{l}-\frac{i}{\sqrt{10}}\bar{\chi
}^{k}\Upsilon_{\mu}\varepsilon_{l}\right)  \phi_{\phantom{al}  k}^{al}%
\phi_{a\phantom{i}  j}^{\phantom{a}  i}e^{\sigma}-\bar{\varepsilon}%
^{k}\Upsilon_{\mu}\lambda_{k}^{a}\phi_{a\phantom{i}  j}^{\phantom{a}
i}e^{\sigma}\right\}  \,, \\
\ell_{I}^{\phantom{I}  \alpha}\delta A_{\mu}^{I}  &  =\bar{\varepsilon}%
^{i}\Upsilon_{\mu}\lambda_{i}^{\alpha}e^{\sigma}\,,\nonumber\\
\delta\ell_{I\phantom{i}  j}^{\phantom{I}  i}  &  =-i\sqrt{2}\bar{\varepsilon
}^{i}\lambda_{\alpha j}\ell_{I}^{\phantom{I}  \alpha}+\frac{i}{\sqrt{2}}%
\bar{\varepsilon}^{k}\lambda_{\alpha k}\ell_{I}^{\phantom{I}  \alpha}%
\delta_{j}^{i}\nonumber\\
&\hspace{0.4cm}  +\frac{\kappa_{7}^{2}}{\la_7^{2}}\left\{  \frac{i}{\sqrt{2}%
}\left[  \bar{\varepsilon}^{k}\lambda_{\alpha l}\phi_{\phantom{al}  k}%
^{al}\phi_{a\phantom{i}  j}^{\phantom{a}  i}\ell_{I}^{\phantom{I}  \alpha
}+\bar{\varepsilon}^{l}\lambda_{ak}\phi_{\phantom{ai}  j}^{ai}\ell
_{I\phantom{k}  l}^{\phantom{I}  k}-\left(  \bar{\varepsilon}^{i}\lambda
_{aj}-\frac{1}{2}\delta_{j}^{i}\bar{\varepsilon}^{m}\lambda_{am}\right)
\phi_{\phantom{al}  k}^{al}\ell_{I\phantom{k}  l}^{\phantom{I}  k}\right]
\right\}  \,,\nonumber\\
\delta\ell_{I}^{\phantom{I}  \alpha}  &  =-i\sqrt{2}\bar{\varepsilon}%
^{i}\lambda_{j}^{\alpha}\ell_{I\phantom{j}  i}^{\phantom{I}  j}+\frac{\kappa
_{7}^{2}}{\la_7^{2}}\left\{  -\frac{i}{\sqrt{2}}\bar{\varepsilon
}^{i}\lambda_{j}^{\alpha}\phi_{\phantom{aj}  i}^{aj}\phi_{a\phantom{l}
k}^{\phantom{a}  l}\ell_{I\phantom{k}  l}^{\phantom{I}  k}\right\}
\,,\nonumber\\
\delta\lambda_{i}^{\alpha}  &  =-\frac{1}{2}\Upsilon^{\mu\nu}\varepsilon
_{i}F_{\mu\nu}^{I}\ell_{I}^{\phantom{I}  \alpha}e^{-\sigma}+\sqrt{2}%
i\Upsilon^{\mu}\varepsilon_{j}p_{\mu\phantom{\alpha j}  i}^{\phantom{\mu}
\alpha j}+\frac{\kappa_{7}^{2}}{\la_7^{2}}\left\{  \frac{i}{\sqrt
{2}}\Upsilon^{\mu}\varepsilon_{j}\phi_{ai}^{\phantom{ai}  j}p_{\mu
\phantom{\alpha k}  l}^{\phantom{\mu}  \alpha k}\phi_{\phantom{al}  k}%
^{al}\right\}  \,,\nonumber \\
\delta A_{\mu}^{a}  &  =\bar{\varepsilon}^{i}\Upsilon_{\mu}\lambda_{i}%
^{a}e^{\sigma}-\left(  i\sqrt{2}\psi_{\mu}^{i}\varepsilon_{j}-\frac{2i}%
{\sqrt{10}}\bar{\chi}^{i}\Upsilon_{\mu}\varepsilon_{j}\right)  \phi
_{\phantom{aj}  i}^{aj}e^{\sigma}\,,\nonumber\\
\delta\phi_{a\phantom{i}  j}^{\phantom{a}  i}  &  =-i\sqrt{2}\left(
\bar{\varepsilon}^{i}\lambda_{aj}-\frac{1}{2}\delta_{j}^{i}\bar{\varepsilon
}^{k}\lambda_{ak}\right)  \,,\nonumber\\
\delta\lambda_{i}^{a}  &  =  -\frac{1}{2}\Upsilon^{\mu\nu}%
\varepsilon_{i}\left(  F_{\mu\nu}^{I}\ell_{I\phantom{j}
k}^{\phantom{I} j}\phi_{\phantom{ak}  j}^{ak}+F_{\mu\nu}^{a}\right)
e^{-\sigma}-i\sqrt
{2}\Upsilon^{\mu}\varepsilon_{j}\mathcal{D}_{\mu}\phi_{\phantom{a}
i}^{a\phantom{i}  j}-i\varepsilon_{j}f_{\phantom{a}  bc}^{a}\phi
_{\phantom{bj}  k}^{bj}\phi_{\phantom{ck}  i}^{ck}\,.\nn
\end{align}
This completes our review of $\mathcal{N}=1$ EYM supergravity in seven dimensions.

\end{document}